\documentclass[nofootinbib,showkeys,floatfix,onecolumn,times]{revtex4}
\usepackage{bm,morefloats}
\usepackage{dcolumn}
\usepackage[T2A,T1]{fontenc} 
\usepackage{fonttable}
\usepackage{graphicx}
\usepackage{balance}
\usepackage[latin1]{inputenc}
        \usepackage{cellspace}
\usepackage{latexsym,color}
\usepackage{centernot}
\usepackage{mathtools}
\usepackage{stmaryrd}
\usepackage{amssymb}
\usepackage[T1]{fontenc}
\usepackage{natbib}
\usepackage{amsmath,amssymb,amsthm,mathrsfs,amsfonts,dsfont}
\usepackage{color}

\usepackage{rotating}
\usepackage[dvipsnames]{xcolor}
\definecolor{LinkBlue}{RGB}{6,69,173}
\definecolor{DarkBlue}{RGB}{11,0,128}
\definecolor{red}{rgb}{1,0.,0.}
\usepackage[colorlinks=true,linkcolor=LinkBlue,urlcolor=magenta,
	citecolor=green,hyperfootnotes=true,urlcolor=red]{hyperref}
 \theoremstyle{definition}

\usepackage{enumitem}
\newcommand{\laa}{\mathcal{L}}
\newcommand{\ba}{\mathcal{B}}
\count\footins = 1000

\newcommand{\mnras}{MNRAS}
\newcommand{\apss}{Ap\&SS}

\begin{document}

\title{General relativistic rotational energy extraction from   black holes-accretion disk systems}
\author{D. Pugliese\&Z. Stuchl\'{\i}k}
\email{d.pugliese.physics@gmail.com;}
\affiliation{
Research Centre for Theoretical Physics and Astrophysics,Institute of Physics,
  Silesian University in Opava,
 Bezru\v{c}ovo n\'{a}m\v{e}st\'{i} 13, CZ-74601 Opava, Czech Republic
}

\begin{abstract}The  determination of mass and spin parameters of the  black holes (\textbf{BHs})
 is  crucial in the analysis of the merger of \textbf{BHs} and \textbf{BHs} formation and evolution, including  accretion.  Here   we constrain   the  \textbf{BH} spin   with  the evaluation of the  dimensionless parameter $\xi$ representing
the total rotational energy extracted  versus   the mass of the \textbf{BH}, following   procedure introduced in    \cite{Daly0} that is independent from the details of  the specific extraction process.  The energy extraction can  power an outflow which can be then observed.
 We relate  the energy extraction to the accreting configurations and the accretion processes occurring in a cluster  of agglomerate  corotating and counter-rotating  tori orbiting one central Kerr \textbf{SMBH}, associating  $\xi$ to the  characteristics of the accretion processes. We
relate  the regions of  tori   parameters  to  features of the energy extraction processes, binding $\xi$  to properties of light surfaces by using the bundles   developed in \cite{remnant}, relating    measures in different  regions of the spacetimes. We evaluate
   properties of the \textbf{BH} accretions disks, and
correlate spacetimes prior and after their transition due to  the energy extraction.  Light surfaces  are related to the generators of Killing   horizons,  proving  limiting frequency of the stationary observers of the geometries.
We consider the photon limiting curves of the stationary observers as constraints for various processes regulated by these  frequencies, to relate different  \textbf{BH} states, prior and after  the energy extraction,   investigating regions close to the \textbf{BH}  horizons and rotational axis.  From methodological view-point we used  a naked singularity -\textbf{BH} correspondence defined with metric bundles to predict  observational characteristics of the \textbf{BH}--accretion disk system.
The analysis  points relevant  \textbf{BH}  spins  $a\approx0.94M$, $a\approx 0.7M$ and
$a\approx0.3M$.  We show the relation between the rotational law of the tori, the characteristic frequency of the  bundle and the relativistic velocity defining the von Zeipel surfaces. The inferior limit on the formation of corotating tori is $\ell/a\geq2$, for counter-rotating tori $\ell/a\leq -22/5$ ($\ell$ is the fluids specific angular momentum).
\end{abstract}
\keywords{Black holes-- Accretion disks--Accretion; Hydrodynamics --Galaxies: active -- Galaxies: jets}
\date{\today}

\maketitle

\newcommand{\ti}[1]{\mbox{\tiny{#1}}}
\newcommand{\im}{\mathop{\mathrm{Im}}}
\def\be{\begin{equation}}
\def\ee{\end{equation}}
\def\bew{\begin{widetext}}
\def\eew{\end{widetext}}
\def\Rem{\textbf{Remark}}
\def\bea{\begin{eqnarray}}
\def\eea{\end{eqnarray}}
\newcommand{\tb}[1]{\textbf{\texttt{#1}}}
\newcommand{\actaa}{Acta Astronomica}
\newcommand{\oft}{\mathcal{o}_T}
\newcommand{\Sie}{\mathcal{S}}
\newcommand{\Mie}{\mathcal{M}}
\newcommand{\ofx}{\mathcal{o}_{\times}}
\newcommand{\mso}{\mathrm{mso}}
\newcommand{\mbo}{\mathrm{mbo}}
\newcommand{\ttb}[1]{\textbf{#1}}
\newcommand{\ctimes}{\overset{\mathbf{\cdot}}{\times}}
\newcommand{\rtb}[1]{\textcolor[rgb]{1.00,0.00,0.00}{\tb{#1}}}
\newcommand{\gtb}[1]{\textcolor[rgb]{0.17,0.72,0.40}{\tb{#1}}}
\newcommand{\ptb}[1]{\textcolor[rgb]{0.77,0.04,0.95}{\tb{#1}}}
\newcommand{\btb}[1]{\textcolor[rgb]{0.00,0.00,1.00}{\textbf{#1}}}
\newcommand{\otb}[1]{\textcolor[rgb]{1.00,0.50,0.25}{\tb{#1}}}
\newcommand{\non}[1]{{\LARGE{\not}}{#1}}
\newcommand{\nnon}[1]{{\Huge{\not}}{#1}}
\newcommand{\parr}[1]{\breve{\partial}}
\newcommand{\body}[1]{\|\mathbf{#1}\|}
\newcommand{\ff}[1]{\left\lfloor #1\right\rfloor}
\newcommand{\Hl}{{\large{ \mathrm{H}}}}
\newcommand{\srt}[1]{{\scriptsize{#1}}}

\newcommand{\bp}[1]{$\eth$}
\newcommand{\Ga}{\mathrm{G}}
\newcommand{\hh}[1]{\left\lceil #1\right\rceil}
\newcommand{\kets}[1]{\ket{#1}}
\newcommand{\kett}[1]{\kets_{#1}}
\newcommand{\ssp}{{\footnotesize{\textsf{\textbf{{S}}}}}}
\newcommand{\cc}{\mathrm{C}}
\newcommand{\il}{~}
\newcommand{\ddp}[1]{\vec{\dda}^{\scriptscriptstyle{#1}}}
\newcommand{\rc}{\rho_{\ti{C}}}
\newcommand{\ex}{\exists}
\newcommand{\vex}{\vec{\exists}}
\newcommand{\hex}{\hat{\exists}}
\newcommand{\oftt}{\mathcal{o}_{\widetilde{T}}}
\newcommand{\ofs}{\mathcal{o_S}}
\newcommand{\la}{\mathcal{A}}
\newcommand{\scr}{${\scriptsize{$\circ$}}$}
 \newcommand{\oo}{\mathrm{O}}
  \newcommand{\Qa}{\mathcal{Q}}
\newcommand{\Sa}{\mathcal{\mathbf{S}}}
\newcommand{\Ca}{\mathcal{\mathbf{C}}}
\newcommand{\Ha}{\mathcal{H}}

\section{Introduction}
We examine the extraction of the  rotational  energy from a Kerr black hole (\textbf{BH})  in the presence of accretion disks. The orbiting  accretion matter is considered in the form of aggregates of pressure supported geometrically thick disks of   perfect fluids, analyzing   purely hydrodynamic models of both tori corotating and counterotating  with respect to the central Kerr \textbf{BH}, and centered on the Kerr \textbf{BH} equatorial plane. This model  of tori aggregate is  known  as \textbf{eRAD}, or \textbf{RAD} when the  toroidal components are  also misaligned with respect to each others and the  rotational axis of \textbf{BH}\cite{ringed,open,dsystem,letter,fi-ringed}.
Our investigation is  centered  on the exploitation of a connection between supermassive black holes (\textbf{SMBHs}) and their  accretion tori and the states of \textbf{BHs} before and after the process of energy extraction.

The \textbf{BH} rotational  energy is  supposed to be   source of  many processes of the high energy astrophysics as the
active galactic nuclei (\textbf{AGN})  large-scale outflows \cite{Daly0,Daly2,Daly3,GarofaloEvans}.
The relation between \textbf{BH} spin--disk rotational law, accretion and   rotational energy  extraction is manifold.  The luminosity intensity of the  \textbf{AGNs}  for example is generally attributed to the fall of gas from the accretion disk on the  \textbf{SMBH}, located at the center of the galaxy, therefore en-powering  the observed jet emission (see for example recent analysis of the  galaxy in the X-band  \textbf{NGC2992} \cite{Marinucci}, and \cite{Mgquasar} for a
discussion of high-resolution image of the interaction between gas clouds and jets of material ejected from a \textbf{SMBH}  at the center of  the  quasar galaxy \textbf{MG J0414+0534}).
However, in different  phases of \textbf{SMBHs} evolution cocoons of matter may envelop the hole \cite{Gilli2}--consequently
\textbf{SMBHs}  can also be darkened by surrounding material  which can be observed on the  X-ray band produced by the accreting fluids\footnote{For example  mission to explore   the X-ray  emission sector:  \textbf{XMM-Newton} X-ray Multi-Mirror Mission)\href{}{http://sci.esa.int/science-e/www/area/index.cfm?fareaid=23},  \textbf{RXTE} (Rossi X-ray Timing
Explorer) \href{}{http://heasarc.gsfc.nasa.gov/docs/xte/xtegof.html} or \textbf{ATHENA} \href{}{http://the-athena-x-ray-observatory.eu/}}.
These plasma concentrations,  can  show a remarkable variation in their luminosity  following the phases of \textbf{BHs} growing
-- \cite{impacting,Ricci}.
 A model of a  cluster of  misaligned (inclined) tori orbiting a central static \textbf{SMBH} has been developed in the \textbf{RAD} frame in \cite{embedded-mnras,embedded-cqg} as an
 orbiting aggregate  composed by tori with  different inclination angles relative to a central \textbf{SMBH}.
  The  \textbf{RAD}  accreting agglomeration can be seen, depending on the     tori thickness, as a (multipole) gobules of orbiting matter, with different toroidal   spin orientations, covering the  embedded central \textbf{SMBH}.
 \textbf{BH} spin and galactic morphology have been  argued to be strongly connected,  for example
 \textbf{SMBHs} in elliptical galaxies possess higher spins than those
in spiral galaxies \cite{apite1}. The spins seem  to be related also to the redshift parameter.

In this regard the  investigation   is increasingly directed to  find  correlations between different phenomena  framed  in  the \textbf{\textbf{BH}}-disk
system: for example in the \textbf{BH} populations and galaxy age correlation and \textbf{BH}--spin shift--accretion disk correlation and
in the jet-accretion correlation--see for example \cite{Hamb1,Sadowski:2015jaa,Madau(1988),Ricci:2017wmr,Narayan:2013gca,Mewes:2015gma,Morningstar:2014hea,Yu:2015lqj,Volonteri:2002vz,Yoshida,Regan:2017vre,Yang:2017slb,Xie:2017jbz}.
The \textbf{BH} spin variation  follows  \textbf{BH} interaction with its environment, therefore
many approaches in the evaluation of the \textbf{BHs} parameters and estimation of \textbf{BH} accretion history are connected to the  accretion disk physics, based  for example on the evaluation of the  disk  luminosity or accretion rates.
It has been  shown   that the dimensionless spin of a central \textbf{BH} ($a/M$) and the morphological  and equilibrium  properties of   its accreting  disks, are strongly related \cite{multy}.
\textbf{SMBHs}  spin depends on  the angular momentum of the infalling materials tracing back a story of \textbf{BH} accretions in diverse epochs of the \textbf{BH} history \cite{LiWangHO}.
However, an  issue of these methods consists also in the fact that the  \textbf{SMBHs} spin
(with mass of $10^6-10^9 M_{\odot}$, $M_{\odot}$
being solar mass)  is strictly  correlated with the ``mass-problem'' and connected with  the evaluation of  the main features   of the accretion processes as the \textbf{BH} accretion rate or the location of the inner edge of the accretion disk. These methods are clearly  model dependent \cite{Capellupo:2017qpt,McClintock:2006xd,Daly:2008zk}, constituting a relevant issue considering that
 even the definition of the inner edge of an  accreting disk  is   controversial -see for example \cite{Krolik:2002ae,BMP98,2010A&A...521A..15A,Agol:1999dn,Paczynski:2000tz,open,long}.
On the other hand gravitational waves  detection from coalescence of  \textbf{BHs} in a binary system  may serve, in   future, as a further possible  method to fix a \textbf{BH} spin parameter  \cite{Farr:2017uvj,vanPutten:2015eda,vanPutten:2016wpa,2016PhRvD..93h4042P,2016ApJ...826...91A}.

In this work we consider the approach introduced in \cite{Daly0}, focused  on the definition of  \textbf{SMBH} (irriducible) mass function and  the  definition of rotational energy, and  the  \textbf{BH} classical thermodynamical law   adapted for the analysis of the
 accretion history of \textbf{SMBHs}  in presence of  episodic accretion and spin shifts.
 This approach bridges the  \textbf{BH} classical thermodynamic laws with the physical processes, leading to a  \textbf{BH} transition through interaction with the surrounding environment.

 \textbf{SMBHs} are  powerful engines  of ejections of matter  and energy where  the  radiative
efficiency of energy extraction from rotational   energy depends on the
 spin.
In  numerous examinations  of  the energy extraction   through the accretion process, the rotational  energy converted into
radiation corresponds to  the binding energy of the fluid,  connecting
the angular momentum of the accreting
material with the radiative efficiency. Furthermore, the \textbf{SMBHs} mass growth rate can be  connected
 with a luminosity function of the host galaxy.
Here we consider  \textbf{BHs}  accreting at (super) Eddington luminosity, and   super Eddington accretion disks which are  geometrically thick and opaque, pressure supported and cooled by advection--\cite{Universe,2005MPLA...20..561S,2009CQGra..26u5013S,2017EPJC...77..860K}.

From methodological view-point, we use geometrically thick, "stationary" disks as  largely
adopted    as the initial conditions in the set up for simulations of the GRMHD (magnetohydrodynamic)  accretion structures for the numerical  analysis--\cite{Fragile:2007dk,DeVilliers,Porth:2016rfi,Shafee}.
The \textbf{RAD} and \textbf{eRAD} models are essentially "constraining-models",   providing  initial data for  dynamical situations.
The location of the inner edge of an accreting torus is indeed  a key element of the \textbf{BHs} spin estimation, and it   is usually identified  with  marginally stable orbit or with a radius in the  region bounded by  marginally bounded circular orbit  and marginally stable circular  orbit according to  the accretion  disk model,  geometrically thick or thin, and the    mass of the central attractor.
Thick (stationary) disks give  a striking good approximation of  several  aspects of accretion also for   more complex dynamical models,   estimating  the   tori  elongation on their symmetry plane, the location of the  inner edge of quiescent and accreting disks, the  location of  critical pressure points, an evaluation of the tori   thickness, their   maximum height.

The inner ringed structure of the \textbf{eRAD}  offers  a set of interesting scenarios that include its unstable states. More points of accretion  can be present in the \textbf{eRAD} {inside} the ringed structure, which may also include inner shells of jets.
Jet in the \textbf{BH}-accretion disk systems can also change the accretion disk inner edge \cite{NL2009,Fender:2004aw,Fender:2009re,Soleri:2010fz,Tetarenko18,AGNnoBH,FangMurase2018,1998NewAR..42..593F,1998MNRAS.300..573F,1999MNRAS.304..865F,Fender(2001)}.
In this article  we    discuss also  the  correlation between the dimensionless spin of the central \textbf{BH} and the specific angular momentum of the fluid in the orbiting tori, characterizing the system with the rationalized specific angular momenta of the fluid as  $\ell/a$ or $\ell/a\sin\theta$, and functions of these variables, supported by considerations related to the general relativistic features  typical of the Kerr geometry and describing the rotational law of the \textbf{eRAD}.

The approach introduced in  \cite{Daly0}
 for the determination of \textbf{SMBHs}  spin, is based on the evaluation of the
 dimensionless ratio, $\xi$, of the released energy  versus the \textbf{BH}  mass, considering the definition of irreducible mass of \textbf{BH} and  the \textbf{BHs} thermodynamical laws  connecting the \textbf{BH}
spin $a$  as function of the  spin energy  and  the \textbf{BH} mass. This approach is quite independent from  the details of the
specific process  of energy extraction\footnote{Note that  the rotation energy  extraction  process is usually  connected with the magnetic fields around the \textbf{BHs}, e.g, the jets are usually connected with the Blandford-Znajek process \cite{BZ}  that could be considered as specific form of the magnetic Penrose process \cite{Penrose,Dad2018} that can demonstrate such extraordinary phenomena as acceleration of protons and ions to high energy  \cite{Universe} due to the so called   chaotic scattering studied in \cite{Stuchlik:2015nlt}. Quite recently a new variation of the Penrose process has been presented in \cite{cts}, that represents new efficient way of extraction of rotational energy related to radiation reaction of charged particles in magnetic fields.}.
Therefore  the quantity $\xi$, which could be measured eventually as energy released from  jet ejection for example, can  provide an indication
of the lower limit on the \textbf{BH}.
The \textbf{BH} spin   can be evaluated by the observation of the  \textbf{BH}  mass $M$
and the energy outflow\cite{Yan-Rong}.
In the non-isolated \textbf{SMBHs} the
outflows, related to the   orbiting accretion matter,
 provide  an indication of features of the \textbf{BH}-accretion disks systems.
In this scenario  a relevant topic is represented by the assumption of counterrotating  material which is an  intrinsic feature of the \textbf{eRAD} model, as  well as the presence of a  magnetosphere.
This estimation  implies the black
hole spin  changes only due to the \textbf{BH} mass  using definition of irreducible mass,  grounded on assumption that  energy  outflow would be  powered
by the  \textbf{BH} rotational  energy only  \cite{Daly:2008zk}.

We constrain the \textbf{BH}  rotational energy
estimating parameters of accretion features in the \textbf{eRAD} frame and the frequencies of light surfaces    related to the analysis of many aspects of \textbf{BHs} physics,   limits of stationary observers frequencies, connecting measures in  different points of spacetime accessible to the observer and  points of different space-times  before and after the transition  induced by the energy release.
We use the concept of metric Killing bundles (\textbf{MBs}) introduced in \cite{remnant}, which  are conformal invariant and can be easy read in terms of the light surfaces.  Metric Killing bundles   define  also  properties of the local causal structure and thermodynamical properties of \textbf{BHs} as the surface gravity, temperature and luminosity\cite{LQG-bundles}.
 Metric bundles can be used to investigate properties of the geometries close to the horizons,  connecting the different metrics of the bundles. The observer could extract information  (locally) of  the region close to the  horizons $r_{\pm}$-- and connecting different geometries of the metric family considering local properties of causal structure with the analysis of  photon-like orbits.
Metric bundles of the Kerr geometry are defined as the  sets of all Kerr spacetimes  having equal limiting light-like (stationary) orbital frequency, which is  the  asymptotic limiting value for time-like stationary observers (as measured at infinity) and also limits of the tori fluid relativistic velocity.
Therefore \textbf{MBs}  satisfy   the condition $\mathcal{L}_{\mathcal{N}}\equiv\mathcal{L}\cdot\mathcal{L}=0$,
where $\mathcal{L}$ is a  Killing field of the geometry   $\mathcal{L}\equiv \partial_t +\omega \partial_{\phi}$-- \cite{remnant,observers,ergon,remnant0,remnant1,inprep} .
The event horizons  of a spinning \textbf{BH}  are    Killing horizons   with respect to  the Killing field
$\mathcal{L}_H\equiv \partial_t +\omega_H \partial_{\phi}$, where  $\omega_H$ is  the  angular velocity of the horizons. Conditions on $\omega_H=$constant represents  the \textbf{BH} rigid rotation.
Characteristic  frequencies $\omega$ of the \textbf{MBs} are also the horizons frequencies\footnote{
In general, a Killing horizon is a  light-like hypersurface, generated by the flow of a Killing vector,
where the norm of a Killing vector is null. The hypersurface {null} generators coincide with the orbits of an
one-parameter group of isometries,  thus there exists a Killing field $\mathcal{L}$, which is normal to the null surface.}.
Metric bundles are defined as curves in an extended plane,  $\mathcal{P}-r$, where $\mathcal{P}$ is the metrics family parameter and $r$ is a radial distance.
In the extended plane  we can consider the horizons confinement and the horizons replicas.
For a   property $\Qa_{\pm}$  of the horizon as distinguished in the extended plane, as the horizon  frequency $\omega$, there is a {replica} of the horizon, in the same spacetime when there is a  orbit (radius) $r_{*}>r_\bullet$  such that
$\Qa(r_\bullet)\equiv \Qa_\bullet= \Qa (r_{*})$, where  $r_{\bullet}$ is a point of the horizon curve in the extended plane.
There are  horizon replicas in different  geometries along the  bundle curves,  i.e.  there is a $p\neq p_{*}$ and a $r_{*}>r_+$, where $p$ and $p_{*}$ are  values of   the extended plane parameter $\mathcal{P}$,  corresponding to two different geometries (distinguished with two  horizontal lines of the extended plane)  such that:
$\Qa(r_\bullet(p),p)\equiv \Qa_\bullet^p=\Qa (r_{*}(p_{*}),p_{*})$. In
 both points, $(r_\bullet,r_{*})$, there is equal  light-like   orbital  frequency.
  The horizon confinement  is viceversa interpreted as the presence of  a region of the extended plane $\mathcal{P}-r$ where \textbf{MBs} are entirely confined-i.e.,  there are no horizon replicas  in any other region of the extended plane, particularly  in any other geometry. However this definition is often  specified  to the  confinement  of the $\Qa$ property in  the same geometry.  In the  Kerr spacetime,  this   region is  upper bounded in the extended plane by the a portion of the horizon curve corresponding to the  set of the inner horizon \textbf{BHs}--\cite{remnant,remnant0,remnant1,inprep}.

In this article we  consider  phenomenology and processes  constrained   and regulated by the characteristic   frequencies of the bundles. We  investigate the extraction of  rotational energy in dependence of the  accretion features  and in relation  to the light surfaces  through \textbf{MBs} by extrapolating information of the physics close to the \textbf{BH} rotational  axis and \textbf{BHs} horizons with the concept of replicas.  We predict the possibility of  observational evidence of the presence of replicas of the horizon, and indications of the replicas along the bundle curves evident in the thermodynamical transitions of the \textbf{BH} after energy release. We claim that it should be possible to observe the replicas  along the axis of the black hole.  \textbf{MBs}  connect  different  \textbf{BH} states, prior and after  the energy extraction.
 We    consider also  the  rotational law of the fluid in dependence on  the central \textbf{BH} spin.

{We note that in more general contexts there can be  a different setup  on the initial and final state of the \textbf{BH} accretion system, a different outflow contribution and the establishment  of special instabilities which can change the \textbf{BH}  disks system    or even  change the  \textbf{BH} spin orientation due to  Bardeen--Petterson effect, where the
\textbf{BH} spin  changes  under the action of the disk torques-\cite{BP75}.
 We base our analysis on the   hypothesis  that
the energy  outflow to be measured is  powered by the \textbf{BH} spin energy only, however
more generally  the  extracted energy can be determined considering the ratio between the  outflow energy  versus   the spin energy while we consider here  the two quantities be coincident.
 The   establishment of  runaway  instability and  the  tori self-gravity are also  factors  neglected  in this model that  may be relevant, requiring  a different characterization of the energy outflow.
Furthermore we assume that, from an initial  Kerr  \textbf{BH}, the final stage of extraction process  is a  static spherically symmetric  Schwarzschild spacetime and we do not consider the contribution of the mass and momentum  to the \textbf{BH}  subsequent  to accretion while   geometrically thick disks have large accretion rates (with super Eddington luminosity).
}

\medskip
The article plan is as follows:
In Sec.\il(\ref{Sec:model})
we introduce the \textbf{eRAD} model.
Sec.\il(\ref{Sec:EE})
 focuses on the energy extraction  processes:
 discussion on the energy-spin relations is in Sec.\il(\ref{Sec:both-s}).
 Metric bundles, horizon replicas and photon frequencies are the focus of Sec.\il(\ref{Sec:MB-K}).
Tori energetics and accretion  is discussed in Sec.\il(\ref{Sec:e-complex-asly}).
Sec.\il(\ref{Sec:lona}) contains notes  on the \textbf{RAD}  and \textbf{BH}-accretion disk spin correlation:
the discussion is deepened  in
Sec.\il(\ref{Sec:barl}), while
special sets of tori are introduced in Sec.\il(\ref{Sec:setsoftori}).
{discussion and concluding remarks} follow in Sec.\il(\ref{Sec:dic-con}).
Four  Appendix sections close this article:
in Sec.\il(\ref{Sec:mislao}) and Sec.\il(\ref{Sec:ls-explicit}) are some  explicit solutions, while in Sec.\il(\ref{Sec:vonZeipel}) the relation between
{Von Zeipel surfaces, metric Killing bundles and  tori} is investigated and
  in Sec.\il(\ref{Sec:adaptedd}) an adapted solution parametrization is discussed.
\section{Fluid configurations on the Kerr spacetime}\label{Sec:model}
We consider geometrically thick tori of perfect fluids orbiting  a central  Kerr \textbf{SMBH}.
 The   metric tensor of the spacetime background can be
written  in Boyer-Lindquist (BL)  coordinates
\( \{t,r,\theta ,\phi \}\)
as follows
\bea\label{alai}&& ds^2=-dt^2+\frac{\rho^2}{\Delta}dr^2+\rho^2
d\theta^2+(r^2+a^2)\sin^2\theta
d\phi^2+\frac{2M}{\rho^2}r(dt-a\sin^2\theta d\phi)^2\ ,
\\&&
\rho^2\equiv r^2+a^2\cos\theta^2,\quad\Delta\equiv r^2-2 M r+a^2,\quad
r_{\pm}\equiv M\pm\sqrt{M^2-a^2};\quad r_{\epsilon}^{+}\equiv M+\sqrt{M^2- a^2 \cos\theta^2},
\eea
where $r_{\pm}$ are the outer and inner Killing horizons respectively, $r_{\epsilon}^+$ is the outer ergosurface.  Here $M$ is the {(ADM  and Komar)}  mass parameter and the specific angular momentum is given as $a=J/M$, where $J$ is the
total angular momentum of the gravitational source.
 There  is $r_+<r_{\epsilon}^+$ on the planes  $\theta\neq0$  and  is $r_{\epsilon}^+=2M$ on the equatorial plane $\theta=\pi/2$.
In the  {region $r\in]r_+,r_{\epsilon}^{+}$[} ({outer \em ergoregion} or simply ergoregion) there  is  { $g_{tt}>0$} and $t$-Boyer-Lindquist coordinate becomes spacelike.
This fact implies that   static observers cannot exist inside
the ergoregion.
  For convenience
  we  introduce %
quantities
\bea
\alpha=\sqrt{(\Delta \rho^2/A)}; \quad \omega_z=2 a M r/A,
\eea
where
\bea A\equiv (r^2+a^2)^2-a^2 \Delta \sigma,\quad \sigma\equiv\sin^2\theta,
\eea
 the lapse function and the frequency of the zero angular momentum fiducial observer (or {ZAMOS}) \cite{observers}, whose four velocity is $u^a=(1/\alpha,0,0,\omega_z/\alpha)$ which is  orthogonal  to the surface   of constant $t$.
Since the metric is independent of $\phi$ and $t$, the covariant
components $p_{\phi}$ and $p_{t}$ of a fluid  four--momentum are
conserved along its  geodesic.
%
\begin{figure}
\centering
  \includegraphics[width=9cm]{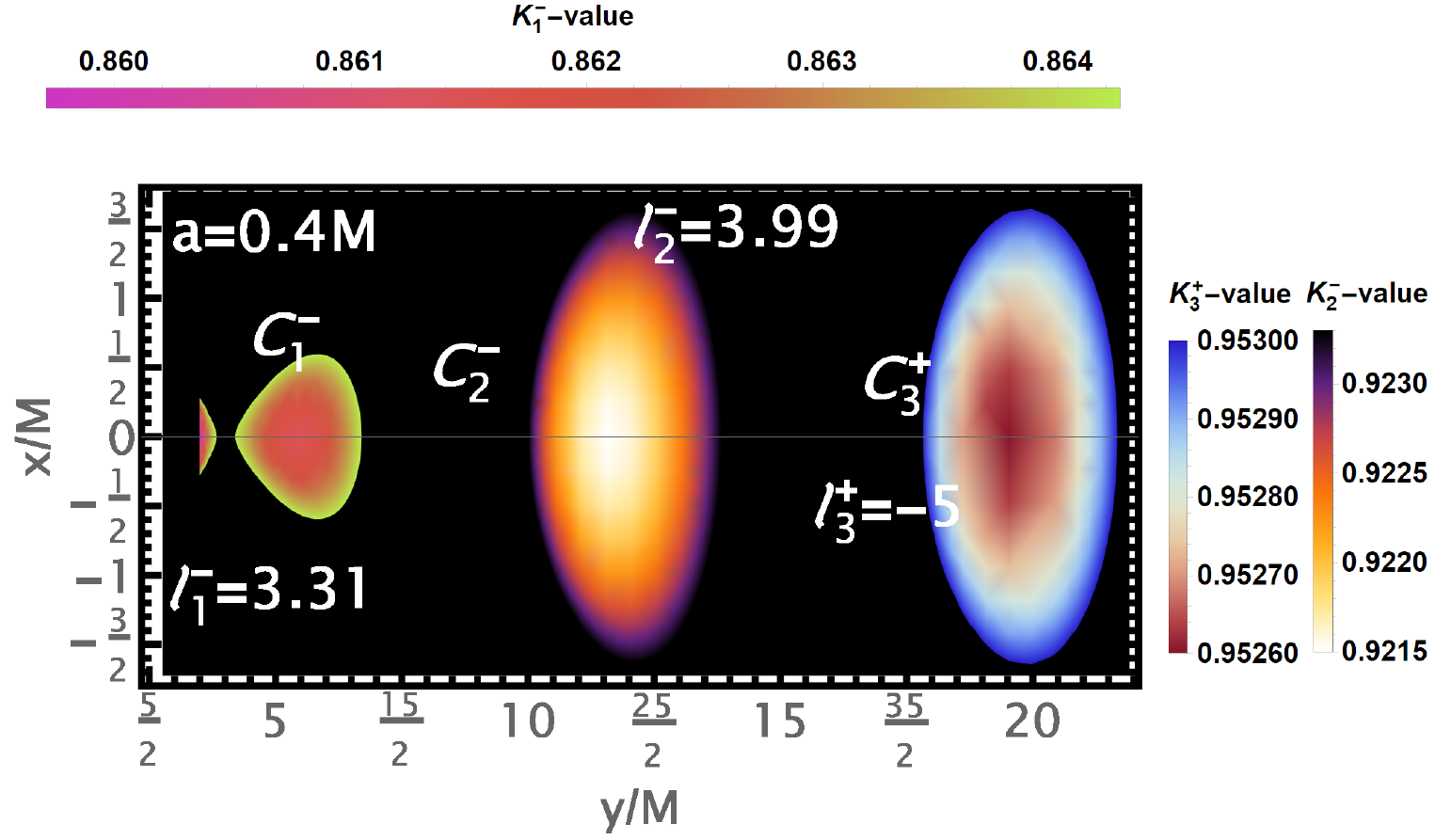}
  \includegraphics[width=4cm]{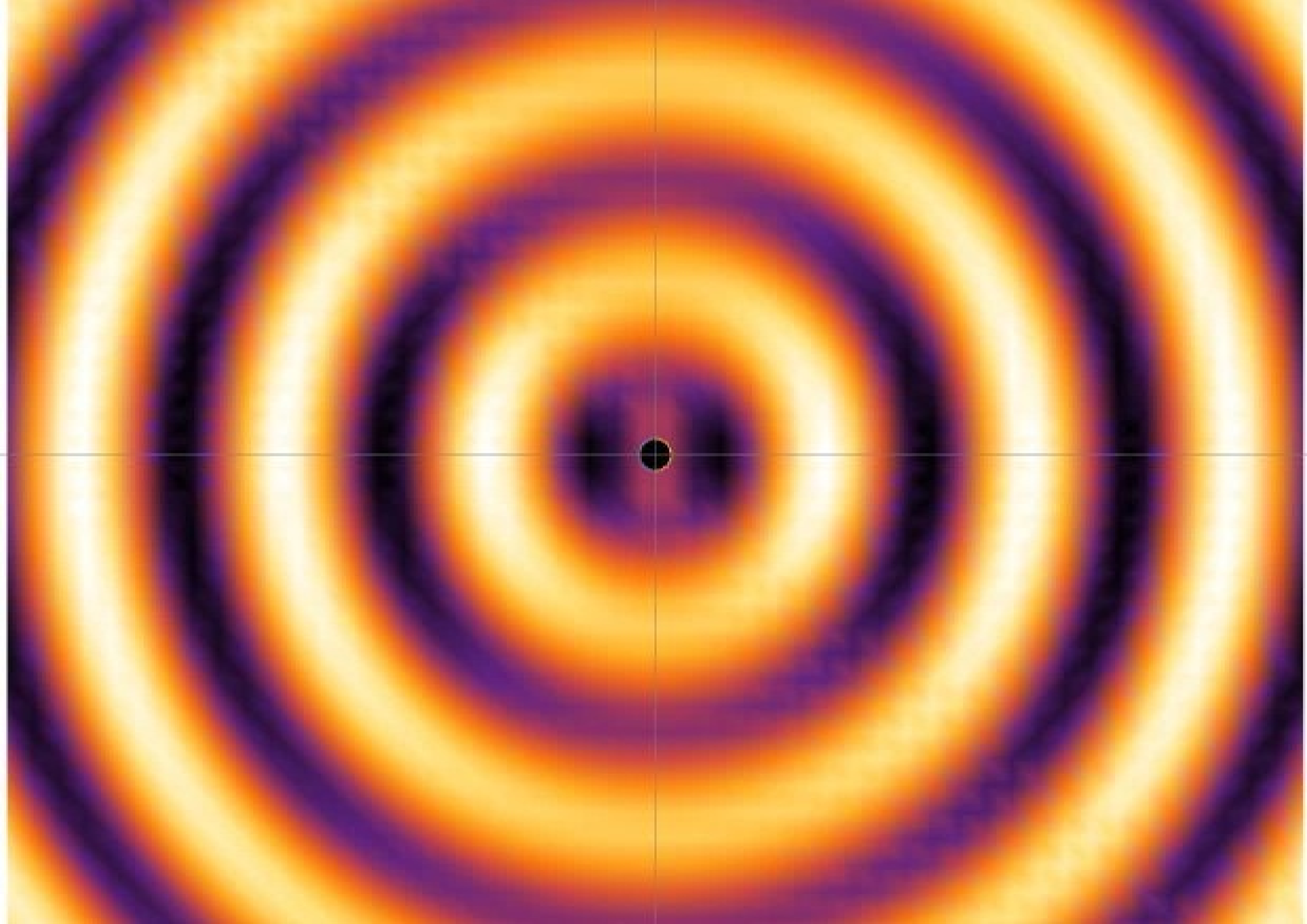}
  \includegraphics[width=4cm]{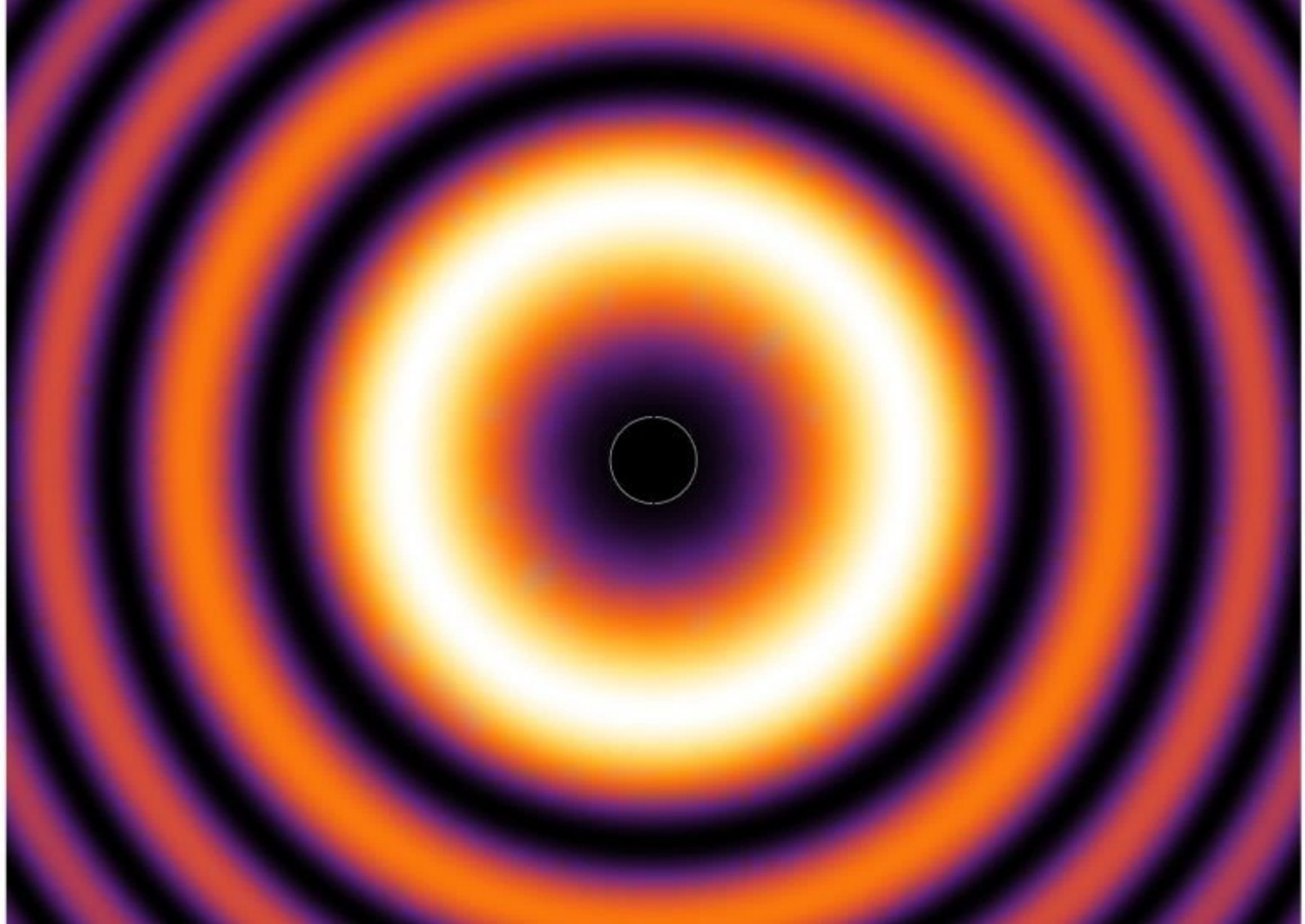}
  \includegraphics[width=8cm]{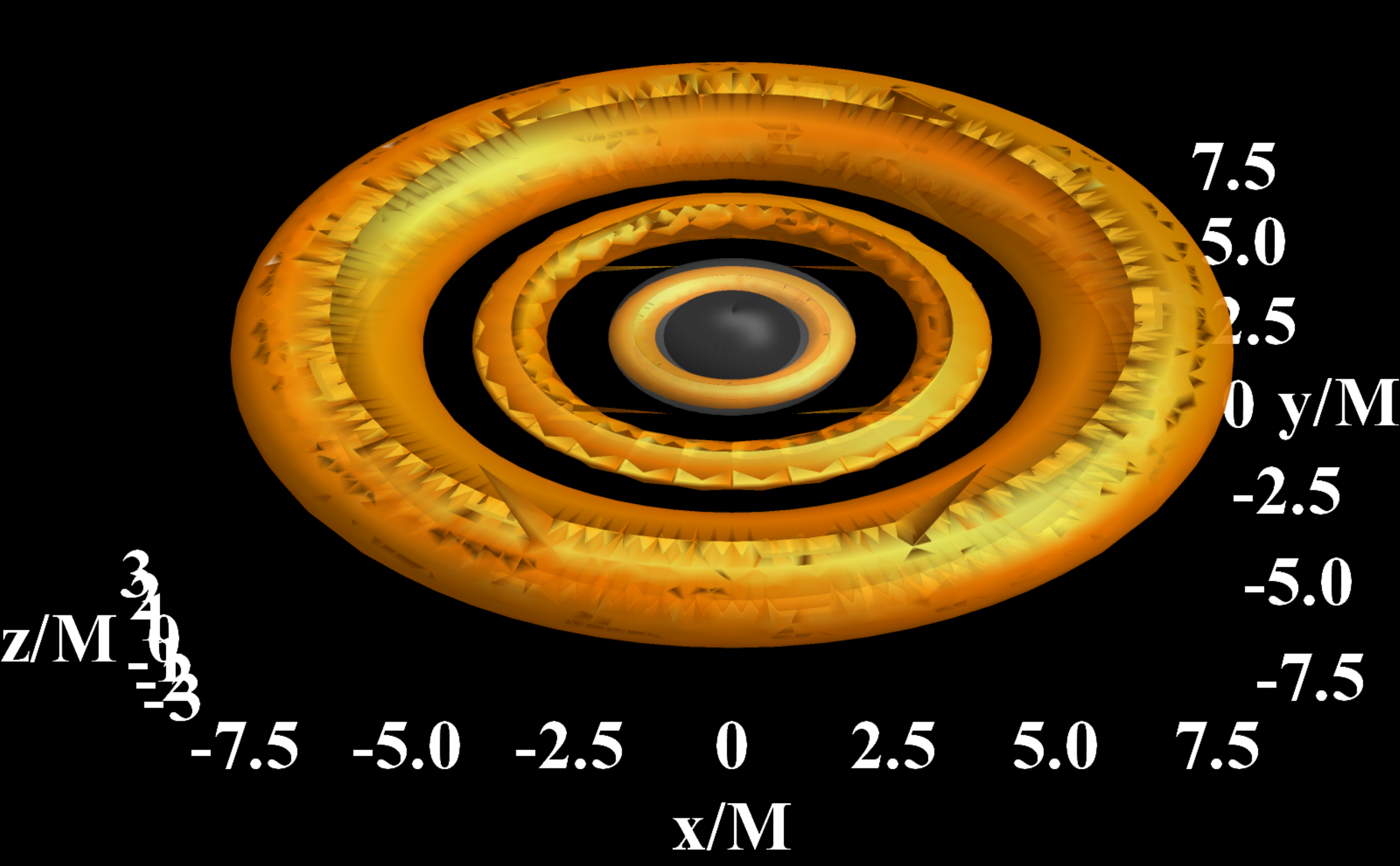}
  \includegraphics[width=8cm]{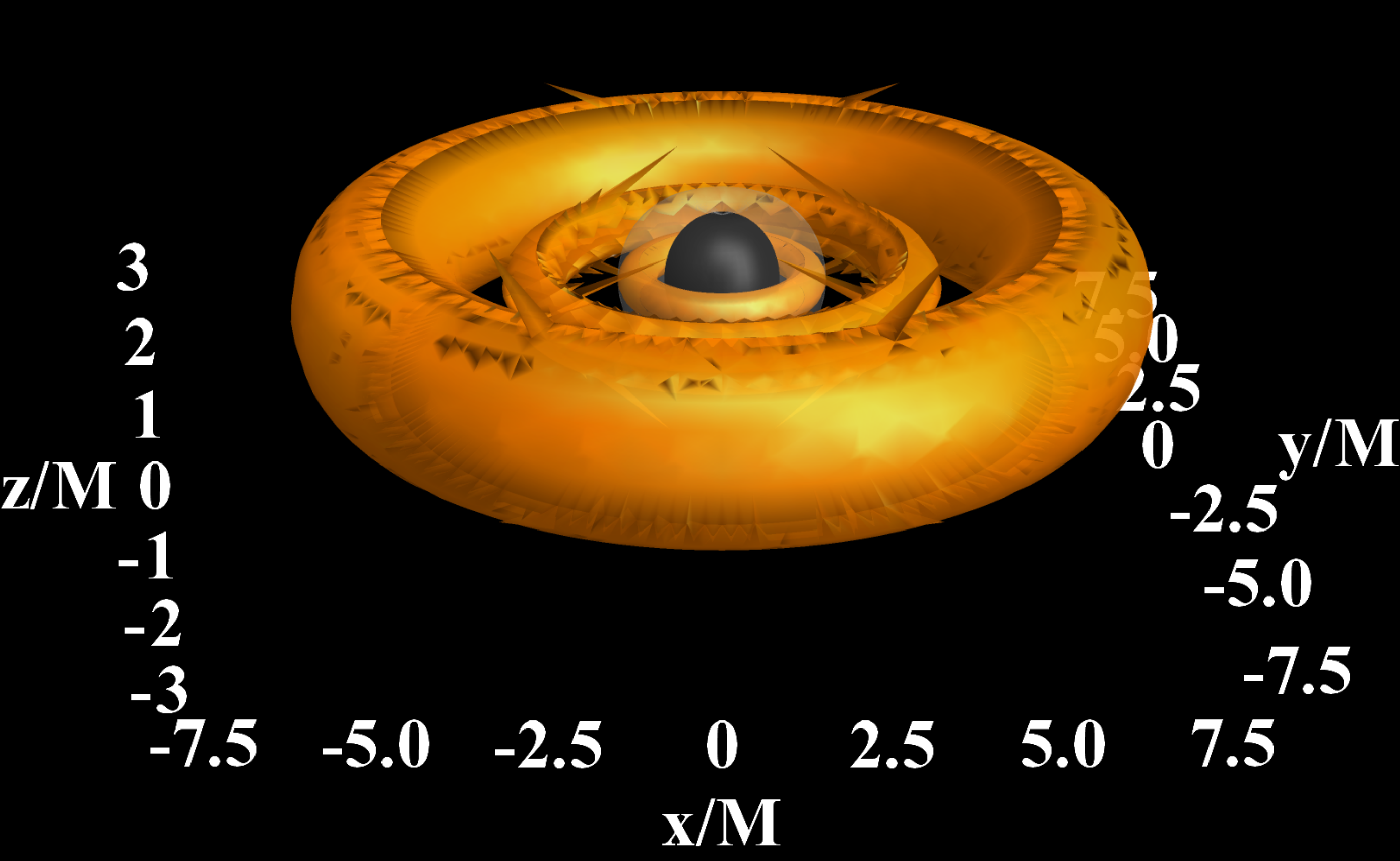}
  \caption{Upper line. Left panel: \textbf{eRAD}--Ringed accretion disk of the order $3$, composed by three tori, inner corotating $C_1^-$ with cusp, middle  quiescent corotating $C_2^{-}$ and the outer counter-rotating  $C_3^+$ torus. The sequence of tori $C_1^-<C_2^-<C_3^+$ is  therefore $\ell$counter-rotating. Values of fluid specific angular momentum $\ell^-_1,\ell^-_2, \ell^+_3$ for the three tori are signed on the panel as the \textbf{BH} dimensionless spin $a/M$.  The values of the $K$ parameters, levels of the tori effective potential, $K^-_1, K^-_2, K^+_3$ are in legends.  Right panels: pictorial representations of a view from above of the ringed structure of  \textbf{eRAD}, the center is noted the presence of a \textbf{BH}. Right panels show  structures of differently thick  tori composing the ringed structure. For this representation we used a wave model for the ringed disk,  adapted to the need for representation it follows the considerations of \cite{ringed}. (The fictiouse density function adopted for the  right plot  is $\phi_{eff}={48 \sin ^2r^2}/{r}+4 \cos ^2r^2$, where $r=\sqrt{x^2+y^2}$, providing minimum and maximum density profile points, in \cite{ringed} there is a discussion on the wavelike solution for a \textbf{eRAD} profile.). Bottom line:  two views of a \textbf{eRAD} composed by three corotating  tori in the \textbf{BH} spacetime with spin $a=0.994298M: r_{(\gamma)}^-=r_{\epsilon}^+$ where $r_{\epsilon}^+=2M$ is the outer ergosurface on the equatorial plane, $r_{(\gamma)}^-: \ell^-(r^-_\gamma)=\ell^-(r^-_{(\gamma)})$ see Eqs\il(\ref{Eq:conveng-defini}).
  {Left panel  from above emphasizing the inner \textbf{eRAD} structure  and right panel  in a front-view to emphasize the thickness of the toroidal components}.  Black center region is the Kerr \textbf{BH}, light gray surface is the outer ergosurface, the parameters for the inner central and outer torus of the \textbf{eRAD} are respectively: ($\ell=2.10728,K=0.74385$), ($\ell=
2.5,K=0.867797$), ($\ell=3,K=0.927362$). }\label{Fig:developedur}
\end{figure}
Consequently
$
{E} \equiv -g_{ab}\xi_{t}^{a} p^{b}$ and $L \equiv
g_{ab}\xi_{\phi}^{a}p^{b}\ ,
$ are  constants of motion for test particle orbits, where  $\xi_{t}=\partial_{t} $  is
the Killing field representing the stationarity of the Kerr geometry and  $\xi_{\phi}=\partial_{\phi} $
is the
rotational Killing field (the vector $\xi_{t}$ becomes    spacelike in the ergoregion).
%
It is convenient to  introduce also   the fluid angular frequency  $\Omega$ and the fluid specific angular momentum $\ell$ as follows
\bea\label{Eq:flo-adding}
\Omega \equiv\frac{u^\phi}{u^{t}}&=&-\frac{{E} g_{\phi t}+g_{tt} L}{{E} g_{\phi \phi}+g_{\phi t} L}= -\frac{g_{t\phi}+g_{tt} \ell}{g_{\phi\phi}+g_{t\phi} \ell},\quad
\ell\equiv\frac{L}{{E}}=-\frac{g_{\phi\phi}u^\phi  +g_{\phi t} u^t }{g_{tt} u^t +g_{\phi t} u^\phi } =-\frac{g_{t\phi}+g_{\phi\phi} \Omega }{g_{tt}+g_{t\phi} \Omega }.
\eea
 The case of an orbiting {circular}  or toroidal configuration is defined by   the constraint
$u^r=0$, as we assume the motion on the  fixed equatorial plane  which is also the plane of symmetry of each toroid of the configurations  ($\sigma=1$), besides  no
motion is  in the $\theta$ angular direction and it is  $u^\theta=0$.
We mainly consider in  our analysis  the case of  positive values of $a$
for corotating  $(L>0)$ and counter-rotating   $(L<0)$ orbits, or for corotating  $(\ell>0)$ and counter-rotating   $(\ell<0)$ fluids. Couple of orbiting toroids, $(i,o)$,  with $\ell_i\ell_o>0$ are said $\ell$corotating while toroids having $\ell_i\ell_0<0$ are said $\ell$counterotating.
%
The geodesic structures, composed by the radii
regulating  the particle dynamics, regulate also large part of the  disks dynamics especially in the case of geometrically thick tori, namely the marginally  circular orbit for timelike particles  $r_{\gamma}^{\pm}$,  which is also the photon orbit, the {marginally  bounded orbit}  is $r_{mbo}^{\pm}$, and the {marginally stable circular orbit} is $r_{mso}^{\pm}$ for corotating $(-)$ and counter-rotating $(+)$ motion.
 Radii of geodesic structures are relevant to the accretion  physics. We consider  also the radius $r_{\mathcal{M}}^{\pm}$   as solution  of  $\partial_r^2\ell=0$, and
the set of radii  $r_{(mbo)}^{\pm}$ and $r_{(\gamma)}^{\pm}$  or more generally $(r_{(mbo)}^{\pm}, r_{(\gamma)}^{\pm},r_{(\mathcal{M})}^{\pm})$ where
\bea&&\label{Eq:conveng-defini}
{r}_{(mbo)}^{\pm}:\;\ell_{\pm}(r_{(mbo)}^{\pm})=
 \ell_{\pm}({r}_{{mbo}}^{\pm})\equiv {\ell_{mbo}^{\pm}},
\quad
  r_{(\gamma)}^{\pm}: \ell_{\pm}(r_{{\gamma}}^{\pm})=
  \ell_{\pm}(r_{(\gamma)}^{\pm})\equiv {\ell_{{\gamma}}^{\pm}},
\quad\mbox{and}\quad
r_{(\mathcal{M})}^{\pm}: \ell_{\pm}(r_{(\mathcal{M})}^{\pm})= \ell_{\mathcal{M}}^{\pm}
\eea
where there is
$
r_{\gamma}^{\pm}<r_{mbo}^{\pm}<r_{mso}^{\pm}<
 {r}_{(mbo)}^{\pm}<
 r_{(\gamma)}^{\pm}
$.

Each toroid is described by a  one-species particle perfect  fluid (simple fluid) energy momentum tensor  where
\bea&&\label{E:Tm}
T_{a b}=(\varrho +p) {u}_{a} {u}_{b}+\  p g_{a b},
\eea
{where from} the conservation of the energy momentum tensor  $\nabla^aT_{ab}=0$ projected along  each fluid  four velocity $u^b$ and the   related  3-sheet spatial  metric tensor projector   $h^{ab}=g^{ab}+ u^a u^b$  we obtain  the  continuity (density $\varrho$ evolution) equation and Euler equation for the pressure respectively
 \bea
&&\label{E:1a0}
u^a\nabla_a\varrho+(p+\varrho)\nabla^a {u}_a=0\, \quad
(p+\varrho)u^a\nabla_au^c+ \ h^{bc}\nabla_b p=0\, ,
\eea
  where $\varrho$ and $p$ are  the total energy density and
pressure as measured by an observer moving with the fluid. Properly chosen boundary conditions determine  the inner  ringed structure of the tori agglomerate. Then model constructed in \cite{ringed,open,dsystem,letter,multy,long,proto-jet,embedded-mnras,embedded-cqg} is known as ringed accretion disk (\textbf{RAD}) for tori with a generic inclination with respect the central
attractor (aggregates of tilted tori). For the case we consider here, where all tori are centered on the equatorial plane of the attractor the model is distinguished as \textbf{eRAD}, Figs\il(\ref{Fig:developedur}).  For the
symmetries of the problem, we always assume $\partial_t \mathbf{Q}=0$ and
$\partial_{\phi} \mathbf{Q}=0$, being $\mathbf{Q}$ a generic spacetime tensor
(we can refer to  this assumption as the condition of  ideal hydrodynamics of
equilibrium).
As consequences of this choice,
the motion of the fluid  is described by  the \emph{Euler equation} only.
We assume moreover a barotropic equation of state $p=p(\varrho)$.
From the Euler  equation (\ref{E:1a0}) we obtain
\be\label{Eq:scond-d}
\frac{\partial_{\mu}p}{\varrho+p}=-{\partial_{\mu}}W+\frac{\Omega \partial_{\mu}\ell}{1-\Omega \ell},\quad W\equiv\ln V_{eff}(\ell),\quad V_{eff}(\ell)^2\equiv (\mathbf{u}_t)^2=\frac{g_{\phi t}^2-g_{tt} g_{\phi \phi}}{g_{\phi \phi}+2 \ell g_{\phi t} +\ell^2g_{tt}}
\ee
where $V_{eff}(\ell)$ is the effective potential    and  the function $W$ is Paczynski-Wiita  (P-W) potential. Assuming  the fluid  is   characterized by the   specific angular momentum  $\ell$  constant (see also \cite{pugtot}),  we consider  the equation for $V_{eff}=K=$constant obtaining the toroidal configurations as
   surfaces of constant pressure  or $\Sigma_{i}=$constant for \(i\in(p,\varrho, \ell, \Omega) \), where it is indeed $\Omega=\Omega(\ell)$ and $\Sigma_i=\Sigma_{j}$ for \({i, j}\in(p,\varrho, \ell, \Omega) \). The toroidal surfaces  are obtained from the equipotential surfaces, critical points of the effective potential  $V_{eff}(\ell) $.
   This set of results is known as von Zeipel solutions.
%
%
%

 The function $V_{eff}(\ell)$ is related to the energy  ${E}$ of the test particle
as it is   $V_{eff}(\ell)^2=L^2/\ell^2= E^2$.
Clearly there is  $
\lim_{r\rightarrow\infty}L(\ell)=\sqrt{\ell^2}$
and  it is $\partial_{\ell} L(\ell)\neq 0$.
 Models of geometrically thick tori  allow the determination of a wide number  of aspects of disks morphology, dynamics and stability. The inner and outer edges  of an equilibrium torus are also strongly constrained.
For each torus, the extrema of the effective potential functions fix   the center $r_{cent}$,  as minimum point of the effective potential and the maximum point for the hydrostatic pressure.
Cusped $\cc_{\times}$ equipotential surfaces are associated to   tori accreting  onto the central \textbf{BH}, due to the   Paczynski-Wiita  (P-W)  hydro-gravitational instability  mechanism at the cusp $r_{\times}$ \cite{Pac-Wii}.
 More specifically, the matter outflows because of  a  violation of mechanical equilibrium of the tori, due to an instability in the balance of the gravitational and inertial forces and the pressure gradients in the fluid.
The cusp  $r_{\times}$,  inner edge  of the  accreting  torus,  corresponds to the maximum  point of  the effective potential and  also the zero point for the hydrostatic pressure.
The distribution of these critical points, fixing the distribution of tori in the \textbf{eRAD}, as well as features of stability properties, are  governed by
 the \textbf{RAD} rotational law $\ell(r,\sigma;a):\partial_r V_{eff}(r,\sigma;a)=0$, regulating  the distribution of tori in the \textbf{RAD}. Function $K(r)\equiv V_{eff}(\ell(r))$ on the other hand  provides  the values of  {$K$} at the critical pressure points inside each torus of the agglomerate.
 Thus the  equipressure surfaces at  $K=$constant,   could be closed, determining equilibrium or quiescent $C$ configurations,  cusped $C_\times$ for the "accreting" tori,  or also  open  $O_{\times}$ for proto-jet configurations  related to  jets \cite{open,proto-jet,Koz-Jar-Abr:1978:ASTRA:,Pac-Wii,abrafra}, constituting funnels of matter with a cusp on the equatorial plane, characterizing
the possible  inter-disk activity in the agglomerate, such as the arising  of proto-jets  shells, collision or double accretion \cite{long,open}.
Each orbiting toroid  is governed by the   general relativistic hydrodynamic Boyer  condition  of equilibrium configurations  of rotating perfect fluid. Within the so  called ``Boyer's condition'', we   can  determine    the boundary of  stationary, barotropic, perfect fluid body as  the
surfaces of constant pressure (also   equipotential surfaces) which are also called Boyer surfaces\cite{Boy:1965:PCPS:}.

\medskip

   We can define the  ranges  of fluids specific angular momentum $(\mathbf{L_1,L_2,L_3})$  governing the fluids and tori topology as follows (we adopt the notation $\Qa_{\bullet}\equiv \Qa(r_{\bullet})$ for any quantity $\Qa$ evaluated on a radius $r_{\bullet}$):

\medskip

\textbf{$\ell\in \mathbf{L_1} $: for $\ell\in \mathbf{L_1} $ there are  quiescent  and cusped tori}--where $
\mp \mathbf{L_1}^{\pm}\equiv[\mp \ell_{mso}^{\pm},\mp\ell_{mbo}^{\pm}[$.
Therefore tori  have
    topologies $(C_1, C_{\times})$;   the accretion point is in    $r_{\times}\in]r_{mbo},r_{mso}]$ and  the center with maximum pressure in $r_{cent}\in]r_{mso},r_{(mbo)}]$ (for  tori with $K<1$).

\textbf{$\ell\in \mathbf{L_2}$: for $\ell\in \mathbf{L_2}$ there are  quiescent  tori and proto-jets}--$\mp \mathbf{L_2}^{\pm}\equiv[\mp \ell_{mbo}^{\pm},\mp\ell_{\gamma}^{\pm}[ $.
Topologies    $(C_2, O_{\times})$ are possible; the    unstable point  is  $r_{j}\in]r_{\gamma},r_{mbo}]$  and  center with maximum pressure $r_{cent}\in]r_{(mbo)},r_{(\gamma)}]$;  For these configurations there is $K_{j}>1$.

\textbf{ $\ell\in \mathbf{L_3}$: for $\ell\in \mathbf{L_3}$ there are only quiescent  tori} $C_3$ -- $d\mp \mathbf{L_3}^{\pm}\equiv\ \ell \geq\mp\ell_{\gamma}^{\pm}$:
and center $r_{cent}>r_{(\gamma)}$.
\begin{figure}
\centering
\includegraphics[scale=.4]{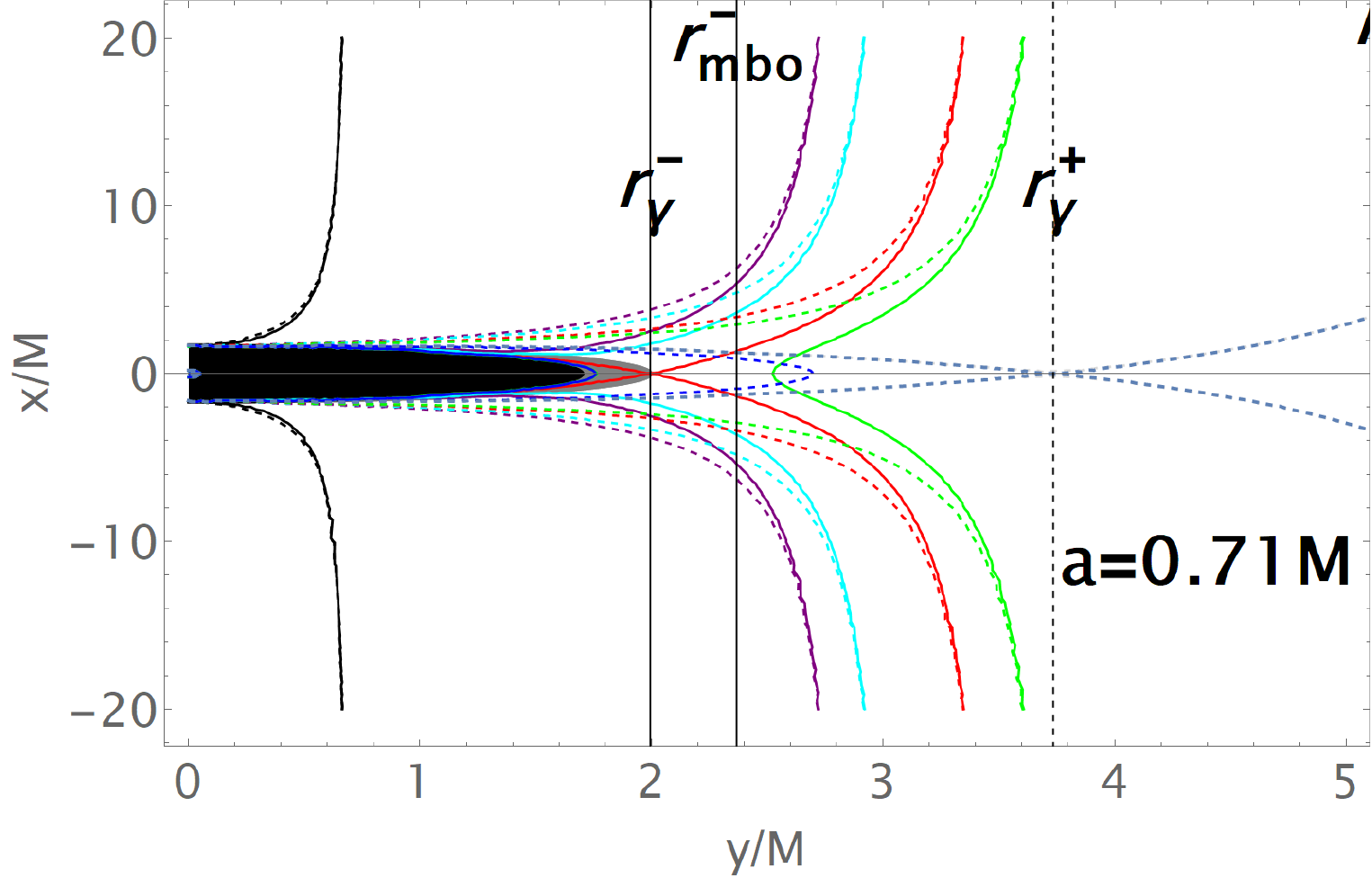}
\includegraphics[scale=.2]{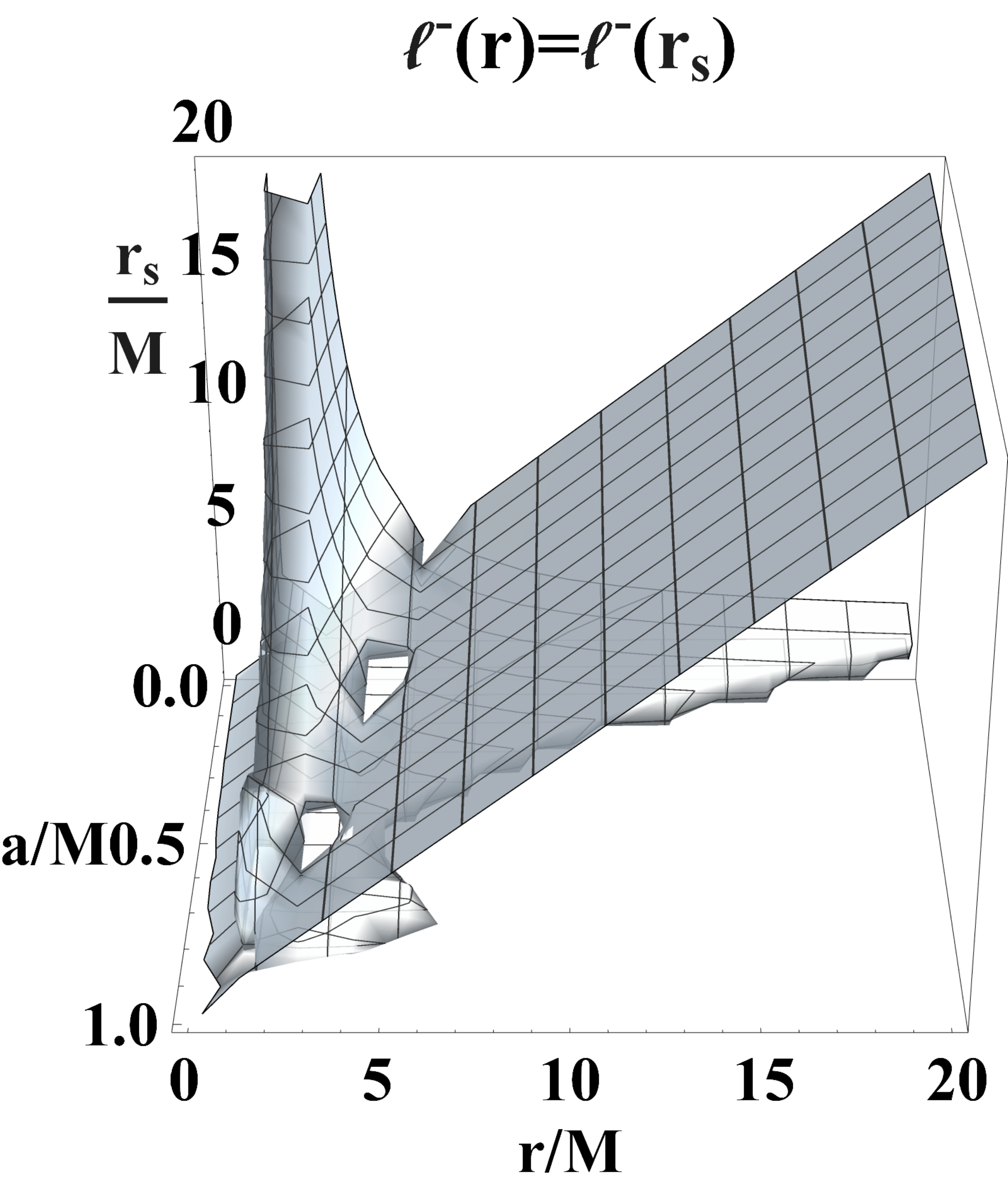}
\includegraphics[scale=.2]{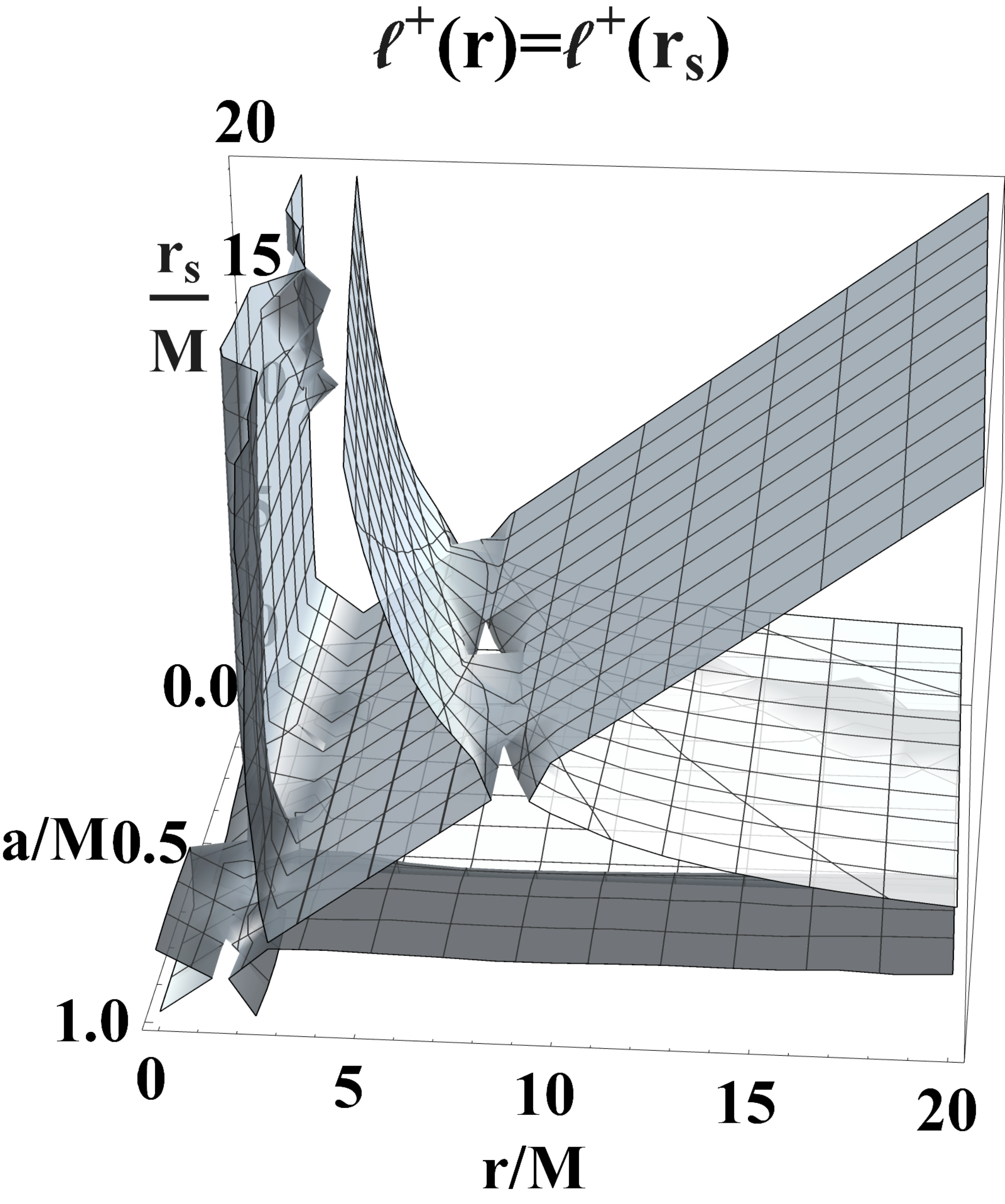}
\caption{Left  panel:   open surfaces, solutions of Euler equations for corotating fluids  ($(-)$ plain) and counter-rotating  fluids ($(+)$ dashed). Black region is the \textbf{BH}, gray region is the outer  ergoregion.  We can note the presence of cusp for the photon orbit location $r_{\gamma}^{\pm}$', $M$ is the \textbf{BH}  mass, $r_{mbo}^-$ is for marginally bounded orbit, curves are open equipotential surfaces  associated to proto-jets, limiting almost collimated funnels at the horizons are present as  limiting surfaces at various constant values of specific fluid  angular momenta:
 black curve is  $\ell=\pm0.7$, purple $\ell=\ell^{\pm}(r_{mso}^-)$,  cyan $\ell=\ell^{\pm}(r_{mbo}^-)$,  light-gray curve is $\ell=\pm6.5$, green curve is $\ell=\pm3.8$, blue curve is $\ell=\pm 8$. Center  and right  panels: solution of condition $\ell^{\mp}(r)=\ell(\mp)(r_s)$ for corotating (-) and counter-rotating (+) fluids respectively. The pair $(r,r_s)$ relates centers and cusps (in general critical points) of family of tori for different  black holes. }\label{Fig:sccreem}
\end{figure}

\medskip

For illustration of the presented definition see Figs\il(\ref{Fig:sccreem}): left panel shows the open surfaces and the limiting cusped surfaces  with cusps at $r_{\gamma}^{\pm}$ for counterrotating and corotating fluids respectively,  enlightening the spatial separations between the two sets of fluids, and the  vertical direction  parallel to the \textbf{BH} rotational axis.
Center and right  panels show radii $(r_s,r)$ solutions of $\ell(r)=\ell(r_s)$  for corotating and counterrotating tori  used in Eq.\il(\ref{Eq:conveng-defini}), the separation $r_s-r>0$ is the distance between the center of maximum pressure and minimum pressure in the disk (cusp) when the minimum is defined. The distance depends on the \textbf{BH} spin and the rotation orientation of the fluid  respect to the central \textbf{BH} (there is $r>r_+$ and $r_s>r$).
\section{Energy extraction}\label{Sec:EE}
In order to look for  global characteristics  of rotational energy extraction from  black holes we have to introduce specific functions of the  black hole spin, which will be addressed in Sec.\il(\ref{Sec:both-s}).
It   is also convenient to introduce here some notable  \textbf{BH} spins:
  \bea
  &&
  a_{mso}^{\epsilon}/M\equiv {2\sqrt {2}}/ {3}\approx 0.942809: r_{mso}^{-} = r_{\epsilon}^+,\quad a _{mbo}^{\epsilon}/M\equiv 2 \left(\sqrt{2}-1\right)\approx 0.828427: r_{mbo}^ -= r_{\epsilon}^+,
 \\
  && a_{\gamma}^{\epsilon}/M\equiv 1/\sqrt {2}\approx 0.707107: r_{\gamma}^ -= r_ {\epsilon}^+,
  \eea
related to the cross of the  radii $\{r_{mso},r_{mbo},r_{\gamma}\}$ with  the $r_{\epsilon}^+=2M$, outer ergosurface on the equatorial plane.
We proceed in Sec.\il(\ref{Sec:both-s}) with the analysis of the \textbf{BH} rotational energy in the \textbf{eRAD} context, while we expand the discussion with  the inclusion of the metric bundles, and the concept of  horizon replicas in Sec.\il(\ref{Sec:MB-K}).
This section closes in Sec.\il(\ref{Sec:e-complex-asly}) with an estimation of   the
mass-flux, the  enthalpy-flux
and  the flux thickness,  determined by the geometric properties of spacetime via the torus effective potential.
\subsection{The energy-spin relations}\label{Sec:both-s}
\begin{figure}
  \includegraphics[width=5.6cm]{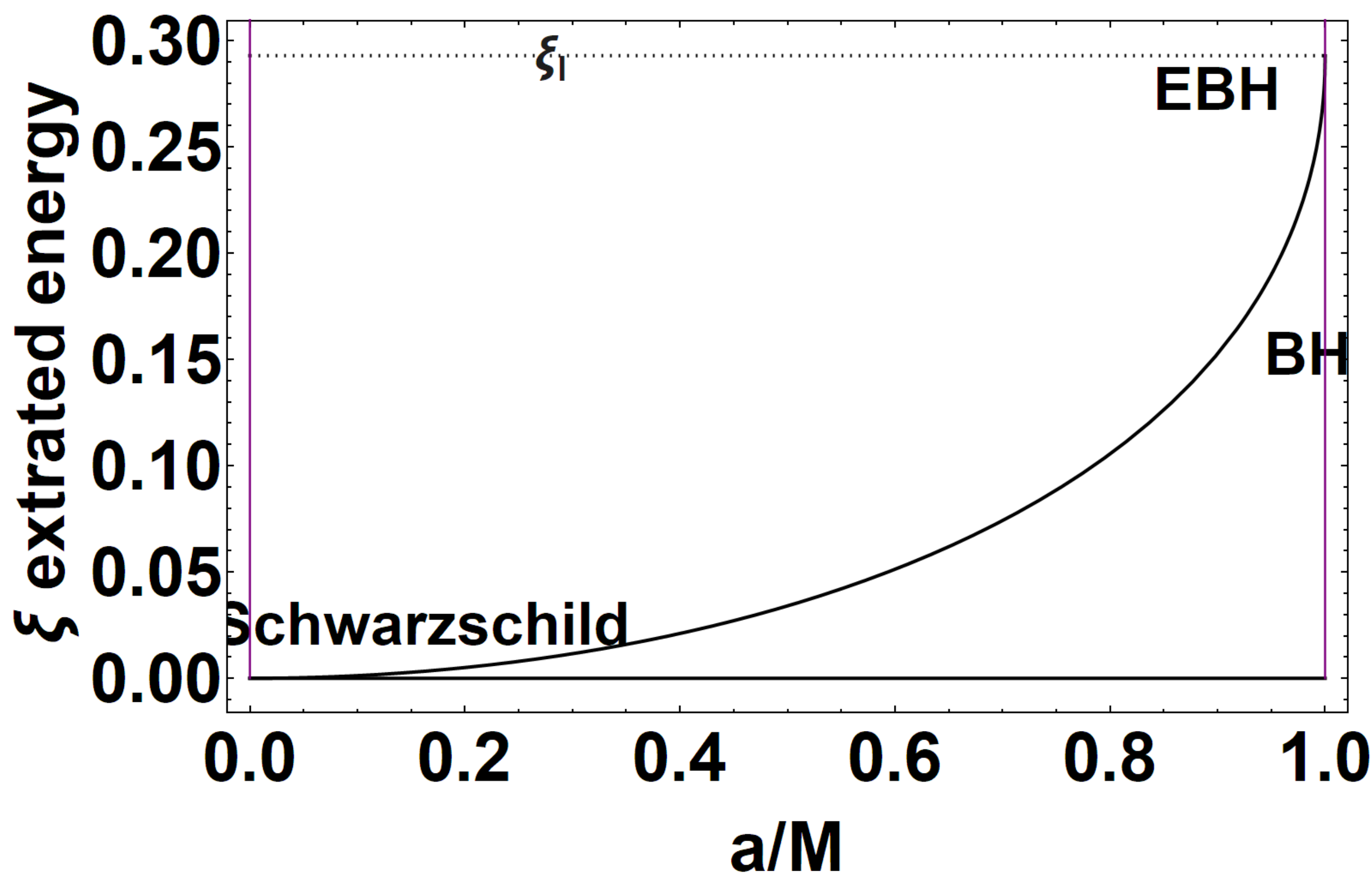}
   \includegraphics[width=5.6cm]{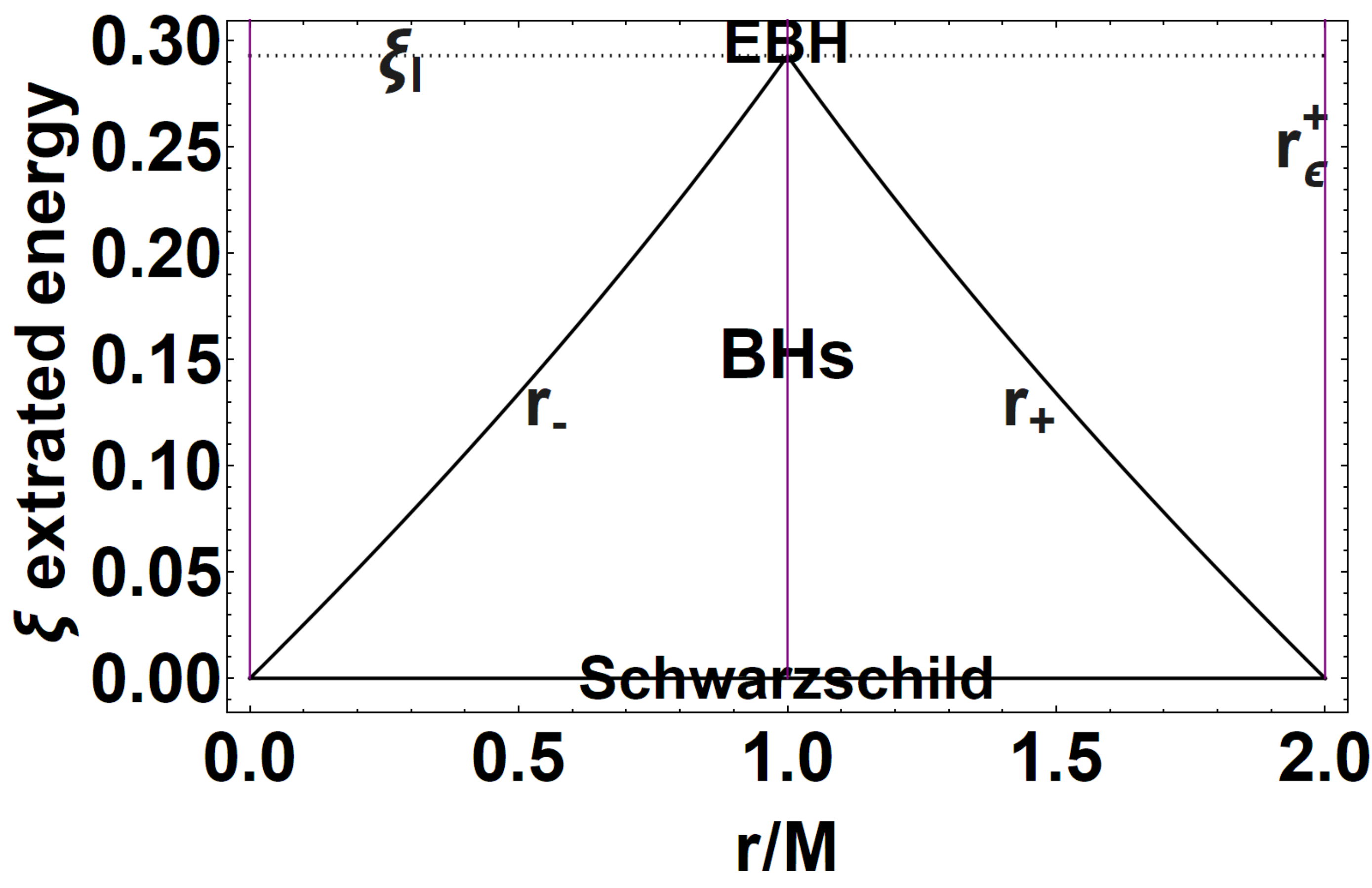}
    \includegraphics[width=5.6cm]{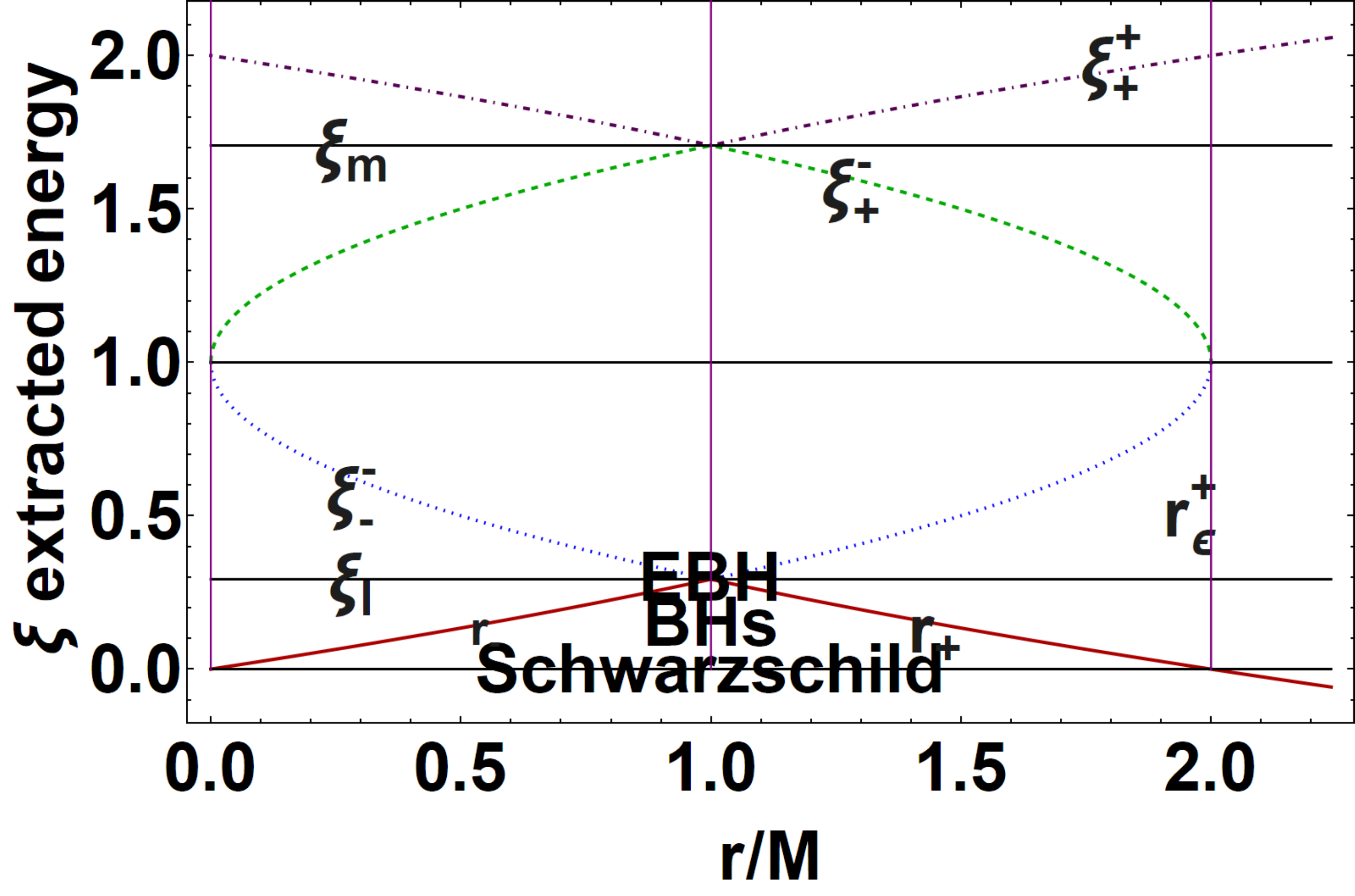}
  \caption{Left panel:  the extracted rotational  energy $\xi$ versus the spin $a/M$. {\textbf{EBH}} is the  extreme Kerr  \textbf{BH}, $\xi_\ell$, maximum extraction, occurs  for total reduction to Schwarzschild attractor from the maximum or nearly  maximum  spin $a=M$.
Center panel:
the extracted energy as parameter  in the extended plane, as function of $r/M$, two Killing horizons $r_{\pm}$ are shown,
$r_{\epsilon}^+=2M$ is the outer  ergosurface  and the outer horizon in the Schwarzschild case.
All the \textbf{BHs} are the boundaries of the inner triangle.
Right panel:  extended plane $\xi-r/M$.
The red triangle  is  the \textbf{BHs} (and represents the inner and outer Killing horizons).
Curves $\xi^{\pm}_{\mp}$ of Eq.\il(\ref{Eq:candd}) (horizons) are shown.  Values $\xi_\ell\equiv  \frac{1}{2} \left(2-\sqrt{2}\right)$ and  $\xi_m\equiv\frac{1}{2} \left(2+\sqrt{2}\right)$, are maxima of $\la(\xi)$ (dimensionless spin of the \textbf{BH}) where $\la(\xi)=1$ and  $r_+=M$(the extreme Kerr \textbf{BH}).}\label{Fig:planeRTexteprop}
\end{figure}

We start by considering the  spin function $\la(\xi)$ :
\bea\label{Eq:exi-the-esse.xit}
\la(\xi)\equiv 2 \sqrt{-(\xi -2) (\xi -1)^2 \xi },
\eea
 relating the  dimensionless \textbf{BH} spin $a/M$ to   the dimensionless ratio $\xi= M_{rot}/M$   representing
the total released rotational  energy   versus \textbf{BH}  mass (measured by an observer at infinity, that is $M_{rot}\equiv M-M_{irr}$ where $M_{irr}$ is the irreducible \textbf{BH} mass),  assuming a  process ending with  the  total extraction of  the    rotational energy of the central Kerr \textbf{BH}--Figs\il(\ref{Fig:planeRTexteprop}). (Here  and in the following  where we do not intend differently, we shall  use dimensionless quantities, viceversa where necessary we will make explicit  the dependence on the mass.).
We can express Eq.\il(\ref{Eq:exi-the-esse.xit}) in the form
\bea\label{Eq:candd}&&
\xi^{\mp}_{\pm}=1\pm\sqrt{\frac{r_{\mp}}{2}},
\eea
relating directly the energy parameter $\xi$ to the horizons--see Figs\il(\ref{Fig:planeRTexteprop}).  However,  as there is  $\xi\propto M-M_{irr}$  (here $\xi$ has units of mass $M$), only solution $\xi_-^+$ has to be considered.
 \mbox{Considering} $\la(\xi)\equiv a_s^{(\pm)}$, \mbox{solving for $\xi$} we find  the   eight functions
$\xi_s^{(\pm)}$
 \bea
&&\xi_s^{(\pm)}=1\flat\frac{\sqrt{1\natural\sqrt{1-(a_s^{(\pm)})^2}}}{\sqrt{2}},\quad\mbox{where}\quad \flat=\pm; \quad \natural=\pm.
\eea
Note
 the  general solution
$\la(\xi)=a_{s}^{(\pm)}$, for a couple of spins $a_{s}^{(\pm)}$,  provides  eight functions
$\xi_s^{(\pm)}$,  four for each spin   $a_{s}^{(+)}$ and $a_{s}^{(-)}$ according to the four (not related) signs $\flat$ and $\natural$ --see Figs\il(\ref{Fig:minneaPlotFlo}).
The surface area $A_{\textbf{BH}}=16 \pi M_{irr}^2$ of the event horizon
identifies the \textbf{BH} irreducible (or rest) mass.
The total  \textbf{BH} mass $M$
 can be decomposed into the  mass $M_{irr}$, the
 rotational energy  and, eventually, the   electromagnetic energy contribution in the Kerr-Newman solution. Therefore  there is  {$M^2= M_{irr}^2+J^2/4M_{irr}^2$} (here $J$, black hole's angular momentum, has { units   of mass $M$ as measured in the asymptotical flat region)}.
  The maximum rotational energy which
can be extracted from the black hole is $(M-M_{irr})$.
A result  of Christodolou and Ruffini  gives an upper
limit on  the   energy  extraction from the   Kerr \textbf{BH}, to the rotational energy   extraction, assuming  a final stage  of the process resulting  a static   Schwarzschild \textbf{BH} \cite{CRR}. The bottom limit of $M$,  at the end of extraction
   process, has to be  $M_{irr}$. More precisely, considering  $M(0)$ and  $ J(0)$ the mass and  angular
momentum of the initial state of the \textbf{BH},   the upper limit for   the energy extracted during  the  stationary process bringing the \textbf{BH} to the state $(1)$ is $M(0) - M_{irr}(M(0), J(0))$. All the quantities are evaluated at the state $(0)$ prior the process, therefore all the quantities evaluated here inform on the status of the \textbf{BH}-accretion disk system at its initial state (0), by evaluating  the related $\xi$ parameter. The \textbf{BH} angular momentum in the final state  $(1)$ is zero. Obtaining the limit of $\xi_{\ell}\equiv \frac{1}{2} \left(2-\sqrt{2}\right)$ of energy extraction (hence $\xi\in[0,\xi_{\ell}]$) where at the state $(0)$ (prior the extraction) there is an extreme Kerr spacetime (having spin $a=M$).
Eq.\il(\ref{Eq:exi-the-esse.xit}) has been found considering the rotational (spin)
energy  $E_{rot}\approx(M - M_{irr})c^2$, where $c$ is the light velocity and   $M_{irr}=\frac{1}{2} \sqrt{a^2+r_+^2}$ is the \textbf{BH} irreducible mass. The rotational energy is  equal to  the extracted  total energy
$E$ of the  outflow $\xi\equiv E_{rot}/Mc^2$.

{Consider the state $(0)$ prior the extraction, there is \bea
M_{irr}^2=\frac{1}{2} \left(\sqrt{M^4-J^2}+M^2\right),\quad\mbox{therefore}\quad
M(0)^2-M_{irr}(0)^2=\left(\frac{J(0) M(0)}{M(0) (2 M_{irr}(0))}\right)^2
\eea
the variation is thus
\bea
\frac{\delta M_{irr}}{M_{irr}}=\frac{\delta M-\delta J(0) \omega_H}{\sqrt{M(0)^2-\frac{J(0)^2}{M(0)^2}}},\quad\mbox{where}\quad
\delta M_{irr}\geq 0, \quad \mbox{thus}\quad (\delta M-\delta J(0) \omega_H)\geq 0.
\eea
($\omega_H$ is the frequency of for the outer Killing horizon.).
From the first law of thermodynamic
$ M^2=\frac{J(0)^2}{4 {M_{irr}}^2}+{M_{irr}}^2
$,
and (extracted rotational) energy emission is essentially\footnote{We note that analogue reasoning guides a first evaluation  of the energy radiated by  gravitational waves following  two
black holes collisions resulting in a merger.
(Particularly considering three  Schwarzschild  \textbf{BHs}  the maximum limit is coincidentally $\xi_{\ell}$ otherwise in case of rotating \textbf{BHs} the maximum values is $\xi=1/2$)}\footnote{Note that the Smarr formulas can be modified considered  an embedded  black hole actually an astrophysical not isolated \textbf{BH} immersed in  material environment, where there is an additional contribution of the  non-vanishing energy--stress tensor due to the matter--see Bardeen in \cite{BARDEEN}.}
}
\bea M_{rot}=M(0)-\sqrt{\frac{1}{2} \left(M(0)^2-\sqrt{M(0)^4-J(0)^2}\right)},\quad \mbox{and}\quad
\frac{M_{rot}}{M(0)}=1-\frac{\sqrt{1+\sqrt{1-\frac{J(0)^2}{M(0)^4}}}}{\sqrt{2}}.
\eea

 In general, there is  $\la(\xi)\in[0,1]$ (dimensionless) and  $\xi\in [0,2]$. However, an immediate calculus from definition of rotational energy, $\xi=1-M_{irr}/M$, with quantities evaluated at the initial state prior the process leads to the restricted range  $\xi\in[0,\xi_{\ell}]$ where $\xi_{\ell}\equiv \frac{1}{2} \left(2-\sqrt{2}\right)$, limiting therefore the energy extracted to a superior of  $\approx 29\%$ of the mass $M$.  %
It is  convenient to view   $\xi$  in its extended range $[0,2]$, and to focus on an extended range  corresponding to the outer and inner horizons in the sense explained below. The extension is intended as a reparametrization of the spin $a(\xi)\in[0,1]$,  since    $\la(\xi)$  shows remarkable properties of symmetries. We shall  see that this extension does  not preserve the symmetry when applied to certain properties of the disk.

As clear from    Figs\il(\ref{Fig:replicas2}),
 functions $\la(\xi)$ and $\xi^{\mp}_{\pm}$ of Eq.\il(\ref{Eq:candd})  are symmetric in $\xi\in[0,2]$.
 For spin value  corresponding to the Scwarzchild
 spacetime,  $\la({\xi})=0$, there is    $\xi=0$ (note in the extended range  $\xi\in[0,2]$,   $\la({\xi})=0$ corresponds also to the values  $\xi=1$   and $\xi=2$). Each curves $\la=\bar{\la}=$constant corresponds  generally  to four  values $\bar{\xi}_1<\bar{\xi}_2<\bar{\xi}_3<\bar{\xi}_4$,  excluding some notable cases as the value  $\la=1$ and $\la=0$.
There is a maximum as function of the spin, $\xi_{\ell}\equiv  \frac{1}{2} \left(2-\sqrt{2}\right)$ (and  $\xi_m\equiv\frac{1}{2} \left(2+\sqrt{2}\right)$) where $\la(\xi)=1$ and  $r_+=M$(the extreme Kerr \textbf{BH}).
Measuring $\xi$ therefore will provide indication of the \textbf{BH} spin. This method,  independent of the specific model of energy extraction,  was introduced   first in \cite{Daly0}, and applied in other analysis  in \cite{Daly0,Daly2,Daly3,GarofaloEvans}.
Here we connect  the quantities $\{\la({\xi}),\xi\}$ with the  tori parameters $\{\ell, K\}$ and fluid angular velocity $\omega$.
The analysis using  function  $\la(\xi)$ in Eq.\il(\ref{Eq:exi-the-esse.xit})  distinguishes  continuum curves of classes of attractors in a plane  $a/M-\xi$. %

The study of the relativistic velocity of the fluid, constrained by the characteristic frequencies of the metric bundle introduced below, allows  to alternatively constrain the  tori fluids  and to constrain structures dependent on the light surfaces defined by the boundary of stationary observers orbits.
The relativistic velocity of the fluid  $\omega\equiv u^{\phi}/u^t$ is related to the function  $1/\ell$  of the fluid specific angular momentum and the conditions of von Zeipel,  governing  the relation between $\omega$ and $\ell$--(see Eqs\il(\ref{Eq:flo-adding})--\cite{zanotti}. Von Zeipel theorem characterizes  any stationary, axisymmetric, non-self-gravitating perfect fluids in circular motion in the gravitational
field of a  central compact object. It
guarantees that the angular velocity $\omega$ depends only on the specific angular momentum $\ell$ and the metric components, therefore
velocity $\omega$ and parameter $\ell$ have common iso-surfaces (assuming a barotropic equation of state).
The von Zeipel
theorem eventually   represents a set of integrability conditions to compute the equilibrium toroidal solutions.
Tori models defined here have four velocity $u^a$   sharing symmetries with the stationary observers, therefore bounded by the limiting  $ \omega_\pm $  relativistic velocities (light-like orbital angular frequencies) of photons defining the light-surfaces regulating many aspects of the \textbf{BH} physics and accretion physics and jet launching or the magnetosphere structure around a \textbf{BH} accretion disk system. Light surfaces are given explicitly in Appendix\il(\ref{Sec:ls-explicit}).  The connection with the von Zeipel surfaces is explored in Appendix\il(\ref{Sec:vonZeipel}).
The fluid specific angular momentum is therefore delimited by the the light-like orbital  frequencies  $\omega_\pm$  defining  the light surfaces and, as particular case the horizon angular velocity $ \omega_H^+=\omega_{\pm}(r_+)$.

 The concept of metric bundles introduced here is based on the classes of all  the geometries having the same limit frequency   $\omega$ in a point; these classes can be represented as curves in a plane called the extended plane, where they are all tangent to the curve defining the Killing horizons of the Kerr geometries. The angular momentum of the fluid and its relativistic velocity are  constrained in a point by the frequency values of the intersection of two bundle curves, at fixed plane. There are situations where these  frequencies are replicated in other portions of the spacetime (the replicas) thus giving the same constraints.
For this reason it is useful to  use the concept of  "extended plane" introduced in  \cite{remnant}  in the definition of metric Killing bundles.
Some bundles on the equatorial plane of the Kerr spacetimes are represented in Figs\il\ref{Fig:Bundleplot}.
The characteristic frequencies $\omega$ are the limiting photon-like frequencies of the stationary observers $\omega_{\pm}$, and also the horizons frequency of the \textbf{BH} identified in the extended plane by the tangent point of the bundle  with the horizon.

 Metric bundles $\ba_{\omega}$ are collections of black holes  or black holes and   naked singularities (\textbf{NSs})   defined generally for any  axially symmetric spacetime with Killing horizons.

 Each geometry of the bundles has  equal "limiting photon  frequency", $\omega_{\pm}$, which is  the characteristic frequency  of the metric bundle,  and it  is also demonstrated be the frequency of a \textbf{BH} Killing horizon as defined   in the extended plane $r/M-a/M$. Metric bundles are represented by curves in the  extended plane  tangent to the horizon curve, which  therefore  emerges as the envelope surface of all the  bundles.
 The metric bundles $\ba_{\omega}$ introduced in \cite{remnant} satisfy the conditions $\mathcal{L}\cdot\mathcal{L}=0$ defining the  null particles frequencies $\omega$  for the vector $\mathcal{L}\equiv\xi_t+\omega\xi_\phi$ which is also an horizon frequency for a certain spacetime with spin  $a$ on the curve    $a_{\pm}\equiv \sqrt{r(2-r)}$  in units of mass. (We note that the curves would be $\pm \sqrt{r(2-r)}$ featuring counter-rotating  orbits.).
 We deepen this definition below.
\subsection{Metric bundles, horizon replicas and photon frequencies}\label{Sec:MB-K}
Metric bundles  (\textbf{MBs}) of the Kerr geometry  were fully characterized in  \cite{inprep}--see also \cite{remnant0,remnant1,remnant,LQG-bundles}.
 This definition concentrates on the  null orbits  frequencies of the   stationary observers on the orbits $r$, and   introduce the concept of replicas for light-surfaces  with equal photon frequencies, which is useful to connect different points of the spacetime, and different geometries following the \textbf{BH} transition for the rotational energy release.
More precisely we   define the replicas in the extended plane $a/M-r/M$,
as a special  set of points  $\{p_i\}_{i=1}^\kappa$ of the extended plane  corresponding to equal   (positive) limiting frequency $\omega>0$ \cite{remnant1,remnant0}. It is proved that there is a  maximum of $\kappa=2$  on the section of the extended plane $a>0$ for fixed plane. In the case considered here,   replicas are a couple  of orbits   $(r,r_1)$ (and planes $(\sigma,\sigma_1)$)   corresponding to   the same value of the  limiting frequency  $\omega$ which is also the   horizon frequency  of the \textbf{BH} defined by the tangency condition of the bundle with the horizon: all the geometries of the bundles have at an point  $r$ and $\sigma$, one equal limiting frequency (the characteristic bundle frequency).
More  generally, we can extend the definition of bundle to group geometries on the basis of an equal value of a property $\Qa$, but the convenience of the choice of the  orbital limiting frequencies for  photon   is manifold: this definition  is naturally related to the geometry symmetries, it has a numerous of relevant astrophysical applications; eventually,  in the extended plane, the \textbf{BHs} horizons emerge as the envelope surface of all the bundle curves, providing therefore also an horizon definition.

The concept of horizon confinement can be introduced by considering that  there is a replica when in the same spacetime it is possible to find at least a couple of points having the same value of properties $\Qa$. There is a \emph{confinement} viceversa, when   that value is not replicated.
In the  Kerr spacetime, part of the inner horizons frequencies are {"confined"}.
The confinement analysis, which is  a study of curves $\ba_{\omega}$ topology in the extended plane and particularly the self-crossing of the bundle curves in the extended plane,   provides the possibility to extract information of  determined local properties of the spacetime in some regions, by their possible replicas in other regions  more accessible  to the observers. Conversely,  it provides  a way to connect measures in different  spacetimes, bundles connect different points in different spacetimes, all characterized by equal value of the property $\Qa$.
Therefore \textbf{MBs} connect   two or more points  in the same spacetime  or different spacetimes and one point of the horizon in the extended plane to other points by equal values of the quantity  $\Qa$.

\begin{figure}
  \includegraphics[width=6cm]{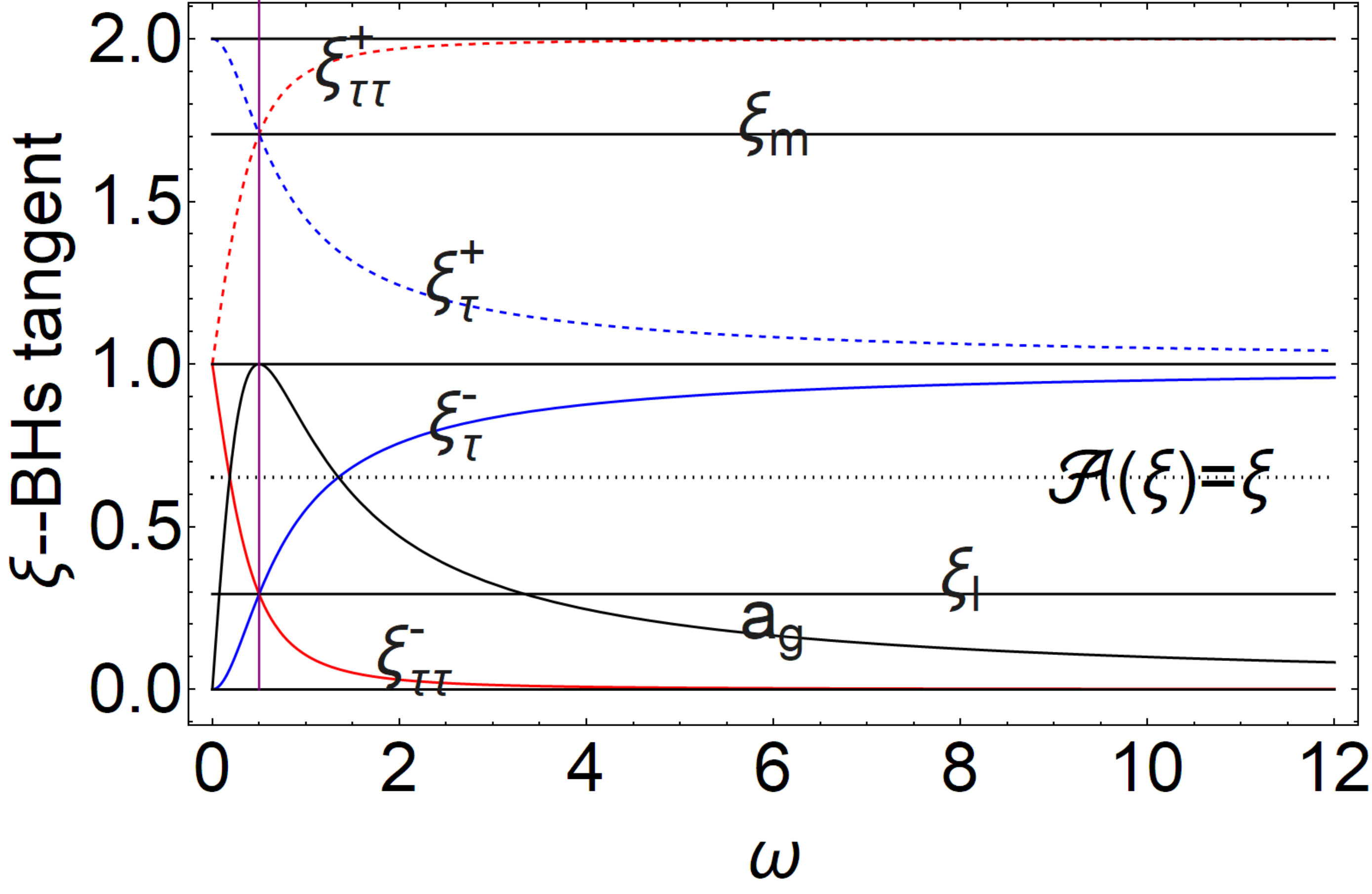}
    \includegraphics[width=6cm]{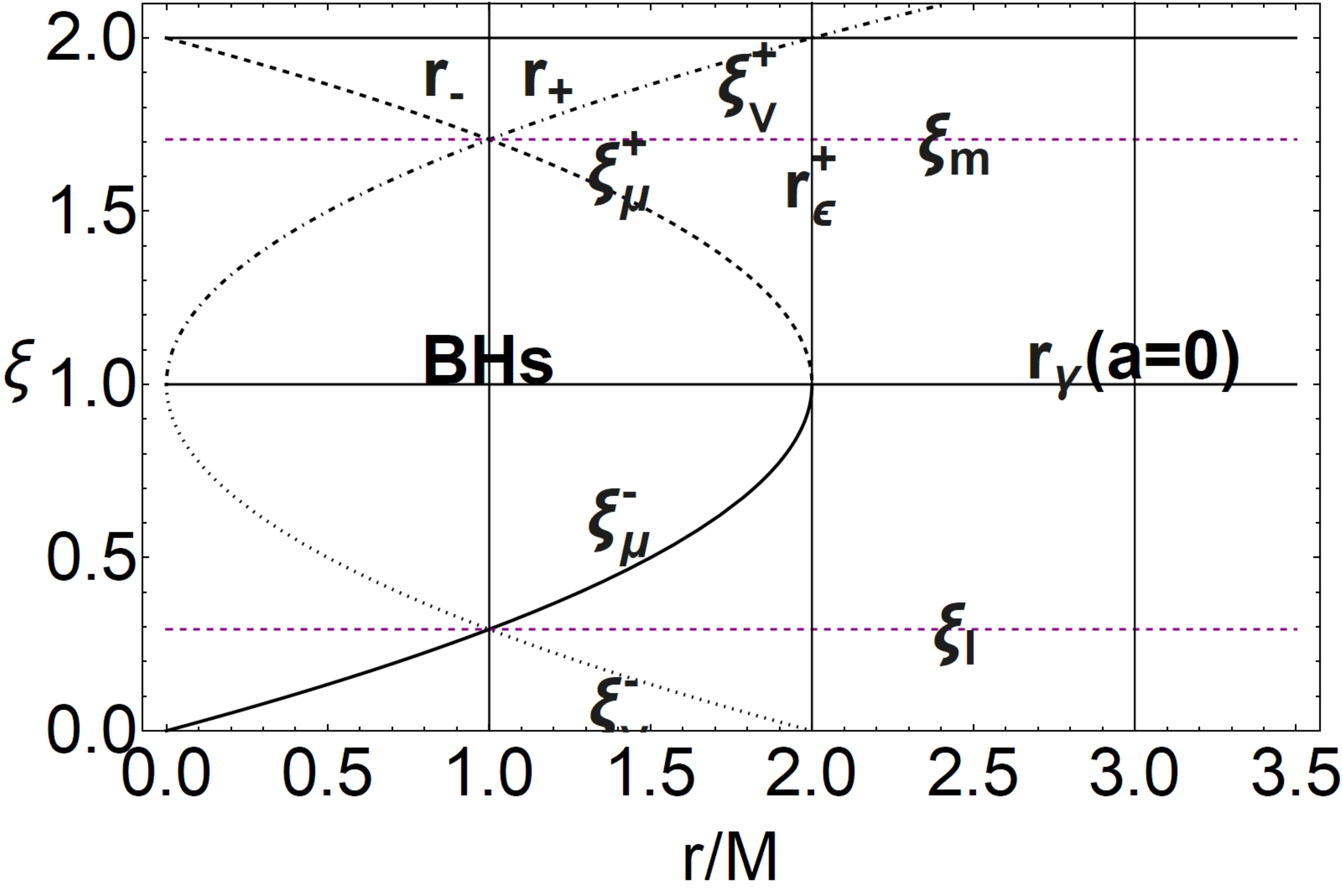}
    \includegraphics[width=5.6cm]{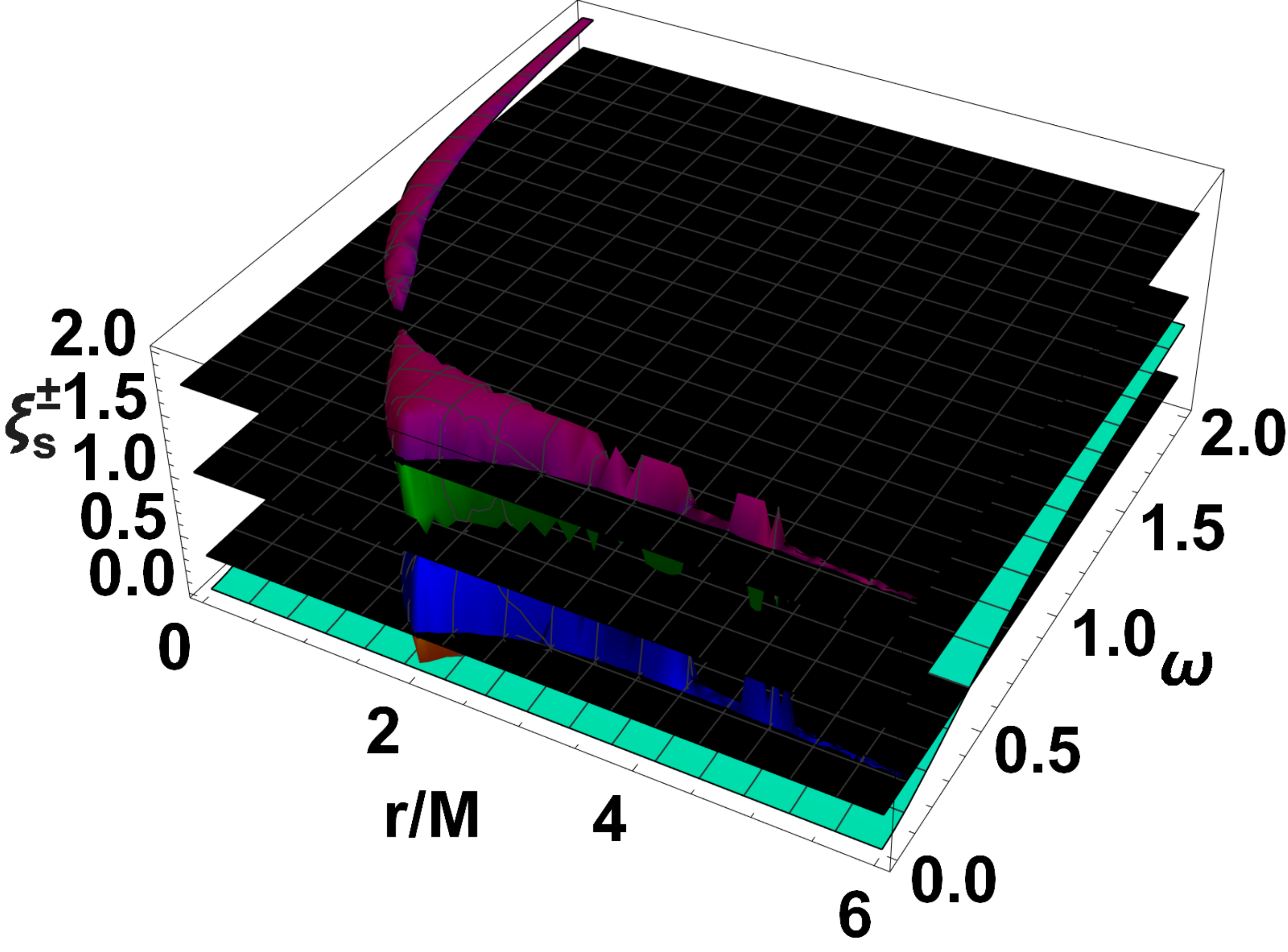}
  \caption{Spin energy extraction.  Left panel: curves $\xi_{\tau}^{\mp}$, $\xi_{\tau\tau}^{\mp}$, defined in Eqs\il(\ref{Eq:aextretautau}),   of tangent curves of the horizons in the extended plane as functions of the horizons frequencies $\omega$. Value $\omega=1/2$ is the frequency of the extreme \textbf{BH} horizon. Spin function $a_g$ is the tangent curve to the horizon as function of the horizon frequency--see Eq.\il(\ref{Eq:show-g}).  Black lines are  $\xi_\ell$ and $\xi_m$ maxima of $\xi$ as functions of the \textbf{BH} spin $a/M$.   Values $\xi_{\ell}$ and  $\xi_m$, are maxima of $\la(\xi)$ (dimensionless spin of the \textbf{BH}) where $\la(\xi)=1$ and  $r_+=M$ (the extreme Kerr \textbf{BH}).  Solution $\la(\xi)=\xi$, dotted line, is the crossing point $a_g=\xi_\tau^-$ and $\xi_{\tau\tau}^-=a_g$ for the outer and inner horizon of the spacetime with spin $a=a_g=\xi_\tau^-=\xi_{\tau\tau}^-$. Spin $\la(\xi)$ is defined in  Eq.\il(\ref{Eq:exi-the-esse.xit}) as function of  dimensionless emission energy $\xi$.  Center panel: plot of the functions $\xi_\mu^{\pm}$ and $\xi_{\nu}^{\pm}$, rotational energy, defined in Eqs\il(\ref{Eq:xtautau}) as function of $r$, horizons points in the extended plane: $r/M=1$ corresponds to the extreme Kerr \textbf{BH} spacetime, $r=0$ is a limiting value correspondent to the Schwarachild singularity with $r=2M$, which corresponds also to radius $r_{\epsilon}^+$ outer ergosurface  on the equatorial plane. $r_{\gamma}=3M$ is the photon orbit on the equatorial plane for the Swarzachild spacetime.  The region $r\in[0,M[$ refers to the inner horizons, while $r\in[M, 2M]$ corresponds to the outer horizons. Right panel: 3D  solutions  $\xi_s^{\pm}$  in Eq.\il(\ref{Eq:candd}), of  $\omega=\omega_{\pm}(a,r)$, for $\sigma=1$  (equatorial plane)  and $a=\la(\xi)$  on the metric bundles  Eq.\il(\ref{Eq:bundle-a-equa})   as function of the horizon frequency.
}\label{Fig:minneaPlotFlo}
\end{figure}

In the Kerr geometry  we restrict the definition to the \textbf{BH} case, therefore the section of the extended plane $a/M\in [0,1]$ or $\la\equiv a\sqrt{\sigma}\in[0,M]$ for $\la-r/M$ (the quantity $\la$ should not be confused with spin function Eq.\il(\ref{Eq:exi-the-esse.xit})), and  applying the symmetries for corotating and counterrotating fluids,  related to the poloidal angle,  connecting a point $p_i$, typically a point of the outer horizon  of the \textbf{BH} geometry,   to a set of points  $\{p_i\}$ all characterized by same value of  photon orbital frequency $\omega_{*}$  of the horizon. It can be proven that in the extended plane $a/M-r/M$,  for fixed plane $\theta=$constant,  it is possible to connect two points at most through the frequency of the outer horizon (existence of at most a replica of the outer horizon in the same geometry). In the case of the inner horizon the situation is more complex and it is not always possible to find a replica of the horizon\cite{remnant}. The dependence of the orbit on  the poloidal angle $ \theta $,  allows a more accurate study of the regions close to the \textbf{BH} rotational axis,  revealing  interesting observational implications\cite{remnant}.
Similarly transformation  from the bundles in $a/M-r/M$ to   $\la-r/M$  has been discussed in \cite{remnant1,remnant0}.
In this context it is clear  that   $ \omega$ is tied to the   rotational energy extraction (we also add further notes on this issue in  Sec.\il(\ref{Sec:barl})).
The process of energy extraction   brings the point on the horizon  from  $p_p$ to a point  $ p_1$ on the horizon curve,   $a_{\pm}$ in the plane $ a/M-r/M$,  and then a rigid rotation in the sense of  \cite{remnant}, a shift from an horizontal line of the extended plane  to another.
\begin{figure}
  \includegraphics[width=5cm]{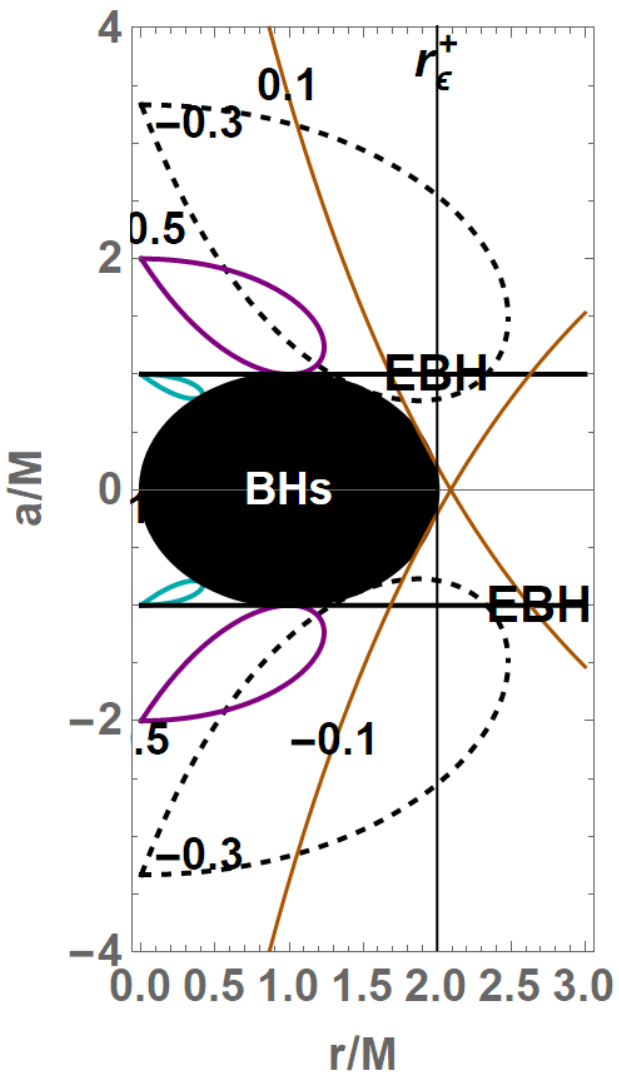}
  \caption{ Metric  Killing bundles of the Kerr spacetimes, on the equatorial plane, in the extended plane $a/M-r/M$ see also Figs\il(\ref{Fig:Bundleplot1}). $\omega$ is the bundle characteristic frequency which is a limiting photon frequency of the  stationary observers, constant values of $\omega$ are signed  on the curves. All the bundles are tangent to the horizon curves (inner and outer horizons) in the extended plane. The central black region is the set of Kerr \textbf{BHs}. Inner and outer Killing horizons are represented. The extreme Kerr black hole (\textbf{EBH}) spacetime $a=\pm1$ is also shown. $r_{\epsilon}^+$ is the outer ergosurface, counter-rotating orbits are also shown ($\omega a<0$). The limit of the Schwarzschild spacetimes is on point $(a=0,r=0)$ and  point $(a=0, r=2M)$.
  for $|a|>M$ there is a naked singularity $(\textbf{NS})$. Details on the structure of the metric bundles are in \cite{observers,remnant,remnant0,remnant1}.}\label{Fig:Bundleplot}
\end{figure}
\begin{figure}
    \includegraphics[width=6cm]{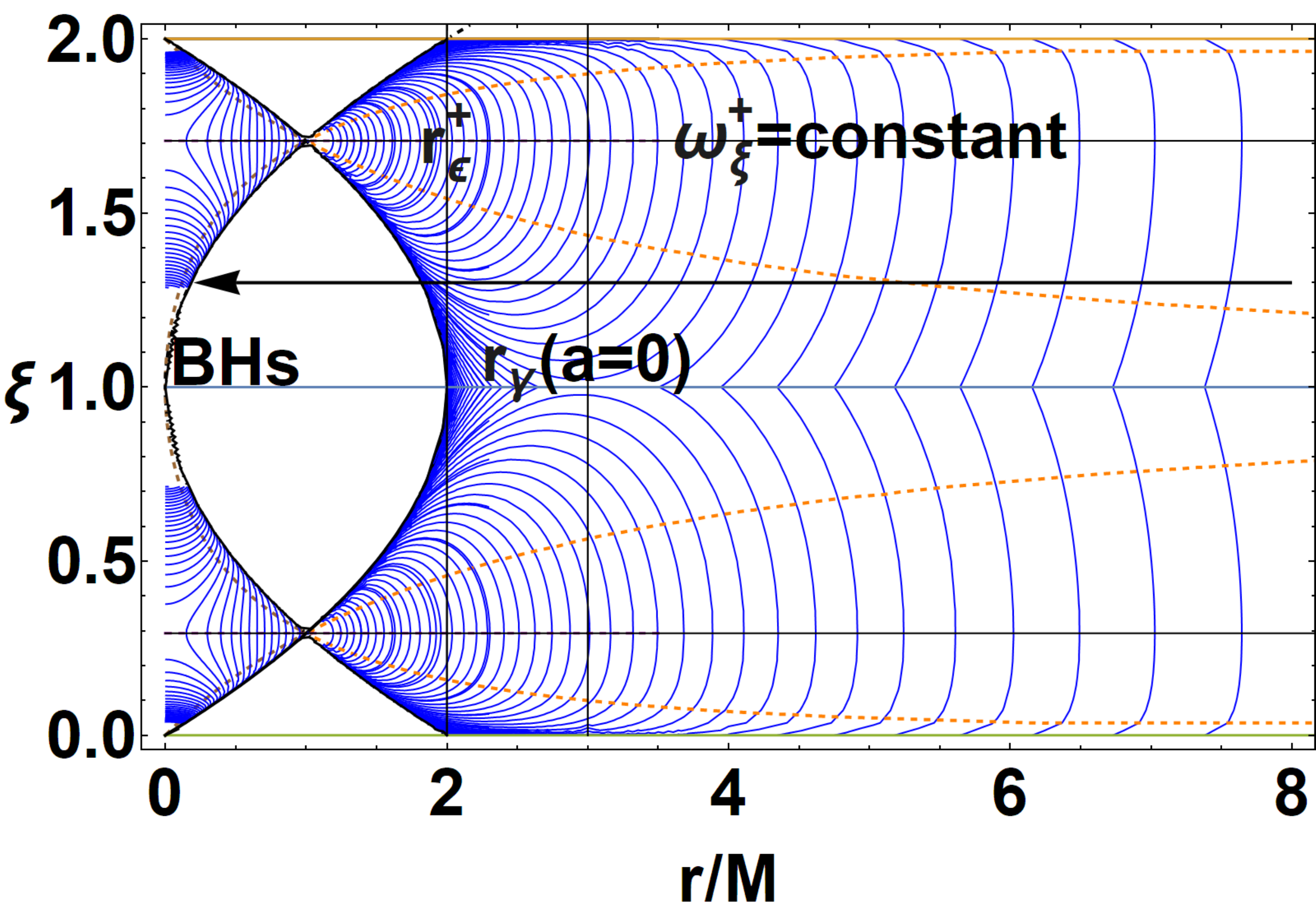}
      \includegraphics[width=6cm]{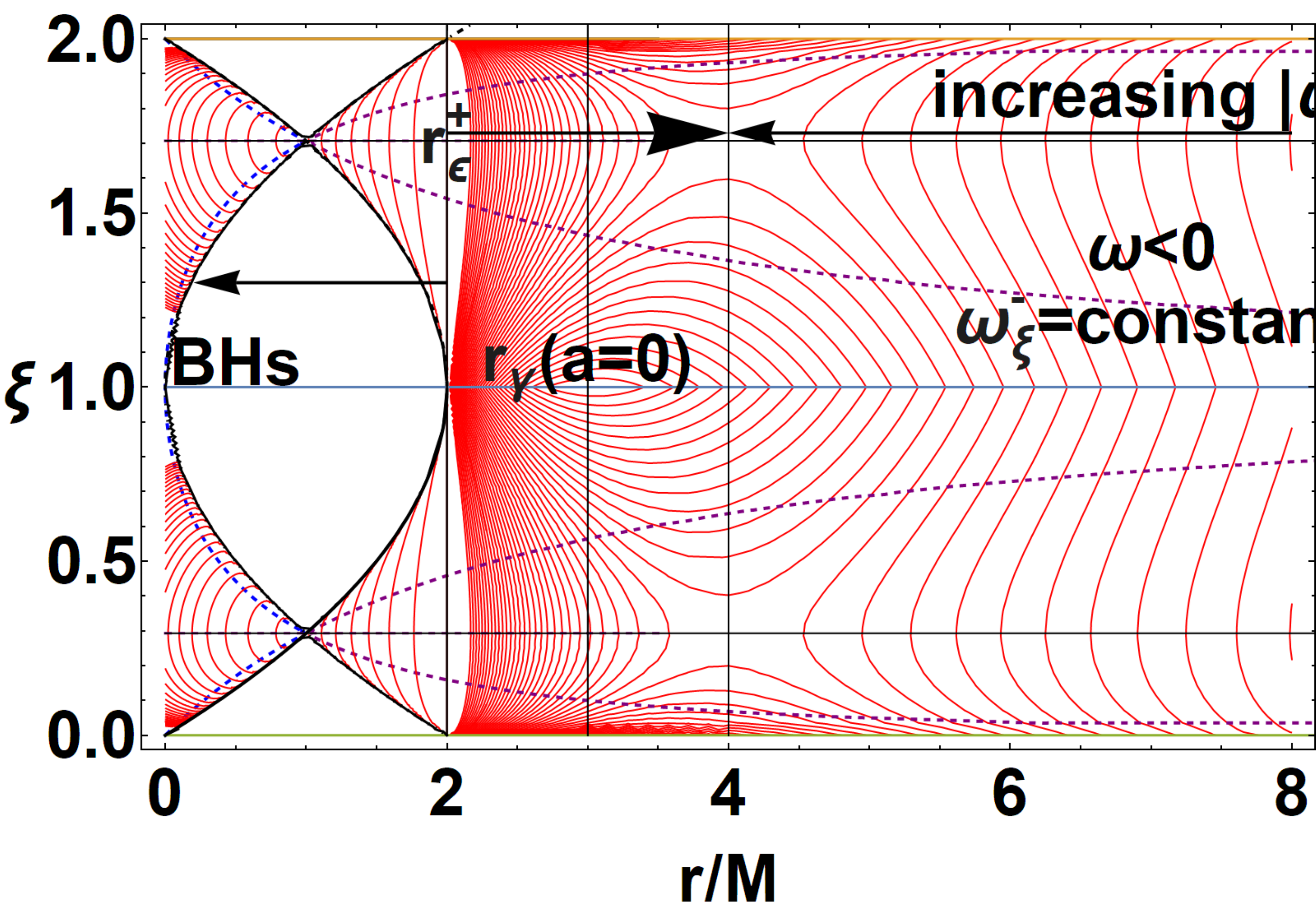}\\
        \includegraphics[width=5cm]{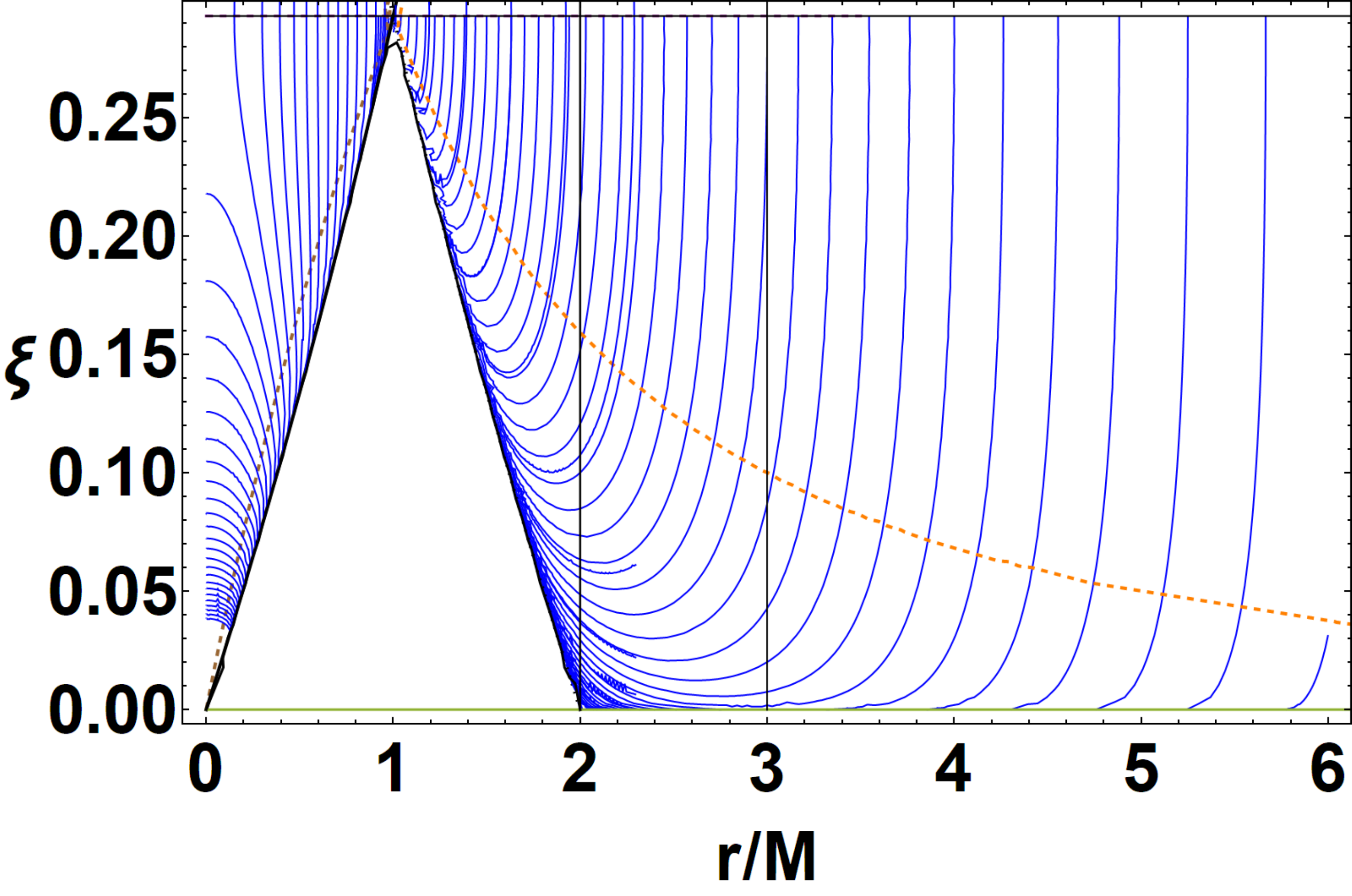}
       \includegraphics[width=5cm]{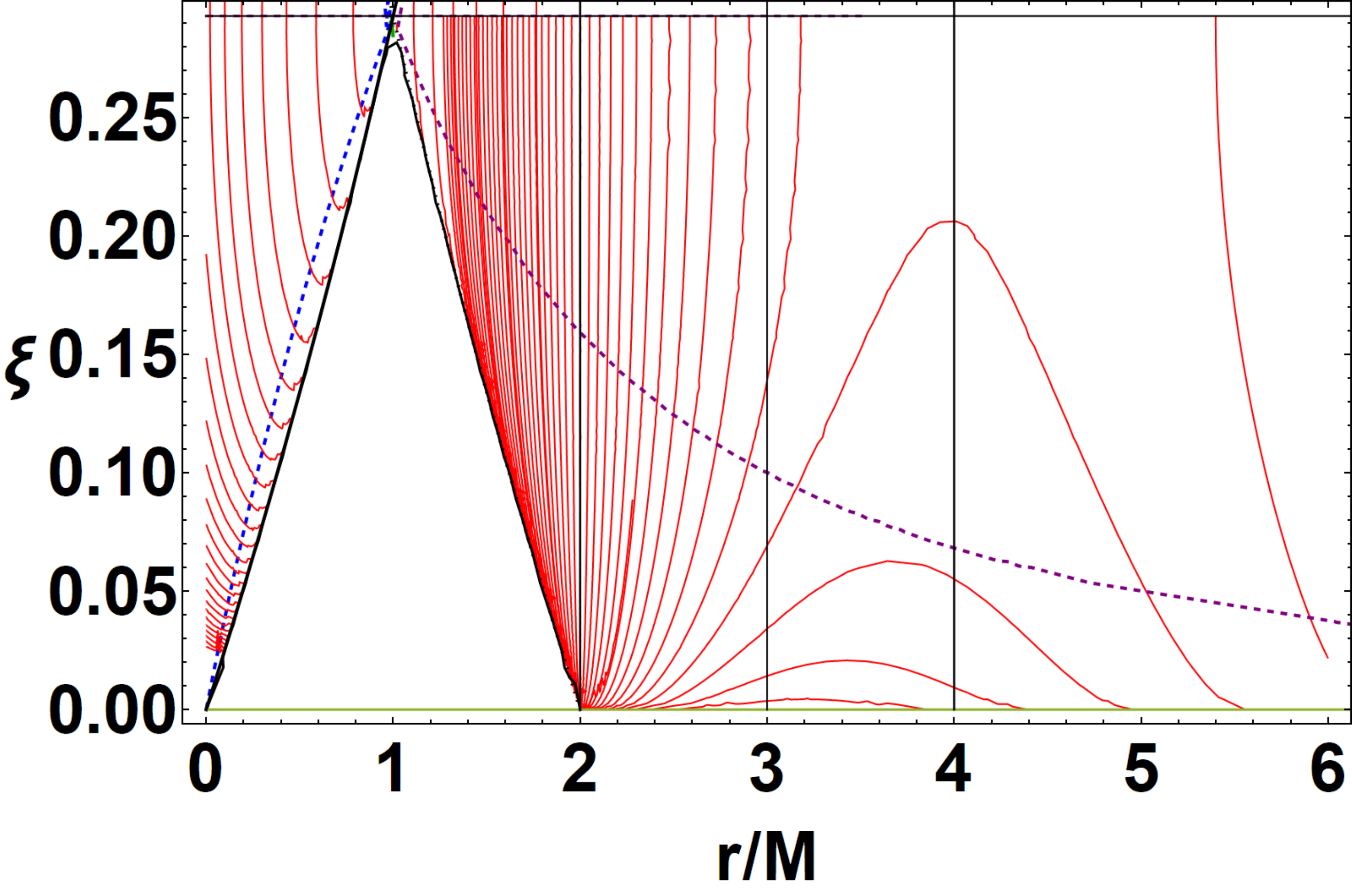}
          \includegraphics[width=6cm]{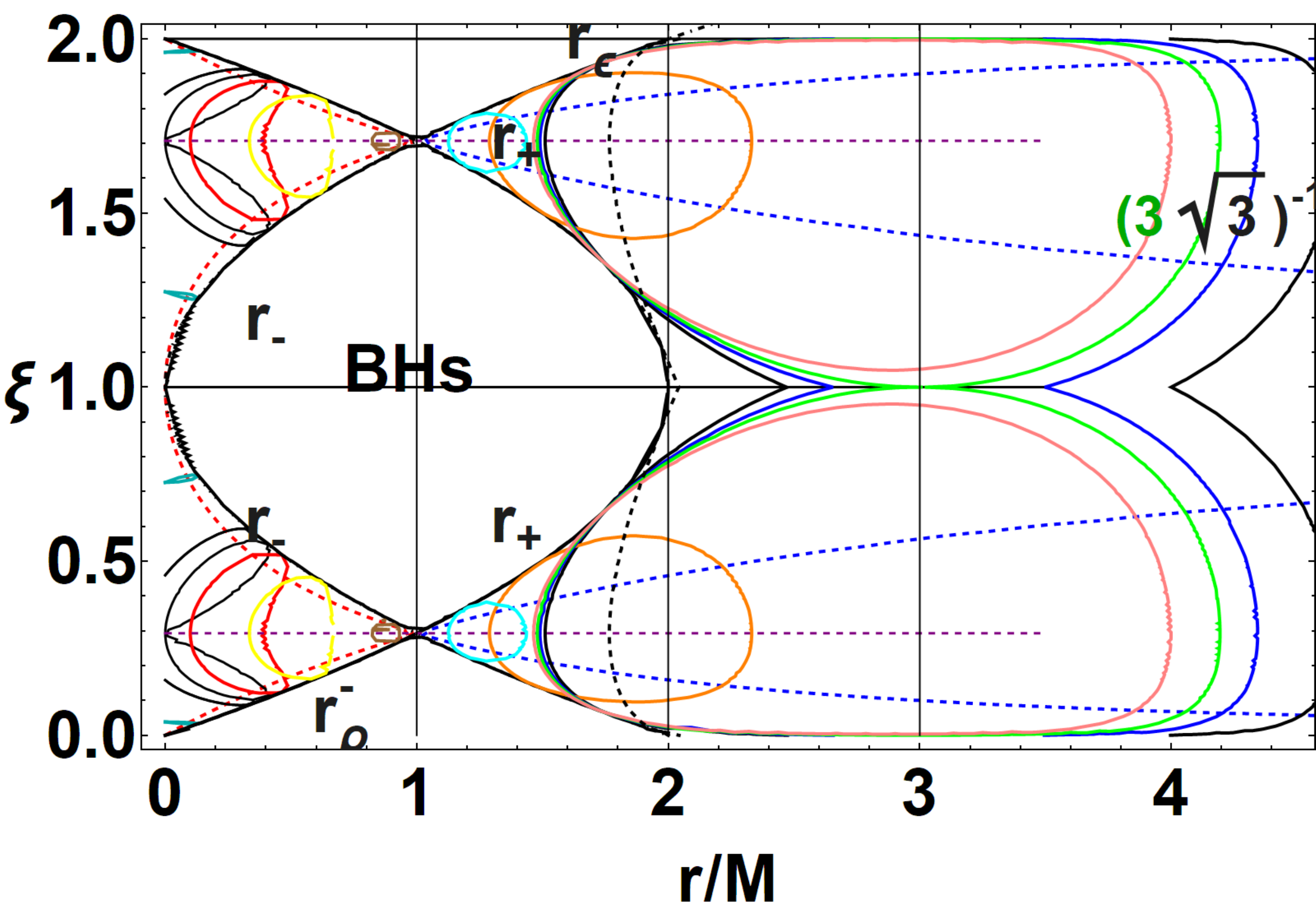}
  \caption{Metric Killing Bundles of the Kerr spacetimes (the equatorial plane) in the extended plane $\xi-r/M$,  $\xi$ is the rotational energy extraction parameter--see also Figs\il(\ref{Fig:Bundleplot}).  The bundle characteristic frequency $\omega$, constant on each bundle is a limiting photon frequency of the  stationary observers. The special frequency $\omega=1/\sqrt{27}$, related to the zeros of the metric bundles (which is the Schwarzschild geometry) and the photon orbit  is signed on the curve. All the bundles are tangent to the horizon curves (inner and outer horizons) in the extended plane. Inner and outer Killing horizons are represented as well as the horizons replicas (dashed curves) $r_{\rho}^{\pm}$: for fixed  spacetimes there are two orbits with equal horizons  frequencies $\omega_{H}^{\pm}$ one on the horizons $r_{\pm}$, the second  on the orbit replica--see also Figs\il(\ref{Fig:replicas2}). The extreme Kerr black hole (\textbf{EBH}) spacetime (for $\xi=0.5$) is also shown. $r_{\epsilon}^+$ is the outer ergosurface, counter-rotating  orbits ($\omega a<0$)  are also shown. The limit represented by the  Schwarzschild spacetime in on points $(a=0,r=0)$ and $(a=0, r=2M)$.
  For $|a|>M$ there are  naked singularities $(\textbf{NS})$.  Arrows set the increasing values of the frequencies magnitude. We note the special role of the photon orbits $r_{\gamma}=3M$ and marginally bounded orbit $r_{mbo}=4M$ for the Schwarzschild spacetime (the limiting case $a=0$ in the extended plane) for each \textbf{BH} spin. Below--left and center panels show a zoom on the region $\xi\in [0,\xi_{\ell}]$.  Below--right panel shows a selection of curves. Some frequencies, for $r_{\gamma}=3$ are shown--see also Figs\il(\ref{Fig:Combplot1}).}\label{Fig:Bundleplot1}
\end{figure}
Specifying
these arguments,  we consider the  \textbf{BHs} horizons frequencies $\omega_{H}^{\pm}$  and the horizons curves   in the extended plane $\xi-r/M$  respectively on the equatorial plane
\bea\label{Eq:xtautau}
&&
\omega_{H}^{\pm}= \frac{a}{2r_{\pm}},\quad \xi_\mu^{\mp}=1\mp\sqrt{1-\frac{r}{2}},\quad \xi_\nu^{\mp}\equiv1\mp\sqrt{\frac{r}{2}},
\eea
see Figs\il\ref{Fig:Bundleplot1},\ref{Fig:minneaPlotFlo},\ref{Fig:HonRingZN} and  Eqs\il(\ref{Eq:candd}) for functions
$\xi^{\mp}_{\pm}$--Figs\il(\ref{Fig:planeRTexteprop}).
 It is shown the  frequency
$\omega=1/\sqrt{27}$ relevant  to the bundle structure for each value of $ a / M $. On the equatorial plane, the zero of the bundles define the  Schwarzschild static case, in this spacetime the frequency is related to photon orbit $r=3M$.

\begin{figure}
  \includegraphics[scale=.52]{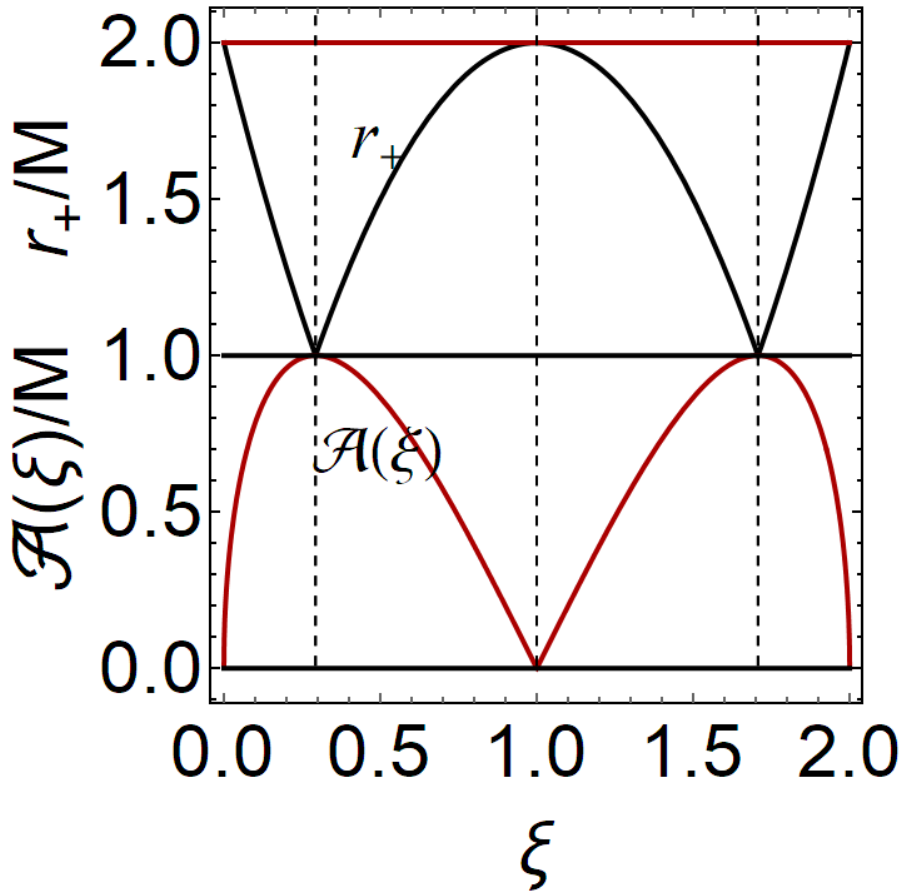}
  \includegraphics[width=6cm]{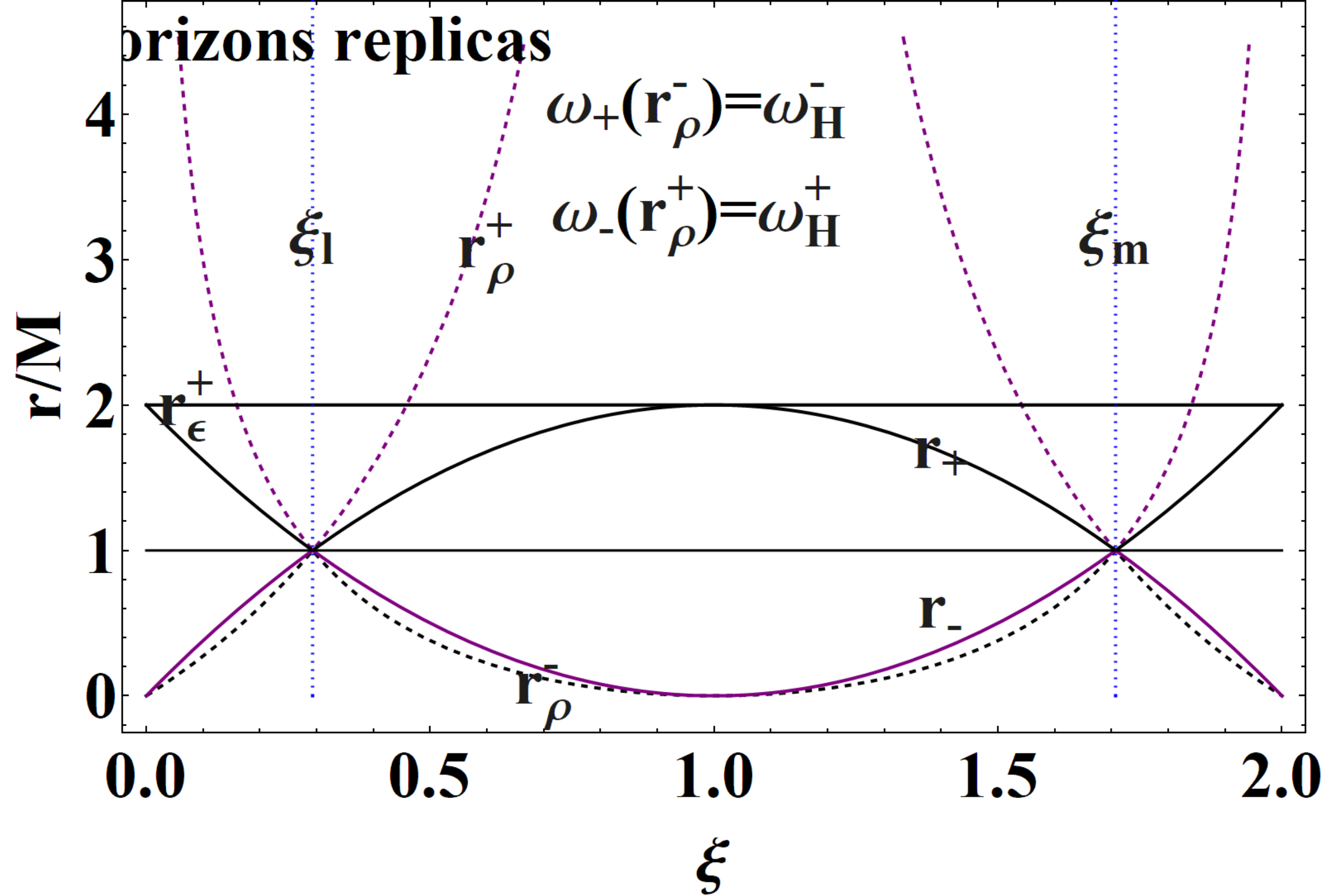}
    \includegraphics[width=6cm]{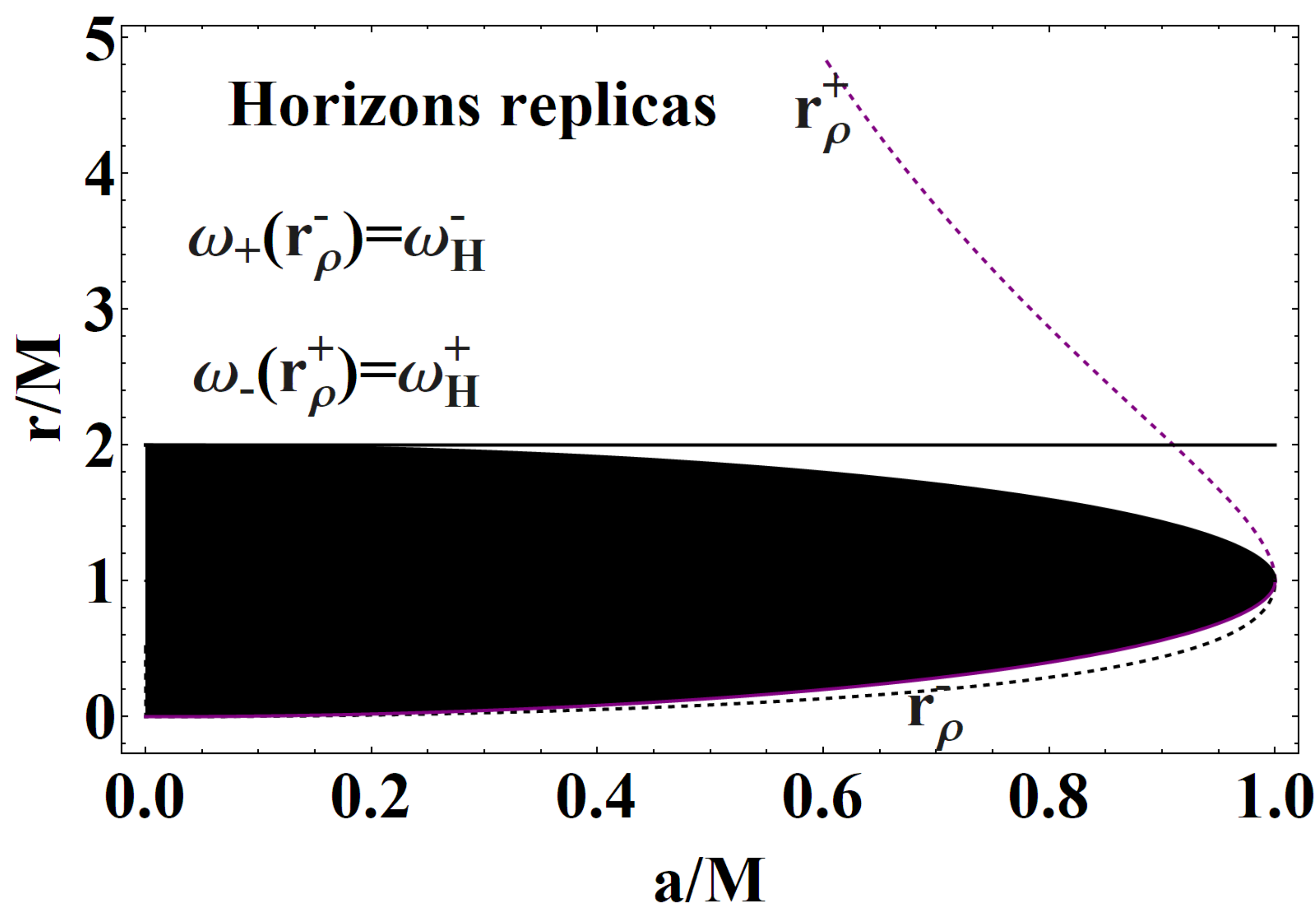}
  \caption{Left panel: horizon $r_+$ and dimensionless spin $\la(\xi)$ of Eq.\il(\ref{Eq:exi-the-esse.xit}) function of the extracted  rotational energy $\xi$. Center and right panel:  Horizons replicas $r_{\rho}^\pm$ on the equatorial plane as functions of the \textbf{BH} spin $a/M$ and of the energy parameter $\xi$. There is $\omega_H^{\pm}=\omega_\mp(r_{\rho}^\pm)$--see Eq.\il(\ref{Eq:replicas2}). Horizons curve $r_{\pm}$ are also shown. Black region is the central \textbf{BH}.  Radius $r_{\epsilon}^+$ is the outer ergosurface, $\xi_m$ and $\xi_{\ell}$ are maxima of the curve $\la(\xi)$.}\label{Fig:replicas2}
\end{figure}

Let us consider now the tangent curves to the horizons in the extended planes. There are the functions of the  frequency
\bea
&&\label{Eq:aextretautau}
a_g=\frac{4 \omega }{4 \omega ^2+1},\quad\mbox{and}\quad
\xi_{\tau}^{\mp}\equiv 1\mp\frac{1}{\sqrt{4 \omega ^2+1}},\quad
\xi_{\tau\tau}^{\mp}\equiv 1\mp\frac{2 \omega }{\sqrt{4 \omega ^2+1}}
\eea
 where  $a_g$ is the curve of tangent points of the  bundles with the horizons, from now on the tangent curve to the horizon, in the extended plane $a/M-r/M$.
 {In this way we  represent the extracted energy in terms of characteristic  frequency  of the bundle and through the tangency  condition of the bundle in the extended plane--Figs\il(\ref{Fig:minneaPlotFlo}).}
Tangent $a_g$ can be  written  in terms of extracted rotational energy as $\xi_{\tau\tau}^{\mp}$,
represented   in Figs.\il(\ref{Fig:minneaPlotFlo}); we note the limiting value  $a=M$. We introduce  the curves $\xi_{\tau\tau}^{\mp}$ and $\xi_{\tau}^{\mp}$ in the extension of the plane for extended values of $\xi$.
In Figs.\il(\ref{Fig:minneaPlotFlo}) we represented the horizons curves of Figs \il(\ref{Fig:replicas2})  parametrized as $\xi_\mu^{\mp}$ in the extended plane,
the inner and outer horizons are shown. The limiting null-like particle frequencies  of the stationary observes, horizons frequencies as characteristic frequencies of the bundles $\ba$ are on the equatorial plane given by
\bea\label{Eq:omegaxi}
&&
 \omega_{\xi}^{\mp}\equiv \frac{4 r\sqrt{-(\xi -2) (\xi -1)^2 \xi } \mp\sqrt{r^4 \left[r-2 (\xi -1)^2\right] [2 (\xi -2) \xi +r]}}{r \left[-8 (\xi -2) (\xi -1)^2 \xi +r^3-4 (\xi -2) (\xi -1)^2 \xi  r\right]},
\eea
see-- Figs\il(\ref{Fig:Bundleplot}).
On the equatorial plane, the bundles are expressed in the form
\bea\label{Eq:bundle-a-equa}
a_{\pm}=\frac{2 \omega \pm\sqrt{r^2 \omega ^2 \left[1-r (r+2) \omega ^2\right]}}{(r+2) \omega ^2}.
\eea
In our case, the bundles in the plane $\xi-r/M$ are evaluated,   i.e. bundles describe   the Schwarzschild \textbf{BH}, the Kerr \textbf{BHs} as well as  Kerr \textbf{NSs}. Considering the counter-rotating orbits, as in \cite{remnant1},  we take into account  also  $a<0$.
Accordingly, bundles  can be given in the extended plane $\xi-r/M$, providing the eight solutions for $a_s^{(\pm)}\equiv a_g^{\pm}$--see  Eq.\il(\ref{Eq:candd}).
Horizons replicas on the equatorial plane are
\bea\label{Eq:replicas2}
r_{\rho}^\mp\equiv \frac{1}{2} \left(\sqrt{\frac{32 r_\mp}{a^2}-a^2\pm6 \sqrt{1-a^2}-22}-r_\mp\right);
\eea
where there is $\omega_H^{\pm}=\omega_\mp(r_{\rho}^\pm)$ respectively--see
\cite{remnant} and Figs\il(\ref{Fig:replicas2}).
 In Figs\il(\ref{Fig:Bundleplot1}) we note the special role of the photon orbits $r_{\gamma}=3M$ and marginally bounded orbit $r_{mbo}=4M$ for the Schwarzschild spacetime (the limiting case $a=0$ in the extended plane) for each \textbf{BH} spin.  Right panel shows a selection of curves, where  frequencies for $r_{\gamma}=3M$ are shown.
Considering the notion of replicas,  an observer on a point  $p$  in the  \textbf{BH}  spacetime with  (dimensionless) spin $a/M$    has orbital frequencies limited by the  photon orbital frequencies  $\omega_{\pm}$, here $\omega_{\xi}^{\pm}$ of Eqs\il(\ref{Eq:omegaxi}),  reachable only in the case  of  null-particles.
 On the outer  horizon of the \textbf{BH} spacetime    there is $\omega_{\pm}=\omega_{H}^{+}$ and the frequency "window" characterizing the stationary observes reduces to  circle  defining the  \textbf{BH} outer  horizon. On the other hand, in the case of   weak naked singularities as studied in \cite{remnant,remnant0,remnant1,1980BAICz..31..129S}, there is a region (of $a/M$ and $\la=\sqrt{\sigma} a/M$) proximate to the singularity  where the frequency range bounded by  $\omega_{\pm}$ is not null  (in this case interpretable as absence of the horizons) but reduced to a minimum.
These singularities have peculiar characteristics emerging very clearly as properties of their bundles $\ba_{\omega}$ in the extended plane.  Bundles corresponding to these \textbf{NSs} are  tangent to a portion of the inner horizon curve which is not confined. This  is related to a region in the plane $r-\omega$ that  can be further reduced  to a minimal  range of frequencies  and spins around the central limiting value $a=M$ (a bottleneck) \cite{ergon}. The issue of naked singularity characterization goes beyond the targets  of this analysis, however we can say that such \textbf{NSs} are characterized by this bottleneck regions   connected to the horizons as their "remnants" or "memory" of  the \textbf{BHs} horizons in the extended plane\cite{remnant}. This region coincides with the region where repulsive gravity effects   appear. \textbf{NSs} region of the extended plane containing parts of the \textbf{MBs} tangent to the horizons have  a role in the \textbf{MBs} analysis of \textbf{BHs} as the origins of the bundles, which  are  points of the axis $r=0$ also in \textbf{NSs}.

Replicas connect the two null vectors $ \mathcal{L}(r_+,a,\sigma)$ and
$ \mathcal{L}(r_p,a,\sigma_p)$ (we also consider the special case $\sigma=\sigma_p$), i.e., $\omega_{H}^{\pm}(a)=\omega_{\pm}$ defining the bundles $\mathcal{B}_{\omega}$ and connecting  points belonging to $\ba_{\omega}$  in the same spacetime or different spacetimes.
 Here we consider  the observers registering  the presence of a replica at the point $p$ of the \textbf{BH} spacetime with spin $a_p$, belonging  to the  Killing bundle $\ba_{\omega}$.  The observer will  find the replica of the \textbf{BH} horizon frequency $\omega_H^+(a_p)$ at point $p$, therefore  her/his orbital stationary frequency will be $\omega_p\in]\omega_{\bullet},\omega_{*}[$ where one of  ($\omega_{\bullet},\omega_{*}$) is the  horizon frequency, i.e. the horizons frequency $\omega_H^{+}$ is replicated on a pair of orbits $(r_+.r_p)$,     the second  light-like frequency  $\omega_\bullet$   is the frequency of a horizon in a \textbf{BH}  spacetime which is correlated by the bundles to $\ba_{\omega}$ in the considered  spacetime.
The relation between the two  frequencies $(\omega_{*},\omega_{\bullet})$ is determined by a characteristic ratio studied in details in \cite{remnant}.

\begin{figure}
  \includegraphics[width=5.4cm]{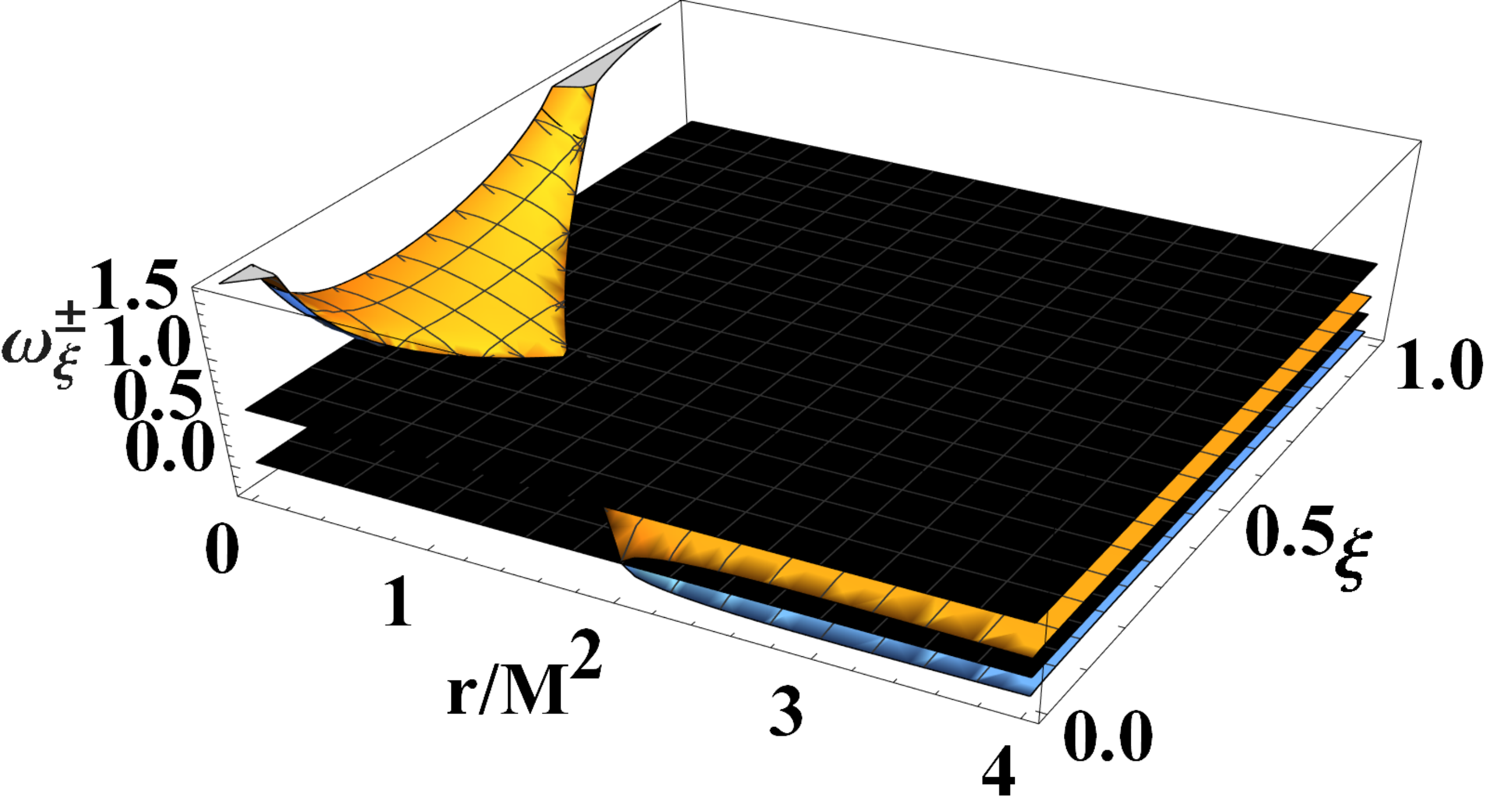}
  \includegraphics[width=5.4cm]{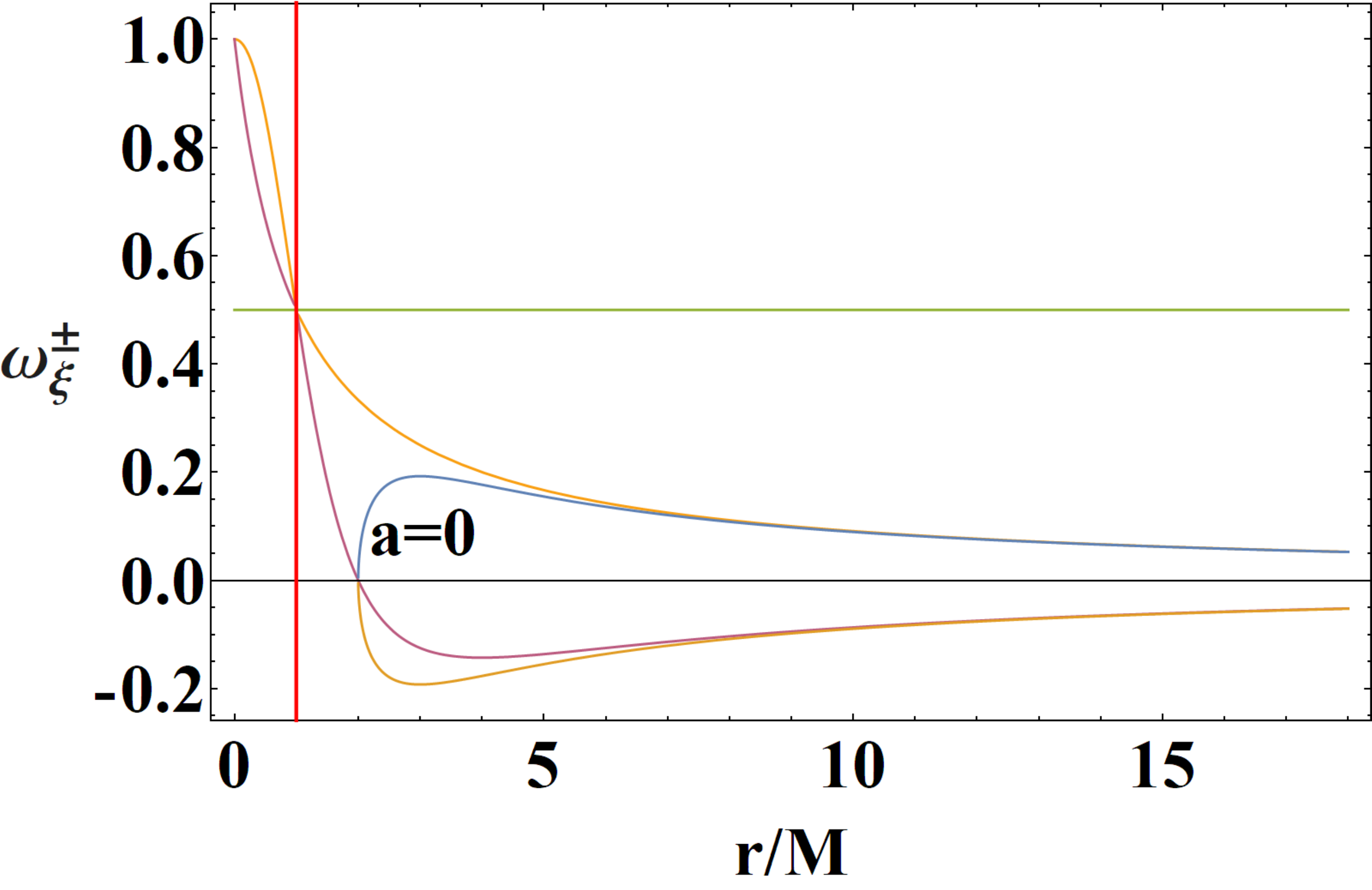}
  \includegraphics[width=5.4cm]{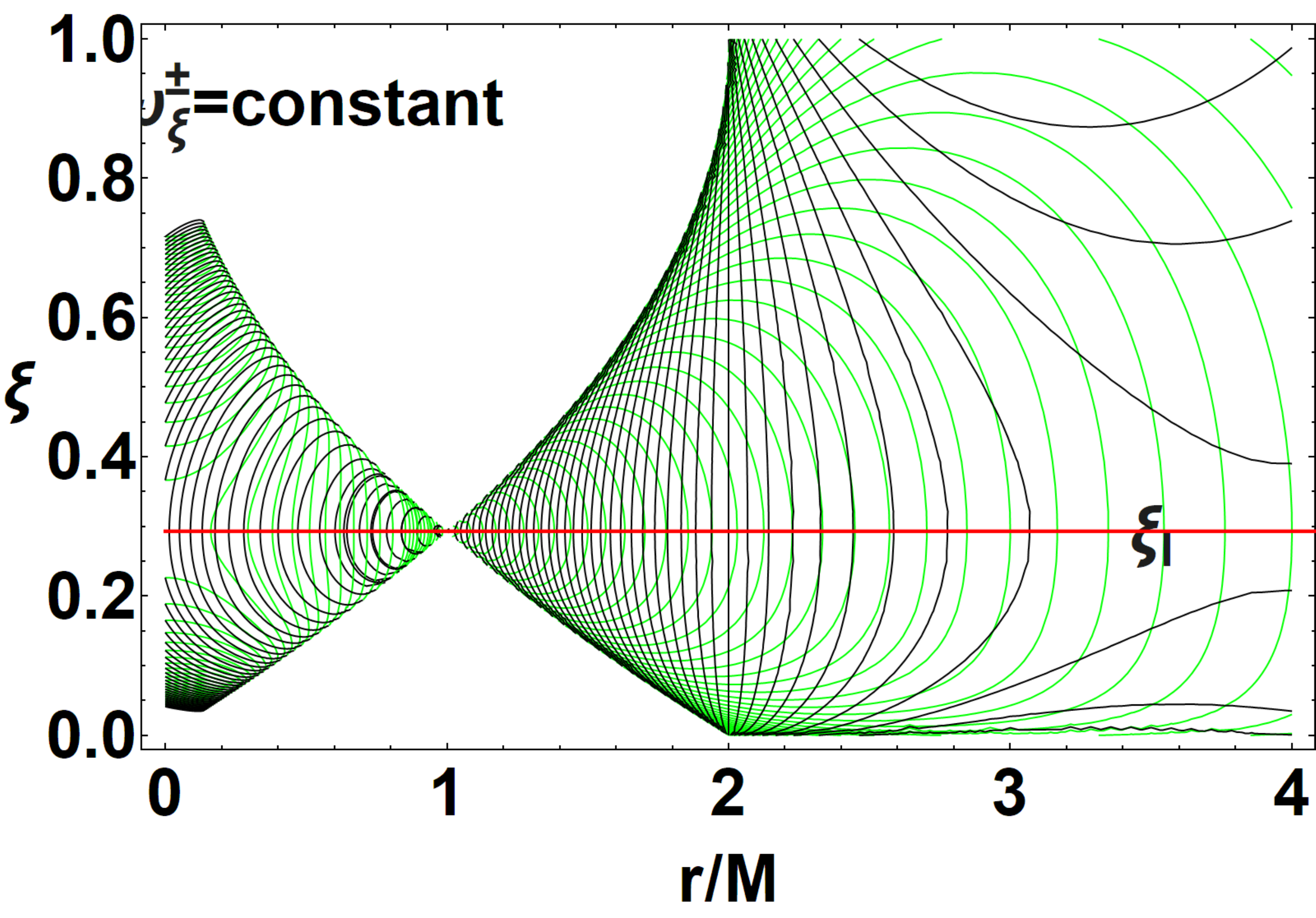}
  \includegraphics[width=5.4cm]{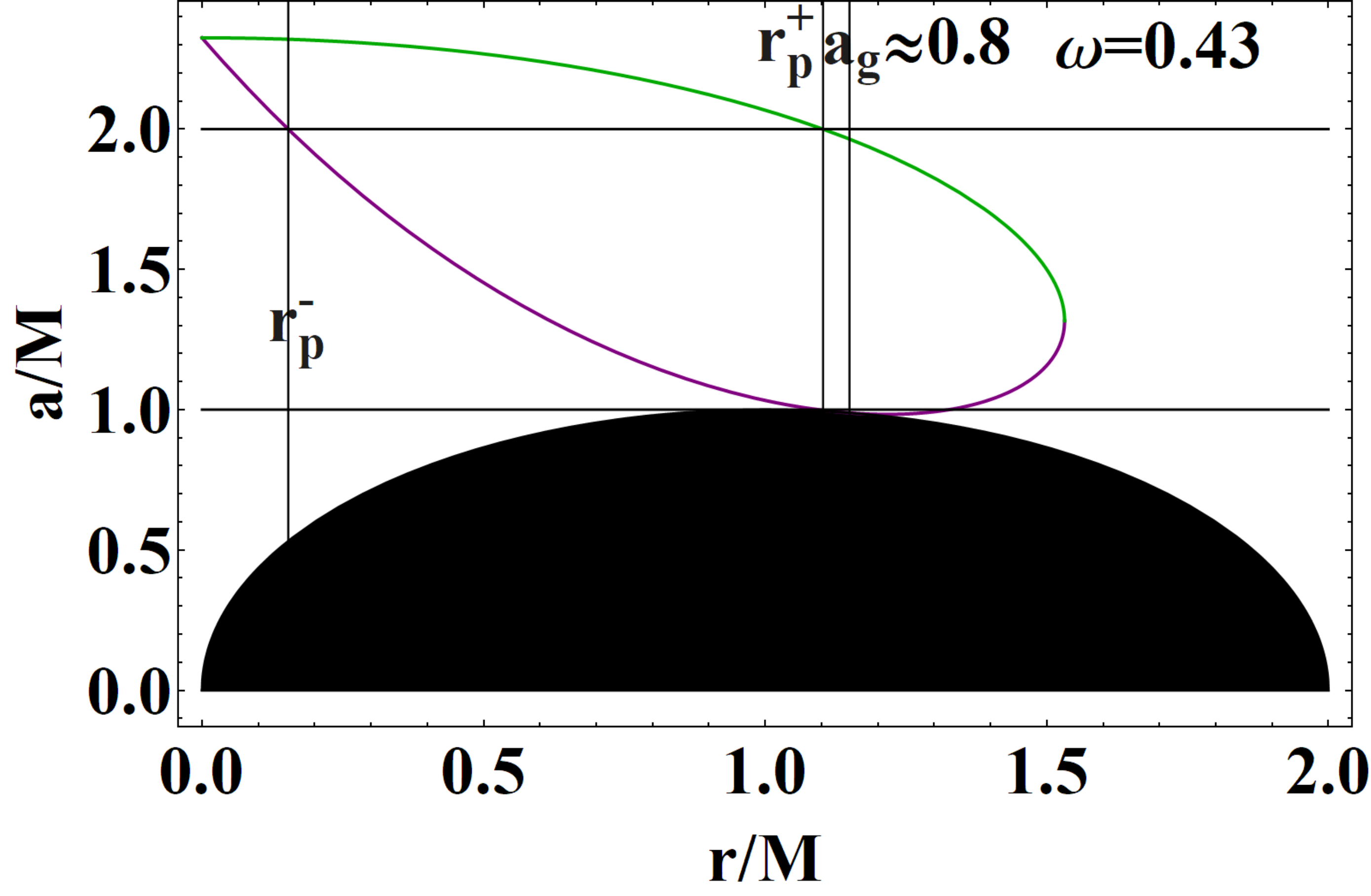}
  \includegraphics[width=5.4cm]{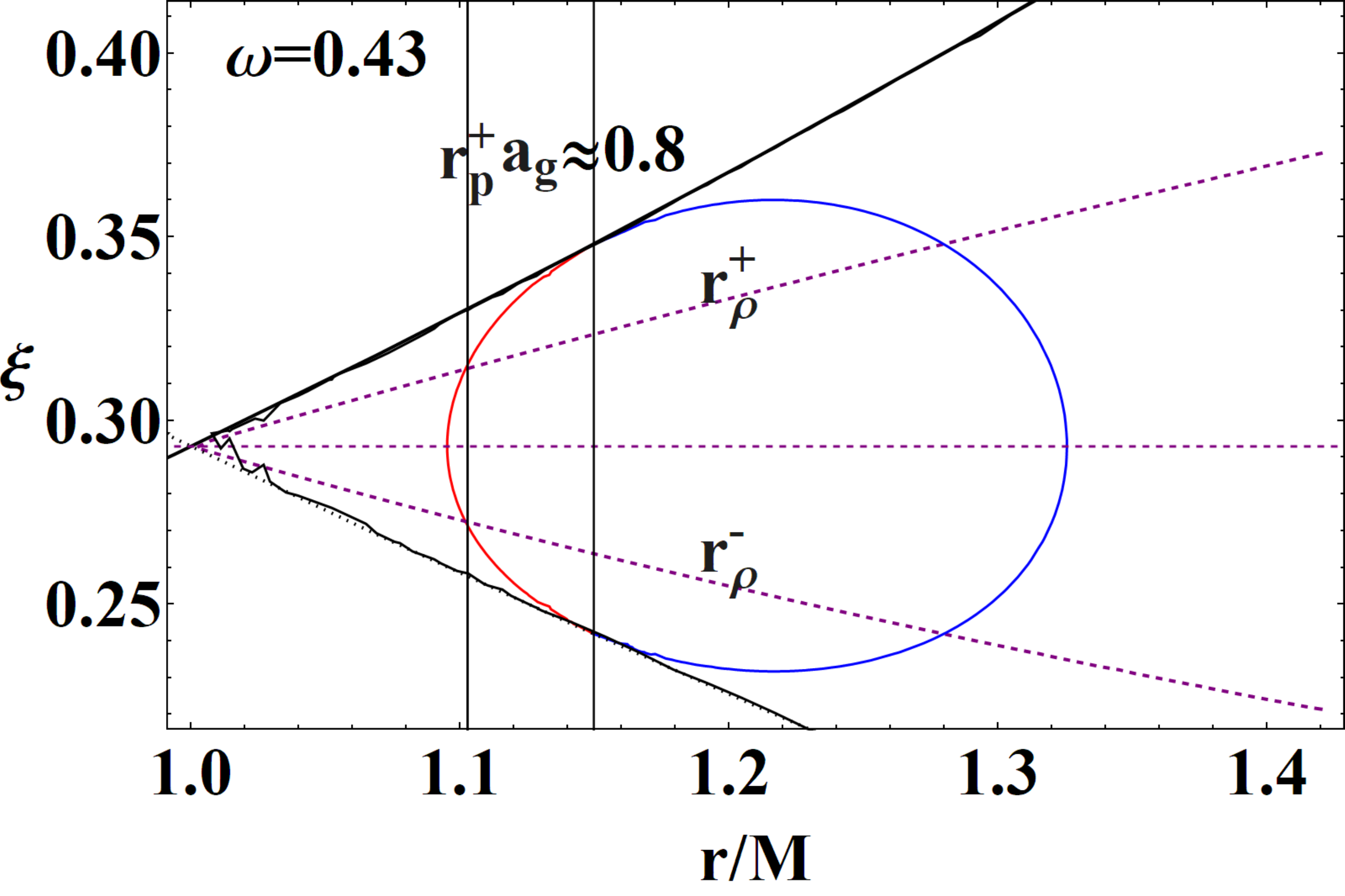}
    \includegraphics[width=5.4cm]{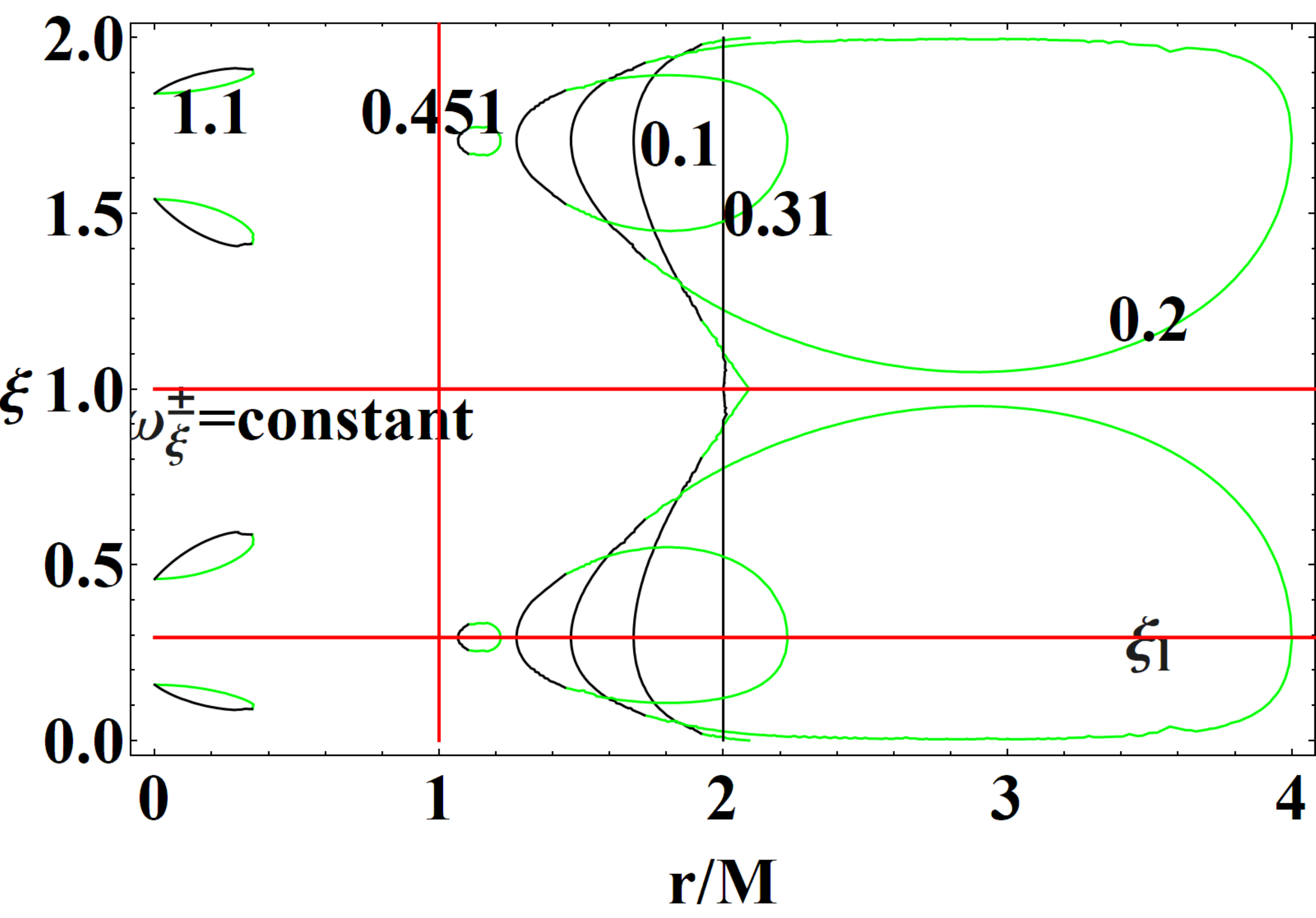}
  \caption{Upper panels: the orbital frequencies $\omega_\xi^{\pm}$  in Eqs\il(\ref{Eq:omegaxi})  of   null particles, limits of stationary observers as functions of the extracted energy $\xi$.
Left upper panel: $3D$ plots of $\omega_\xi^{\pm}$ a functions of $\xi$, rotational energy and the radius $r/M$, $M$ is the \textbf{BH} mass (as measured at infinity).
 Planes $\omega=0$ related to the Schwarzschild solution
and $\omega=1/2$ frequency of the extreme   Kerr \textbf{BH} horizon are represented. Note the presence of negative frequencies at $r>2M$, outer stationary limit and horizon of the limiting Schwarzschild spacetime.
Center panel: curves for different $\xi$.
The limiting case of the Schwarzschild  solution $a=0$ is shown.
The extreme Kerr \textbf{BH} case $a=M$ is also shown. (Note that curves do not represent  \textbf{NSs} solutions.) Curves provide the inner and outer Killing horizons.
Right upper  panel:  curves $\omega_{\xi}^{\pm}$  in a portion of the extended plane $\xi-r/M$
 Bottom left panel:
 bundle at fixed frequency. Black  region is the \textbf{BH} in the extended plane, spin
$a_g$  is the tangent point between the  bundle and the horizon, $r_p^{\pm}$ are points of replicas  the \textbf{NS} at spin
$a=2M$.
Center panel:bundle in the plane  $\xi-r/M$,
replicas $r_\rho^{\pm}$ for fixed frequency  and   spin $a_g$ and  the radius $r_p^+$  are shown.
Right panel: different  bundles in the plane $\xi-r/M $ at fixed frequency, the limit  case, $r=M$,  for the extreme \textbf{BH}  is also shown. See also Figs\il(\ref{Fig:Bundleplot1}).}\label{Fig:Combplot1}
\end{figure}

In the extended plane,  $\xi-r/M$,  the extracted rotational energy is  $\xi=1-(\sqrt{\sqrt{1-(2-r) r}+1})/({\sqrt{2}})$ see Figs\il(\ref{Fig:planeRTexteprop}).
Clearly, the extended plane  $\xi-r/M$  does not describe  bundles,  or portions of bundles $\ba_{\omega}$,   in the  \textbf{NSs} region similarly to the plane  $a/M-r/M$, particularly in the case fully  contained in \textbf{NS}  region which is defined by the  frequency $\omega=1/2$, corresponding to the Kerr extreme \textbf{BH} as tangent point bundle-horizon\footnote{
Exploring the role of the \textbf{NSs} as  considered in the extended plane, we note that the three solutions  $\xi(r)/M$ not  verifying the condition
$1-M_{irr}/M\in [0,\xi_{l}]$  could provide an indication of the portion of the bundles  contained in the \textbf{NSs} region. We note that considering the characteristic frequency $\omega=0.43<0.5$ of the  bundle on the equatorial plane tangent to  the curve of the outer horizon in the extended plane, thus, $\omega=0.43=\omega_H^+$, there is a replica in the spacetime at $a=2M$, not shown  however in the plane $\xi-r/M$--as clear from Figs\il(\ref{Fig:Combplot1}).
It is also evident that these bundle structures are very similar to those obtained for the Schwarzschild case.}.
Thus it cannot  describe energy extraction from a \textbf{NS} although we  use  a \textbf{NS} to construct the bundles on the \textbf{BH} region.

Function $\la(\xi)$ links the former state spin $a_0$ to the rotational energy extraction in the subsequent phase  where the \textbf{BH} is settled in a Schwarzschild \textbf{BH}.
Therefore analysis of a quantity  $\Qa(\xi)$ relates quantities $\Qa(0)$, before the transition, to the energy extraction measured by $\xi$--
 Figs\il(\ref{Fig:Combplot1})--\cite{Camezin,Punsly,Meier}.

\begin{figure}
  \includegraphics[width=7cm]{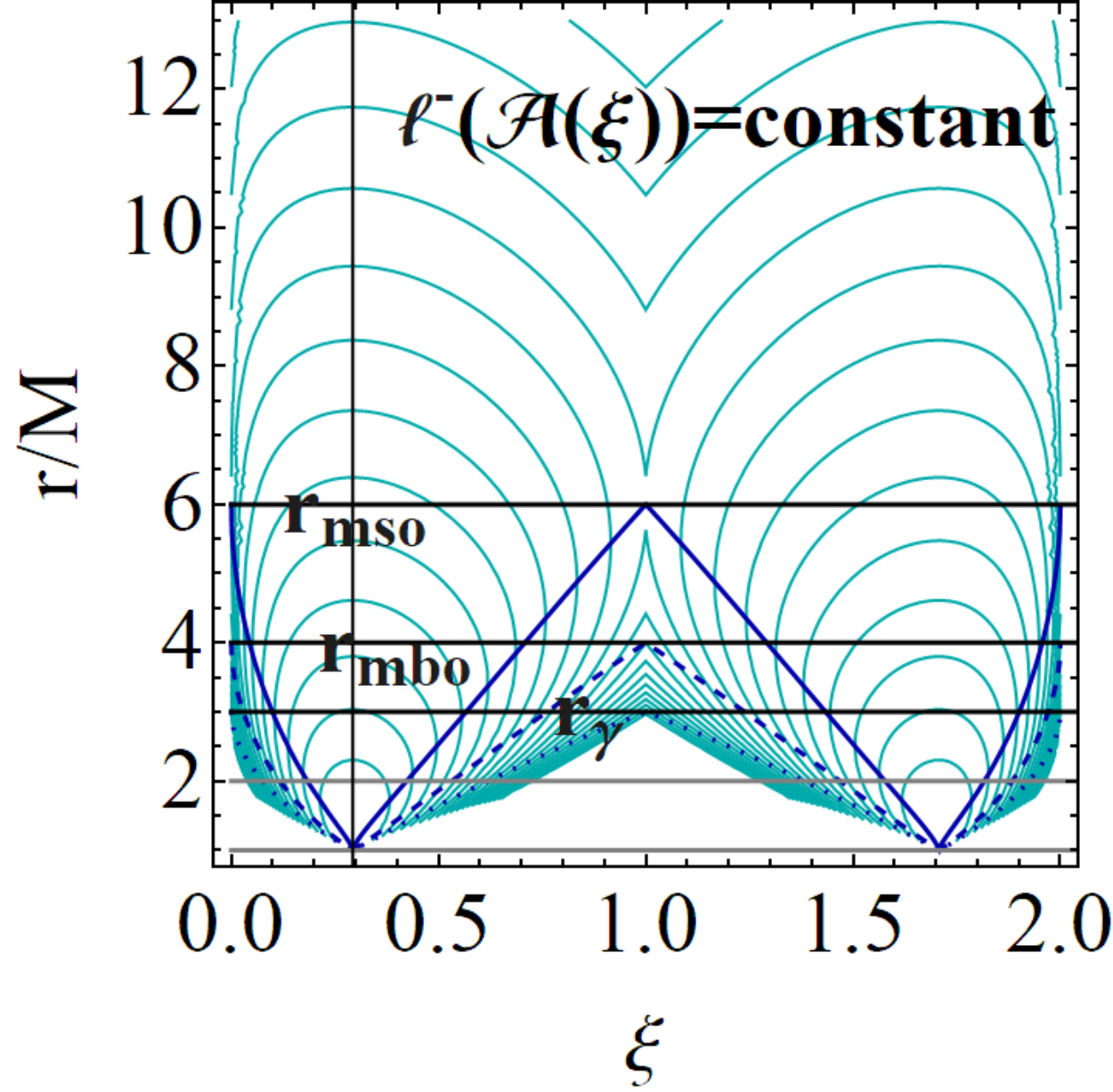}
    \includegraphics[width=7cm]{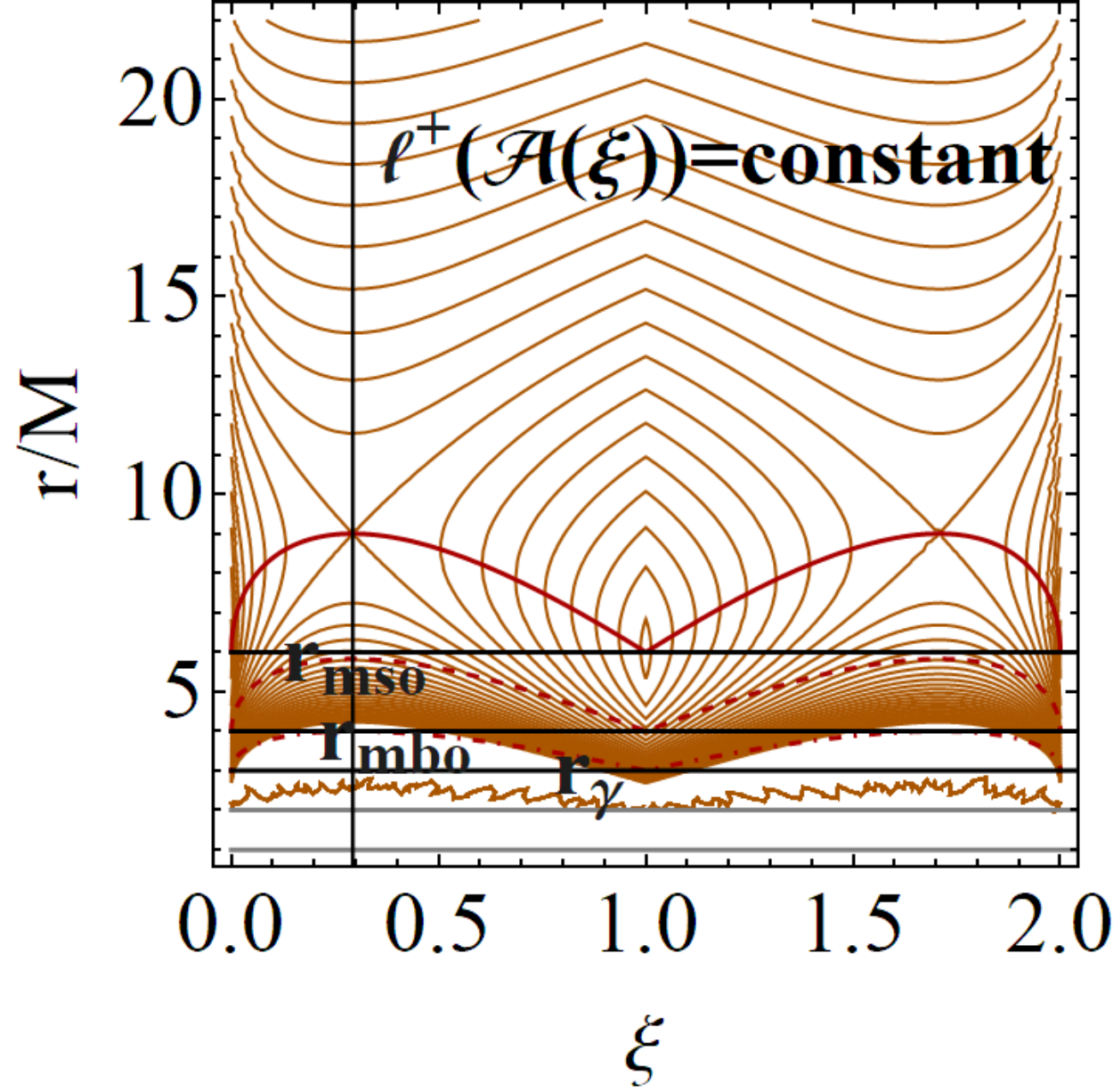}\\
     \includegraphics[width=8cm]{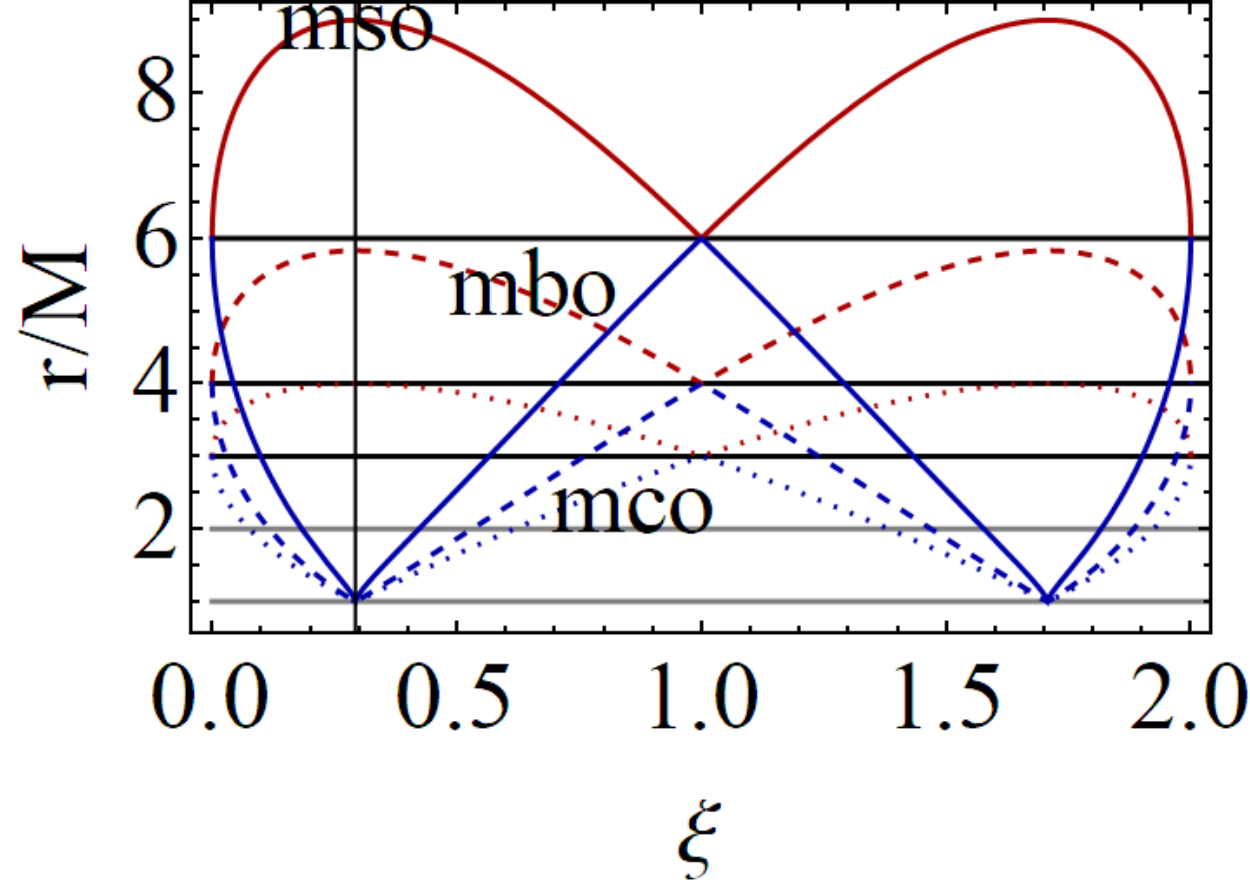}
      \includegraphics[width=8cm]{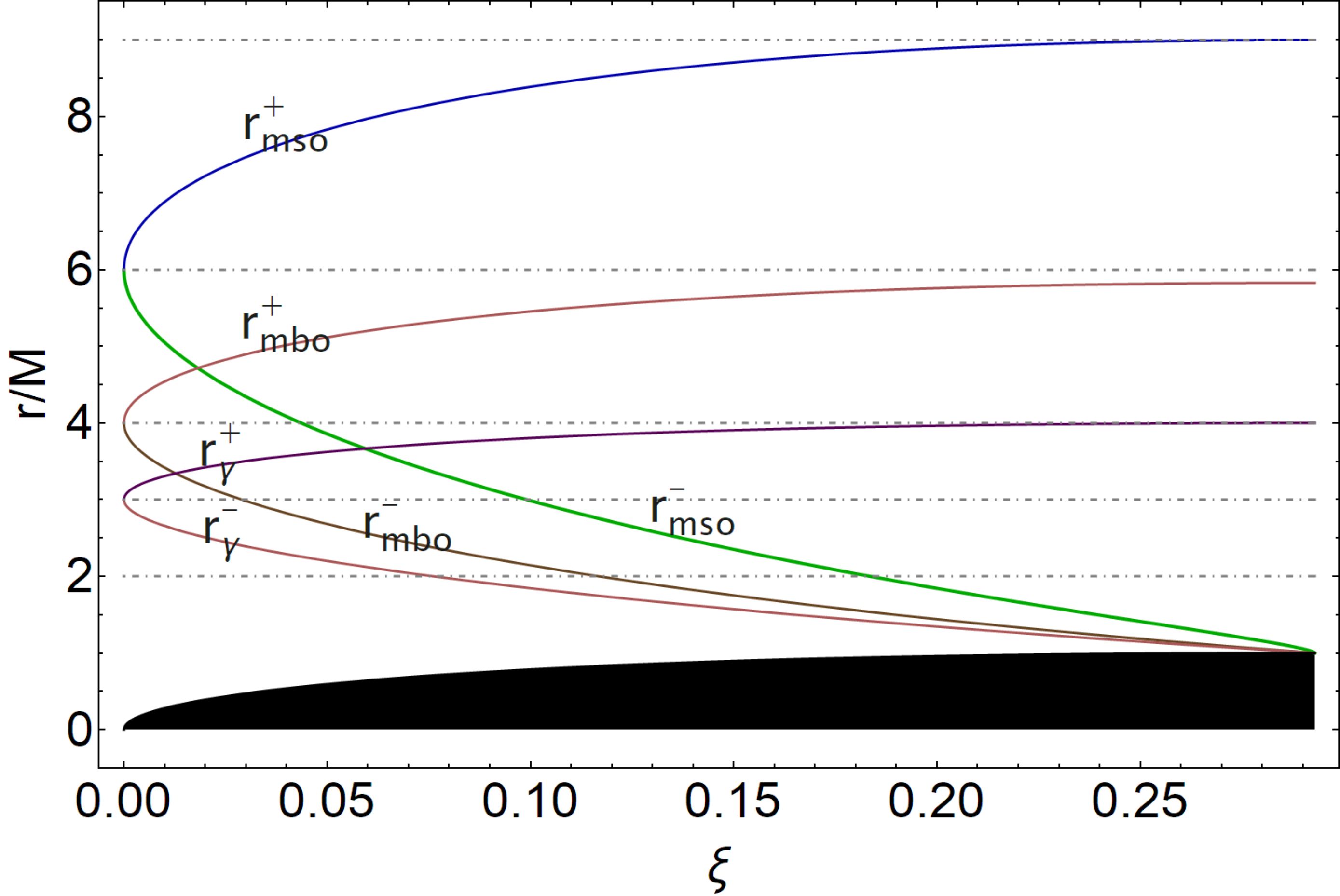}
  \caption{ Plane $r/M-\xi$, extracted rotational energy parameter  of  Eqs\il(\ref{Eq:exi-the-esse.xit}).
Upper line  panels and bottom right panel: Black lines are the marginal bounded orbits, $r_{mbo}$, marginal stable orbit $r_{mso}$ and $r_{mco}=r_{\gamma}$ marginal circular orbit and photon orbit for the  Schwarzschild   spacetime $(a=0)$.
Red curves are for counter-rotating fluids, blue curves for  corotating fluids.  Right bottom panel: Curves   $\ell(\xi;r)=$constant evaluated on $r_{mso}$ (plain curve),
$r_{mbo}$ (dashed curve), and $r_{mco}$ (dotted  curves). Therefore each curve represents a torus.
 Upper  panels represent curves $\ell^\mp (a_{\xi})$ in the plane $r/M$-$\xi$ for   corotating and counter-rotating   curves respectively. The maximum $\xi_\ell$ is the vertical black line. Bottom right panel shows  the geodesic structure as function of $\xi$, black region is the central \textbf{BH}.}\label{Fig:rightse-s}
\end{figure}

 In  the processes of   accretion and classical interaction  between \textbf{BHs} and the surrounding matter and fields,  the horizon area, $A_{\mathbf{BH}}$, can remain constant or  can increase.
Assuming that all the (corotating) fluid provides an initial contribution through the cusp
$r_{\times}$, from its specific angular momentum and a specific internal
energy,  the mass and angular momentum of the \textbf{BH}  grow by the corresponding  amount
given by $dM_{BH}=E_{\times}d M_0$,
and $dJ_{BH}=L_{\times} dM_0$ ($[J]=M^2$).
Next section focuses on the analysis of quantities  relative to the disks as functions of the extracted energy measured  by the observers.
In Figs\il(\ref{Fig:rightse-s}) we represent the geodesic structure $\{r_{mso},r_{\gamma},r_{mbo}\}$ as functions of
the rotational energy parameter, including the curves $\ell=$constant which define each torus of the \textbf{RAD} agglomeration  as functions of the extracted rotational energy in  the range $\xi\in [0,2]$ and  for corotating and counter-rotating fluids.
The ranges defined by values $\ell_{mso}$, $\ell_{mbo}$ and $\ell_{\gamma}$ define the topology of the critical configurations. In Figs\il(\ref{Fig:ManyoBoFpkmp1})   the curves $K(r)\equiv K_{crit}(r)$ are shown as functions of $\xi$ providing the values of the $K$ parameter of each toroidal configuration (at $\ell=$constant) at the critical points of pressure inside the configurations.  (Note that   the functions in  Figs\il(\ref{Fig:ManyoBoFpkmp1})    are not symmetric in the left and right range of the extreme value $\xi_\ell$)
Therefore, in this frame we solve equation   $\ell^\pm=\ell$  obtaining the curves:
\bea\label{Eq:chi-repo-try}
\tilde{\tilde{a}}_{\mp}^{\pm}\equiv \frac{1}{2} \left(\pm\sqrt{\ell^2-4 \ell (r-1) \sqrt{r}-4 (r-1) r}+\ell\mp2 \sqrt{r}\right),
\eea
representing the families of  tori-\textbf{BHs} systems. Note that solutions of $\ell^-=\ell$ or $\ell^+=\ell$ provide the solutions of Eq\il(\ref{Eq:chi-repo-try}), for $\ell^\pm=\ell$ where $\ell$ is positive or negative, if fluids are corotating or counter-rotating. (Alternatively solutions are given  in Sec.\il(\ref{Sec:mislao}).)
Considering the geodesic structure in the  extended plane, we can find classes of spin-radius
\bea\label{Eq:travseves}
a_{\gamma}^-\equiv-\frac{1}{2} (r-3) \sqrt{r},\quad
a_{\gamma}^+\equiv\frac{1}{2} (r-3) \sqrt{r},\quad \frac{\ell^{\pm}(a_{\pm})}{a_{\pm}}=-\frac{2}{r-2}
\eea
connecting  the geodesic structures of different  spins  of the Kerr \textbf{BHs}, where $a_{\gamma}^{\pm}$
refers to the photon orbits $r_{\gamma}^{\pm}$, and  $\ell^{\pm}(a_{\pm})$ refers to the horizons. Similar definitions can be used for  $a_{mso}^{\pm}$, showed in   Figs\il(\ref{Fig:rightse-s}).
Explicit solutions  in different parameterizations are in Sec.\il(\ref{Sec:mislao}).
Solving the problem   $\ell^\pm(\xi,r)=\ell$, we find  functions $\xi(r,\ell)$ connecting the tori, defined by
values $\xi=0$, and locations of maximum and minimum pressure points $(r)$, with the  rotational energy $\xi$, which now can be read in the extended plane.
Figs\il(\ref{Fig:ManyoBoFpkmpMig}) and Figs\il(\ref{Fig:quesa}) show the results of $\ell(\xi,r)=\ell$ for corotating and counter-rotating tori and   the location of  the inner edge of accretion tori  as function of the rotational energy $\xi$.
In Figs\il(\ref{Fig:sccreem}) we give the spread between the rotational law curves which demonstrates  the ranges of parameters  for the existence of tori, tori extension  on the equatorial plane and maximum possible  distance  between  tori in the \textbf{eRAD}.
  We relate in this way  centers and cusps (in general critical points) of family of tori for different  black holes, determined as solutions of the problem $\ell(r)=\ell(r_s)$ on the  {rotational curve} of the \textbf{eRAD}.
The distance  relates  the point of maximum and minimum pressure  inside each torus, enlightening the outstretching of the torus on its equatorial plane and the location of the inner edge (cusp).
The parameter
$\xi$ on the horizon at the extreme case $a =
 M $ where the frequency is $\omega= 1/2$ and  $ \xi=1\pm {1}/ {\sqrt {2}}$.
The characteristic frequencies of the bundles, seen as horizons frequencies in the extended plane, are
\bea\label{Eq:show-g}
\omega_g^ {\pm} \equiv \frac {2  \pm \sqrt {16 (\xi -
            2)\xi (\xi - 1)^2 +
        4}} {8\sqrt {-(\xi - 2) (\xi - 1)^2\xi}};
\eea
$\omega_g^ {-} $ has a saddle point for the extraction parameter $\xi=\left (3\pm \sqrt {6} \right)/3$ correspondent to the spin $ a =a_{mso}^{\epsilon}$ and frequency $\omega=({1}/{2\sqrt {2}} ,{1}/{\sqrt {2}}$)----see Figs\il(\ref{Fig:fas2}) and Sec.\il(\ref{Sec:barl}). (The spin  $a_{mso}^{\epsilon}$,  corresponds to a \textbf{BH} with $\xi= \frac{1}{3} \left(3-\sqrt{6}\right) \approx 0.183503$ and  \textbf{BH} with spin
$a=a_\gamma^{\epsilon}$ to $\xi \approx0.0761205$.)
\subsection{Energetics and accretion}\label{Sec:e-complex-asly}
Our analysis focuses on  a state $(0)$ prior the extraction process, considering  the  \textbf{RAD} tori     of polytropic fluids,
with pressure given by polytropic equation of state
 $p=\kappa \varrho^{1+1/n}$, here $\kappa$ is a polytropic constant, $\gamma\equiv(1+1/n)$ is the polytropic index. The limiting  case of  polytropic index  $\gamma=0$  would correspond to the case of zero pressure represented  by gravitating  dust  of test  particles.
 We    estimate   the
mass-flux, the  enthalpy-flux (which is related to the temperature parameter),
and  the flux thickness, which are important for the evaluation of the  energy release from \textbf{RADs}--see \cite{abrafra,Japan}. In details, definitions of these  quantities  are listed in Table\il(\ref{Table:Q-POs}).
\begin{table*}[ht!]
\caption{Quantities $\mathcal{O}$ and $\mathcal{P}$.  $\mathcal{L}_{\times}/\mathcal{L}$ stands for  the  fraction of energy produced inside the flow and not radiated through the surface but swallowed by central \textbf{BH}. %
Efficiency
$\eta\equiv \mathcal{L}/\dot{M}c^2$,    $\mathcal{L}$ representing the total luminosity, $\dot{M}$ the total accretion rate where, for a stationary flow, $\dot{M}=\dot{M}_{\times}$,  $W=\ln V_{eff}$ is the potential  of Eq.\il(\ref{Eq:scond-d}), $\Omega_K$ is the Keplerian (relativistic)  angular frequency,
$W_s\geq W_{\times}$ is the value of the equipotential surface, which is taken with respect to the asymptotic value, $ W_{\times}=\ln K_{\max}$  is the  function at the cusp (inner edge of accreting torus), $\mathcal{D}(n,\kappa), \mathcal{C}(n,\kappa), \mathcal{A}, \mathcal{B}$ are functions of the polytropic index and the polytropic constant.
}
\label{Table:Q-POs}
\centering
\begin{tabular}{|l|l|}
 \mbox{\textbf{Quantities}}$\quad  \mathcal{O}(r_\times,r_s,n)\equiv q(n,\kappa)(W_s-W_{\times})^{d(n)}$ &   $\mbox{\textbf{Quantities}}\quad  \mathcal{P}\equiv \frac{\mathcal{O}(r_{\times},r_s,n) r_{\times}}{\Omega_K(r_{\times})}$\\\hline\hline
$\mathrm{\mathbf{Enthalpy-flux}}=\mathcal{D}(n,\kappa) (W_s-W)^{n+3/2},$&  $\mathbf{torus-accretion-rate}\quad  \dot{m}= \frac{\dot{M}}{\dot{M}_{Edd}}$  \\
 $\mathrm{\mathbf{Mass-Flux}}= \mathcal{C}(n,\kappa) (W_s-W)^{n+1/2}$& $\textbf{Mass-accretion-rates }\quad
\dot{M}_{\times}=\mathcal{A}(n,\kappa) r_{\times} \frac{(W_s-W_{\times})^{n+1}}{\Omega_K(r_{\times})}$
 \\
 $\frac{\mathcal{L}_{\times}}{\mathcal{L}}= \frac{\mathcal{B}}{\mathcal{A}} \frac{W_s-W_{\times}}{\eta c^2}$&     $\textbf{Cusp-luminosity}\quad  \mathcal{L}_{\times}=\mathcal{B}(n,\kappa) r_{\times} \frac{(W_s-W_{\times})^{n+2}}{{\Omega_K(r_{\times})}}$
 \\
\hline\hline
\end{tabular}
\end{table*}
All these quantities   can be written in general form $\mathcal{O} (r_\times,r_s,n)=q(n,\kappa)(W_s-W_{\times})^{d(n)}$,  and $\mathcal{P}=\mathcal{O} (r_\times,r_s,n)/r_\times\Omega_K(r_\times)$,  considering that  $\Omega_K(r_\times)$ is the Keplerian frequency  of the accreting  tori cusp  $r_\times$ (the inner edge of accreting disk), where $\{q(n,\kappa),d(n)\}$ are    functions of the polytropic index  for each torus. Parameters  $(\kappa,n)$ within the constraints  $q(n,\kappa)=\bar{q}=$constant,  fix  a  polytropic-family, while   $r_s<r_\times$ is related to thickness
 of the accreting matter flow and the  potential
  $W=\ln V_{eff}$, thus
$W_s(W_{\times})$ denotes, for a torus with fixed specific angular momentum $\ell$,  the (constant) value of the  potential of  the $p=$constant surface corresponding to radius $r_s$ ($r_\times$).
We consider also  the quantity
$W-W(r_j)$,
     where $r_j\approx r_{\times}\approx r_{mbo}$ is a limiting  case  corresponding to a large  centrifugal component of the disk  tending to balance the gravity and pressure force components in the torus.
In this case the  inner edge of  the accreting tori  $r_{\times}^\pm\approx r_{mbo}^{\pm}$, there is   $W_\times\equiv W(r_{\times})=\ln K_\times\approx0$. In our analysis, for this first evaluation, we define the accretion point
$\widehat{r}_\times(a)$ %
and  {$\widehat{r}_s(a)$   as in Figs\il(\ref{Fig:ManyoBoFpkmpMig}) (there is therefore $1>K_s>K_{\times}$).
As the cusp approaches the limiting radius  $r_{mbo}$,  the potential  $W_\times\approx0$,  which is also the limiting asymptotic value for very large  $r$ as well as for the emergence of the proto-jets for $\ell\in \mathbf{L_2}$.
The couple of parameters  $\{r_s(a),r_{\times}(a)\}$
 has been fixed, to simplify the  comparison of the $\mathcal{O}^{\pm}$ and $\mathcal{P}^{\pm}$ quantities in the  corotating and counter-rotating tori,  and to characterize the dependence of these quantities on the \textbf{SMBH} spin-to mass ratio $a/M$.

 We examine the  fraction of energy produced inside the flow and not radiated through the surface but swallowed by central \textbf{SMBH}, the  efficiency
$\eta\equiv \mathcal{L}/\dot{M}c^2$,   together with   the total luminosity $\mathcal{L}$,  the total accretion rate, $\dot{M}$,   and  accretion for a {stationary flow}, $\dot{M}=\dot{M}_{\times}$.
 We  examine also $\mathcal{P}$-quantities--Figs\il(\ref{Fig:ManyoBoFpkmp1},\ref{Fig:ManyoBoFpkmp},\ref{Fig:ManyoBoFpkmpMig}) ,  for $\kappa\equiv n+1=4 (n=3)$, with the new variables independent from details of the selected specific polytrope,  $\Psi _*^{\pm }\equiv \mathcal{O} (r_\times,r_s,n)/q(n,{K})$  for $\mathcal{O} $-quantities  and
 $\mathrm{N}_*^{\pm}=\mathcal{O}(r_\times,r_s,n) r_\times/q(n,K)\Omega_K(r_\times)$  for $\mathcal{P}$-quantities.
Considering therefore the
rate of the  thermal-energy    carried at the  cusp and  the {disk accretion rate }      $\dot{m}= \dot{M}/\dot{M}_{Edd}$,
as well as the  mass flow rate through the cusp (i.e., mass loss accretion rate).
\begin{figure}
          \includegraphics[width=8cm]{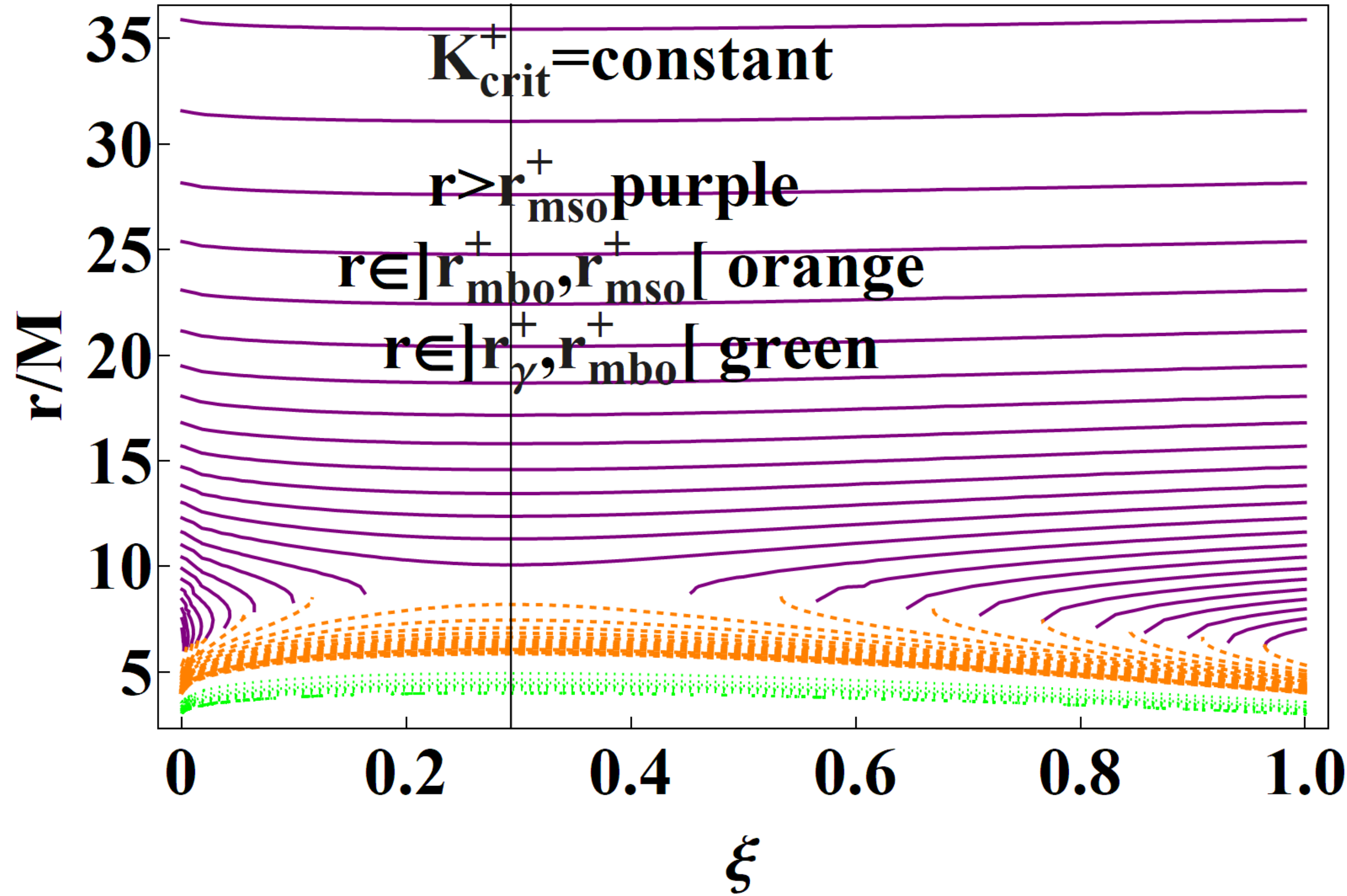}
            \includegraphics[width=8cm]{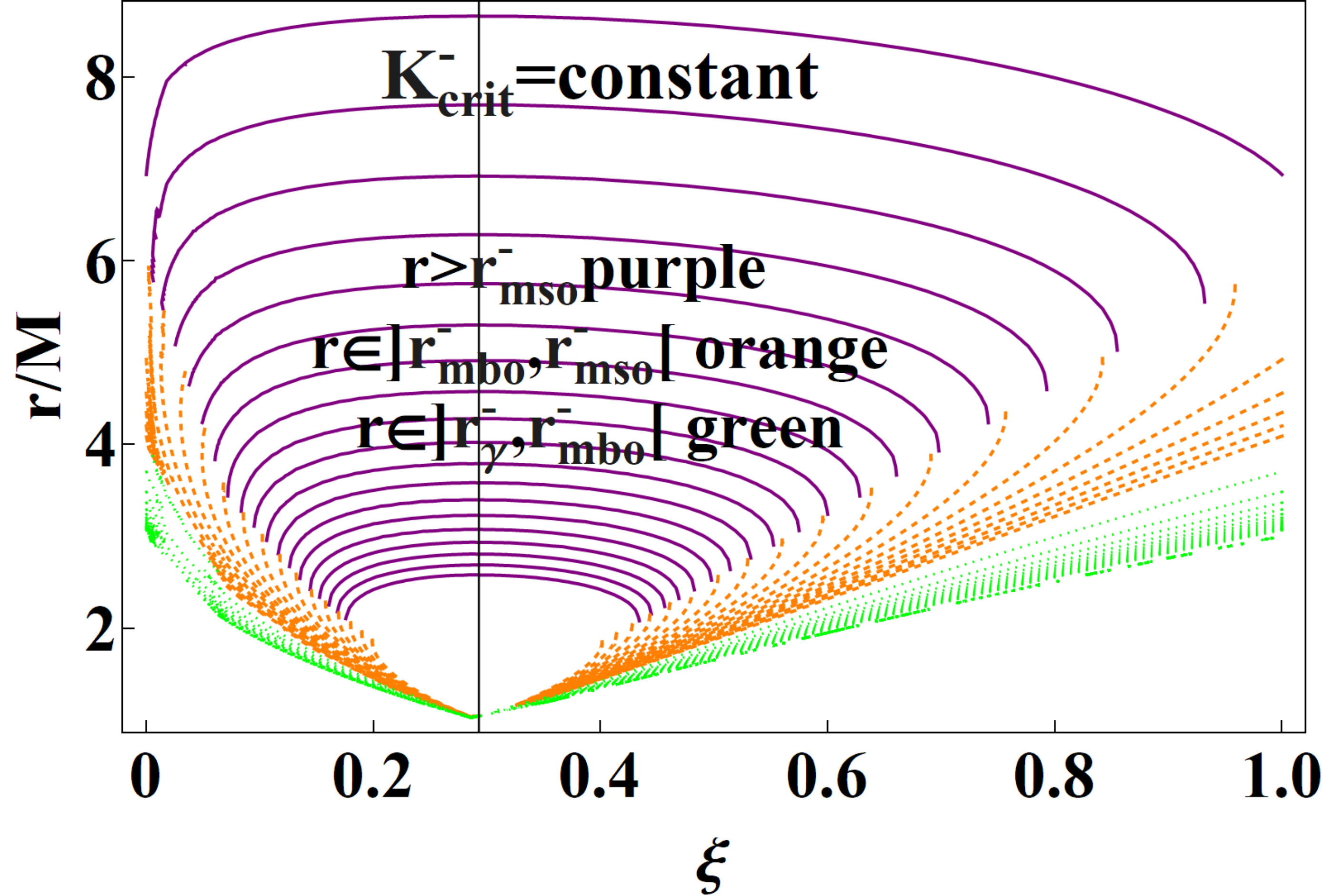}\\
                \includegraphics[width=8cm]{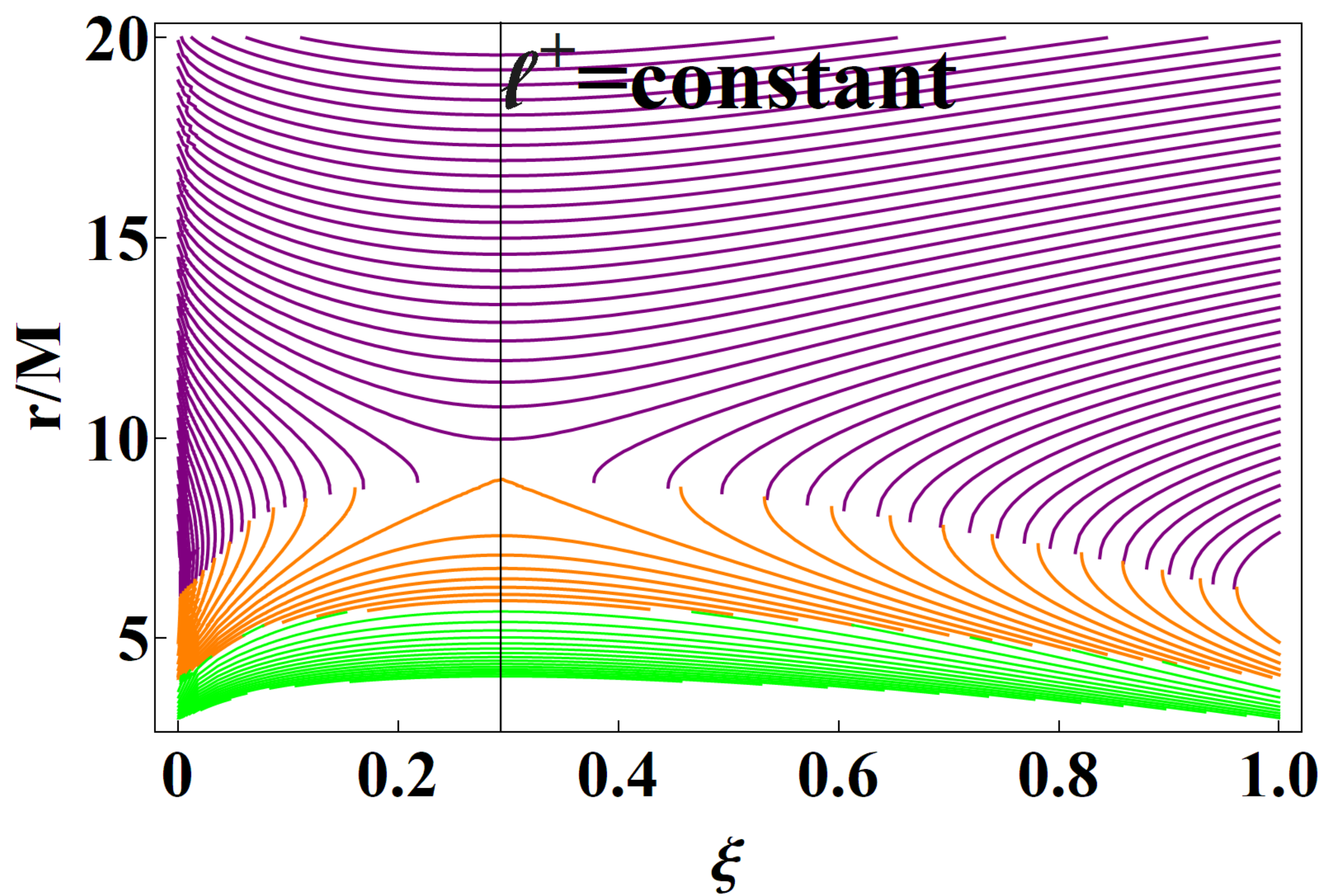}
  \includegraphics[width=8cm]{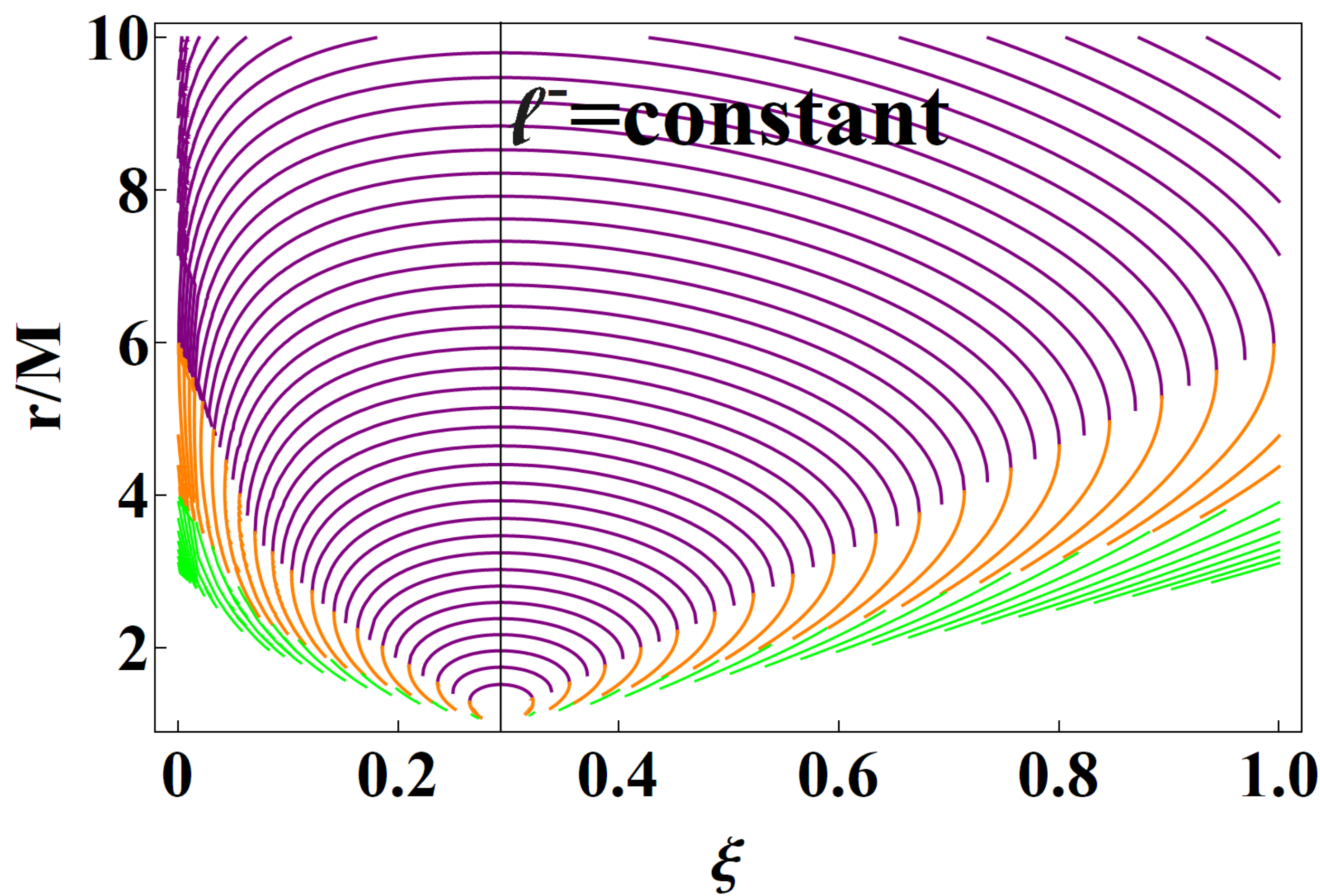}
  \caption{$K_{crit}^{\pm}$ is $K^{\pm}(r;a)\equiv V_{eff}(\ell(r),a)$  on the equatorial plane,  there is  spin  $a=\la({\xi})$, see Eq.\il(\ref{Eq:exi-the-esse.xit}), $\xi$ is the  rotational energy parameter.
$\ell(r)$ is the fluid specific angular momentum. Panels show $K^{\pm}(r;\xi)=$constant  and $\ell^{\pm}(\la_{\xi})=$constant in $(\xi,r/M)$ plane for corotating $(-)$ counter-rotating tori $(+)$ tori, $M$ is the central \textbf{BH} mass.  It is clear the role of $\xi_{\ell}$, maximum for $\la(\xi)$ (black vertical line).}\label{Fig:ManyoBoFpkmp1}
\end{figure}
\begin{figure}
\includegraphics[width=8cm]{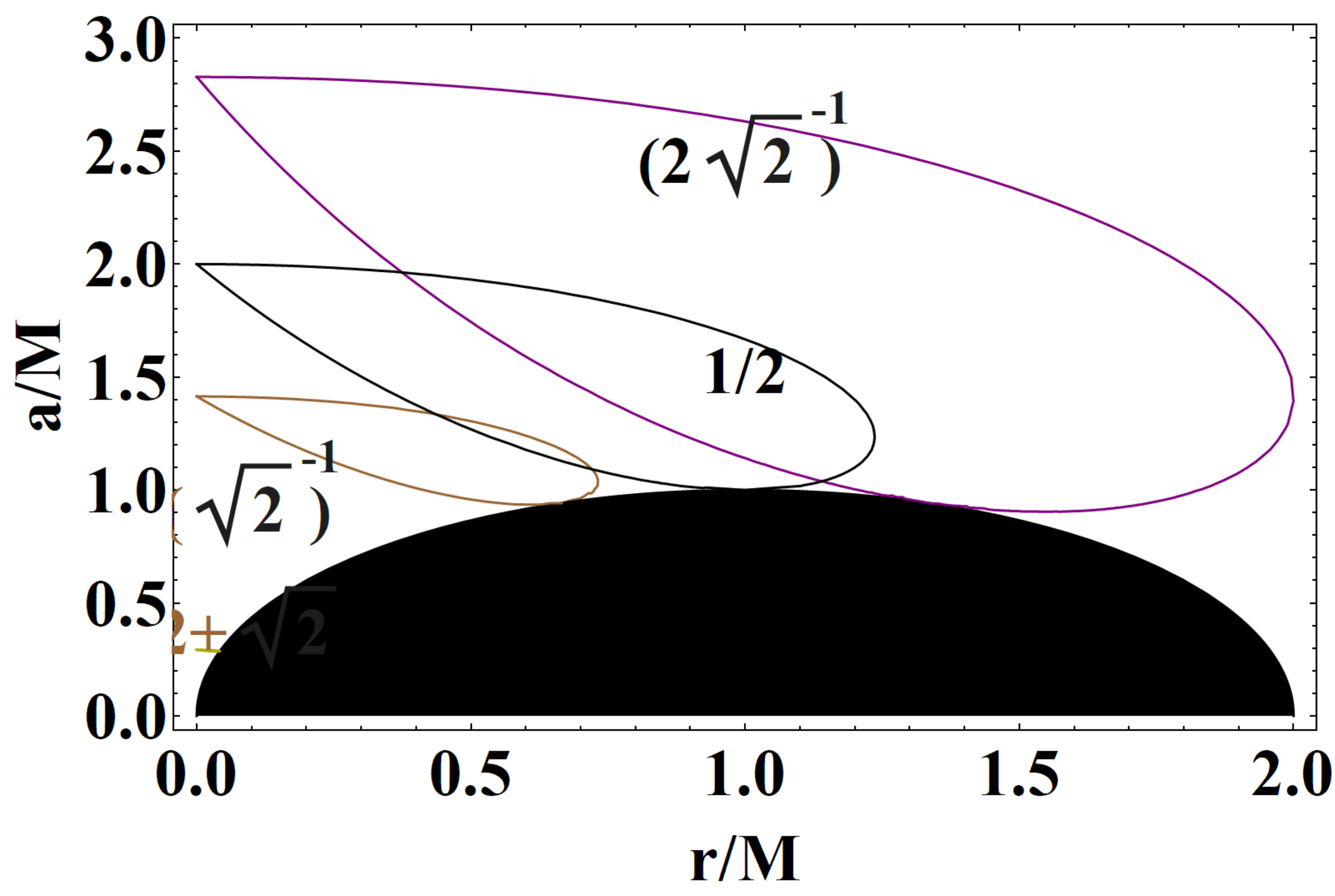}
\includegraphics[width=8cm]{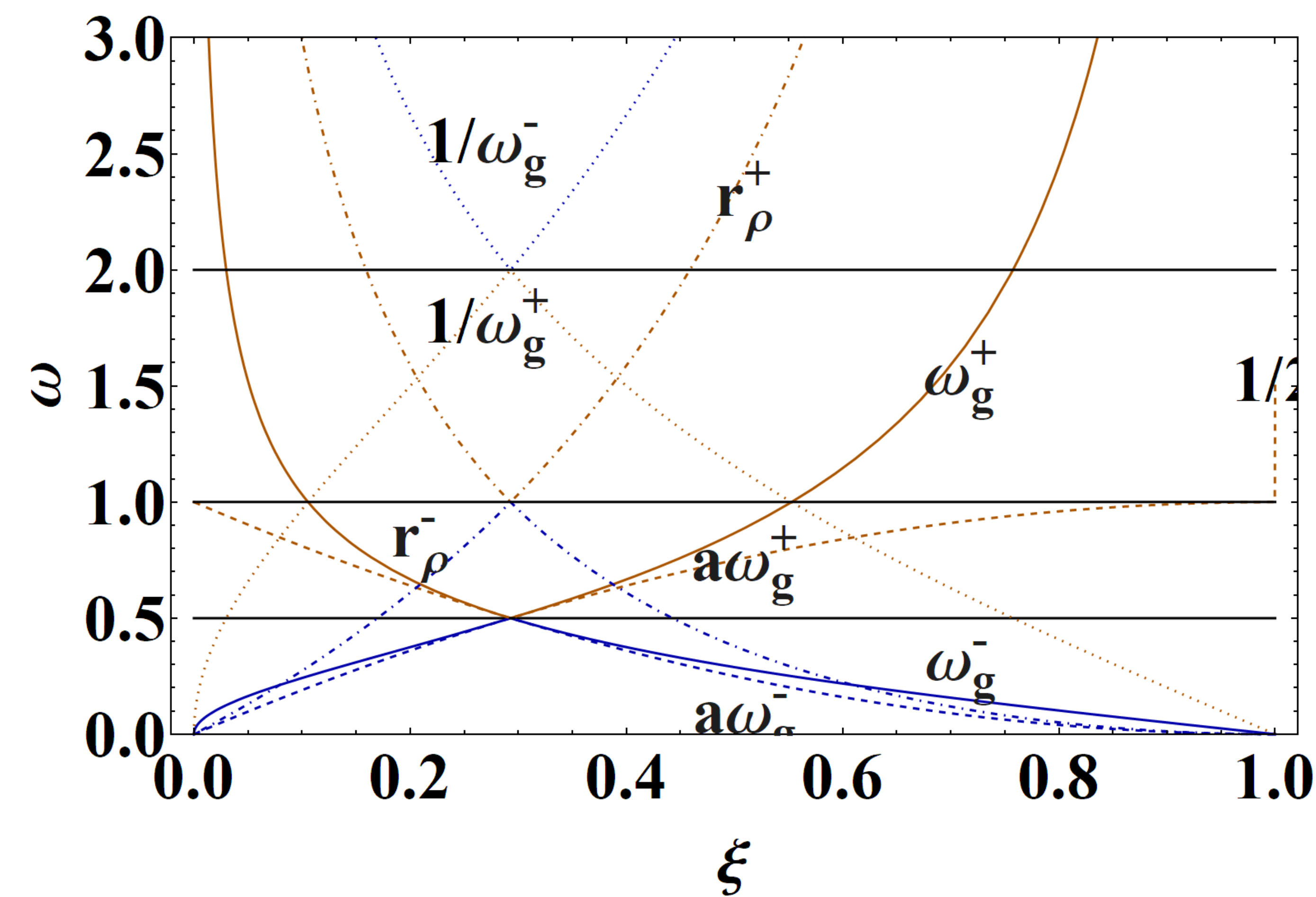}
  \caption{ Left panel: metric Killing  bundles  on the equatorial plane for selected values of the characteristic bundle frequencies $\omega$ signed on the panel, $\omega=1/2$ is the bundle  tangent to the extreme Kerr  spacetime. Black region is the \textbf{BH} in the extended plane, $r=2M$ is the  schwarzschild horizon and the outer engosurface. Brown and purple curves  are correspondent bundles  to one tangent  spin $a_g/M= a_{mso}^{\epsilon}/M={2\sqrt {2}}/{3}$ see discussion in Sec.\il(\ref{Sec:both-s}) and Sec.\il(\ref{Sec:barl}).
Right panel: $\omega$ is the bundle characteristic frequency and light particle  orbital frequency, $\xi$ is the  parameter for the maximum rotational energy  extracted. $r_{\rho}^{\pm}$ are horizon replicas on the equatorial plane see Eq.\il(\ref{Eq:replicas2}).
$\omega_g^{\pm}$ are defined in  Eq.\il(\ref{Eq:show-g}) and are the  horizons frequencies.
Quantities  $a\omega_g^{\pm}$ refer to the limiting condition on the energy process
and $\ell=L/E=\/\omega_H^{\pm}$ it refers to  quantity  $\bar {\ell} _{\max}$ from analysis of Sec.\il(\ref{Sec:barl}).
$\xi_\ell$ is the maximum energy extracted.
}\label{Fig:fas2}
\end{figure}
\begin{figure*}
        \includegraphics[width=8cm]{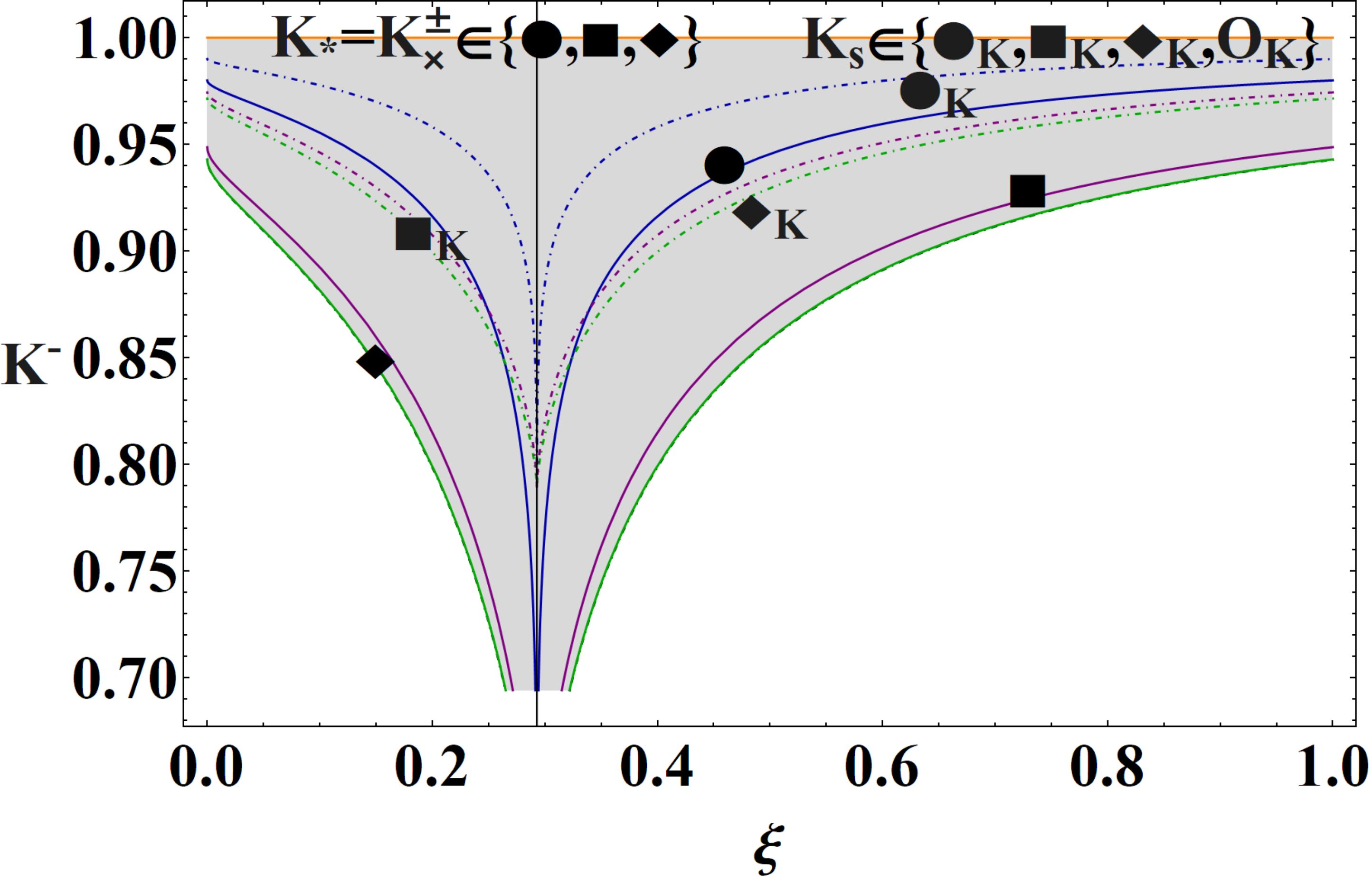}
           \includegraphics[width=8cm]{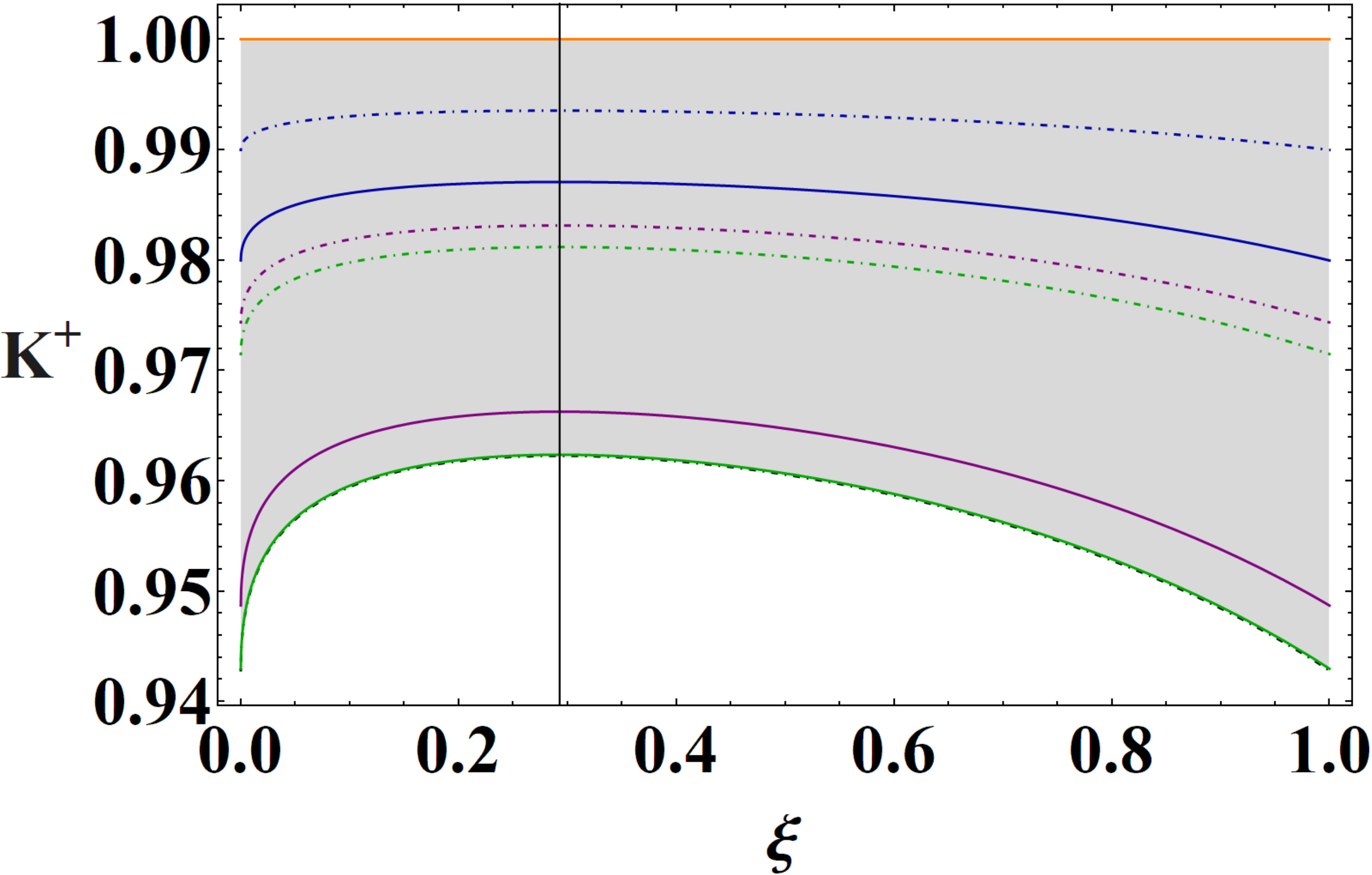}
                  \includegraphics[width=8cm]{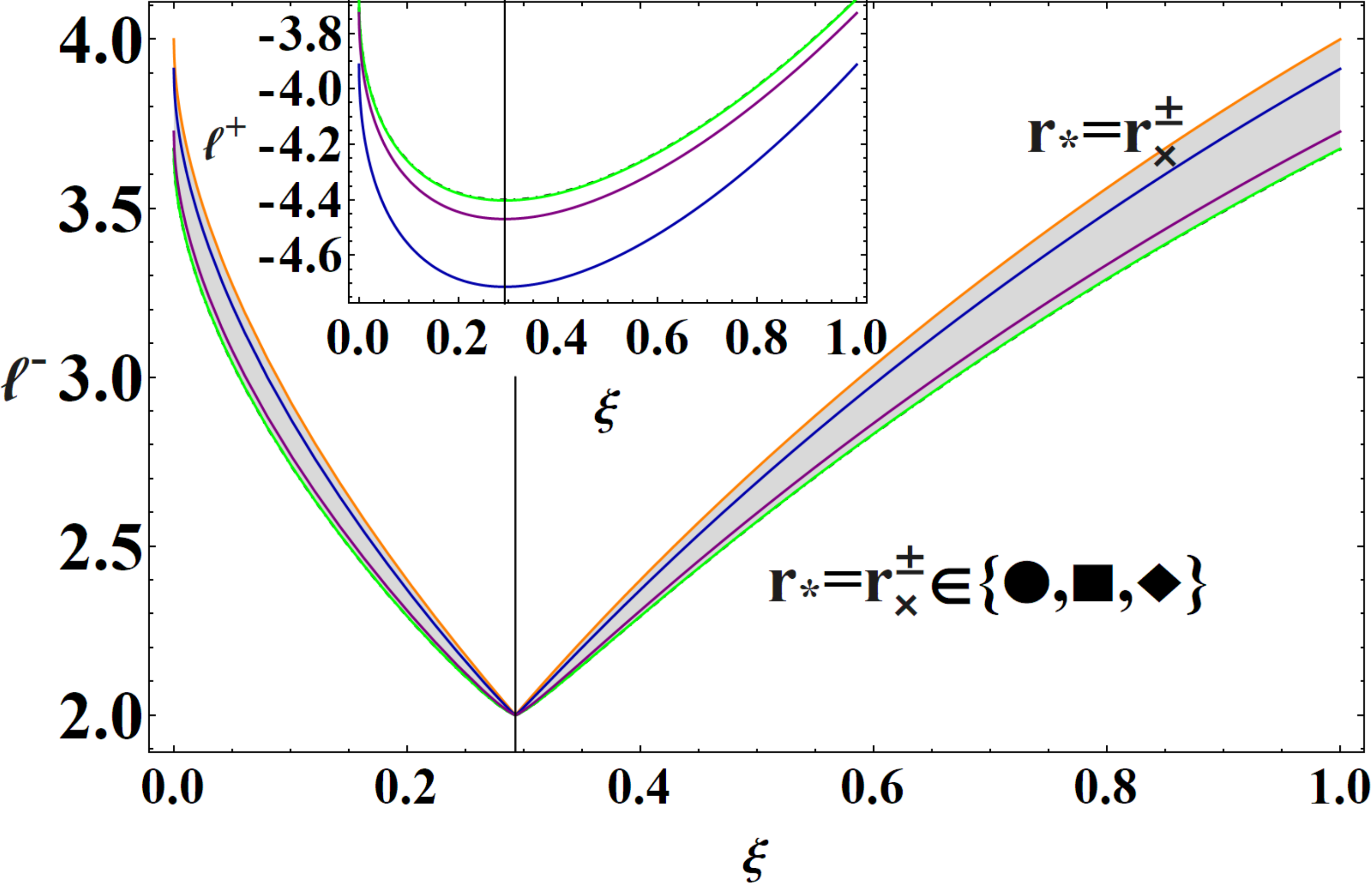}
          \includegraphics[width=8cm]{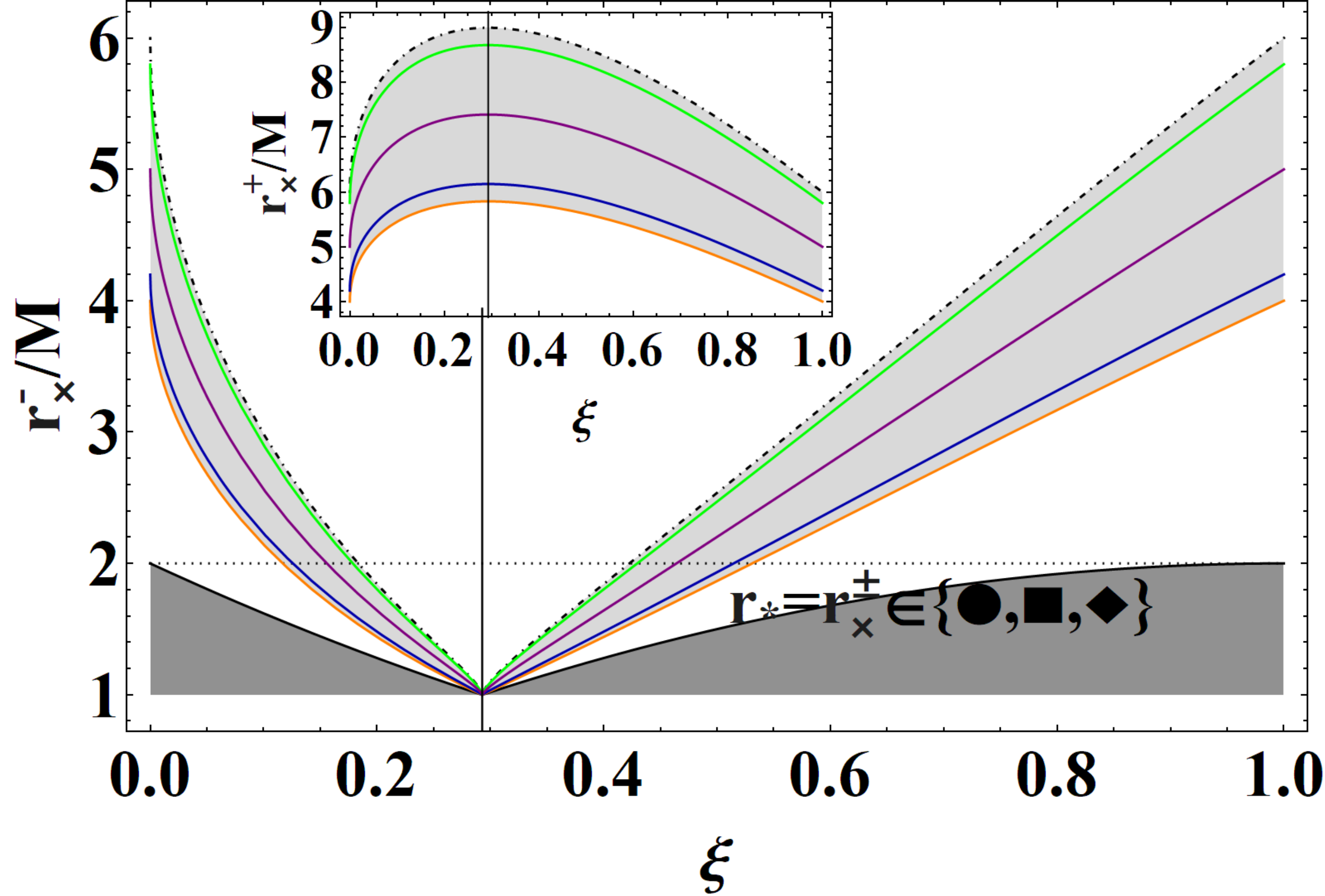}
  \caption{Analysis of  the energetics  of  \textbf{BH}-accretion disks  systems in Sec.\il(\ref{Sec:e-complex-asly}).  Constant specific angular mometum $\ell$ (left bottom panel)  and constant function $K(r)$ (upper  panels) and location of the inner edge of accreting tori (right bottom panel)  for counter-rotating fluids $(+)$,  and  corotating fluids $(-)$, $\xi$ is the extracted rotational energy parameter.
It refers to analysis in Figs\il(\ref{Fig:ManyoBoFpkmpMig}).
 Radii $(r_*,r_s)$ and  the associated    angular momentum  $\ell$ and $K$ parameters  are shown with $\{\bullet,\blacksquare,\blacklozenge,\bullet_K,\blacksquare_K,\blacklozenge_K,\mathrm{O_K}\}$.  $\Omega_K$ is  the Keplerian angular velocity, $r_{\times}$ is the accreiting tori cusp (inner edge of accreting torus), $r_s$ is related to thickness of the accreting  matter flow.  $r_{mbo}$ is the marginally bounded orbit. $\Omega_K^{+}$  has been considered for the counter-rotating fluids. The maximum location of inner edge is  $r_{\times}\lessapprox r_{mso}$. Vertical black line is the maximum value of the rotational energy parameter $\xi_\ell$, corresponding to the extreme Kerr \textbf{BH}.}\label{Fig:ManyoBoFpkmp}
\end{figure*}
\begin{figure*}
  \includegraphics[width=7cm]{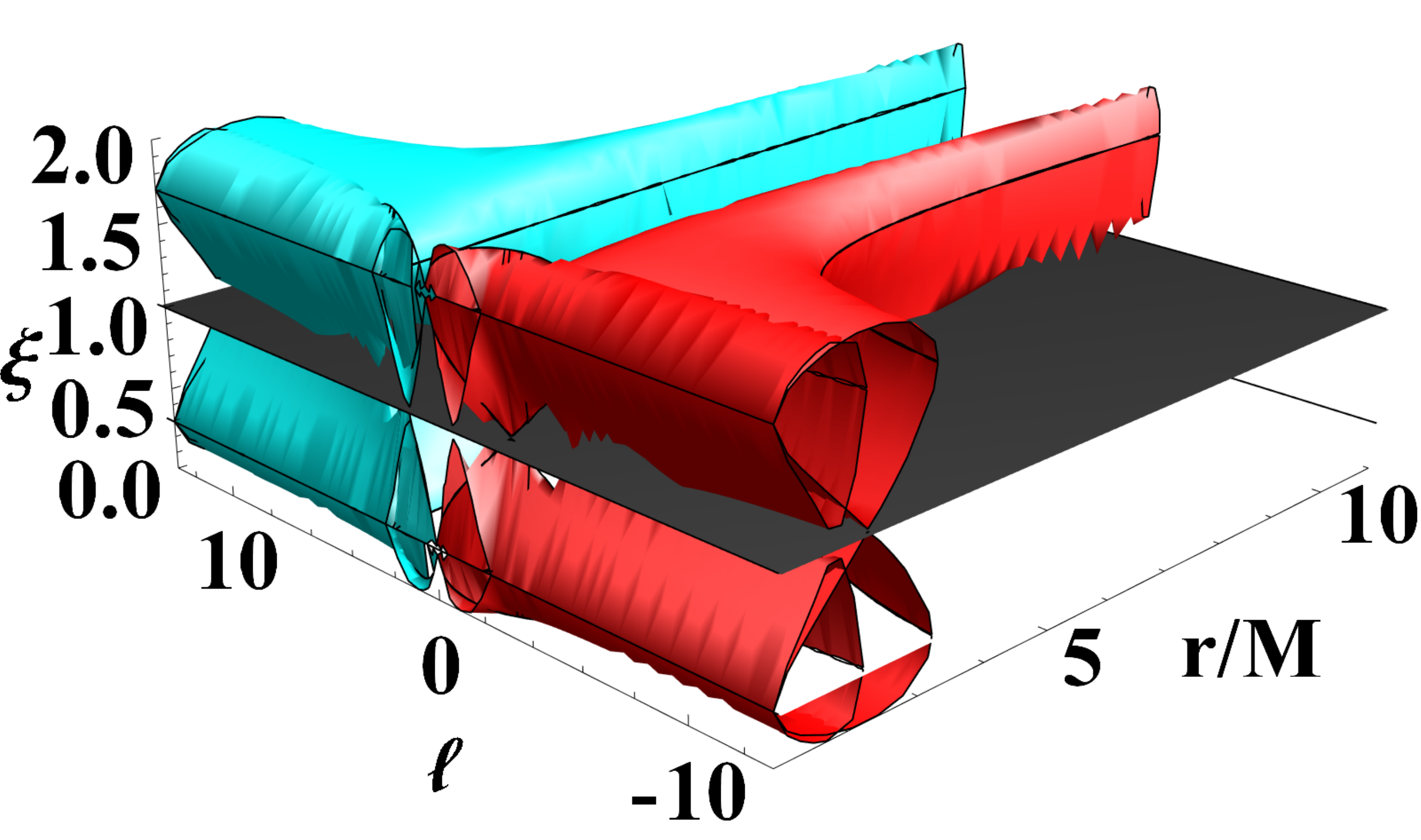}
            \includegraphics[width=7cm]{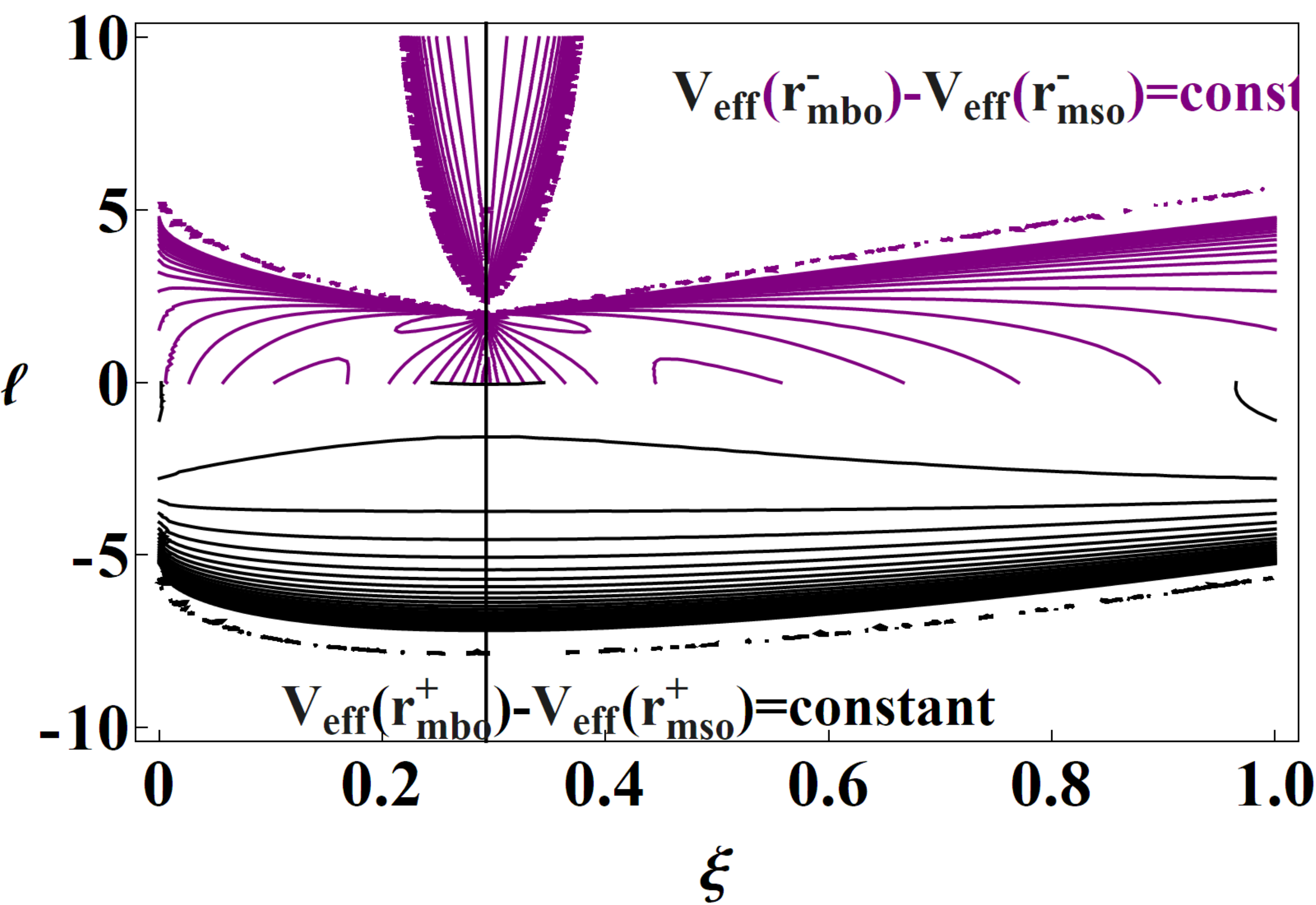}\\
                 \includegraphics[width=8cm]{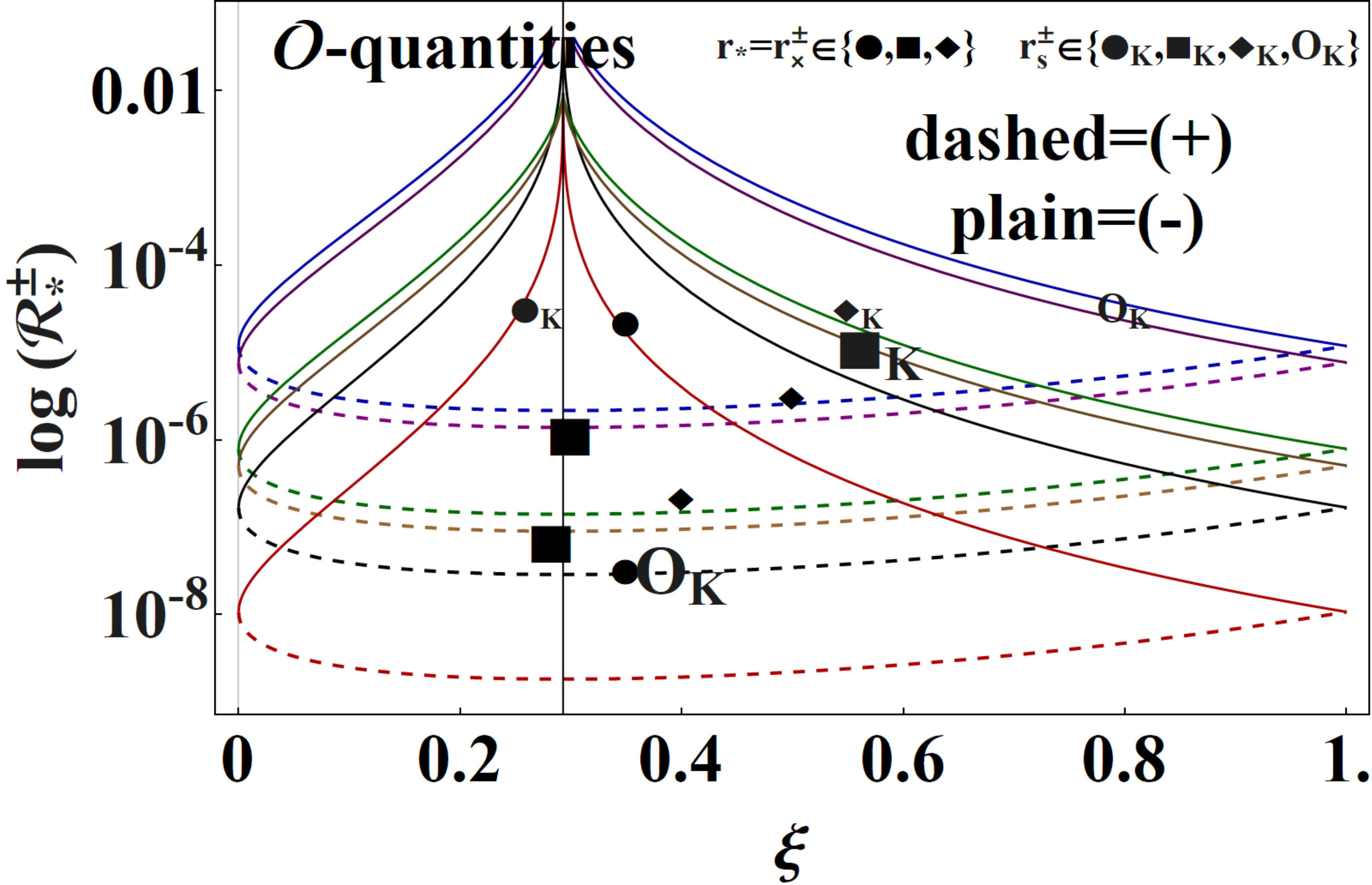}
      \includegraphics[width=8cm]{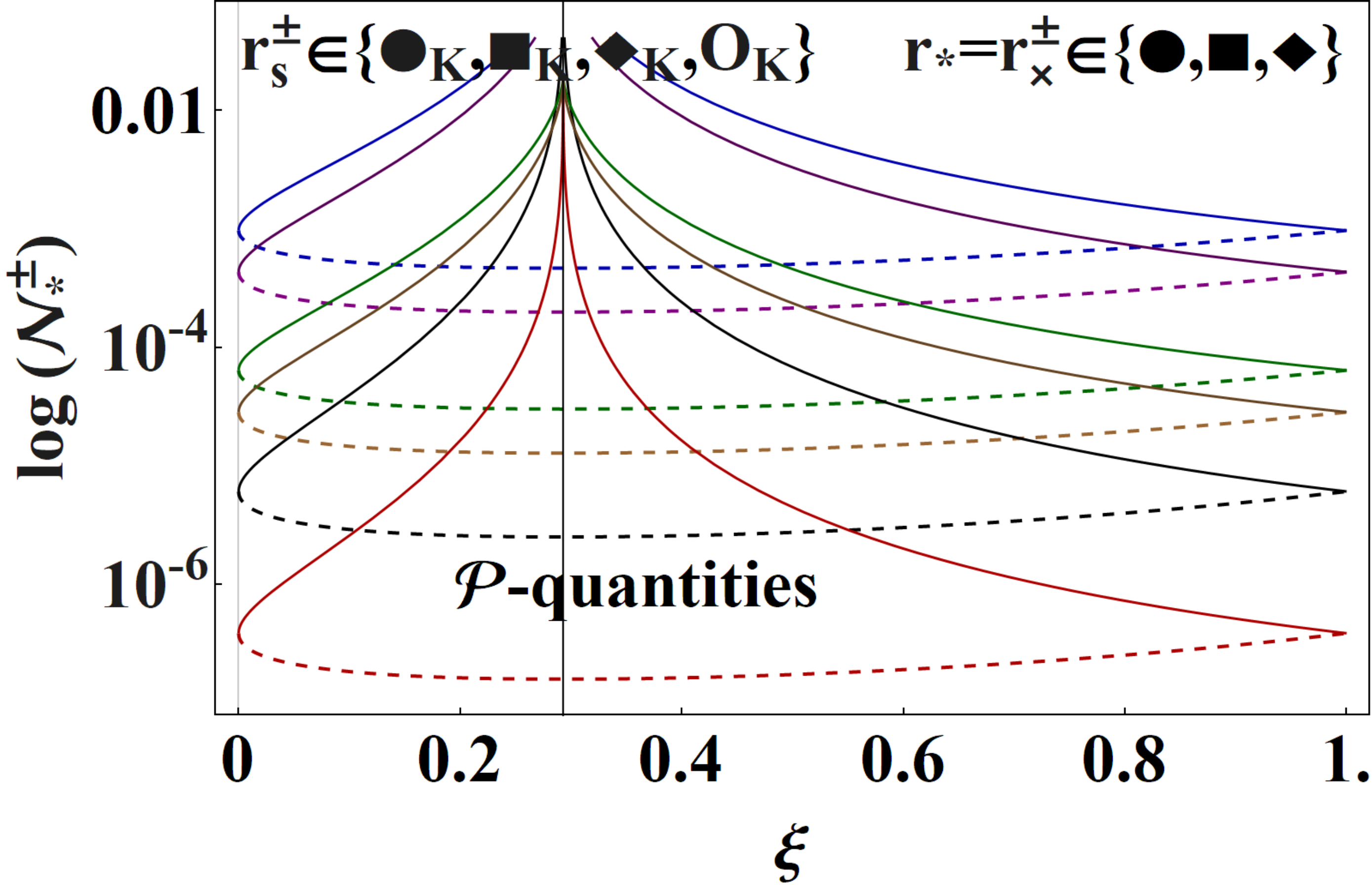}
  \caption{ Analysis of  the energetics  of  \textbf{BH}-accretion disks  systems in Sec.\il(\ref{Sec:e-complex-asly}).  Upper left
  panel: Solutions for the  extracted rotational energy parameter $\xi$: $\ell^{\pm}(r,\xi)=\ell$  for corotating $(-)$ (cyan) and counter-rotating $(+)$ (red) tori respectively, of Sec\il(\ref{Sec:mislao}) at different angle view, where  the \textbf{BH} spin is $a=\la({\xi})$ of Eq.\il(\ref{Eq:exi-the-esse.xit}), function of $\xi$.    Functions $\xi({\ell})$ are represented for corotating and counter-rotating fluids as functions of $(\ell,r/M$), $\ell$ is a constant value of  the  distributional law of the \textbf{eRAD} and specific angular momentum of the torus. Exact expression of solutions $\ell^{\pm}(r,\xi)=\ell$ are also in Sec.\il(\ref{Sec:mislao}). Upper right panel: $(V_{eff}(r_{mbo}^{\pm})-(V_{eff}(r_{mso}^{\pm})=constant$ respectively where  $a=\la({\xi})$  in the plane $(\xi,\ell)$, $\xi$ is the extracted rotational energy parameter and $\ell$ is the fluid specific angular momentum, $(+)$ is for counter-rotating fluids, $(-)$ is for corotating fluids.
Right  and  left bottom panel: evaluation of $\mathcal{P}$- and $\mathcal{O}$-quantities of Table\il(\ref{Table:Q-POs})  for corotating and counter-rotating tori, versus  $\xi$.     Plots of $\mathcal{N}_{*}^{\pm}\equiv{r_*} (W^{\pm}(r_{s})-W^{\pm}_{*})^\kappa(\Omega_K(r^{\pm}_*))^{-1} $ for $\mathcal{P}$-quantities analysis and  $\mathcal{R}_{*}^{\pm}\equiv(W^{\pm}(r_{s})-W^{\pm}_{*})^\kappa$ for $\mathcal{O}$-quantities analysis  for corotating ([\textbf{-}]--continuum curves) and counterrotating  ([\textbf{+}]--dashed curves)   tori at  different $r_*=r_{\times}^{\pm}\in\{\bullet,\blacksquare,\blacklozenge\}$ and $r_{s}\in\{\bullet_K,\blacksquare_K,\blacklozenge_K,\mathrm{O_K}\} $ where  $\varpi=n+1$, with  $\gamma=1/n+1$ is the polytropic index. %
 Radii $(r_*,r_s)$ and  the associated    angular momentum  $\ell$ and $K$ parameters  are shown with $\{\bullet,\blacksquare,\blacklozenge,\bullet_K,\blacksquare_K,\blacklozenge_K,\mathrm{O_K}\}$.  $\Omega_K$ is  the Keplerian angular velocity, $r_{\times}$ is the accreting tori cusp (inner edge of accreting torus), $r_s$ is related to thickness of the accreting  matter flow.  $r_{mbo}$ is the marginally bounded orbit. $\Omega_K^{+}$  has been considered for the counter-rotating fluids. The maximum location of inner edge is  $r_{\times}\lessapprox r_{mso}$--see Figs\il(\ref{Fig:ManyoBoFpkmp}). Vertical black line is the maximum value of the rotational energy parameter $\xi_\ell$, corresponding to the extreme Kerr \textbf{BH}. }\label{Fig:ManyoBoFpkmpMig}
\end{figure*}
The \textbf{eRAD}  is in fact a geometrically thin disk with a complex inner ringed structure composed by both corotating and counter-rotating  geometrically thick accretion disks having many features common with the thick and opaque disks as the expected super Eddington luminosity.
This evaluation,
 leading to the  results shown in Figs\il(\ref{Fig:ManyoBoFpkmp}), connects  diverse states of the  \textbf{BH} and its \textbf{RAD}  system,   more specifically, different initial states $(0)$,  prior  the rotational energy extraction. The procedure  is based on the   geometric considerations derived from the geometrically thick  torus model considered here:  each  \textbf{RAD}  component  is  pressure supported  and subjected to conditions of  the  von Zeipel  theorem, ensuring the integrability of the Euler equation and assumptions on the boundary conditions (which are assumptions on the pressure at  the center, point of the maximum pressure, and torus edge, equipressure surface).
We limited the description   of the  situation of \textbf{BH}, neglecting  the contribution of  mass feeding processes, leading likely  to a multi-stage evaluation, as an interactive process may be engaged   between geometry and disk similar to the  runaway instability.
 For corotating  and counter-rotating fluid characteristics,  fluid momenta and location of cusps have  different role  for energy extraction parameter $\xi$.  For corotating tori  the  variation of the location of the minimum points of the pressure, the instability points of the toroidal configurations, changes  minimally with  respect to counter-rotating tori with the parameter $\xi$.
  A  relevant  aspect concerning accretion is whether the presence of double accretion phase due to the doubled inner ringed structure of the \textbf{RAD}, enhances \textbf{BH}  accretion, considering also the possibility  of interrupted phase of accretion due to the inner screening torus of the couple.  There is  the maximum of two accreting tori  in the \textbf{eRAD}, with the outer torus being  counter-rotating, and inner torus, being  corotating. This scenario is then enriched by  the possibility of inter disk emission jets and obscuring tori, which are  typical aspects of the \textbf{RAD} presence around the central \textbf{SMBH}, and shells of jets.
The processes related to counter-rotating fluids may however be drastically different from the corotating case  especially   due to the influence of the  ergosurface.
There are indeed  corotating solutions in the ergoregion  for these tori  stretching with the inner quiescent or cusped tori down  to very close vicinity   of  the horizons for large \textbf{BH} spins\cite{next}. This is essentially due to the fact that the fluid in the tori could be considered stationary in the sense  that the fluid four velocity is $u=\xi_t+\omega\xi_\phi$, where $\omega$ is the relativistic angular momentum. On the other hand, a cusped (corotating) torus can stretch towards the horizon enhancing  the so called runaway instability-- see Figs\il(\ref{Fig:travseves}).
\begin{figure}
 \includegraphics[width=8.cm]{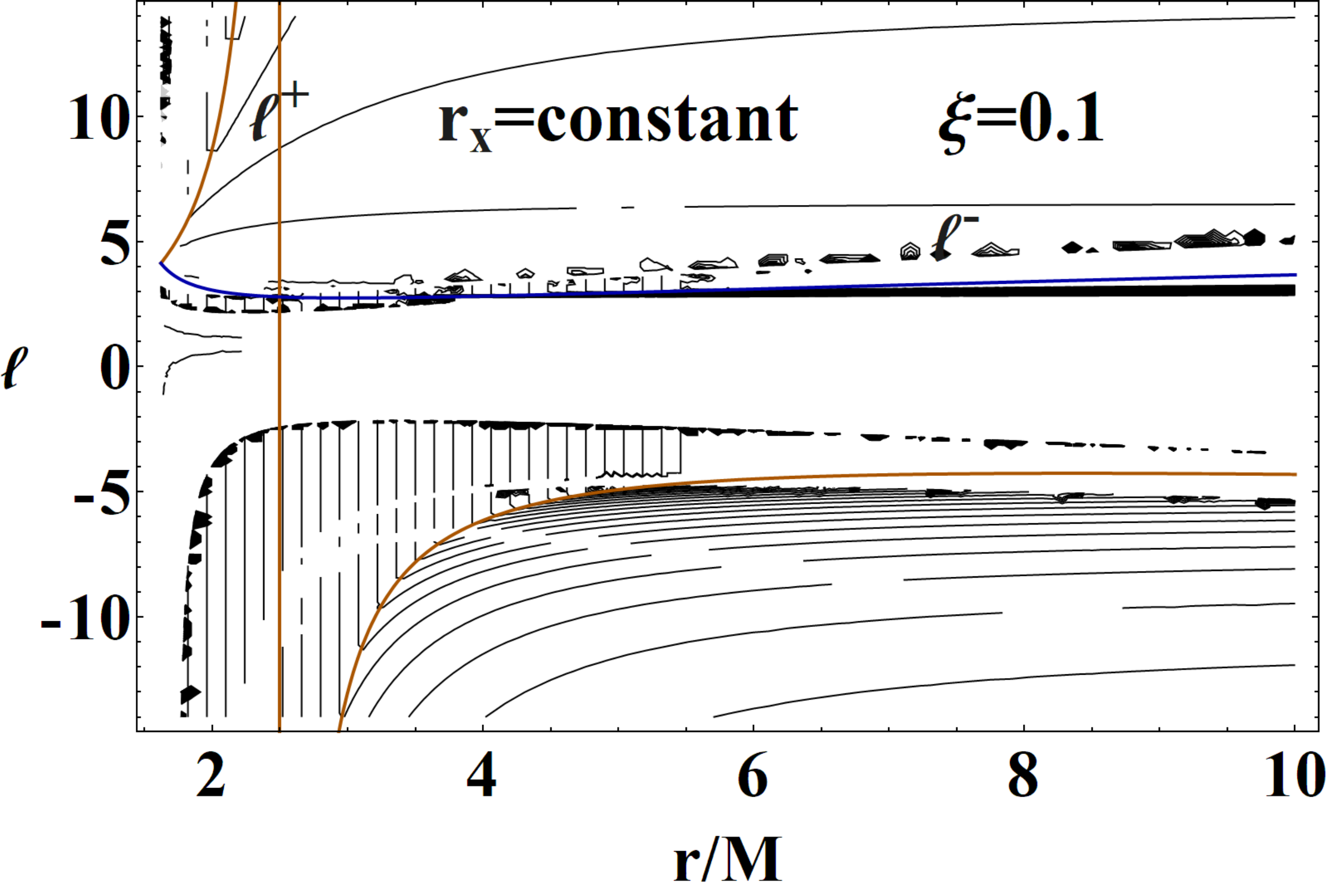}
 \includegraphics[width=8.cm]{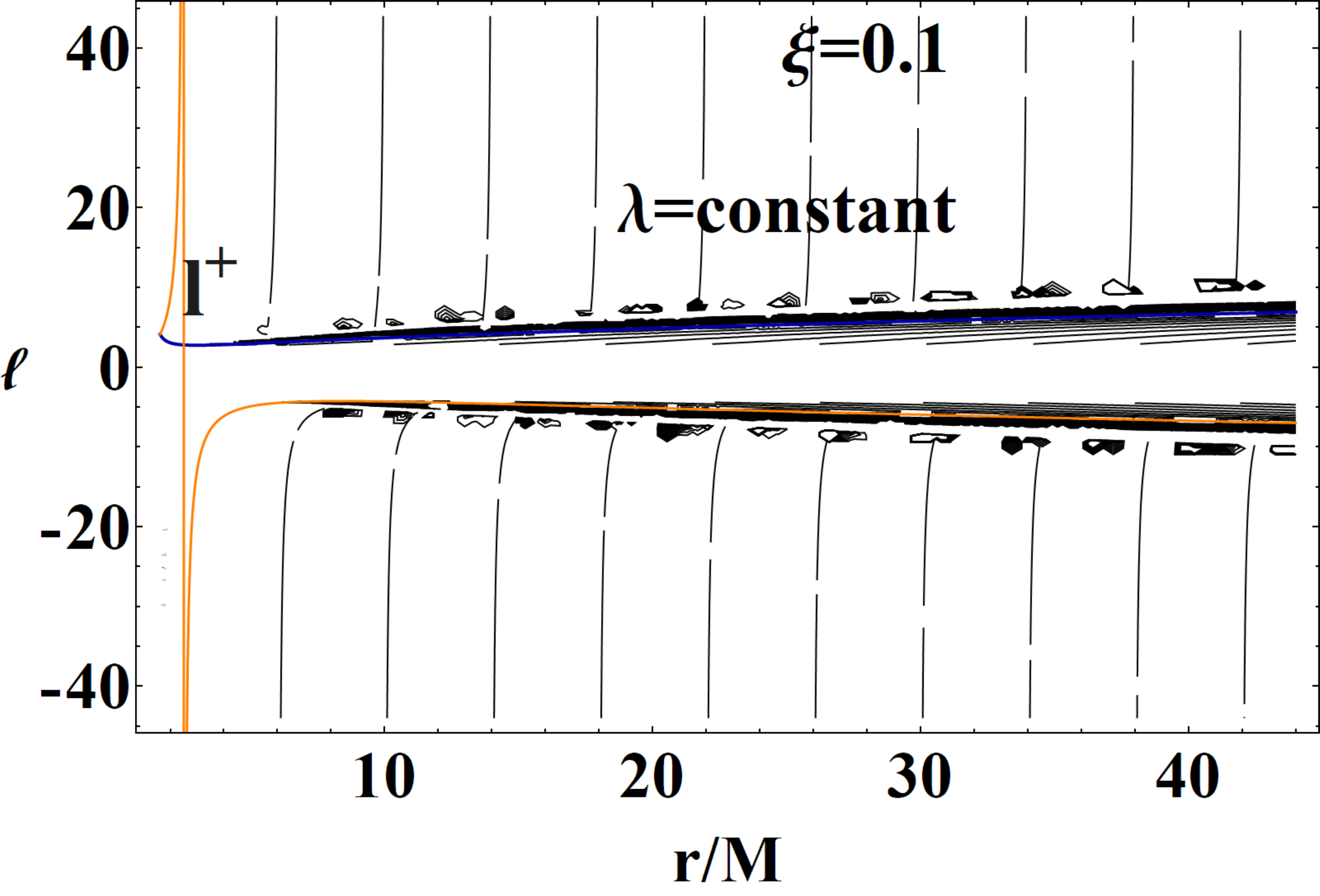}\\
 \includegraphics[width=8.cm]{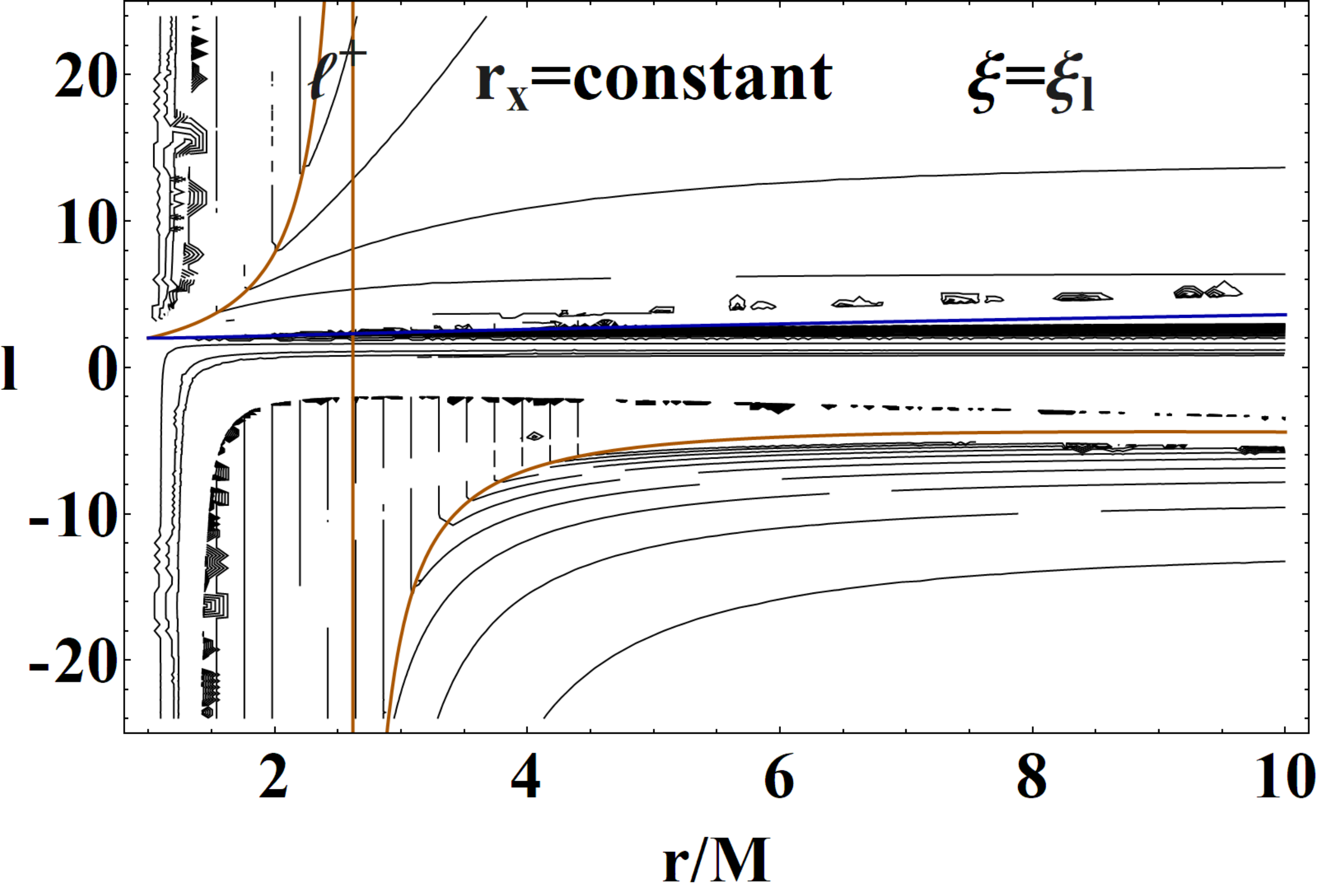}
\includegraphics[width=8.cm]{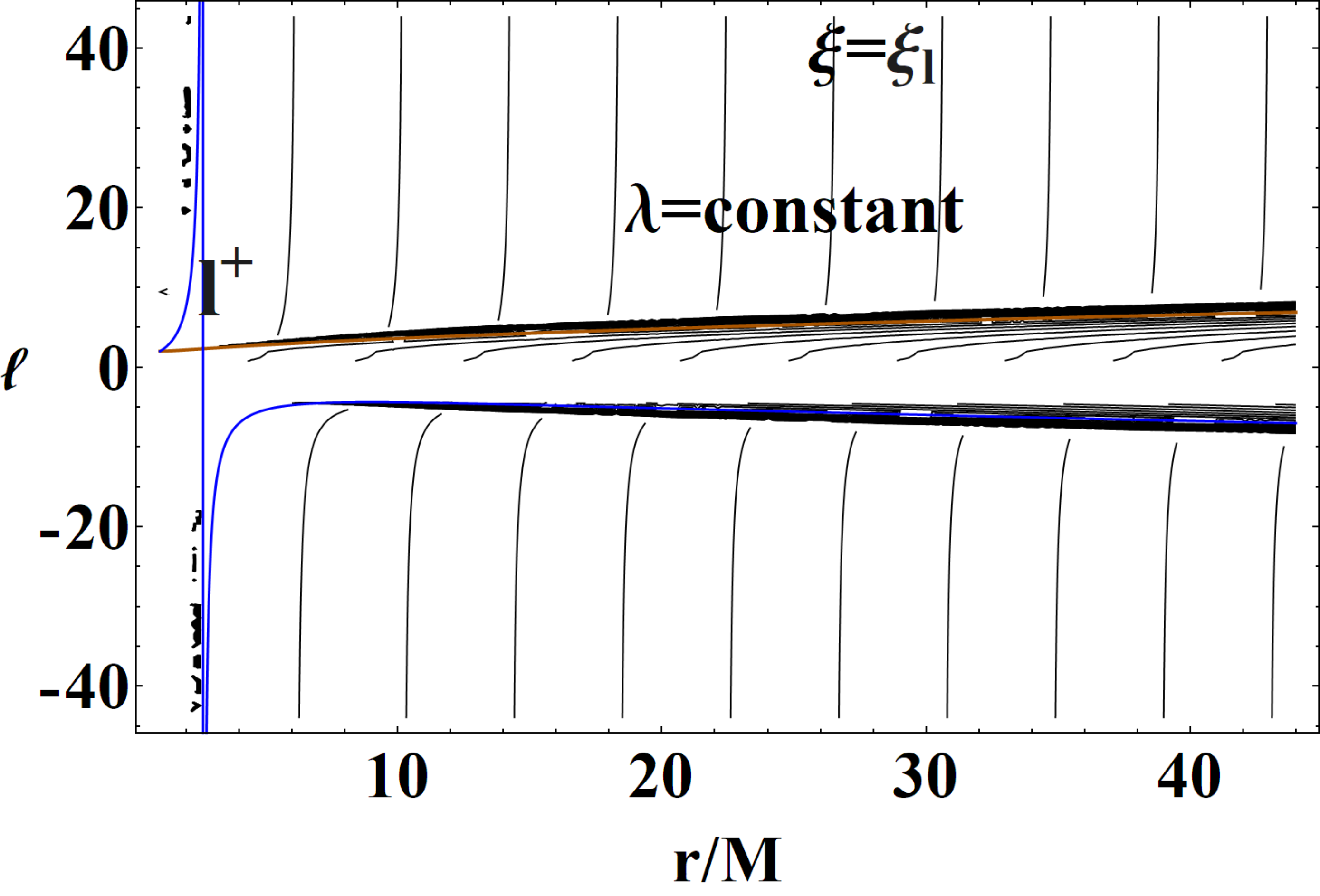}
\caption{Inner edge $r_{\times}=$constant  of the cusped tori or the cusp location of a proto-jet in the plane $(\ell,\xi)$, $\ell$ is the fluid specific angular momentum and $\xi$ is the energy parameter, for corotating $(-)$ and counter-rotating  $(+)$ tori.  $\lambda$ is the elongation $r_{out}-r_{in}$, evaluated considering $K=V_{eff}(r,a,\ell)$;  $r_{\times}(a,r,K,\ell)$ is evaluated considering functions $K=V_{eff}(\la_{\xi},\ell,r)$, (where  $a=\la({\xi})$). Central curve is  the fluid specific angular momentum $\ell^{\pm}(\xi,r)$ as function of $r/M>r_+(\xi)$ for selected values of $\xi$, $r_+$ is the \textbf{BH} horizon as function of extracted rotational energy.}\label{Fig:quesa}
\end{figure}
We conclude  from Figs\il(\ref{Fig:ManyoBoFpkmpMig})  that the functions are not symmetric in the left and right range  of the limiting value $\xi_\ell$,
we note the different behaviour of the curves for corotating and counterrotating fluids.
\section{On the \textbf{RAD}  and \textbf{BH} spin--accretion disk  correlation}\label{Sec:lona}
We conclude this analysis    exploring  tori characteristics in terms of dimensionless \textbf{BH} spin.
Here we investigate in detail  the dependence of the \textbf{RAD} rotational law and the \textbf{BH} spin.
It was proved in \cite{dsystem,letter,multy} that there is a relation between the specific angular momentum of the fluid  $(\ell_i,\ell_o)$ of the inner $(i)$ and outer $(o)$  torus of an \textbf{eRAD}  couple and  the central \textbf{BH} spin  $a$,
different  for  $\ell$corotating and $\ell$counter-rotating couples.
We investigate a simple correlation within the limits of the  range of values of  the dimensionless spin $a/M$ of the \textbf{BH}  ($M$ is the \textbf{BH} ADM mass) and the fluid specific angular momentum $\ell$, here used to parameterize the torus of the orbiting aggregate, through the value assumed along the distribution curve $ \ell(r)$ (the \textbf{eRAD} rotational law) at the  point of maximum density and pressure inside each toroidal configuration of the agglomeration.
Arguments in support of the existence of such correlation in the form of superior and inferior boundaries of the range, have been discussed in  \cite{pugtot, multy}, while in \cite{ergon}  discussion on the possible geometrical origin of this limit has been highlighted. In this section we resume this topic considering the ergoregion as stationary (corotating) disk solution can also be considered  for   sufficiently large values of the central \textbf{BH} spin.
Therefore, the model is also studied in the regions close to the static limit

Function $\ell(r)$   is    considered  as a possible reference distribution  on the fluid angular momentum in an extended region  of  any ("fast" rotating) accretion disk. The condition of almost spherical accretion   provides a first limit on the rotational  law of the fluid in the accretion disk.
Sometimes this limit  is know as    "Bondi regime".
Thick
disks considered in this article are regulated  by a significant  centrifugal force which is more generally assumed   superior or equal the Keplerian specific  angular momentum $\ell_K=L/E$.
In this sense the "slow rotation" case is referred to as "Bondi flows"  (being  the limit of  free fall accretion disks) \cite{Bondi}.
In these  (quasi)spherical "Bondi" accretion conditions,  the angular momentum is not  relevant in the dynamical  forces balance,  i.e.,  the (specific) angular momentum  in the disk  is     smaller
than the Keplerian one. On the other hand,   an accretion  disk must have  an extended region
   where matter has a  large   centrifugal component ($\ell\geq \ell_K$)--\cite{abrafra}.
 It will be    convenient to analyze the  accretion disk properties in terms of the  ratio $\ell/a$ or $\ell-a$ as  an important  parameter  for these  models\footnote{As discussed in Sec.\il(\ref{Sec:model}) there are closed toroidal configurations  for $\pm\ell^{\mp}\geq\pm\ell_{mso}^{\mp}$  and $K^{\mp}\in [K_{mso}^{\mp},1]$. However  in \cite{pugtot} orbiting stationary configurations  for different values of $ K$ and  $\ell$ have been constructed.
 {At lower momentum  (in magnitude) there are  closed surfaces very close to the outer horizon, and related to the inner  Roche lobe of the closed tori. The  stability of these  configurations   which have not cusp has however still  to be assessed \cite{pugtot}}}.
\subsection{On the \textbf{RAD} rotational law  $\ell(r)$,  torus specific angular momentum $\ell$ and \textbf{BH} spin $a/M$.}\label{Sec:barl}
In \cite{proto-jet,pugtot}, variables    $(\ell\pm a)$ and $(\ell\pm a\sigma)$  have been considered to construct the accreting tori around  a Kerr \textbf{SMBH}.
 Many   properties of the fluid effective potential are determined  by  the  quantity   $\bar{\ell}\equiv\ell/(a\bar{\bar{\sigma}}^2)$ %
 where there are the limiting values
$
\ell=-a\bar{\bar{\sigma}}^2$ and $\ell=
a\bar{\bar{\sigma}}^2$
($\bar{\bar{\sigma}}=\sin\theta$ being off function of the polar angle $\theta$).
The origin of this quantity can be understood by considering the following fact.
For simplicity we use here all dimensionless quantities,
 we introduce the rotational version of the Killing vectors $\xi_t$ and $\xi_{\phi}$, i.e., the
canonical vector fields
$\tilde{V}\equiv(r^2+a^2)\partial_t +a\partial_{\phi}$
 and $\tilde{W}\equiv\partial_{\phi}+(a \bar{\bar{\sigma}}^2) \partial_t$.

 Then the contraction of
 the geodesic four-velocity with $\tilde{W}$ leads to the (non-conserved) quantity
$L-E (a \bar{\bar{\sigma}}^2)$,
 function of the conserved quantities $(E,L)$,
  the spacetime parameter  $a$ and the polar coordinate $\theta$;
   on the equatorial plane it then reduces on $ L-E a$.
  We note the existence of two limiting values related to two bundles frequencies and limiting momenta: $\omega=a\sigma$ and $\omega=1/(a\sqrt{\sigma})$.
When we consider the principal null congruence,
$
\gamma_{\pm}\equiv\pm\partial_r+\Delta^{-1} \tilde{V}$,
the angular momentum $L=(a \bar{\bar{\sigma}}^2)$ that is $\bar{\ell}=1$
(and $E=+1$, in proper unit), every principal null geodesic is then characterized
by\footnote{{Here we consider the quantity $\bar{\ell}= \ell/(a\sigma^2)=L/(Ea\sigma^2)$; then if $
L=a\sigma^2$ we have
$\bar{\ell}=1/E$ and therefore for $E=+1$ there is $\bar{\ell}=1$.}} $\bar{\ell}=1$.

\begin{figure}
  \includegraphics[width=6.5cm]{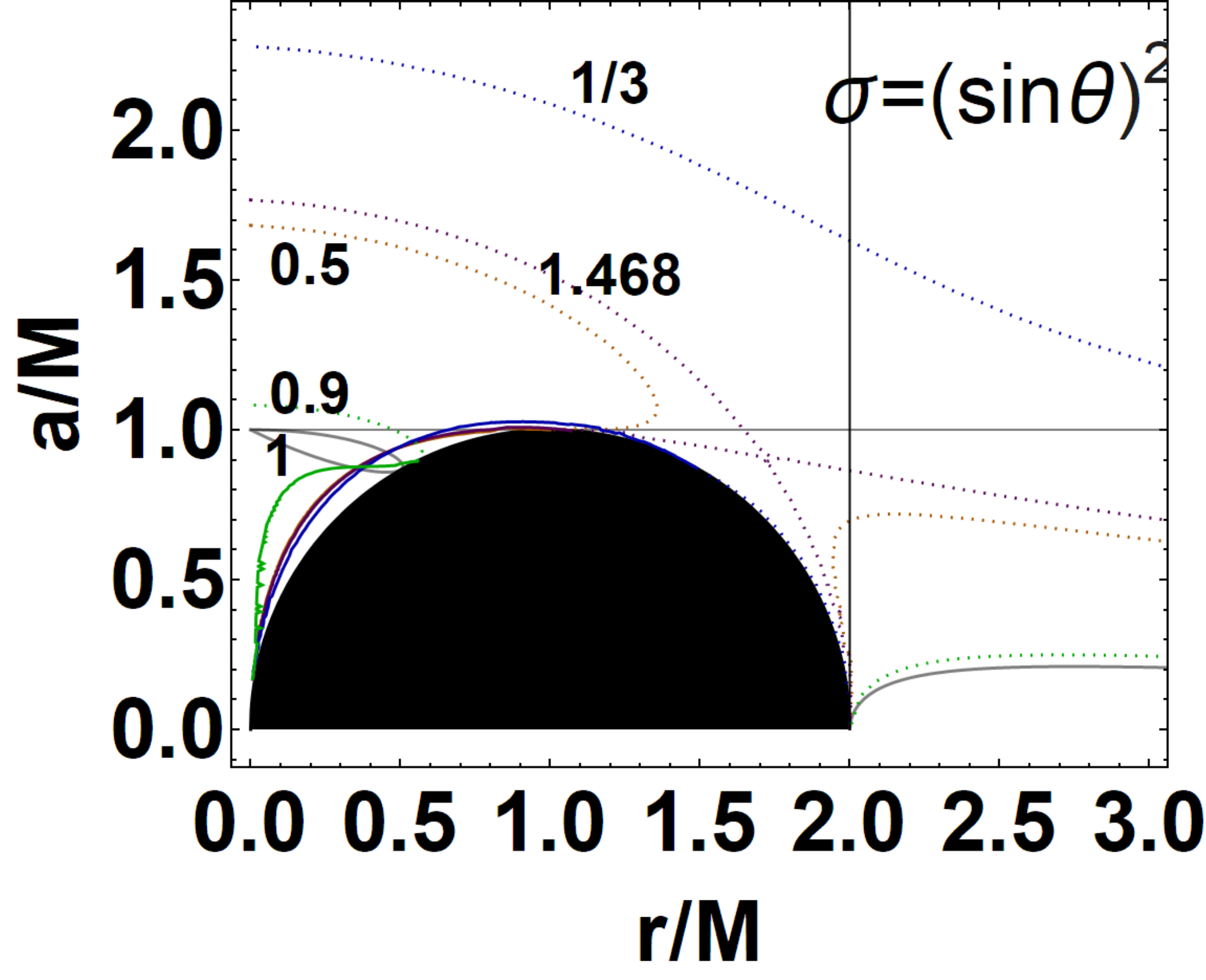}
  \includegraphics[width=5.cm]{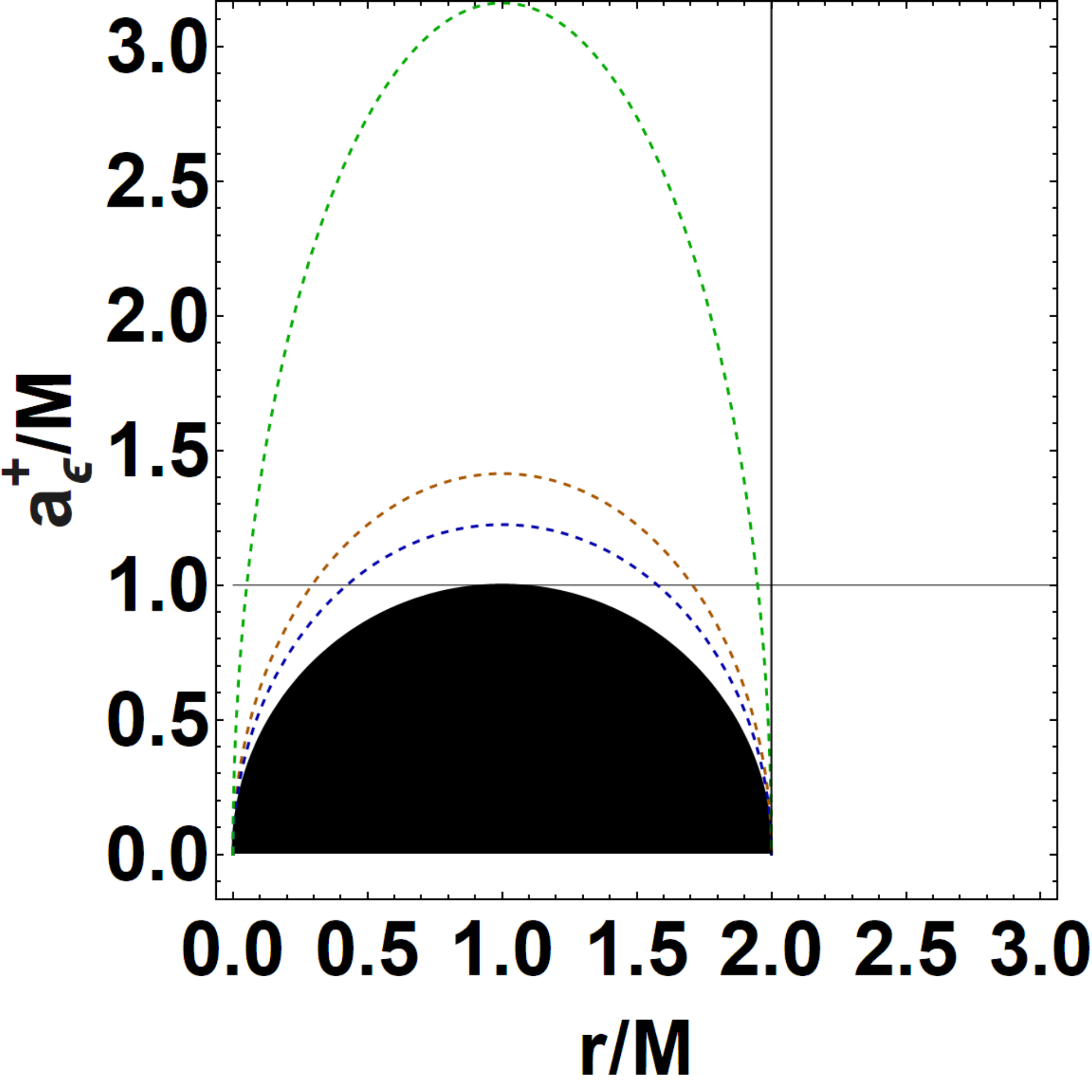}
   \includegraphics[width=6cm]{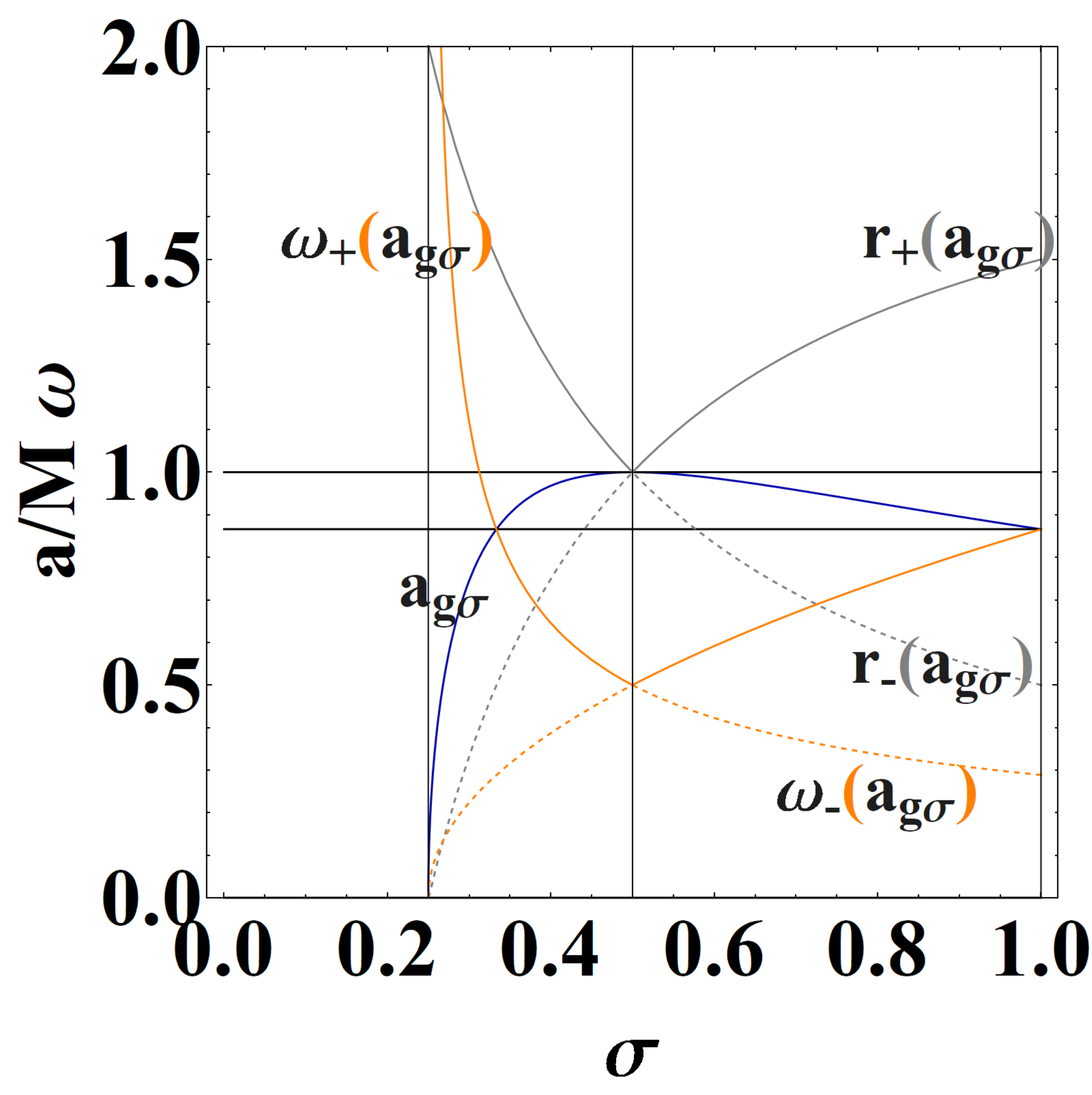}
  \\
 \includegraphics[width=6cm]{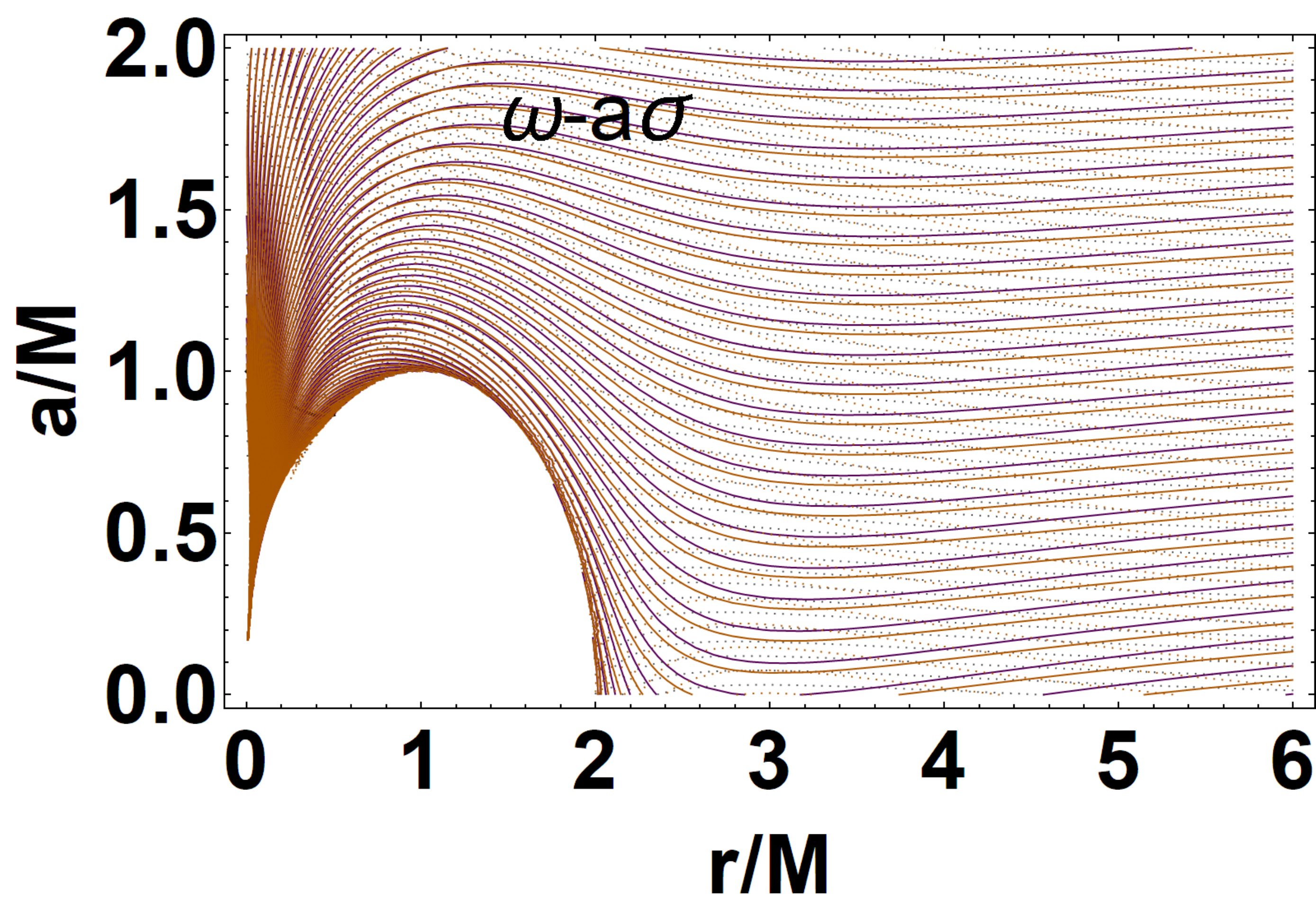}
   \includegraphics[width=6cm]{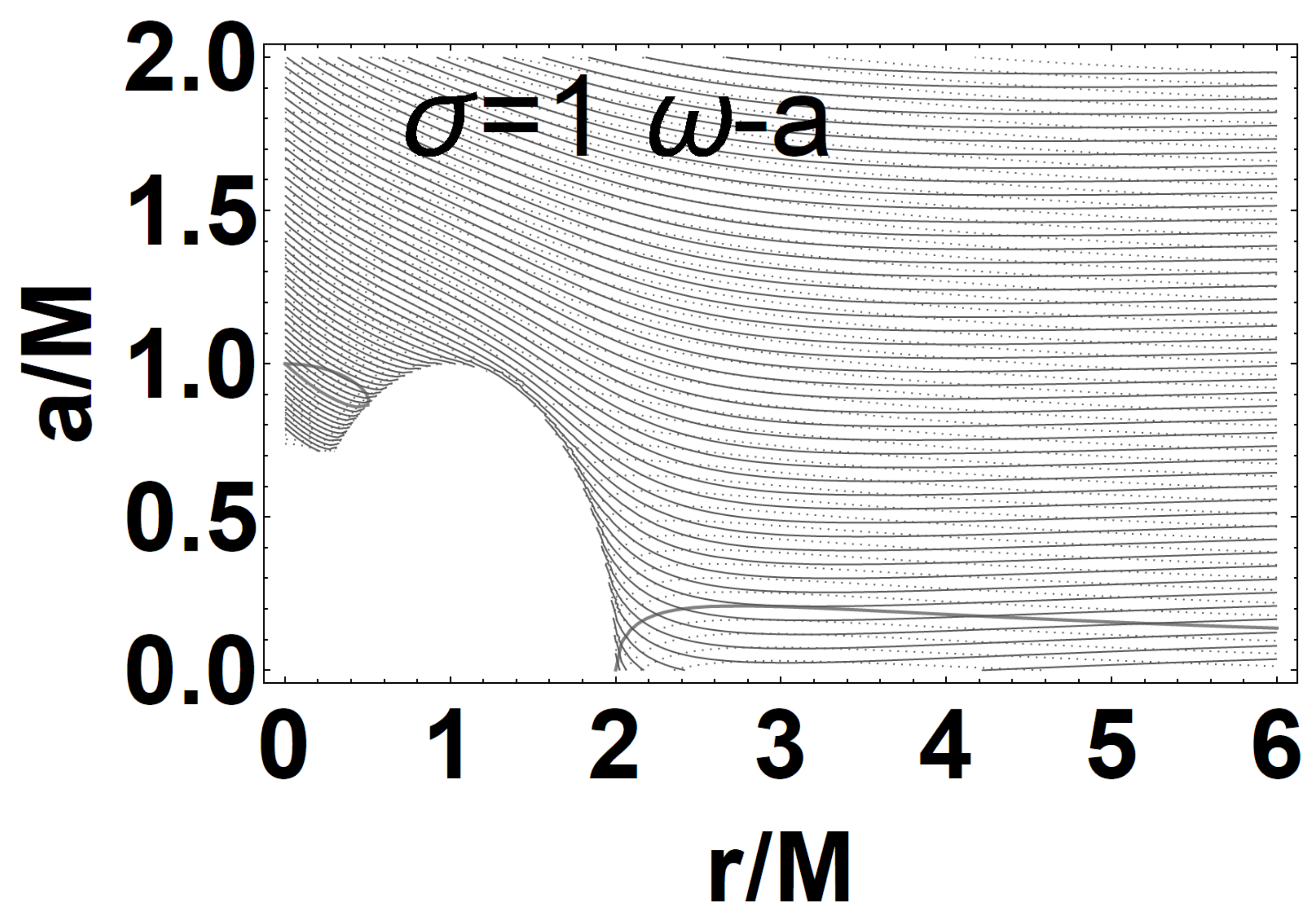}
  \caption{Black region is the \textbf{BH} in the extended plane, the boundary is the horizons curve in the extended plane. There is $\sigma=\sin^2\theta$. Numbers of the left panels are values of constant plane $\sigma$. \emph{Upper line.}  Left panel shows metric Killing bundles $\mathcal{B}_{a\sigma}$ at constant $\sigma$, defined by condition $\omega-a\sigma=0$. Central panel shows the ergosurfaces curve $a_{\epsilon}^+$  in the extended plane which is a spin function, for different $\sigma$s following colors relative to the left panel. Full characterization is in \cite{remnant,remnant0,remnant1}. \emph{Bottom line.} Curves of constant $\omega-a\sigma$ for different $\sigma$s according to colors in upper line panels. Bottom right panel shows the curves on the equatorial plane $\sigma=1$. It can be noted the horizon curves.  Upper right  panel shows notable spins $(a/M)$, radii $(r/M)$ and  characteristic frequencies of the bundles  $ \omega$ as function of  the plane  $\sigma=\sin^2 \theta\in [0,1]$.
$a_{g\sigma}$ is the tangent curve to the horizons of the bundles $\mathcal{B}_{a\sigma}$
$r_{\mp}(a_{g\sigma})$ are the horizon curve for tangent spin, depending on the  plane $\sigma$ from definition of $a_{g\sigma}$.
$\omega_{\pm}$ are limiting frequencies for stationary observers.}\label{Fig:effct}
\end{figure}
With reference to Figs\il(\ref{Fig:effct}) we considered the "bundles-like" solutions  $\mathcal{B}_ {a\sigma}$ with Killing vector
  $\mathcal{L}$ with  $\omega=a \sigma^2$. We note that these particular bundles solutions appear tangent to the horizon curve in the extended plane $a/M-r/M$. The study of these special solutions is relevant considering the following relations hold:
there is  $a_0\equiv 1/(\omega \sqrt{\sigma})$ where  $a_0$  is the origin of the bundles $\mathcal{B}_{\omega}$ that is at $\omega=$constant in the extended plane $a/M-r/M$, the origin is at $r=0$\cite{remnant,remnant0,remnant1}.
It is worth noting  that in the extended plane at fixed spacetime (an horizontal line), on a fixed plane these curves are bundles at constant frequency. Eventually, we can evaluate the gradient  of  $\omega-a \sigma^2=$constant.
  The following relations also hold:   $E-\omega_H L>0$, thus
$\ell<1/\omega_H$, $\omega_H$ is the outer horizon frequency in the extended plane.
Therefore, featuring an adapted process in terms of the variation of the moment of the \textbf{BH}, there is
$\delta J/\delta M<1/\omega_H$, which is expression of the condition
 $\delta M_{irr}>0$
(here $J$ has using of mass square). Note that  there is the limit
$\ell_{\max}=1/\omega$ independently from  $\sigma$ in the spherically symmetric Schwarzschild spacetime.
Condition  $1/\omega_ {\pm} =
 a\sigma$ has no solution, but there is  the singular surface of the ergosurfaces, $a_ {\epsilon}^+\equiv\sqrt { {(r -
        2) r} /({\sigma -
      1}})$,  in the extended plane
 as reported in Figs\il(\ref{Fig:effct}).
   The condition $1/\omega_ {\pm} =
 a\sqrt {\sigma} $ is realized only at the origin  $r =
  0 $, particularly for $\sigma = 1 $ there is
$\omega_ {\pm} = 1/a$ (then $\ell = a$).

On the other hand,  considering the coincidence condition with the horizon curves, the condition   $ \omega_ {\pm} (r_ {\pm}) = a \sigma$ for the bundles  $\mathcal{B}_ {a\sigma}$
is solved for  $a_ {g\sigma}\equiv({\sqrt {4\sigma -
     1}})/{2\sigma} $ which, evaluated on the horizons $r_ {\pm}$, gives respectively
$r_ {\pm} (a_ {g\sigma}) = \pm\sqrt { {(2\sigma -
          1)^2}/ ({4\sigma^2})} + 1 $.
The condition implies $\ell (r_ {\pm}) =
  1/(a\sigma)$.
    The tangency condition of the metric bundles are given in  \cite{remnant} with curve $a_g$ for $\mathcal{B}_{\omega}$. The solution of the condition   $a_g =
  a_{g\sigma}$  for the frequency $\omega$ is $\omega_{g\sigma}^\mp\equiv ({2\sigma \mp \sqrt {(2\sigma - 1)^2}})/({2\sqrt {4\sigma - 1}})$. The horizon curve is clearly delineated as an asymptotic limit of the bundles.

 The characteristic frequencies of the bundles, seen as horizons  frequencies, are the functions $\omega_g^{\pm}$ of Eq.\il(\ref{Eq:show-g}).
$\omega_g^ {-} $ has a saddle point for  $\xi=\left (3\pm \sqrt {6} \right)/3$, correspondent to the spin $\la(\xi) =a_{mso}^{\epsilon}$ and frequencies $\omega_g^{\pm}=({1}/{2\sqrt {2}} , {1}/{\sqrt {2}}$)----see Figs\il(\ref{Fig:fas2}).
The maximum extractable energy   $\xi_\ell =\left (2 - \sqrt {2} \right)/2= 1/\bar {\ell} _ {\max} =2+\sqrt {2}$, confirming the relation $ \bar {\ell} _{\max} =1/\omega$--see Figs\il(\ref{Fig:fas2}).
The  two  bundles whose tangent spin on the equatorial plane is
 $a_g/M =a_{mso}^{\epsilon}$,  are tangent to the  horizon curve in the extended plane on $r = 2/3, r=4/3$, with characteristic frequencies $\omega=\{1/(2 \sqrt{2}), 1/\sqrt{2}\}$ and origin  dimensionless spins in the naked singularity regions,  $a_0/M=(2\sqrt{2},\sqrt{2})$ respectively on the equatorial plane.
Obtaining the limit of $\xi_{\ell}\equiv \frac{1}{2} \left(2-\sqrt{2}\right)$
in the static case $a=0$,  the bundle frequencies  have an extreme as function of $r/M$ for the orbit of photon $ r=3M$
where  $\omega_{\pm}=1/\sqrt{27\sigma}$.

Spin $a_{\gamma}^{\epsilon}=M/\sqrt{2}$ is solution  of  $\partial_a\partial_a\ln s=0$ where $s=\omega_H^+/\omega_H^-$.
Another relevant spin  is   $a_g=\sqrt{3}/2$ which  solves the problem for $\partial_a^{(2)}a_0=0$
and where $s=1/3$, a second saddle  point is $a_g=a_{\gamma}^{\epsilon}=1/\sqrt{2}$,
where  $\omega_H^{\mp}=1/2\pm1/\sqrt{2}$--\cite{remnant0}.

A different adapted  solution parameterizations  is discussed in Appendix\il(\ref{Sec:adaptedd}).

\subsection{Sets of tori}\label{Sec:setsoftori}
 An example of distributions of momenta and radius as functions of the spin in the considered  parametrization is   shown in Figs\il(\ref{Fig:Boundaryaloudfilla},\ref{Fig:travseves}). In Fig.\il(\ref{Fig:sccreem})  we show the dispersion in the corotating and counter-rotating distribution of tori in the \textbf{RAD}.

 Considering the situation on the equatorial plane $(\sigma=1)$, there are lower bounds in magnitude of the rationalized momentum  $\ell/a$. It is easy to see that this limit is always given
as inferior limit (in magnitude) of  the   extreme rotating \textbf{BH} case.
Therefore,   corotating  fluid configurations are formed for $\ell/a\geq2$, where the limiting value increases  with decreasing spin.
In the counter-rotating  case,
fluid  configurations can  form with $\ell/a<-22/5$. The  detailed situation is shown in Figs\il(\ref{Fig:sccreemzero}), which considers also the limits for the formation of stationary configurations in the  ergoregion.
For the  formation of counter-rotating accreting tori and proto-jets there are the limiting ratios $\ell/a=-2 (1 + \sqrt{2})$, and $\ell/a=-7$.
In \cite{pugtot} it was shown  that the limiting cases $\bar{\ell}=\pm 1$ do not admit any toroidal Boyer configurations, $\ell=\bar{\ell}_{\theta=\pi/2}=\pm a$.

A more accurate description is therefore shown in Figs\il(\ref{Fig:travseves}), curves of $\ell$=constant show one torus evolution. The analysis  points out two relevant spins: the first at  $a= 0.7M$, clear also from the Figure\il(\ref{Fig:Boundaryaloudfilla}), and a second relevant spin
is $a\approx 0.3M$. The crossing of areas and curves
show the regions of \textbf{RAD} tori parameter where collision is more probable to occur.
This behavior has also a role in the density   seeds formation. Considering  different regions of the parameters,
the separation (corotating-counter-rotating) is larger for the accretion than the proto-jets parameter space, indicating that at high fluid specific angular momentum  the corotating and counter-rotating fluids may have not been differentiated in the formation of the early phases of  jets, and the separation remains in the range from  $2M$ to  $4 M $ independently from the spin. The model reveals also the possibility of jet shells of corotating or  counter-rotating fluids.
\begin{figure}
\centering
\includegraphics[scale=.2]{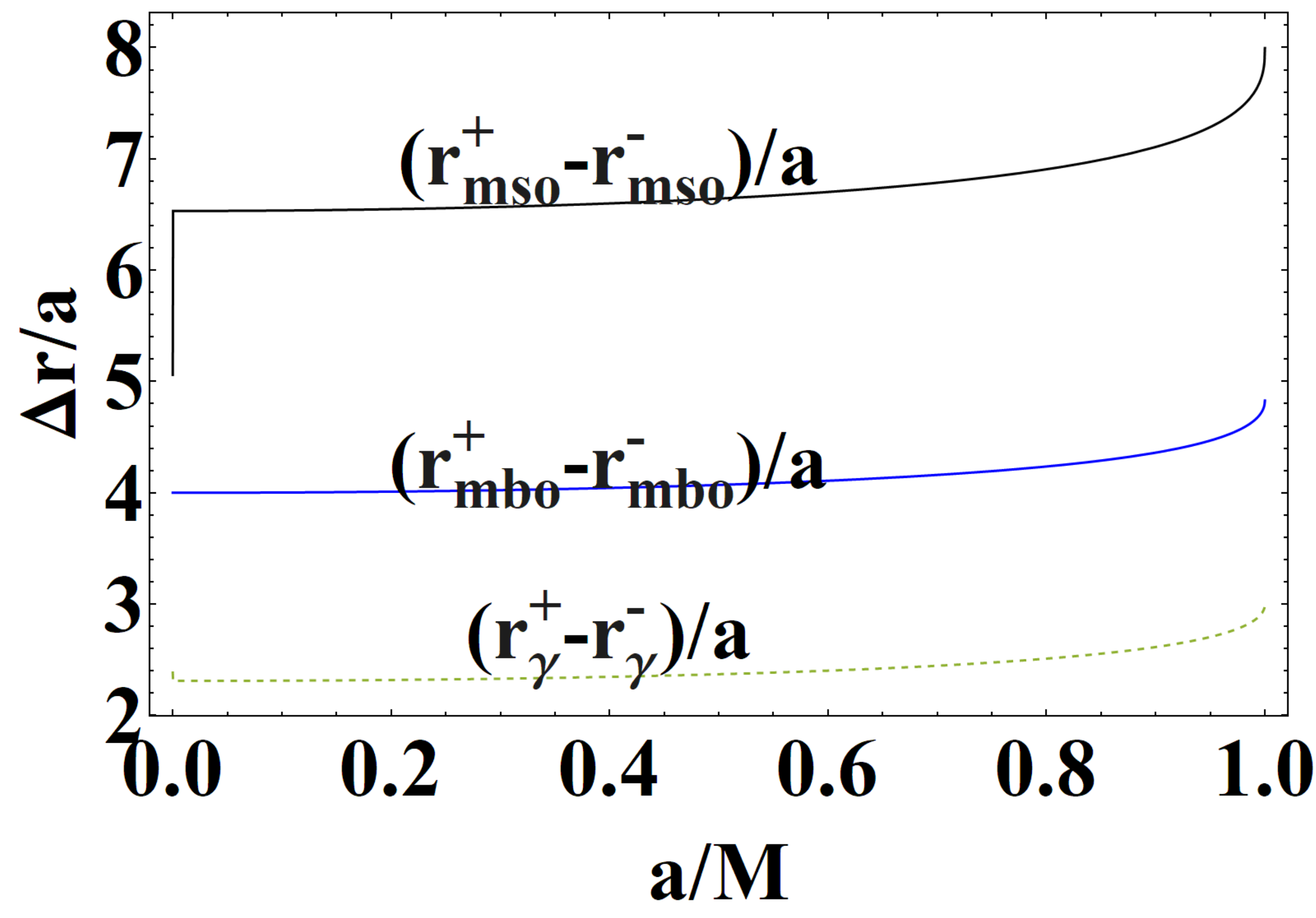}
\includegraphics[scale=.2]{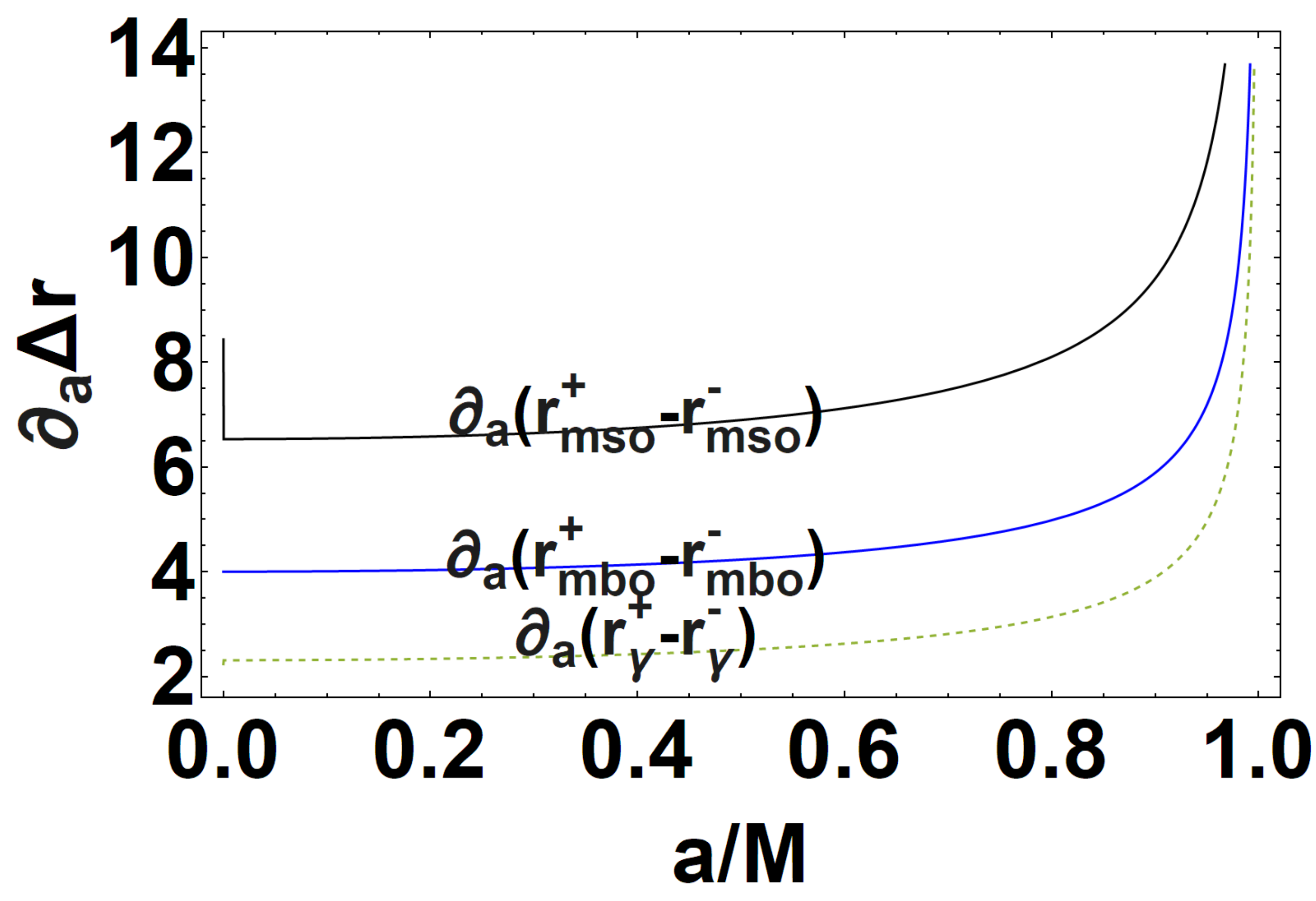}
\includegraphics[scale=.2]{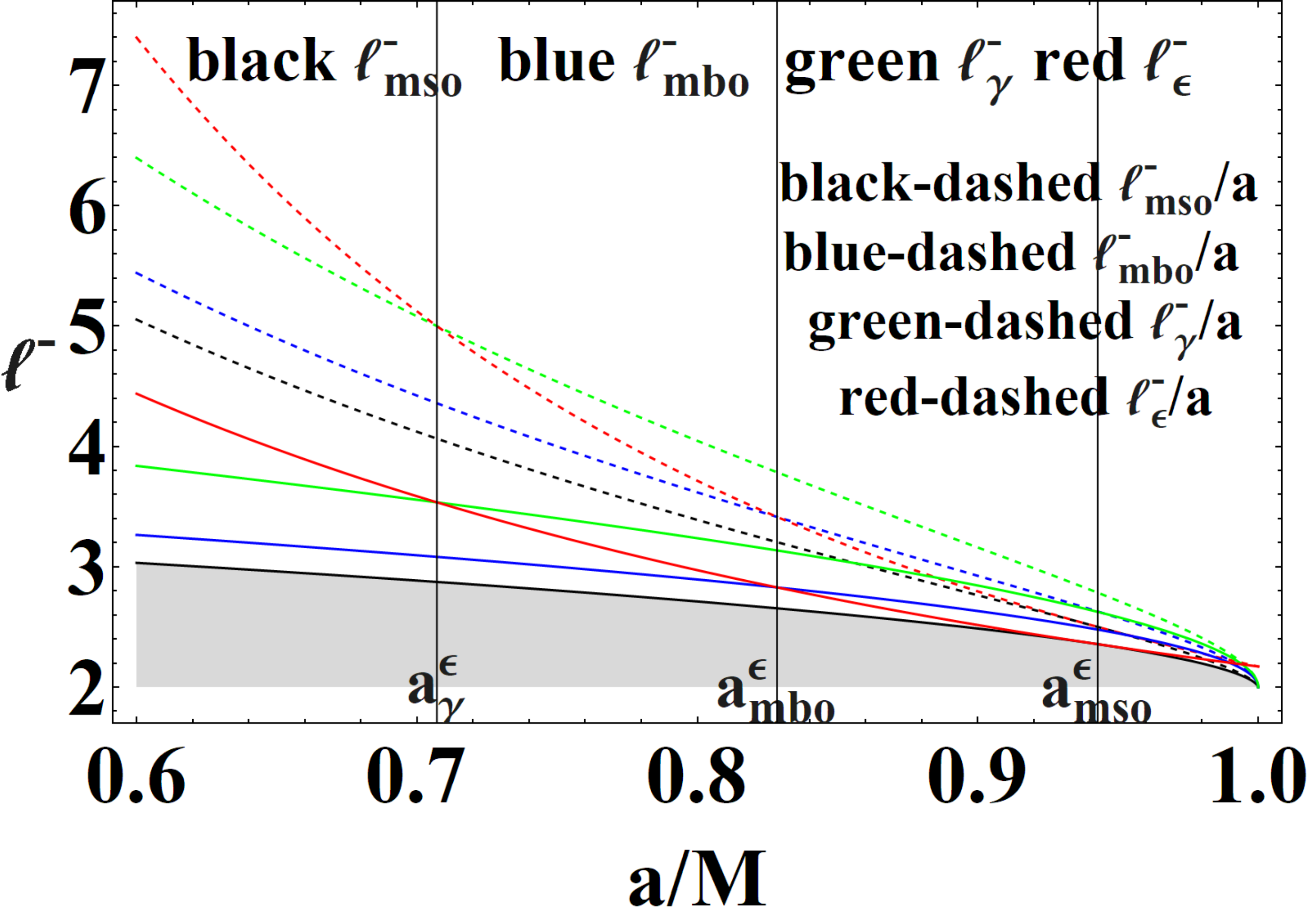}
\caption{Corotating fluids  ($(-)$) and counter-rotating  fluids ($(+)$). The photon orbit location is $r_{\gamma}^{\pm}$, $M$ is the \textbf{BH}  mass, $r_{mbo}^\pm$ is for marginally bounded orbit, $r_{mso}^{\pm}$ is the marginally stable orbits. Left: increments  versus spin of the \textbf{BH}  for marginally  stable orbits, marginally  bounded orbits and photon  orbits. Central panel: rate of   increments  versus spin of the \textbf{BH}. These are dimensionless quantities, the increment of gradients is larger for  the parameter range determining  formation of quiescent tori with limiting unstable  proto-jets. Right panel:  $\ell$ fluid specific angular momentum and spin $a/M$ of the \textbf{BH}, $\ell_{\epsilon}^-\equiv \ell^-(r_{\epsilon}^+)$ where $r_{\epsilon}^+=2M$ is the outer ergosurface. In right panel  role of spins $\{a_{mso}^{\epsilon},a_{mbo}^{\epsilon},a_{\gamma}^{\epsilon}\}$ are shown. No tori are defines for fluid specific angular momentum in the region $\ell<\ell_{mbo}$ (gray).}\label{Fig:sccreemzero}
\end{figure}

A detailed  analysis points out the following spins :
\bea\nonumber&& a_{\flat}=0.638285M: r_ {mso}^ -= r_ {\gamma}^+,\quad
   a_{\beta}= 0.372583M: r_ {mso}^ -= r_ {mbo}^+,\quad
a_*= 0.313708M:
       r_{mbo}^ -= r_ {\gamma}^+
       \\\nonumber
       &&
      a_{h1}=0.172564M: \ell_ {mbo}^ -= -\ell_ {mso}^+,\quad
  a_{h2}=0.390781M: \ell_ {\gamma}^ -= -\ell_ {mbo}^+,\quad
      a_{h3}= 0.508865M: \ell_ {\gamma}^ -= -\ell_ {mso}^+,
      \\
      &&\label{Eq:oper-econcibr}
    \mbox{where}\quad   a_{h1}< a_*<a_\beta< a_{h2}<a_{h3}<a_\flat.
	\eea
	  There is a discriminating value in two spin classes, determined by the value  $a\approx 0.3M$,
	see  Figs\il(\ref{Fig:Boundaryaloudfilla})
	 \cite{dsystem,multy}
	We distinguish the set of spins $\{a_{*},a_{\flat},a_{\beta}\}$ defined by the cross of the radii of the corotating and counterrotating  geodetic structures, and the spins $\{a_{h1},a_{h2},a_{h3}\}$ defined by the cross of the corotating and counterotating fluid angular momentum curves as functions of the \textbf{BH} dimensionless spin.   Spins  $\{a_{*},a_{\flat},a_{\beta}\}$ regulate the  \textbf{eRAD} inner structure associated to classes of \textbf{SMBHs} defined according to their spins.  This can be seen by considering   the role of the geodetic structures in the torus model construction, fixing  location of  maximum and minimum of pressure in the torus, respectively the center and inner edge of the torus, and the  related orbital regions shown in   Figs\il(\ref{Fig:Boundaryaloudfilla}). The inner  \textbf{eRAD}  structure is defined by the number of tori, their  locations  and relative rotation, if  corotating or  counterrotating with the central \textbf{BH},  and their topology.
Spins   $\{a_{h1},a_{h2},a_{h3}\}$ define regions of the rationalized specific angular momentum, $\ell/a = L/(aE)$, for  corotating  and counterrotating  fuids, regulating also  condition on the colliding tori
 \cite{multy,long}.
 Furthermore, in  Figs\il(\ref{Fig:Boundaryaloudfilla})  a quasilinear relation $\ell/a$ as function of $a$ is shown, in logarithm scale:
 $\ell \sim
   a^{(b_1 + 1)} +
   e^{b_0}$ where  $b_1$  and $b_0$ are constant,  and it can be explained by an analysis for small dimensionless spin $a/M$.
It is clear from Figs\il(\ref{Fig:rightse-s}) how the functions of the rotational energy parameter  $\xi$  determining the inner structure of the  \textbf{eRAD} in terms of the relative location of  corotating or counterrotating tori, is more articulated for  small $\xi\in]0,\xi_\ell]$,  that is for geometries $a<a_{\beta}\approx 0.37M$.

On the equatorial plane, the frequency has an extreme for  $a=\sqrt{2} $ where   $\omega=1/(2\sqrt{2})$.
In Figs\il(\ref{Fig:sccreemzero}) we represent  the ratio $\delta r_{\star\bullet}^{\pm}=(r_{\star}^{+}-r_{\bullet}^{-})/a$ and the increment
 $\partial_a\delta r_{\star\bullet}^{\pm}$, for $\star$ and $\bullet$ denoting    geodetic limiting orbits. The differences  $a \delta r_{\star\bullet}^{\pm}$ provide a first indication on the construction of the couples of $\ell$counter-rotating orbits and therefore the possibility of formation and evolution under spin transition, remaining almost constant for $a\in[0,M[$.
\begin{figure}
  \includegraphics[width=6cm]{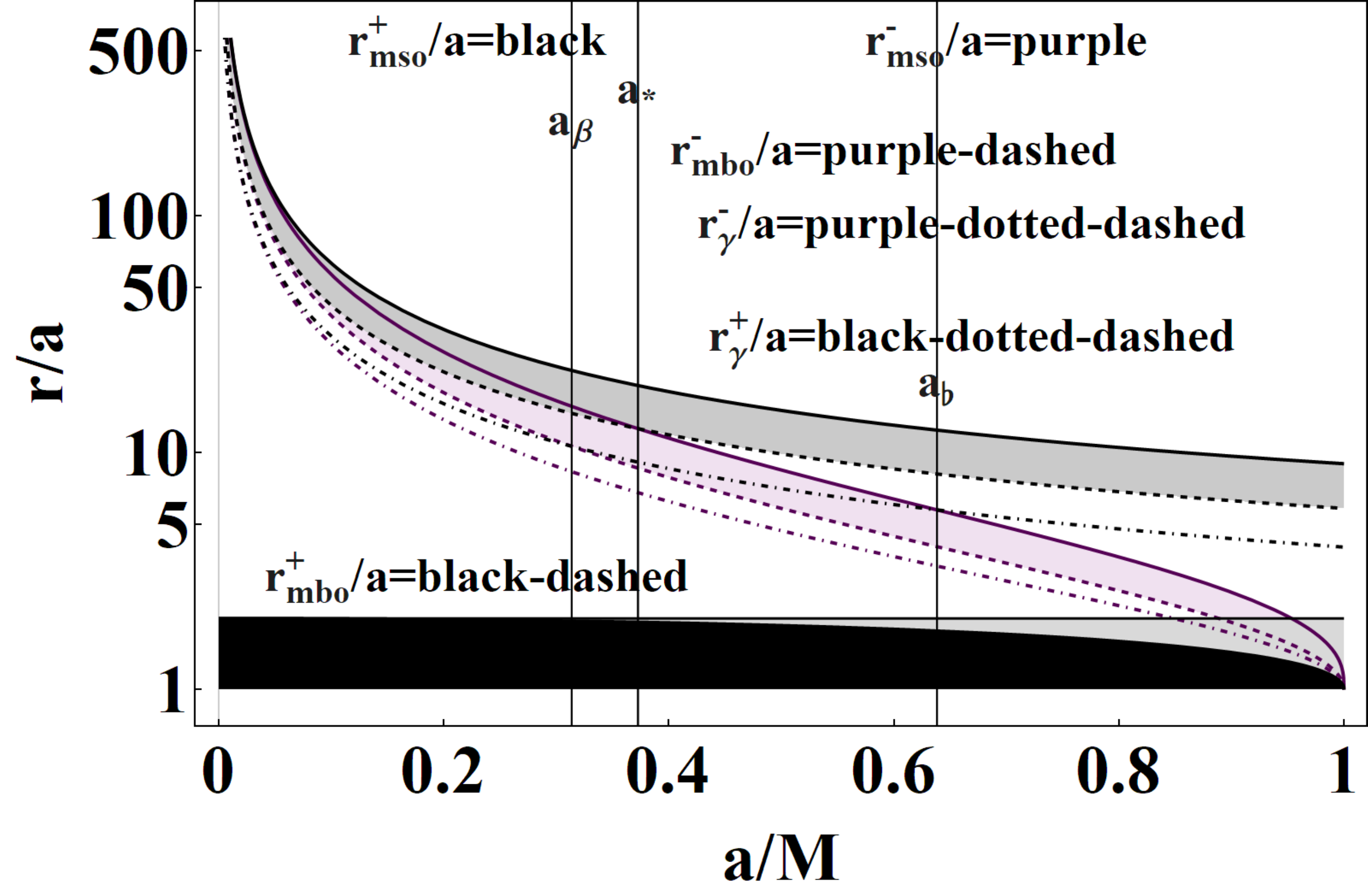}
   \includegraphics[width=5.6cm]{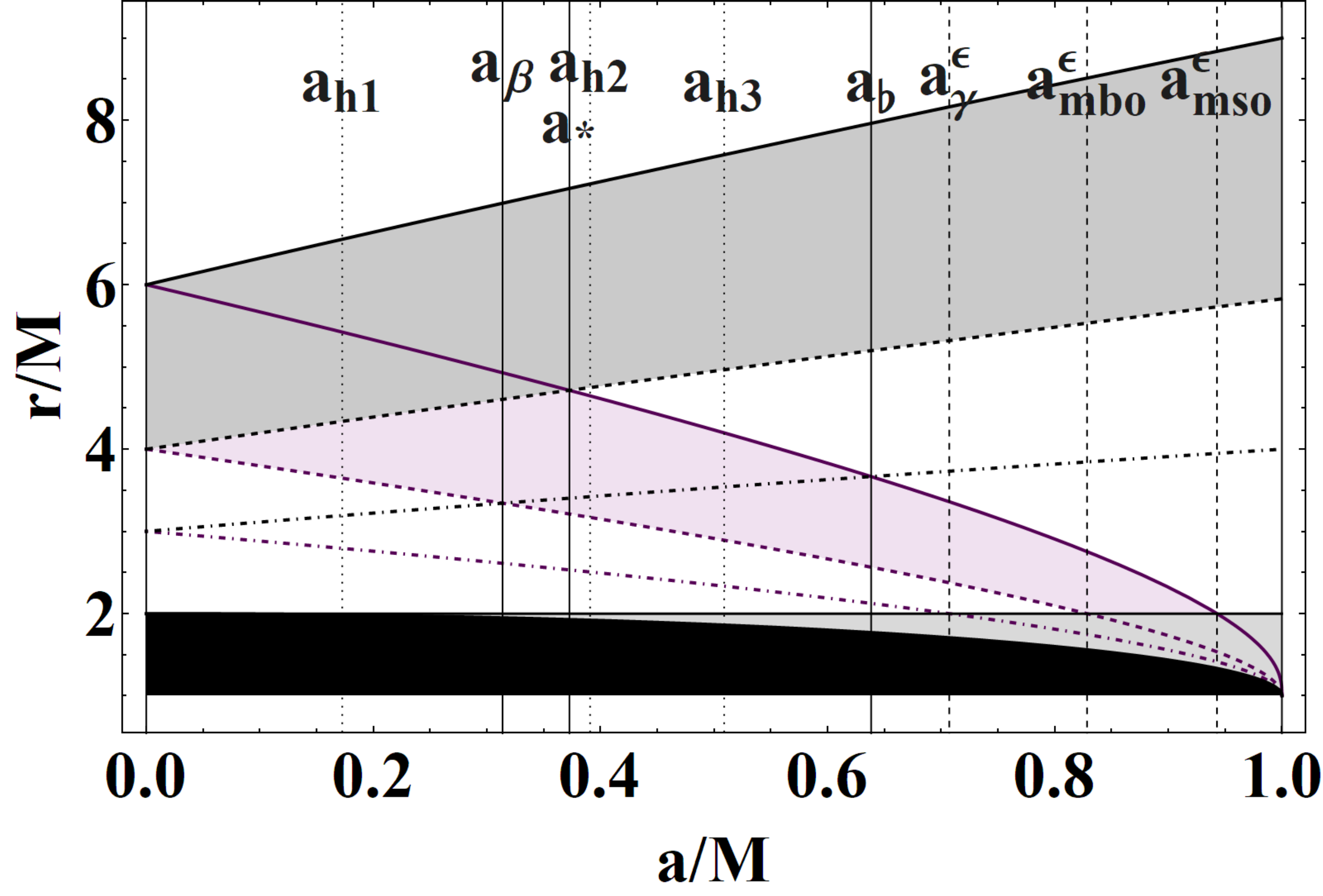}
\includegraphics[width=5.6cm]{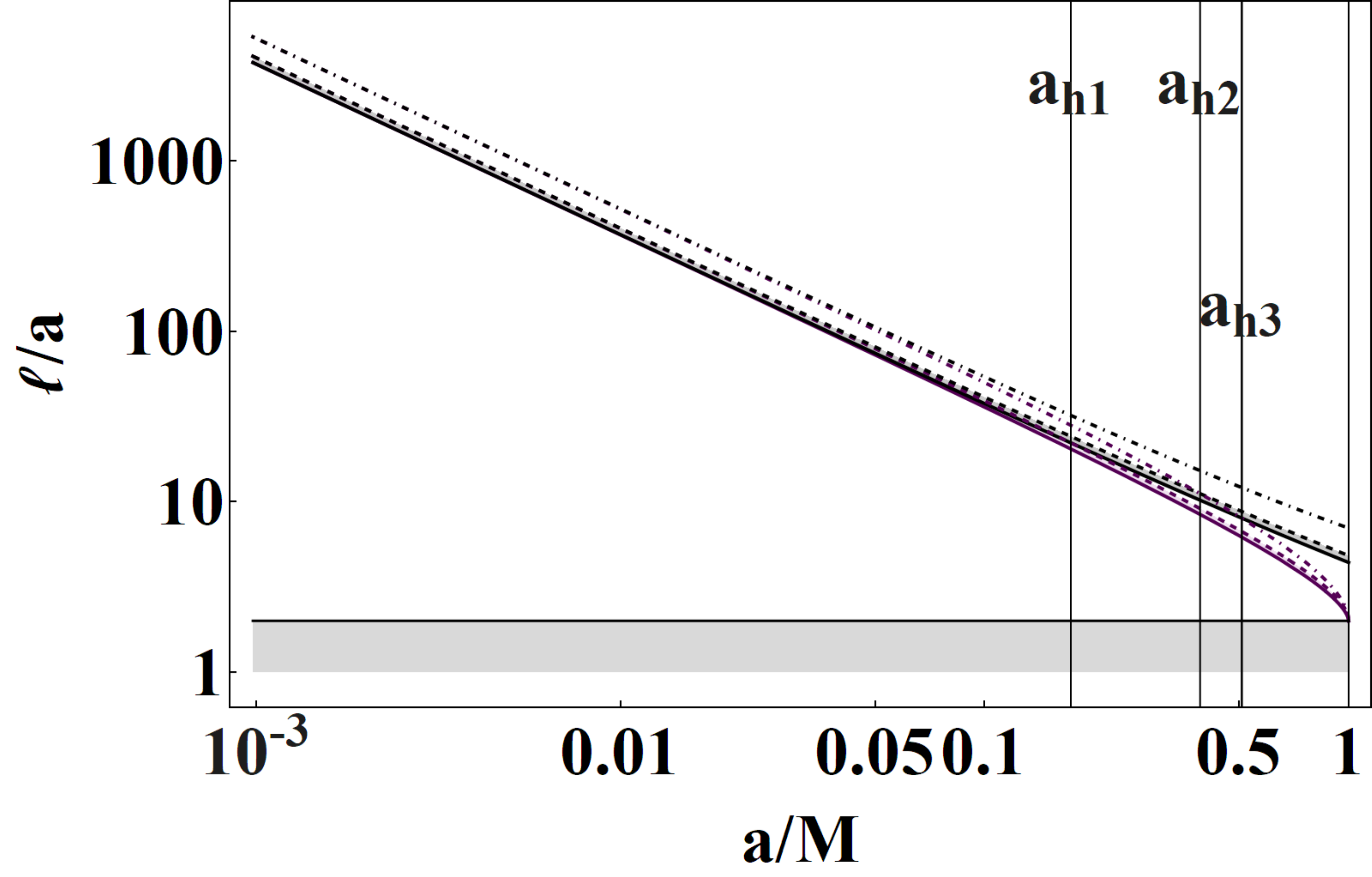}\\
\includegraphics[width=5.6cm]{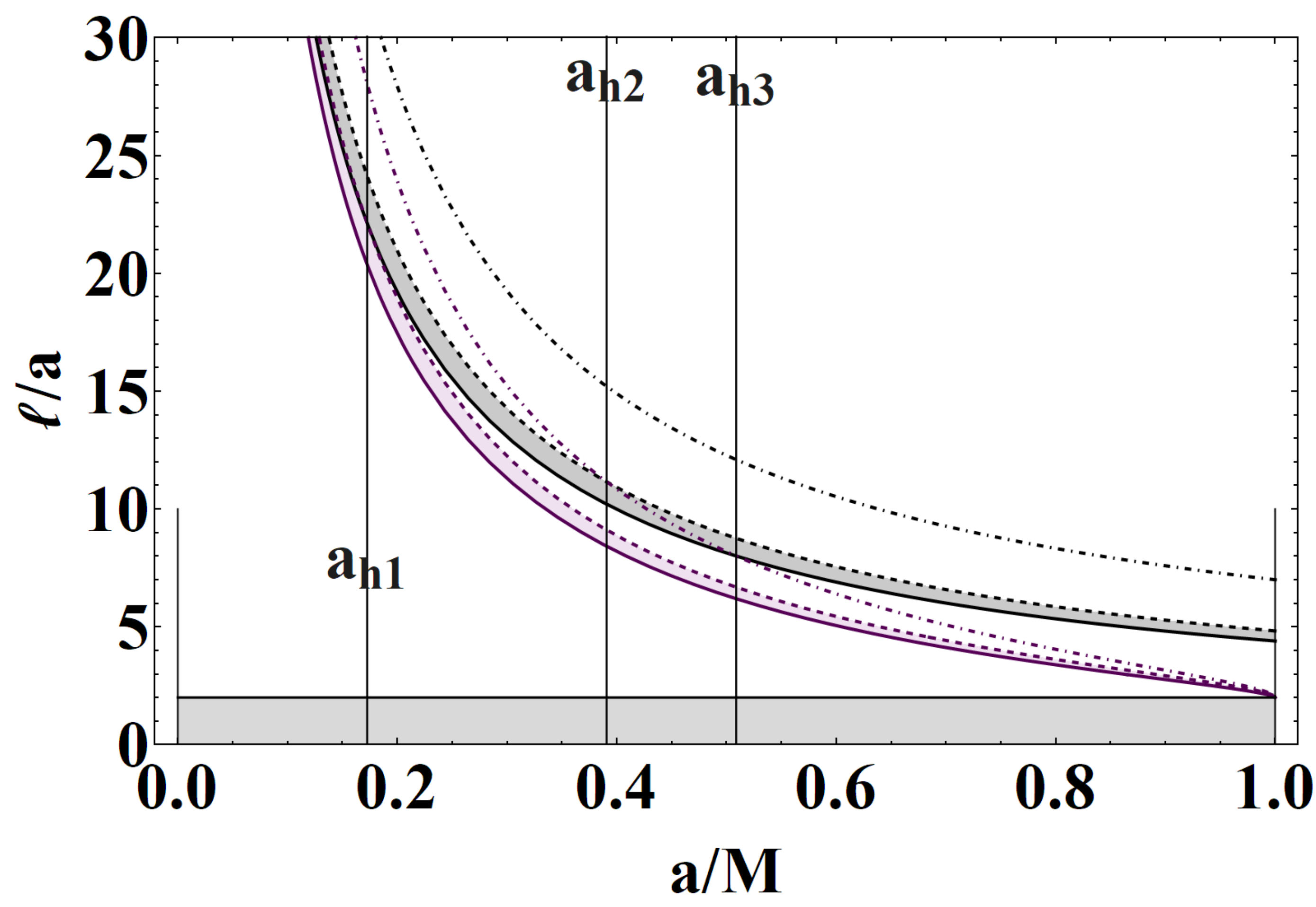}
\includegraphics[width=5.6cm]{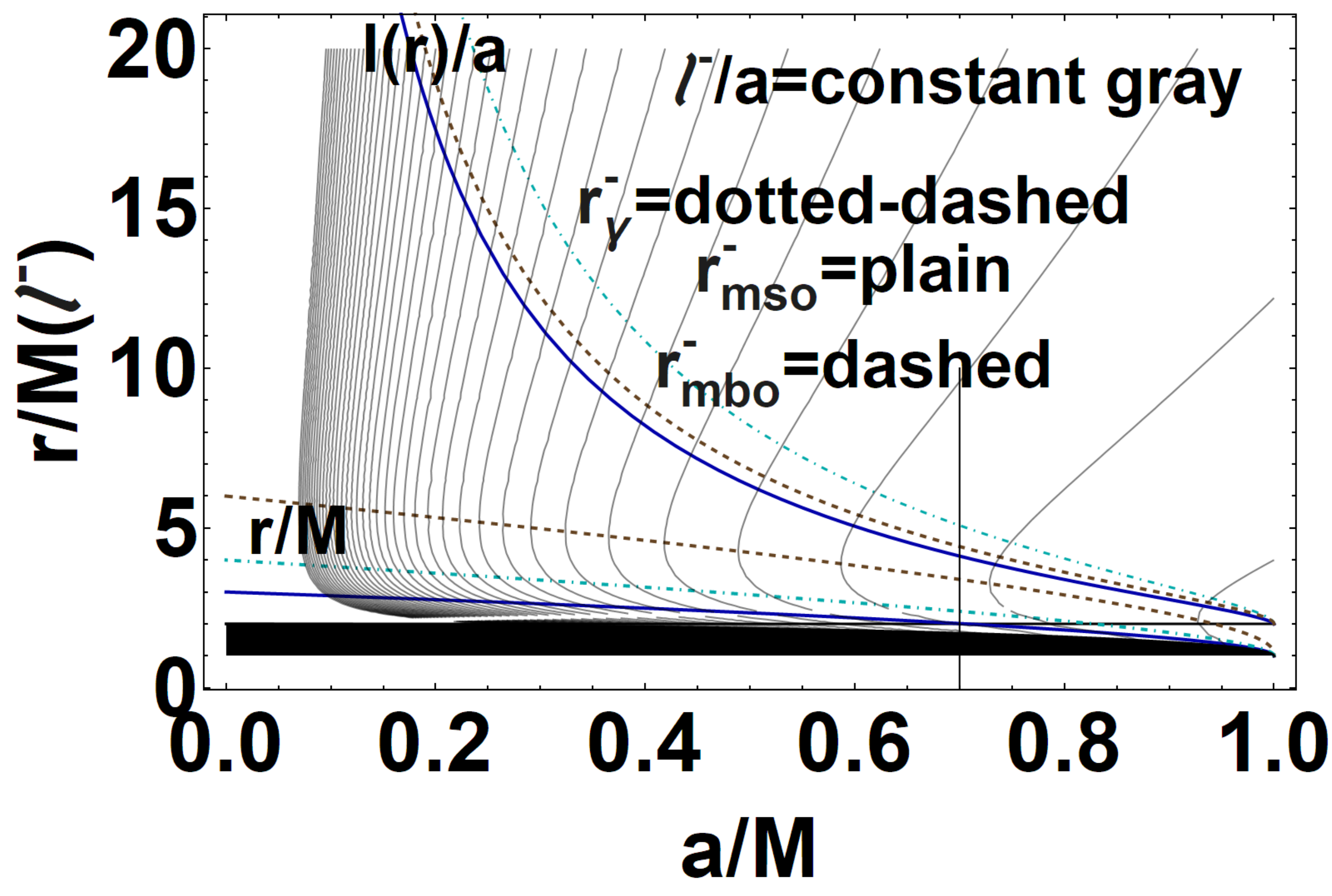}
\includegraphics[width=5.6cm]{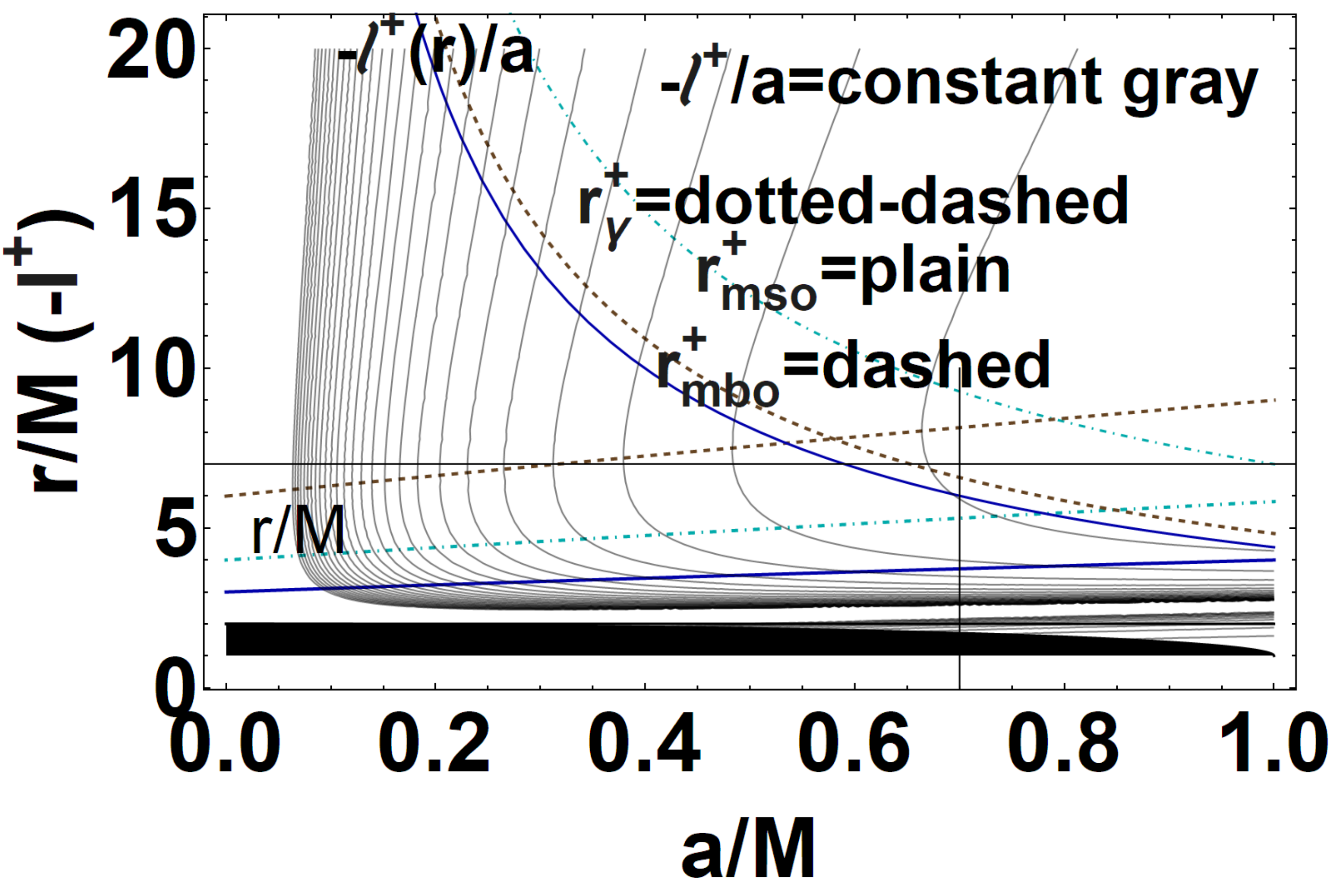}
  \caption{Plot of the geodesic structure  $r/a$ (left upper  panel) and $r/M$ (center  upper  panel) as function of the \textbf{BH} dimensionless spin $a/M$, where $r$ is the marginally stable orbit (mso), marginally bounded orbit  (mbo), and the photon orbit (or marginally circular orbit) $r_{\gamma}$ for corotating $(-)$  and counter-rotating $(+)$ fluids. Black region is the \textbf{BH} $r<r_+$ where $r_+$ is the outer horizon.
Gray strip  is the outer ergoregion.
Purple region is region where cusps of accreting corotating  tori are located. Note that some cross  the outer   ergosurface. Similarly    gray region refers to counterrotating fluids.
Upper right panel  and bottom left panel  show the  specific angular momentum  $\mp\ell^{\pm}/a$ for countterotating and corotating  fluids according to the notation  used in other panels. Here   there is $\ell_{\bullet}\equiv \ell(r_{\bullet})$ for any radius $r_{\bullet}$.
Spins $\{a_{h1},a_{h_2}, a_{h3}\}$   (from the crossing of $\ell^{\pm}$ curves) and $\{a_{\beta},a_{\flat},a_{*}\}$  (from the crossing of the geodetic radii curves).
 Bottom center panel and right panel describe corotating and counter-rotating  fluids respectively.
Curves
of specific  angular momentum versus  spin $a$ constant in the plane $r/a$, gray curves, and evaluated on the geodesic curves as signed in the panel, curves
lines on the geodesic radii are also  plotted according to the  notation on the panel.
Each curve is a class of torus/attractor.
}\label{Fig:Boundaryaloudfilla}
\end{figure}
Curve representing classes of torus/attractor according to the ratios $r/a$ and $\ell/a$  are shown in Figs\il(\ref{Fig:Boundaryaloudfilla}).
\begin{figure}
   \includegraphics[width=6cm]{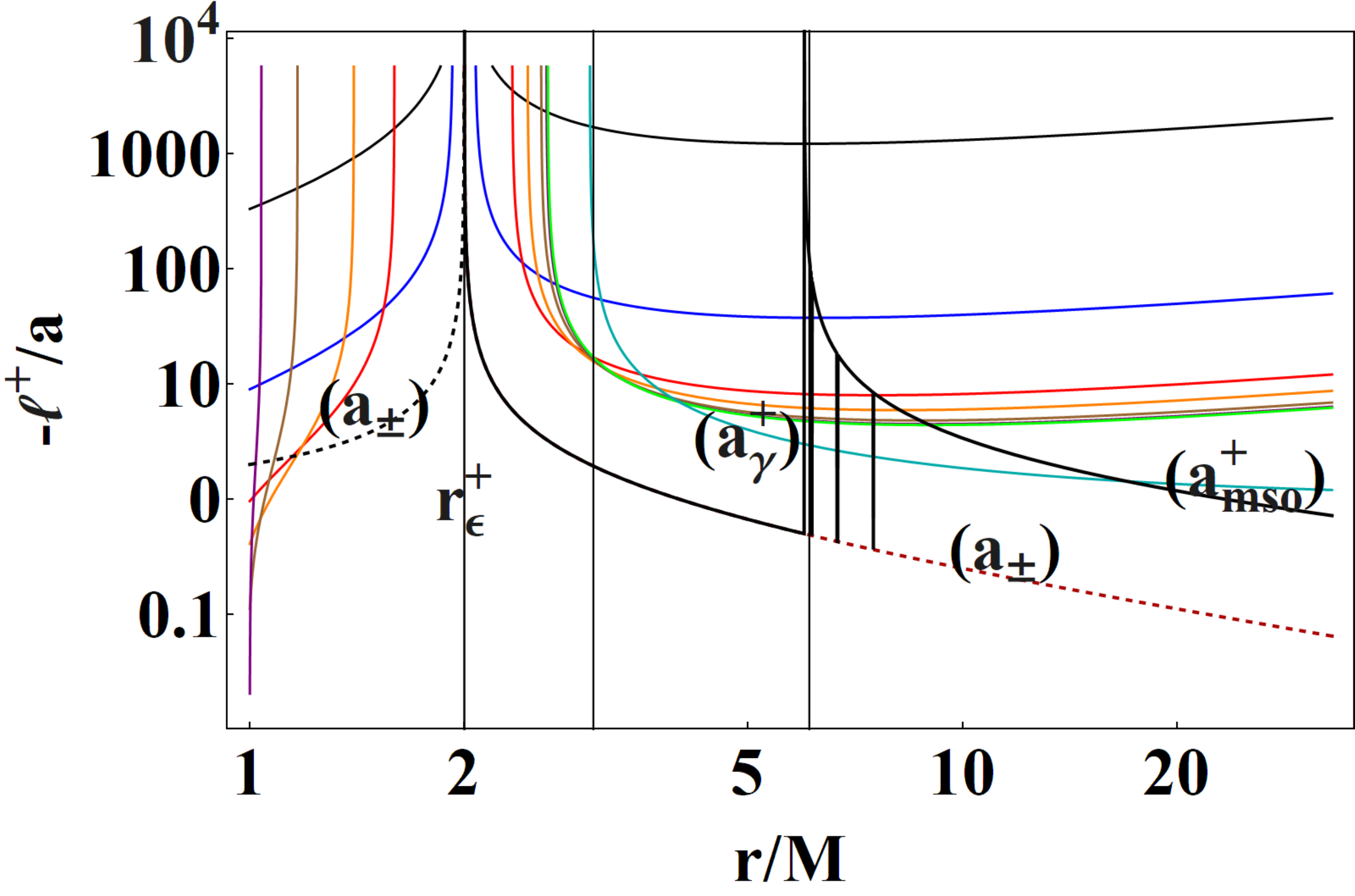}
   \includegraphics[width=6cm]{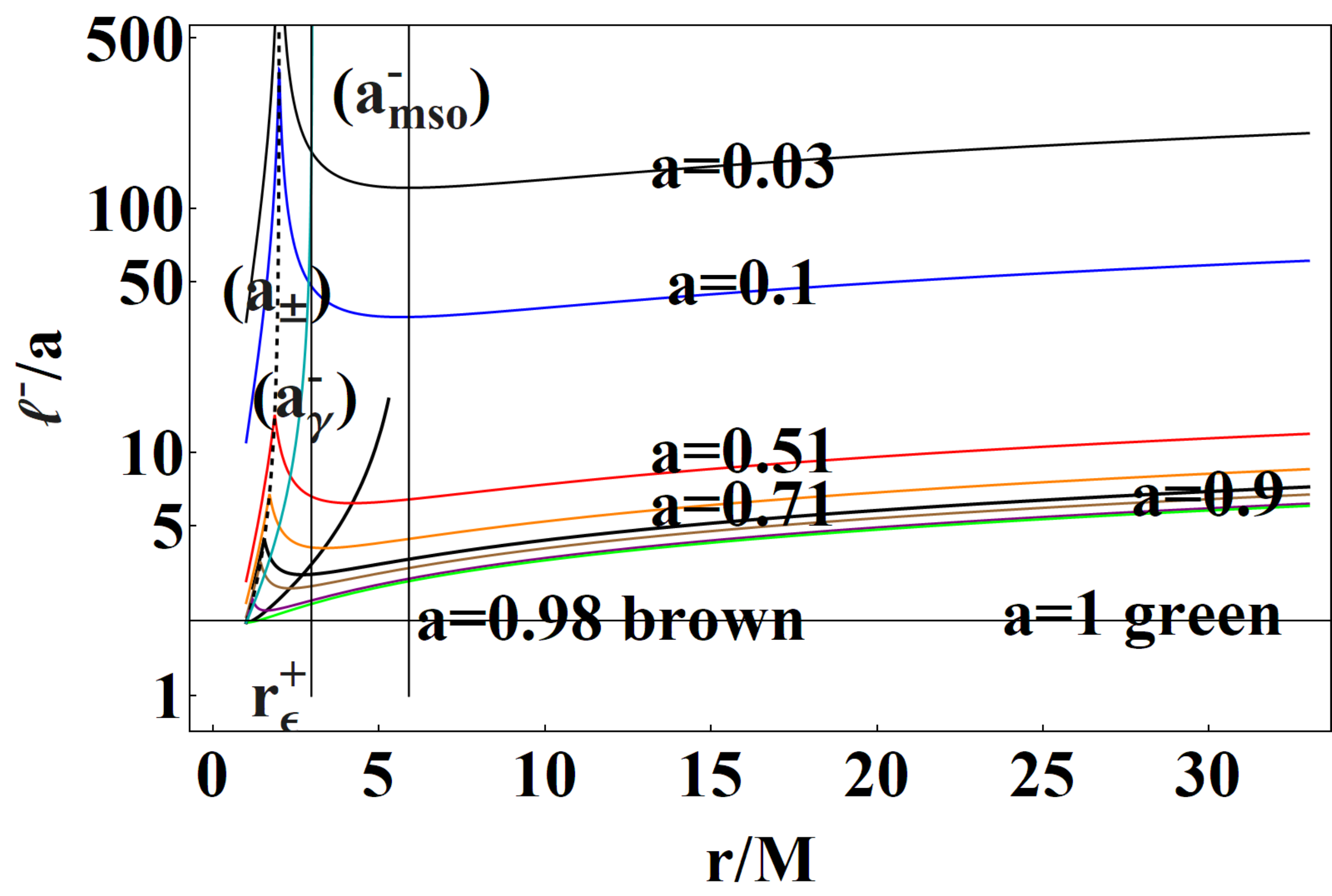}
     \includegraphics[width=5.6cm]{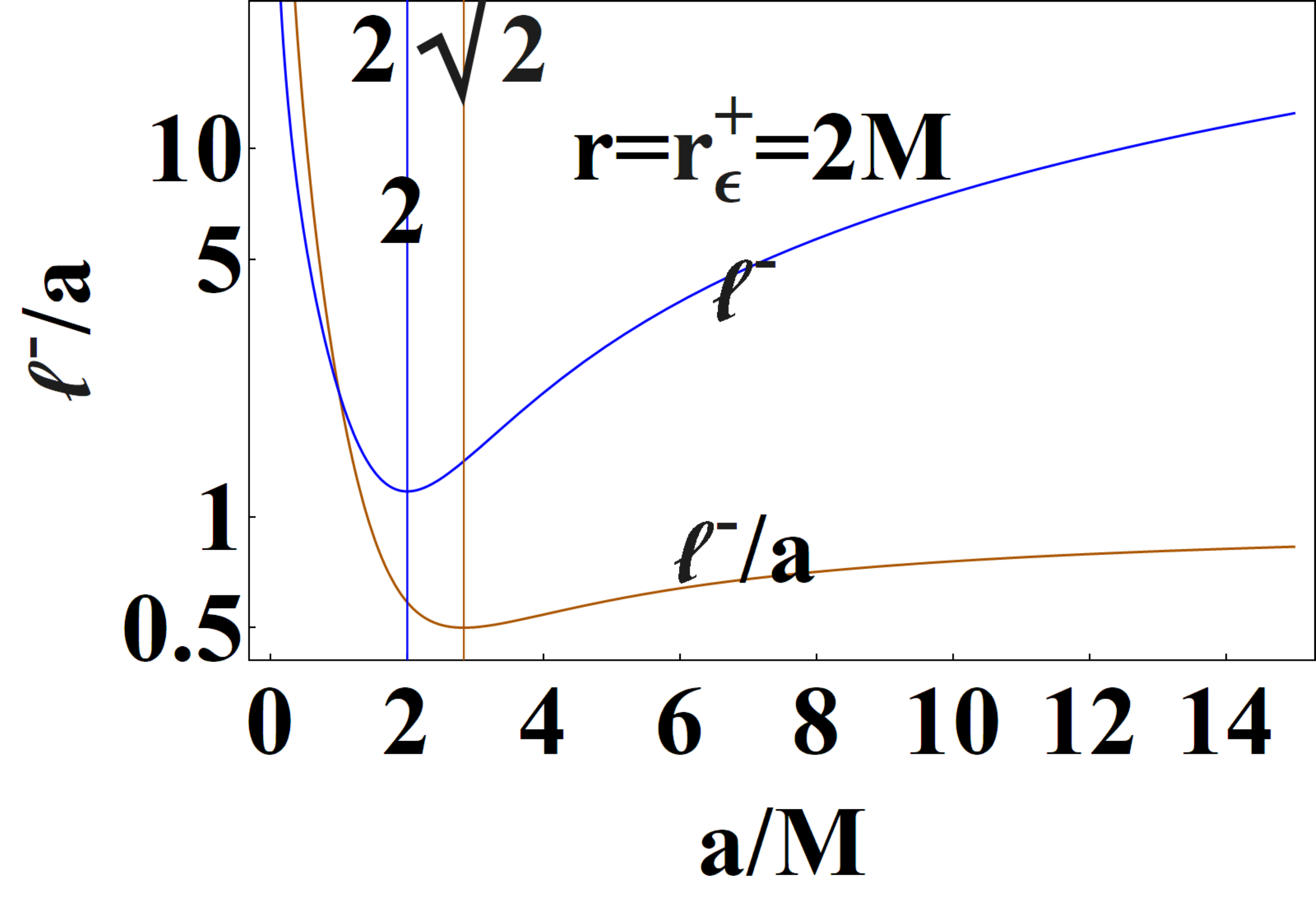}
  \caption{Curves $\mp\ell^{\pm}/a$ for counter-rotating $(+)$ and corotating (-) fluids respectively. Values of spin, signed in the panel close to each curve are  both cases considered in the two panels.  $r_{\epsilon}^+=2M$ is the outer ergosurface on the equatorial plane. $a_{mso}^{\pm}$ and  $a_{\gamma}^{\pm}$, defined in     Eq.\il(\ref{Eq:travseves}), refer to the marginally stable orbits and marginally circular (photon) orbits, $a_{\pm}$ is the horizon curve in the {extended plane}.Where $ \ell(r)$ is the rotational law of the  \textbf{eRAD} and curve of maxima and minima of pressure and density in each torus and distribution of constant $\ell$ fluid specific angular momentum identifying the torus.
Left panel follows the colors association of the  right panel.
Ratio  $\ell/a $  for $r/a$  for different  spins values.
In the counter-rotating  case   the inner ergoregion is shown in the extended plane.
Right panel:  $\ell/a$ and $\ell$,  on the ergosurface $r_{\epsilon}^+=2M$ on the equatorial plane $\sigma=1$. Minimum are for spins  $a=2M$  and $a=2 \sqrt{2}M$.}\label{Fig:travseves}
\end{figure}
We mention here
the extreme case of ``steady'' fluid  with respect to the central object, in other words the cases $\bar{\ell}=1$, or the counter-rotating  case  $\bar{\ell}=-1$. With respect to these limiting conditions, the  corotating and  counter-rotating configurations are constrained by the  range  $\bar{\ell}<1$ as  shown in \cite{pugtot}. The critical points for the pressure can be only at $\bar{\ell}<-1$, therefore only counter-rotating  tori configurations are allowed with $\bar{\ell}<1$.
As the ergosurface, independently from the spin, is located at $r_{\epsilon}^+= 2 M$,  the ratio $\ell^-/a$ has a maximum on the stationary limit on $a =2M$, correspondent to a \textbf{NS}  where  there is limiting $L = a E = 2 E$ (see analysis  \cite{observers,ergon}). The  bundle  with origin in $a_0=2M$ on the equatorial plane is tangent to the extreme Kerr \textbf{BH} on the extended plane, where the characteristic frequency is $\omega=1/2$,
 corresponding to the maximum rotational energy  $\xi=\xi_\ell$, while maximum for
 ratio  $\ell/a$  is at $\ell/a=2 \sqrt{2}$.


\section{discussion and concluding remarks}\label{Sec:dic-con}
Black holes are the powerful  engines of  the active galactic nuclei.  A very relevant question of the  \textbf{BH} accretion physics in this context  is  determination of the characteristic  \textbf{BH} parameters, its spin $a/M$ and mass $M$. This issue is  certainly a preponderant aspect of the physics of accretion and jet emission, such that the \textbf{BH} accretion  history  can be traced in the  evolution of its  spin.

Variation  of \textbf{BH} mass and spin parameters can be the results of   combination of   phases of mass growth  from
 accretion and  spins-up processes,  or rotational energy  extraction, eventually constituting  a spin-down mechanism. The \textbf{BH} spin evolution may consist in several
periods of  episodic accretions  or  of a continuum  prolonged accretion regime.
The spin of \textbf{SMBHs} traces the angular momentum of the
accreted material, tracing consequently the  \textbf{SMBH} feeding.
The accretion processes   may be evident  also by the presence of accretion disk misalignment.
In these processes, the
 presence of fluid viscosity and the  Lense-Thirring effect enhanced by the  \textbf{BH} rotation affect the disks alignment evolution, empowering the
\textbf{BH}  spin evolution (see for example role of the Bardeen--Petterson
effect), and constituting  also a possible mechanism   for  the  origin  of the  inner discrete  inhomogeneous \textbf{eRAD}   structure.

Methods to  provide an evaluation of mass and spin of a \textbf{BH} are usually  highly model (and process) dependent.
   The   establishment of the location of inner edge of accretion disks, or  the analysis of the   jet/radio power of radio-loud
\textbf{AGNs}\cite{Daly0,Daly2,Daly3,GarofaloEvans}, based on a correlation between the beam power and spin are  reliable   methods.
As discussed in Sec.\il(\ref{Sec:e-complex-asly}), we can detect  the \textbf{BH}  spin evolution
through the  efficiency of the energy extraction.

 The approach   we consider here relates  the \textbf{BH} spin "states" prior and after a (complete) rotational energy extraction \cite{Daly:2008zk}. A mayor  characteristic of this approach is  the fact that it   is essentially  model-independent, in the sense that it does  not depend  on the  details of the process of the  energy
extraction, considering exclusively the   final and initial \textbf{BH} states where its  spin  is \textit{only} affected by the  extraction of the reducible
\textbf{BH} mass, according to the \textbf{BH} thermodynamical laws. The rotational
energy extraction parameter $\xi$ is defined by the ratio between the total energy ($M$-ADM mass) and the \textbf{BH} irreducible mass $M_{irr}$.
Therefore, we proceed to study the  \textbf{BH} thermodynamical processes related to  a series of high energy  phenomena affecting  the \textbf{BH} accretion system, combined in the \textbf{RAD} model of tori aggregates.
Studying the \textbf{BH}  mass and
the outflow energy,  we obtain
a lower limit on the  \textbf{BH} spin.

In this   analysis we related three diverse  topics of \textbf{BH} physics: accretion physics, the \textbf{BH} extraction process energy  and the metric bundles concept grounded on the light surfaces definition.  Features of the  extraction processes with \textbf{BH} thermodynamics   introduced in \cite{Daly0},  have been discussed and revisited through metric  bundles  in  Sec.\il(\ref{Sec:both-s}),  and related to  accretion physics  in Sec.\il(\ref{Sec:e-complex-asly}).  The dimensionless  $\xi$   is evaluated for the \textbf{BH} initial  ADM mass, relating the spin  to the mass evaluation for the initial  state of  the  \textbf{BH}.
In this analysis we made use of some simplificative assumptions which we  plan to  focus on in future analysis. Firstly, we assumed a process  leading to \emph{total} rotational energy extraction; on the other hand, contribution of  mass feeding processes has been neglected   especially in the evaluation of the accretion process, eventually not considering the contribution of accretion to the \textbf{BH} mass, or  the  so-called runaway instability that  is essentially an  iterative process  between \textbf{BHs} and disks.
We showed the relation between the rotational law of the tori, the characteristic frequency of the  bundle and the relativistic velocity defining the von Zeipel surfaces, in Section\il(\ref{Sec:lona}).

Black hole  luminosity is another  relevant variable for the \textbf{BH} spin evolution, specifically, if the \textbf{BH} is   accreting at near Eddington luminosity, which may imply  super Eddington accretion rates for  the accreting disks,  as  it is the case for the thick tori considered here.
     Luminosity and accretion rates  should reflect the  \textbf{\textbf{BH}} spin-up  evolution:   the  \textbf{BH}
may have an  initial state of super-Eddington luminosity,  followed by   a phase  characterized by lower luminosity.
The standard accretion disk theory prescribes that the radiative
efficiency of energy conversion in such processes  is  related  to the \textbf{BH}  spin, because it is  related to the location of inner edge, established by the presence of a  tori  cusp of the accreting configuration. In the \textbf{eRAD}  model we discussed  the connection between the specific angular momentum of the disk, which is a parameter related to the point of maximum pressure in the torus, and the \textbf{BH} spin.

The  binding energy   of the  orbiting matter is radiated leaving the  inner edge,  matter freely falling  into the hole, assuming  the total amount of energy converted into
radiation being  binding energy.
There is therefore a relation between the   accreting material angular momentum and  the radiative efficiency. We specialize this relation in Sec.\il(\ref{Sec:e-complex-asly}).

In Sec.\il(\ref{Sec:lona}), we discussed  for  \textbf{RAD}  the  \textbf{BH} spin-accretion disk  correlation.  The    attractor-ringed-disk system shows various  symmetry properties  with respect to the quantities  $\bar{\ell}=\ell/(a \bar{\bar{\sigma}})$ and  $R=r/a$, where $\bar{\bar{\sigma}}=\sin \theta$ \cite{pugtot}.  Noticeably, many  properties of the \textbf{RAD} depend mainly on the rationalized specific angular momentum  $\bar{\ell}$.
 There are  relevant spins  $a\approx 0.94M$, $a\approx 0.7M$ and
 $a\approx 0.3M$. We show from the analysis of ratios $\ell/a$,  the formation of corotating tori is bounded to  $\ell/a\geq2$,  counter-rotating tori to  $\ell/a\leq -22/5$.

At fixed specific angular momentum $\ell$,  the zeros of the $\mathcal{L}\cdot\mathcal{L}$, as functions of $\ell$,  define  \emph{limiting surfaces} of  the fluid configurations (light surfaces for solutions in terms of $\omega$).
For fluids with specific angular momentum $\ell \in \mathbf{L_3}$, the limiting surfaces are  the cylinder-like  surfaces  crossing the equatorial plane on a point which is  increasingly far from the attractor with increasing  specific angular momentum magnitude. A second surface, embracing the \textbf{\textbf{BH}}, appears,  matching  the outer surface at the cusp $r_{\gamma}$.
For  $\ell=\ell_{\gamma}$, a cusp appears together  with an  inner surface closed on the \textbf{BH}.
A further remarkable aspect of the application of the metric bundles  approach to the analysis of the \textbf{BH} energetics, consists in the use a  \textbf{NS}-\textbf{BH} correspondence to predict  observational characteristics of the \textbf{BH} accretion disk systems, through the extension of the geometry in the naked singularity region in the extended plane, since  the metric bundles, tangent to the horizons curve in the extended plane, are in part contained in the \textbf{NS} region.
Results of the analysis of metric bundles are  in Figs\il(\ref{Fig:effct},\ref{Fig:fas2},\ref{Fig:Combplot1},\ref{Fig:replicas2},\ref{Fig:minneaPlotFlo},\ref{Fig:Bundleplot1},\ref{Fig:Bundleplot}).
Metric bundles are defined in the extended plane, considered in Figs\il(\ref{Fig:planeRTexteprop}) in terms of the rotational energy. Metric bundles are curves tangent to the horizon curve in the extended plane. The tangent point changes  along the horizon curve during the process of rotational energy extraction. The characteristic frequencies of the bundle provide limits to the angular momentum of the fluid, and can be measured with the replicas orbits.

A further question regarding the measurement of the \textbf{BH} spin emerges from the fact that  the applied  methods are related  to phenomena in the very
vicinity of the horizon, where the general relativistic effects are  predominant. We explored these conditions adapted to the metric bundles  concept  related to definition of the light surfaces.

{This analysis informs of the overall situation before any occurring process involving disks and central \textbf{BH}.
 All the quantities are evaluated at the state $(0)$ prior the process, therefore all the quantities evaluated here inform on the status of the \textbf{BH}-accretion disk system at its initial state (0), by evaluating  the related $\xi$ parameter.
We proceed with the characterization  of the state  $(0)$  in terms of the  \textbf{eRAD} structure in relation with the rotational energy extracted, characterized by a determined inner composition of tori, relatively to  fluid corotation or counter-rotation, or the associated topological state with the instauration of  certain unstable phases or the position of the  inner edge and  certain characteristics of the energetics of the accretion.
However,  the analysis describes  the state prior  the extraction processes,   considering the rotational energy of the  \textbf{BH} unaltered.
Results of the analysis  of the tori morphological and topological characteristic have been   analysed in Fig\il(\ref{Fig:rightse-s}), (\ref{Fig:ManyoBoFpkmp1}), and (\ref{Fig:quesa}).
The
energetics is considered in Figs\il(\ref{Fig:ManyoBoFpkmp}), Figs\il(\ref{Fig:ManyoBoFpkmpMig}).  The conditions for an \textbf{eRAD}  with $\ell$counterrotating tori is explored in  Sec.\il(\ref{Sec:setsoftori}), and in Figs\il(\ref{Fig:sccreemzero}), Figs\il(\ref{Fig:Boundaryaloudfilla}). This analysis  individuates the spins $\{a_{*},a_{\beta},a_{\flat}\}$ and $\{a_{h1},a_{h2},a_{h3}\}$  significant for the conditions on the fluids in the initial configurations of tori.
Concerning how the accretion disks and particularly the part of inner disk and inner edge   affects this process we refer for example to \cite{Camezin}.  The inner edge (cusp) is constrained in Figs\il(\ref{Fig:sccreem}) in terms of the spin, as  the center of maximum pressure in the  torus for corotating and counterotating fluids,  in
Figs\il(\ref{Fig:rightse-s}) in terms of the parameter $\xi$.  The cusp is in the range $r_{\times}\in]r_{mbo}^{\pm},r_{mso}^{\pm}[$ and it is constrained in Figs\il(\ref{Fig:ManyoBoFpkmp1})- Figs\il(\ref{Fig:Boundaryaloudfilla}),
in terms of the specific angular momentum. In these processes  we should consider for the thick disks the instauration of the runaway process, and in the case of \textbf{RADs} the runaway-runaway process.
 Runaway   instability affecting thick tori and their \textbf{BH} attractors implies large accretion rates  typical of these tori-- the \textbf{BH} mass  increases changing  the spacetime properties, and  in turn  affects   the orbiting accreting   disk. In this process,  the inner edge moves inwards,   or  the cusp moves
 inside the disk inducing  increase of mass transfer.
  In the \textbf{RAD} system one may consider the possibility of Runaway-Runaway instability, originating  when the Runaway instability affecting the  \textbf{BH} and the inner accreting  torus of the agglomeration,  is accompanied by the consequences of the background modification on the outer tori of the \textbf{RAD} which can collide, accrete or stabilize in dependence  on  initial conditions \cite{dsystem}. It is    the combination of runaway instability, involving the inner edge of the inner accreting torus of the \textbf{RAD}, with the consequent destabilization of the aggregate.
The accretion   modifies the  inner torus morphology and the background geometry having  repercussions in the whole  \textbf{RAD} structure establishing a sequence of events having different possible outcomes.}

{More specifically, characteristic features of different accretion disk models   outlining    the  torus interior structure,  the mechanism of  energy transport  in the disk,  and more generally  the different accretion dynamics  might affect  the  energy extraction  processes  from the central \textbf{BH}, constraining eventually  the  assumptions on the tori-\textbf{BH} model adopted here.
The magnetic fields and, in  particular contexts, the quantum effects of extraction in the   \textbf{BH}   vicinity are among the most relevant mechanisms that can interact in the extraction process, requiring  a different characterization of the energy outflow.
 Together with these effects however, there can be  a different setup  on the initial and final state of the \textbf{BH} accretion system, a different outflow contribution and the establishment  of special instabilities which can change the \textbf{BH}  disks system as the runaway instability   or the change in spin orientation due to  Bardeen--Petterson effect.
 More specifically, in  the jets emission, characterizing \textbf{SMBH} also in \textbf{AGNs},
the frame dragging for the spinning  attractor is thought to play an important role, the   Lense--Thirring effect induced by the  central attractor  can engine also
  the Bardeen--Petterson effect  (a process resulting in the  tearing up  of the orbiting disk)--\cite{BP75}). The change in the disks structure can  induce a spin shift in magnitude and orientation of the \textbf{BH}.
The warping and twisting of the disks,   the Bardeen--Petterson  effect, depends  on the   frame-dragging  and the disk   viscosity, the presence of magnetic fields,    and  the original disk inclination with respect  to the  orbital plane of a star companion.  There can be a    final steady state of the  initially misaligned  torus   in a   coplanar accreting  inner  torus  with the outer part of the original   disk
  aligning  more slowly on the longer timescales.
  These  misaligned disks are expected to be characteristics of the transient periods of the \textbf{BH}--accretion disk evolution.
Bardeen-Petterson effect is also  proposed as a possible  cause of  the counter-alignment of \textbf{BH} and
disk spins. In fact   \textbf{BH} can grow rapidly  if
they acquire most of accreting mass  in a sequence of randomly oriented accretion episodes
\cite{King:2006uu,King:2013vva}.
The chaotical
accretion  in \textbf{AGNs} could produce counterrotating accretion
disks or  strongly misaligned disks with respect to the central \textbf{SMBH} spin, and
  misaligned disks evolution orbiting a Kerr
\textbf{BH} might lead to a tearing up of the  disk into several   planes with different inclinations with a precession of the BH spin.
  The
\textbf{BH} spin  changes  under the action of the disk torques, as the disk, being
subjected to Lense-Thirring precession,  becomes  twisted
and warped-- \cite{Lodato:2006kv}, therefore it is possible to study the aligning  of Kerr \textbf{BH}s and accretion disks  \cite{King:2005mv}.
or the
effects of \textbf{BH} spin on misaligned accretion disks in relation  to  the role of the disk inner edge and the alignment of the angular momentum  with the \textbf{BH} spin-- \cite{Nealon:2015jya}.
The method used here of measuring \textbf{BH}  spin is largely not  model dependent, however it is based on the  hypothesis  that
the energy  outflow to be measured is  powered by the \textbf{BH} spin energy only, therefore
the outflow energy is taken as a lower bound on the \textbf{BH} spin energy.
More generally  the  extracted energy can be determined considering the ratio between the  outflow energy  versus   the spin energy. In here we considered the two quantities be coincident, being based on the  suggestion  that the jets  beam power can be
directly related to  \textbf{BH} spin, linking
jet power and  accretion tori with  jet mechanism  supposed to be  originated in the accretion disk with  magnetic fields, and
extraction of the \textbf{BH} spin energy powering  the  outflow from the hole.
More precisely we consider the  dimensionless ratio  between the released energy  (coincident with the rotational energy of the initial \textbf{BH} state) to the \textbf{BH}  mass (ADM total mass of the initial \textbf{BH} state).
We assume that all the   \textbf{BH} rotational   energy is extracted,  and the spin changes only by extraction of the reducible
\textbf{BH} mass.
The energy outflow can be empowered however  from the
 the accretion of disk material  and separately  the rotational energy of the black hole itself.
We assume that, from an initial  Kerr  \textbf{BH}, the final stage of extraction process  is a  static spherically symmetric  Schwarzschild spacetime.
The hypothesis is that all the outflow is induced by the rotational energy extraction leaving constant (and not increased) the \textbf{BH} irreducible mass.
 A divergence from this hypothesis  can  alter  the analysis of
 Figs\il(\ref{Fig:ManyoBoFpkmpMig}) which do not consider the contribution of the mass and momentum  to the \textbf{BH}  subsequent  to accretion and in the case of runaway- instability.
In fact  at the  initial and final state the
 \textbf{BH}-accretion disk system which  is stable  (at least for few seconds), the runaway instability  typical of \textbf{SMBHs} and thick tori  could undermine this situation, the
runaway- instability affecting  the  structure of the torus inner  part can be considered as a
 purely
dynamical effect, meaning that  time-scale is extremely short
 occurring therefore before the start of processes as the viscous transport of angular momentum in the disk in the accretion model supported by such processes.
   To consider properly these aspects, having fixed the details of the process, a sequence of steady states should be considered, re-evaluating  the disks mass and momentum (consequently its inner edge)  and the \textbf{BH} spin and mass altered by the accretion.
When an accretion disk  is subjected to the runaway instability, a large portion of its
mass  fall into the \textbf{BH} within a few dynamical time.
Furthermore  geometrically thick disks, which are particularly subjected to runaway- instability, have large accretion rates (with super Eddington luminosity) and especially for the largest tori (great corotating specific angular momentum or for the counterotating tori far from the \textbf{BH}) may also have a relevant self gravity. For small \textbf{BH} spin the corotating tori matter inflow may have also an initial specific momentum relatively large.
The divergence of the radial mass transfer  expected as an outcome of  the runaway- instability clearly can lead to the disk destruction.
The runaway instability of such systems  is based on the following mechanism:  accretion from the disk induces an increase of  the \textbf{BH} mass, consequently the  geometry background is affected and resettles in a new state. As a consequence of this, the
accretion disk can also  never reach a  completely steady state.
Considering therefore  the case of disks increasing the  \textbf{BH} mass two faces for critical and overcritical accretion from thick disk are possible  i.e.  the disk  cusp can move inwardly,
slowing the mass transfer and consequently reaching eventually a
 stable situation or,  alteratively, the cusp can move outwardly   in
 the disk, the mass transfer increases  (in velocity),
leading to the runaway instability.
Further factors  have been neglected  in this model and may be relevant also for the combinations with the  runaway  instability the establishment   of such  instability and  its outcomes, as the  tori self-gravity seeming to favor the instability, the black hole rotation which can  have  a stabilizing effect on the runaway instability, and
finally a different rotational law for the torus with a non-constant distribution of the angular momentum (which might have a stabilizing effect in relation to runaway instability).}

\acknowledgments
The authors acknowledge the  Institute of Physics of Silesian University in Opava.
ZS acknowledges the Czech Science Agency grant No 19-03950S.

\appendix
\section{Explicit solutions   $\ell^\pm(\xi,r)=\ell$}\label{Sec:mislao}
A convenient parametrization of solutions  $\ell^\pm(\xi,r)=\ell$ are as follows.
See also Figs\il(\ref{Fig:ManyoBoFpkmpMig}).
For corotating fluids, there is for $\ell^-(\xi,r)=\ell$
\bea
\\&&
\xi = \frac{1}{2} \left(2\mp\sqrt{2\natural\sqrt{\Psi^{+}_{\beta}-2 \Psi^-_{\alpha}+4}}\right),
\quad
\xi = \frac{1}{2} \left(2\mp\sqrt{2\natural\sqrt{\Psi^{-}_{\beta}-2 \Psi^+_{\alpha}+4}}\right),
\\
&&
\xi = \frac{1}{2} \left(2\mp\sqrt{2\natural\sqrt{\Psi^{+}_{\beta}+2  \Psi^-_{\alpha}}}\right),
\quad
\xi = \frac{1}{2} \left(2\mp\sqrt{2\natural\sqrt{\Psi^{-}_{\beta}+2  \Psi^+_{\alpha}}}\right),
\eea
where $\natural=\pm$ and
\bea
&&
 \Psi^{\pm}_{\alpha}\equiv\sqrt{\pm4 \ell^3 r^{3/2}-16 (r-1) r^2+4 \ell^2 (3 r-2) r+\ell^4},\quad   \Psi^{\pm}_{\beta}\equiv\pm4 \ell r^{3/2}+4 (r-2) r-2 \ell^2.
\eea
For counter-rotating orbits $\ell^{+}(\xi,r)=\ell$, we have
\bea
\\&&
\xi = \frac{1}{2} \left(2\pm\sqrt{2\natural\sqrt{\Psi^{+}_{\beta}-2  \Psi^{-}_{\alpha}+4}}\right),
\quad
\xi = \frac{1}{2} \left(2\pm\sqrt{2\natural\sqrt{\Psi^{-}_{\beta}-2 \Psi^{+}_{\alpha}+4}}\right),
\\&&
\xi = \frac{1}{2} \left(2\pm\sqrt{2\natural\sqrt{\Psi^{+}_{\beta}+2  \Psi^{-}_{\alpha}}}\right),
\quad
\xi = \frac{1}{2} \left(2\pm\sqrt{2\natural\sqrt{\Psi^{-}_{\beta}+2 \Psi^{+}_{\alpha}}}\right).
\eea
(See Figs\il(\ref{Fig:ManyoBoFpkmpMig}).).
 Using the approach introduced with the bundle on the extended plane, featuring classes of \textbf{BH} solutions, we found the spin functions for this problem for corotating and counter-rotating fluids,
\bea
&&a_1\equiv\frac{1}{2} \left[-y_{i}^{-}+\ell-2 \sqrt{r}\right],\quad a_2\equiv\frac{1}{2} \left(y_{i}^{-}+\ell-2 \sqrt{r}\right),\quad a_3\equiv\frac{1}{2} \left(-y_{i}^{+}+\ell+2 \sqrt{r}\right),\quad a_3\equiv\frac{1}{2} \left(y_{i}^{+}+\ell+2 \sqrt{r}\right),
\eea
 where there is
\bea
&&
y_i^{\pm }\equiv\sqrt{\left[\ell^2\pm 4 \ell(r-1) \sqrt{r}\right]-4 (r-1) r},\quad y_{ii}^{\mp }\equiv\sqrt{\left[4 \ell r^{3/2}\mp 2 y_{i}^{-} \left(\ell+2 \sqrt{r}\right)\right]-2 \ell^2+4 (r-1)^2};
\\&& y_{iii}^{\mp }\equiv\sqrt{4 (r-1)^2-2 \left[\left(\ell^2\mp \ell\left[y_{i}^{+}-2 r^{3/2}\right]\right)-2 \left(\sqrt{r} y_{i}^{+}\right)\right]}.
 \eea
 These solutions are for $r=y$, $y$ being the Cartesian flat coordinate on the equatorial plane of the Kerr \textbf{BH} coincident with the symmetry plane of the torus.
Therefore, rearranging the terms to make explicit  for corotating $(-)$ and counterotating $(+)$ tori we have
\bea
&& a_1:\quad (+)\quad\xi =1\pm\frac{1}{2} \sqrt{2-y_{ii}^{+}};\quad(-)\quad \xi =1\pm\frac{1}{2} \sqrt{y_{ii}^{+}+2},
\\%
&&a_2:\quad (+)\quad\xi =1\pm\frac{1}{2} \sqrt{2-y_{ii}^{-}};\quad(-)\quad \xi =1\pm\frac{1}{2} \sqrt{y_{ii}^{-}+2},
\\
&&a_3:\quad (+)\quad\xi =1\mp\frac{1}{2} \sqrt{2-y_{iii}^{-}};\quad(-)\quad\xi =1\pm\frac{1}{2} \sqrt{y_{iii}^{-}+2},
\\
&&a_4:\quad (+)\quad \xi =1\pm\frac{1}{2} \sqrt{2-y_{iii}^{+}};\quad(-)\quad\xi =1\pm\frac{1}{2} \sqrt{y_{iii}^{+}+2}.
\eea
\section{Explicit solutions for light-surfaces}\label{Sec:ls-explicit}
The zeros of $\laa\cdot \laa=0$, for the Killing field $\laa=\xi_t+\omega\xi_{\phi}$, defines  the light surfaces $r(\sigma; a,\omega)$ when solved for the radius $r$ and the metric bundles $a(\omega;r,\sigma)$.
Explicit solutions for the light surface at any plane $\sigma\equiv \sin^2\theta$ are the four radii:
\bea&&\label{Eq:lad}
r_{\mathbf{u}_\bullet}^{\pm}\equiv\frac{1}{2} \left(\mathbf{u}_\bullet\pm \sqrt{-\frac{4 (a \sigma  \omega -1)^2}{\sigma  \mathbf{u}_\bullet \omega ^2}-\mathbf{bb}_{\pi }-c_{\bullet}}\right),\quad \mathbf{u}_\bullet\equiv\{+\mathbf{uu}_\bullet,-\mathbf{uu}_\bullet\},\quad \mathbf{uu}_{\bullet}\equiv\sqrt{c_{\bullet}-\mathbf{bb}_{\pi}},
\\&& c_{\bullet}\equiv \frac{1}{3} \left[2 \sqrt{\mathbf{bb}_{\pi }^2-3 \mathbf{cc}_{\pi }} \cos \left(\frac{1}{3} \cos ^{-1}\left(\mathbf{dd}_{\pi }\right)\right)+\mathbf{bb}_{\pi }\right],\\&&
\mathbf{dd}_{\pi }\equiv\frac{36 a^2 \mathbf{bb}_{\pi } (\sigma -1) \sigma  \omega ^2 \left(a^2 \sigma  \omega ^2-1\right)+54 (a \sigma  \omega -1)^4+\mathbf{bb}_{\pi }^3 \sigma ^2 \omega ^4}{\sigma ^2 \omega ^4 \left(\mathbf{bb}_{\pi }^2-3 \mathbf{cc}_{\pi }\right){}^{3/2}}.
\\
&&\mathbf{cc}_{\pi }=\frac{4 a^2 (\sigma -1) \left(a^2 \sigma  \omega ^2-1\right)}{\sigma  \omega ^2},\quad\mathbf{bb}_{\pi}\equiv\frac{-a^2 (\sigma -2) \sigma  \omega ^2-1}{\sigma  \omega ^2},
\eea
where $a_0\equiv 1/\omega \sqrt{\sigma}$ is the bundle origin spin--Figs\il(\ref{Fig:HonRingZN}).
On the equatorial plane, $\sigma=1$, light surfaces are the radii
\bea&&
r_1\equiv\frac{2 \mathbf{u}_k \cos \left[\frac{1}{3} \cos ^{-1}\left(-\mathbf{u}_x\right)\right]}{\sqrt{3}},\quad
r_2\equiv\frac{2 \mathbf{u}_k \sin \left[\frac{1}{3} \sin ^{-1}\left(\mathbf{u}_x\right)\right]}{\sqrt{3}}
\quad\mathbf{u}_k\equiv\sqrt{\frac{1}{\omega ^2}-a^2},\quad \mathbf{u}_x\equiv
\frac{3 \sqrt{3} \omega ^2 \mathbf{u}_k}{(a \omega +1)^2}.
\eea
 The analysis through bundles  connects classes of spacetimes with equal limiting photon frequencies in some points, replicas in the same spacetimes, with the classes of \textbf{RAD} disks which we can expect in different phases of their evolution.
Any point of the light surface is related to an horizon  in the extended plane. Excluding  the counter-rotating orbits,  for any point $(r,\sigma)$ there is a maximum of two frequencies. One point is  the  outer horizon of the \textbf{BH}   geometry, the second  frequency is also a frequency of the horizon on a geometry  connected by  the bundle curve in the plane, crossing this point, according to the analysis of Sec.\il(\ref{Sec:both-s}).
A detailed study of the  bundles $\ba_\omega$  at fixed frequency  is in \cite{remnant,remnant0,remnant1};  for fixed frequency and plane $\ba_{\omega \sigma}$ is  prevalently studied in
Sec.\il(\ref{Sec:both-s}) on the equatorial plane  $\sigma=1$ relevant for the \textbf{eRAD} case.
There is a  detailed discussion of the class of bundles  $\{\mathcal{B}_{\omega}\}_{\sigma}$ with frequencies $ \omega \in [\omega_H^-,\omega_H^+] $ and $\sigma \in [0,1]$
(note there is also  the range    $\sigma=[-1,0]$ considered  within the axial symmetry of the metric in \cite{remnant,remnant0,remnant1}).  We considered  $\omega$ as the  horizons frequencies for   $ a>0$, including limiting points of the  extended plane on line  $a=0$  then   ($r=0$, $a=0$)  and ($ r=2M, a=0$).
Transformations from $\mathcal{B}_{\omega,\sigma}$  to a  $\mathcal{B}_{\omega,\sigma_1}$  relate different (couples of)  points of the same spacetime  (horizontal line of the extended plane) and different spacetimes related by transformations  along the   bundle curve,
 at fixed initial state  $0$ (with parameters  $a_0, M_0$)  of final of a  \textbf{BH} transition
   (with parameter  $a_1, M_1$). In general, at fixed $a=\bar{a}$, the light surfaces are given by the collections of points crossing  the horizontal line  $a=\bar{a}$ in the classes
    $\mathcal{B}_{\omega,\sigma}$ for all values of $\sigma$ and $\omega$. Not all the bundles intersect  a fixed horizontal line (there are frequencies not accessible in a given spacetime, depending on the polar angle),  some are confined in a region of the extended plane as it occurs for a portion of the inner horizon tangent bundles.
	Other bundles are partially contained in the naked singularity region of the extended plane, the bundle tangent  to the extreme Kerr \textbf{BH} spacetime for $\sigma=1$ is entirely contained in the \textbf{NS} sector apart for the tangent point.
 Some bundles cross each other, for corotating photons there is a maximum of two bundles at fixed plane.
The observation of the \textbf{BH}-accretion disks and \textbf{RAD} associated phenomenology  should show traces of the  presence of replicas, Eq.\il(\ref{Eq:replicas2}),
and the relation with the states before and after the rotational  energy extraction, described as   a shift on the horizon curve. Main characteristic quantities can be considered as  function of the energy parameter $\xi$ which can then be detected by the measure of energy outflow.

Under spin transition this light cylinder is deformed. There is a fixed point $p$  on the external horizon in the extended plane  which shifts rigidly (in the sense of \cite{remnant})  on the curve  $a_{\pm}$.  (The \textbf{BHs} horizon is  independent of $\theta$  and thus it  rotates rigidly).
The analysis of bundles in the external plane shows the existence of a collection of points with the same frequency of the horizon
$	\omega_{H}^+$ for $ r_+ $  and $\{(r_s,\sigma_s)\}$   which could serve to relate magnetospheres in the two consequential states\footnote{Concerning the proto-jets emission,  in \cite{proto-jet,open,impacting} the set of \textbf{RAD}  open-cusped  configurations and  open   configurations which are   not related to  the critical points of the effective potential were studied. These   correspond to the solutions  $\Pi=0$  of:
$
\mathbf{(a)}\quad\Pi(\ell)={g_{\phi \phi}+2 \ell g_{\phi t} +\ell^2g_{tt}}$,  $
\mathbf{(b)}\quad \Pi(u^t, u^\phi)=g_{tt} (u^t) ^2+2 g_{\phi t} u^tu^\phi +g_{\phi\phi}(u^\phi)^2=(u^t)^2(g_{tt} +2 g_{\phi t} \Omega +g_{\phi\phi}\Omega^2)[\approx(u^t)^2\mathcal{L}\cdot\mathcal{L}],
$ and  $
\mathbf{ (c)}\quad\Pi(L, E)= {E}^2 g_{\phi\phi}+2 {E} g_{\phi t} L(\ell)+g_{tt}  L(\ell)^2$.
These quantities  are  of course related to the equation  $\laa\cdot \laa=0$  providing solutions of stationary observers.
Thus,  $\Pi$ is related to the normalization factor  $\gamma$ for the stationary observers,  establishing thereby the light-surfaces.  The effective potential,  related to the four-velocity component  $\mathbf{u}_t=g_{\phi t} u^\phi+g_{tt} u^{t}$,
is  not well defined on the zeros of $\Pi(\ell)$.}
 \begin{figure}
\centering
\begin{tabular}{cc}
\includegraphics[width=9cm]{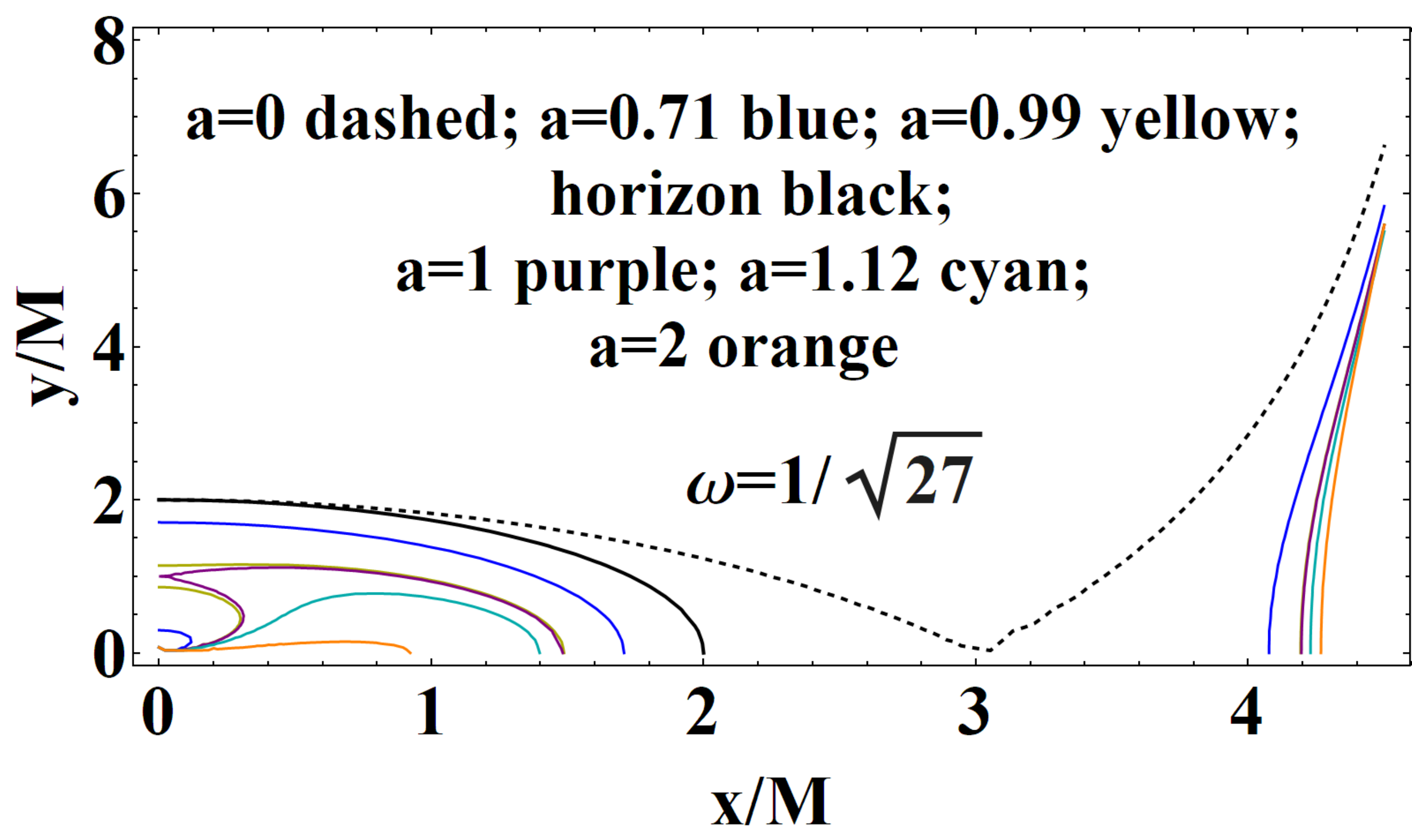}
\end{tabular}
\caption{
Solution of equation $\laa\cdot\laa=0$  for the light surfaces of Eq.\il(\ref{Eq:lad}))   in the plane $(x,y)$, of the central \textbf{BH}, for different \textbf{BH} spin $a$ and for a fixed frequency (different plane $\sigma=\sin^2\theta$, where $r\to \sqrt{x^2+y^2}$, and $\sin (\theta )\to {x}/{r}$ in units of \textbf{BH} mass $M$.).   Different spacetimes (for different dimensional spins of the central \textbf{BH}) are shown from the Schwarzschild $a=0$ to the extreme rotating \textbf{BH} $a=M$ and naked singularity $a>M$. The  frequency is  fixed at $\omega=1/\sqrt{27}$, the collections of all point of the curves is a Metric Killing bundle  corresponding to
$\omega=1/\sqrt{27}$ for different planes $\sigma$. Note the role of photon circular orbit $r=3M$ for the Schwarzschild spacetime, and the double curves at different spins for inner and outer curves for light surfaces. In the extended plane, the class of bundles with frequencies $\omega=1/\sqrt{27}$ because related to the axis $a=0$ and independent  from the parameter  $\sigma$ because of the spherical symmetries of the Schwarzschild metric--see also Figs\il(\ref{Fig:Bundleplot1}).}\label{Fig:HonRingZN}
\end{figure}
\section{Von Zeipel surfaces, metric Killing bundles and  tori}\label{Sec:vonZeipel}
 Here we explore the connection between the von Zeipel surfaces, defined as  $\Omega=$constant, metric bundles, which are  curves with light-like orbital frequency   $\omega=$constant on the plane $a/M-r/M$, and  \textbf{eRAD} tori defined by condition $\ell=$constant.
From  Eq.\il(\ref{Eq:flo-adding}) we have  $\Omega(\ell)=\Omega$, or  viceversa the explicit expression  $\ell(\Omega)=\ell$.
 It is convenient to introduce the rotational law of the \textbf{eRAD} with explicit dependence on the polar parameter $\sigma$:
 \bea&&
 \ell^\mp_{\sigma}\equiv\frac{\text{\emph{\textbf{Cr}}}\pm\text{\emph{\textbf{Cr}}}_1}{\text{\emph{\textbf{Cr}}}_2},\quad\mbox{where}
 \\
 &&\text{\emph{\textbf{Cr}}}\equiv a^3 r^2 (\sigma -2) \sigma +a^5 [-(\sigma -1)] \sigma +a r^3 (4-3 r) \sigma;
 \\
  &&\text{\emph{\textbf{Cr}}}_1\equiv\sqrt{r \sigma  \left[a^2+(r-2) r\right]^2 \left[r^2-a^2 (\sigma -1)\right]^2 \left[a^2 (\sigma -1)+r^2\right]};
  \\
  &&\text{\emph{\textbf{Cr}}}_2\equiv a^2 r^2 [-2 r (\sigma -1)+3 \sigma -4]+a^4 (\sigma -1) [r (\sigma -1)-\sigma]+(r-2)^2 r^3;
 \eea
 with  $\ell^\mp_{\sigma}=\ell^\mp$ on the equatorial plane ($\sigma=1$).
 Important is the special case $a=0$ where $\Omega=\pm\frac{1}{\sqrt{r^3 \sin ^2(\theta )}}$   (therefore the limit $\Omega \sqrt{\sigma}=2/r^{3/2}$) using Eq.\il(\ref{Eq:flo-adding}).
In the Kerr (\textbf{BH}) spacetime, for any spin $a/M$ on the equatorial plane $\sigma=1$, there is
 \bea\label{Eq:lobb-parli}
 \Omega^{2}_{\pm}\equiv\mp\frac{1}{r^{3/2}\mp a},\quad \Omega ^{2}_\mp=\Omega\left(\ell_\mp\right),\quad (\sigma=1),
 \eea
 ($+$ is for counter-rotating and $-$ is for corotating).
 Example of curves $\ell=$constant (tori)  in the extended plane are in Fig.\il(\ref{Fig:capPamericp}).
\begin{figure}
  \includegraphics[width=8cm]{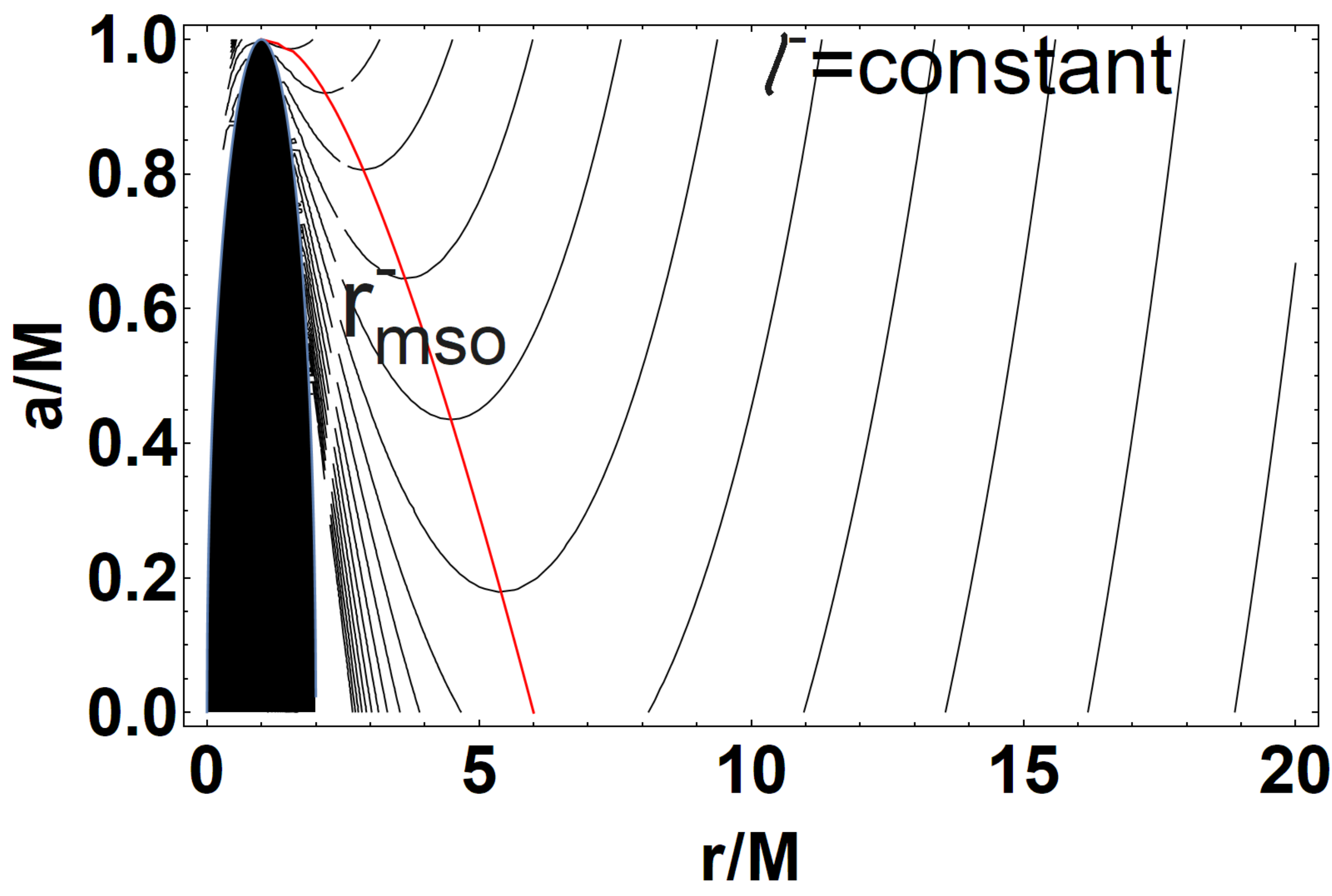}
  \includegraphics[width=8cm]{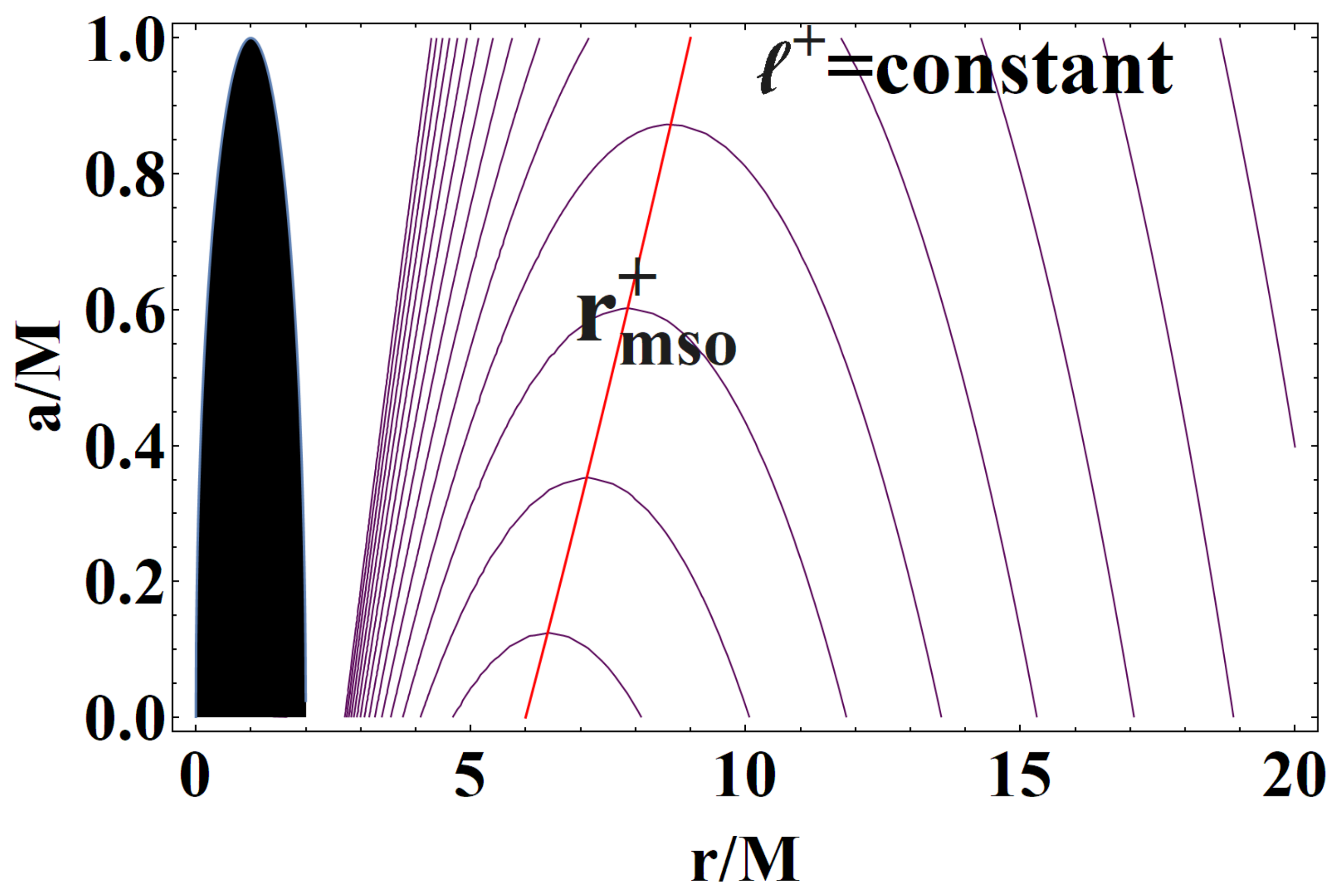}
  \caption{Surfaces $\ell=$constant in the extended plane. Each curve define an \textbf{eRAD} torus. Black region is the \textbf{BHs}. $a/M$ is the dimensionless \textbf{BH} spin, $\ell=$constant is the fluid specific angular momentum and $\ell(r)$ is the \textbf{eRAD} rotational law, $\ell^-$ is for corotating fluids and $\ell^+$ for counter-rotating fluids. Marginally stable circular orbits radii are respectively $r_{mso}^{\mp}$.}\label{Fig:capPamericp}
\end{figure}
Figs\il(\ref{Fig:capPlotitano07}) show the relation between $\omega_{\pm}$ (bundle characteristic frequencies and limiting photon stationary orbits) and $\Omega$.
Surfaces of constant $\Omega^2$ are the von
Zeipel surfaces, surfaces of constant $ \ell$ are the matter configurations considered here, surfaces of constant
$\omega_{\pm}$ are light surfaces defined by stationary observers that  define metric bundles--see Appendix\il(\ref{Sec:ls-explicit}) and \cite{zanotti}.
Clearly  $\partial_{r}\omega_{\mp}=0$ is solved for $r_{\gamma}^{\mp}$, where there is also $\Omega_\pm = \omega_ {\pm}$ respectively, a further solution is for the horizons $r_{\pm}$.
We solve the problem $\Omega(\ell)=\omega _{\pm }$ for $\ell$, obtaining the solution
\bea&&
\ell_{\Theta}^{\pm}\equiv\frac{-g_{t\phi}\mp\sqrt{g_{t\phi}^2-g_{\phi\phi} g_{tt}}}{g_{tt}}=\frac{g_{\phi\phi} \omega^\pm_{\sigma}}{g_{tt}}\quad (\omega_{\pm}=\Omega(\ell)),
\\
&&\omega^\pm_{\sigma}\equiv\frac{-g_{t\phi}\mp\sqrt{g_{t\phi}^2-g_{\phi\phi} g_{tt}}}{g_{\phi\phi}},\quad\mbox{where}\quad
\omega^\pm_{\sigma}=\omega_{\pm} \quad on \quad \sigma=1.
\eea
\begin{figure}
   \includegraphics[width=5.6cm]{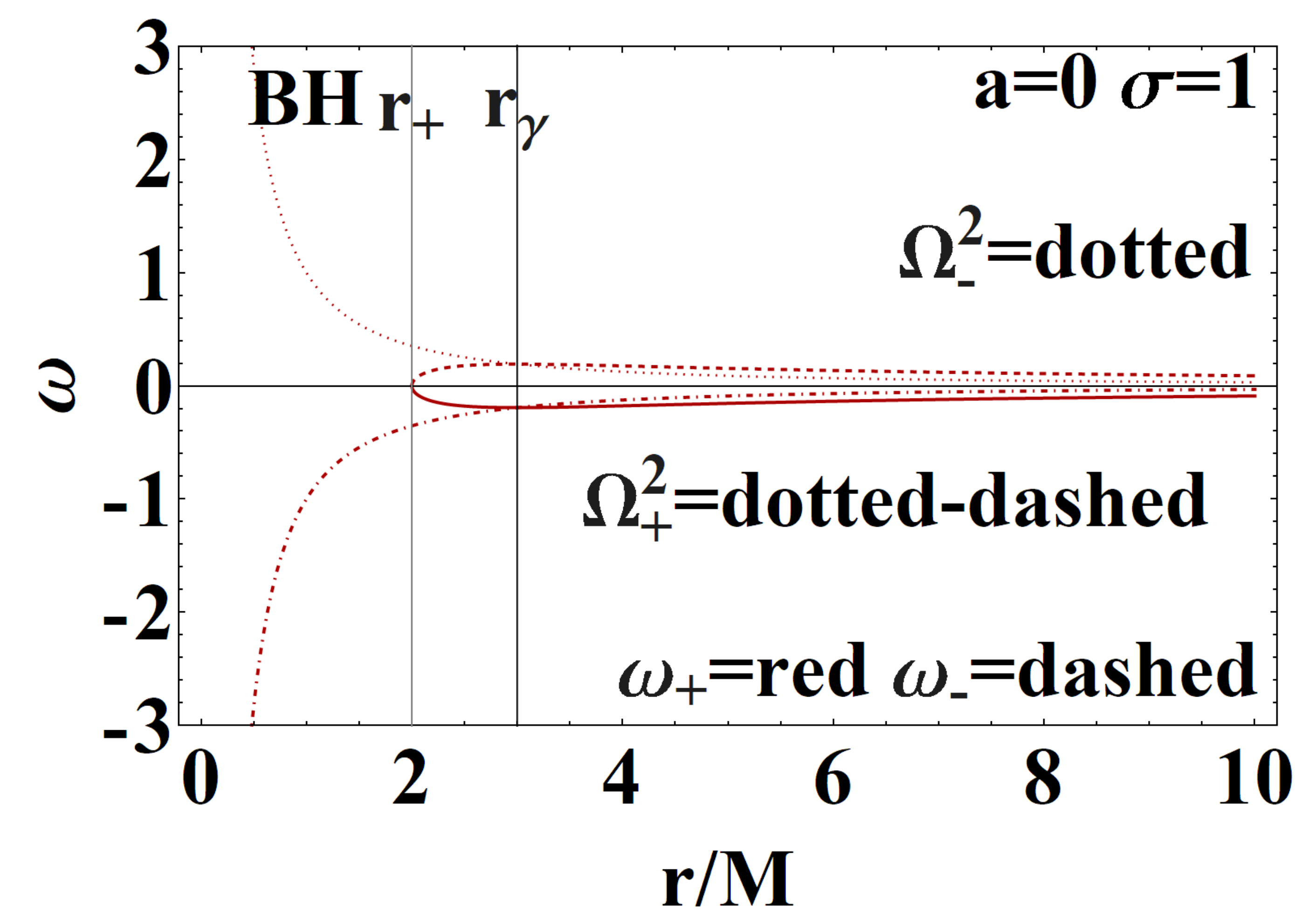}
  \includegraphics[width=5.6cm]{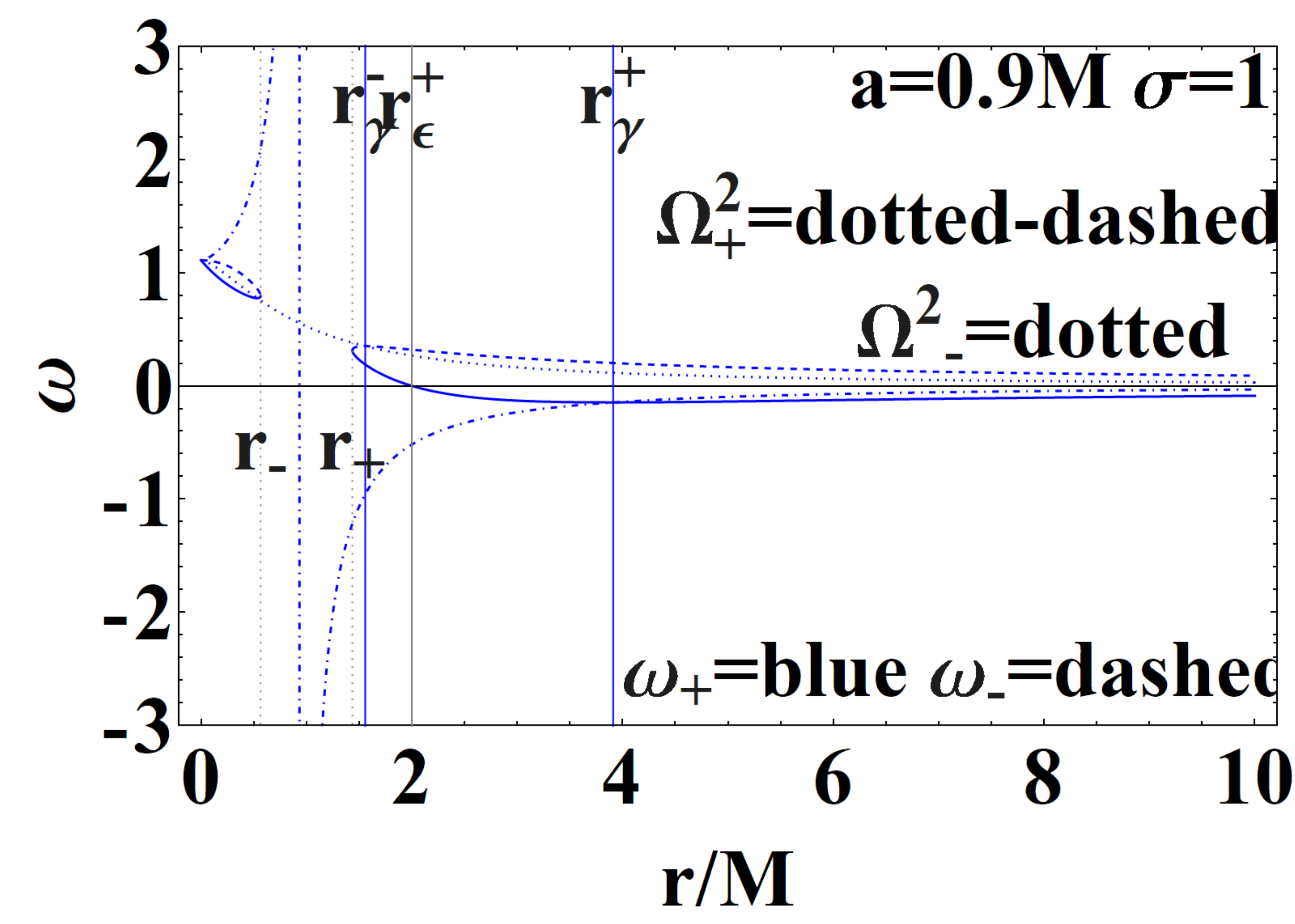}
   \includegraphics[width=5.6cm]{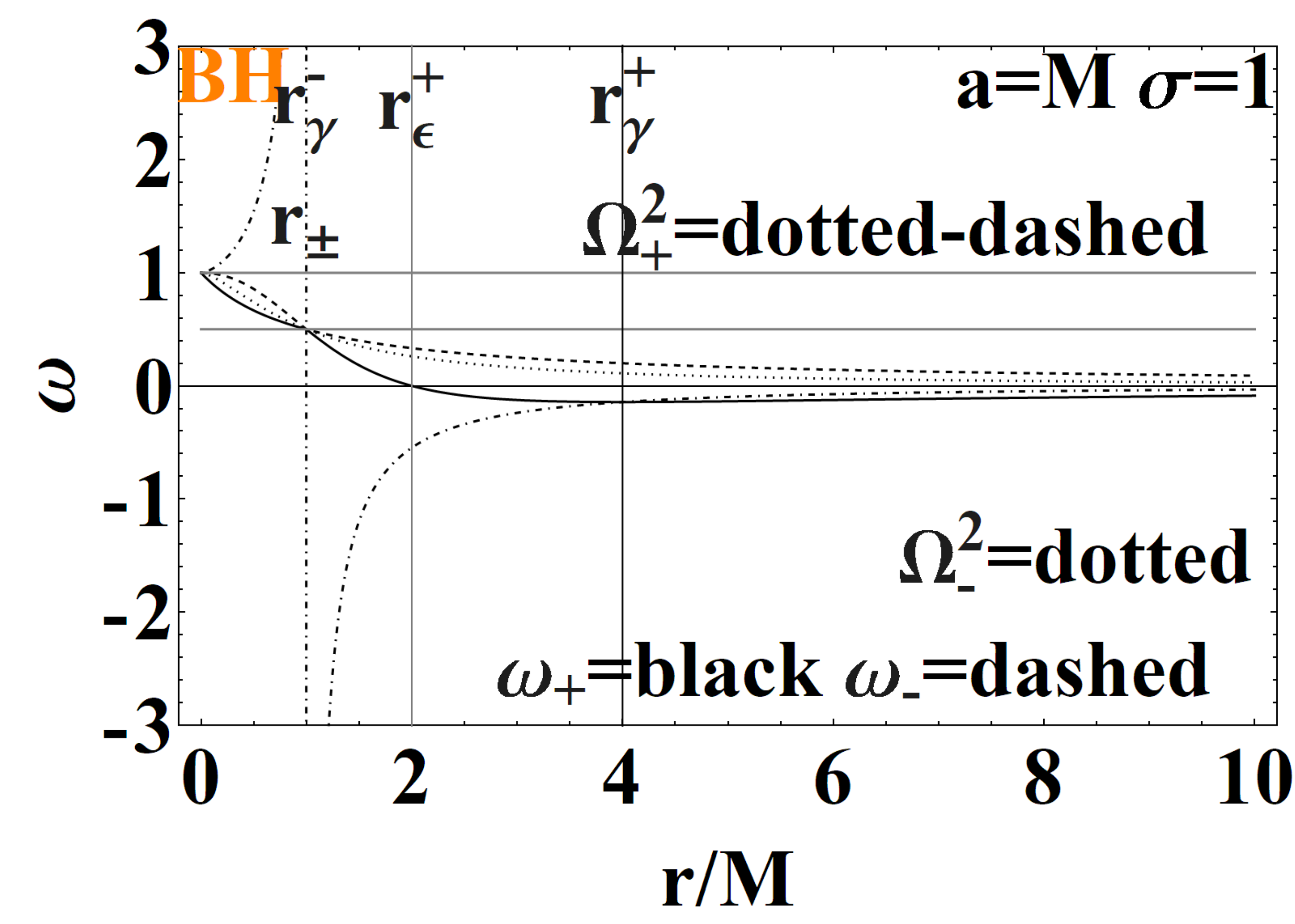}
  \caption{On the equatorial plane $\sigma=1$, $r_{\pm}$ are the outer and inner horizons of the \textbf{BH}. $r_{\epsilon}^+$ is the outer ergosurface. $r_{\gamma}^{\pm}$ are the photon orbits for the counter-rotating and corotating motion respectively. $a=0$ is the Schwarzschild static geometry. $a=M$ is the extreme Kerr spacetime. $\omega_{\pm}$ are the limiting photon stationary orbit frequencies and metric bundles frequencies (for $\omega_{\pm}=$constant for varying $a/M$). Quantity $\Omega_{\pm}^\pm$ is the relativistic velocity in Eq.\il(\ref{Eq:lobb-parli})}\label{Fig:capPlotitano07}
\end{figure}

\begin{figure}
   \includegraphics[width=5.6cm]{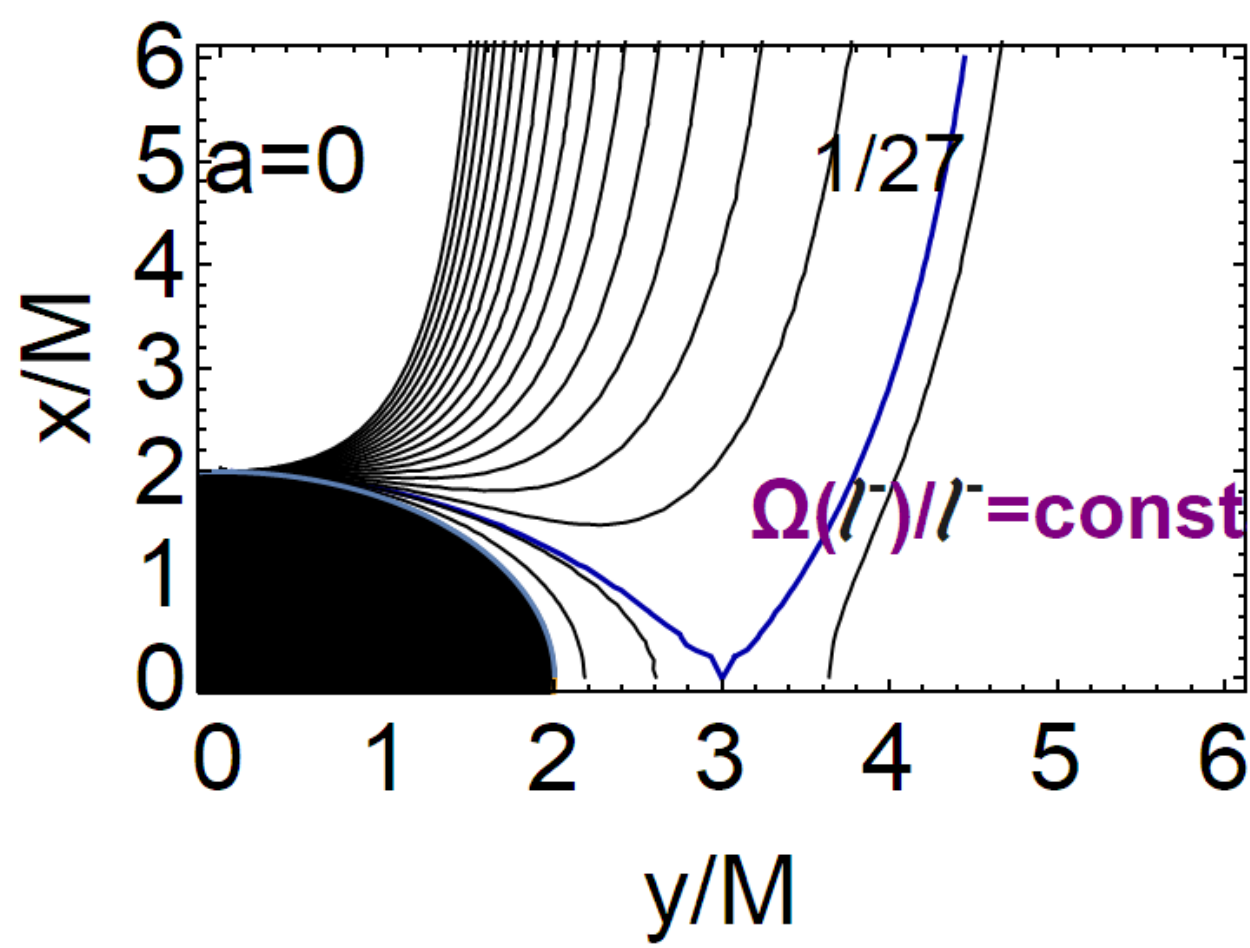}
  \includegraphics[width=5.6cm]{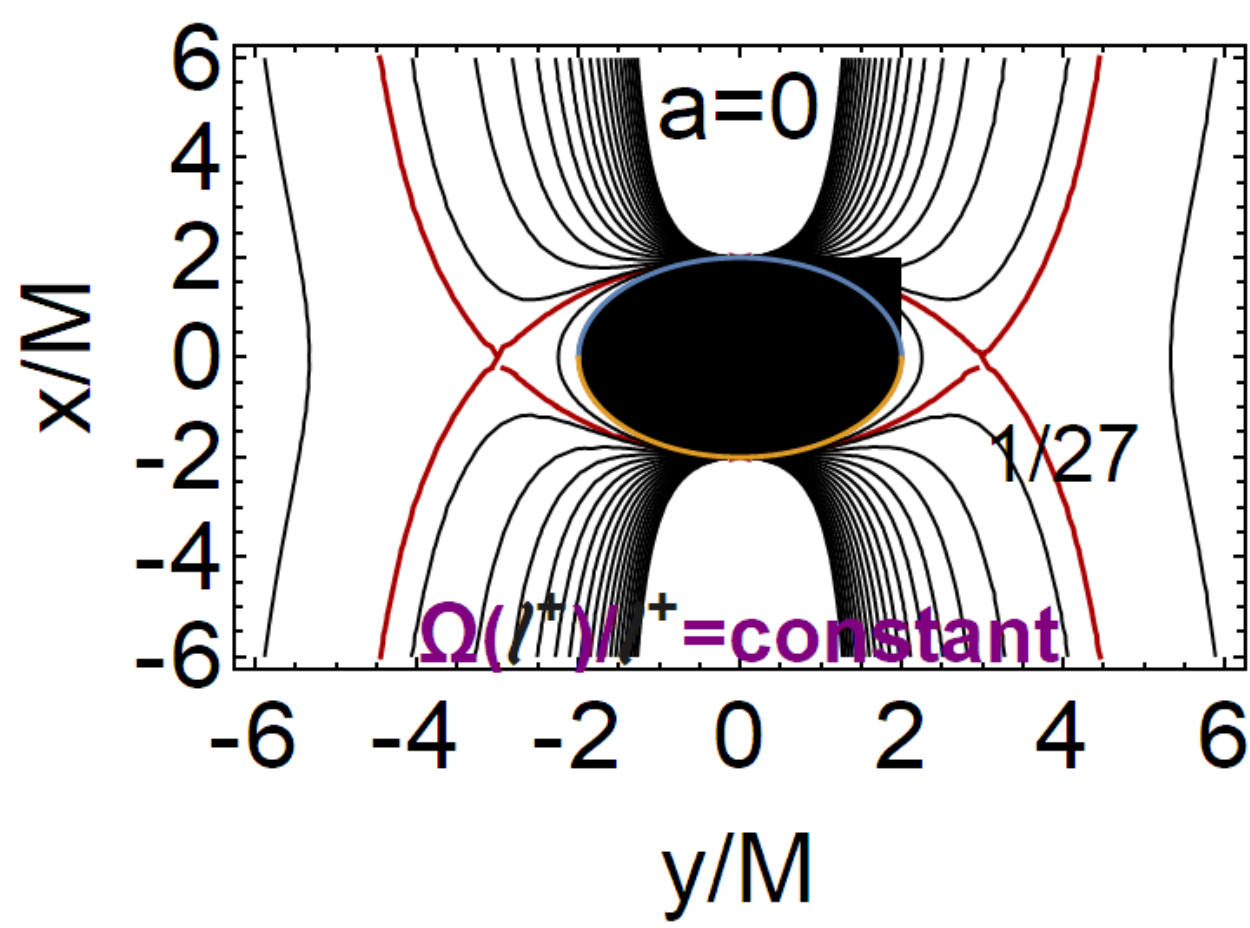}
   \includegraphics[width=5.6cm]{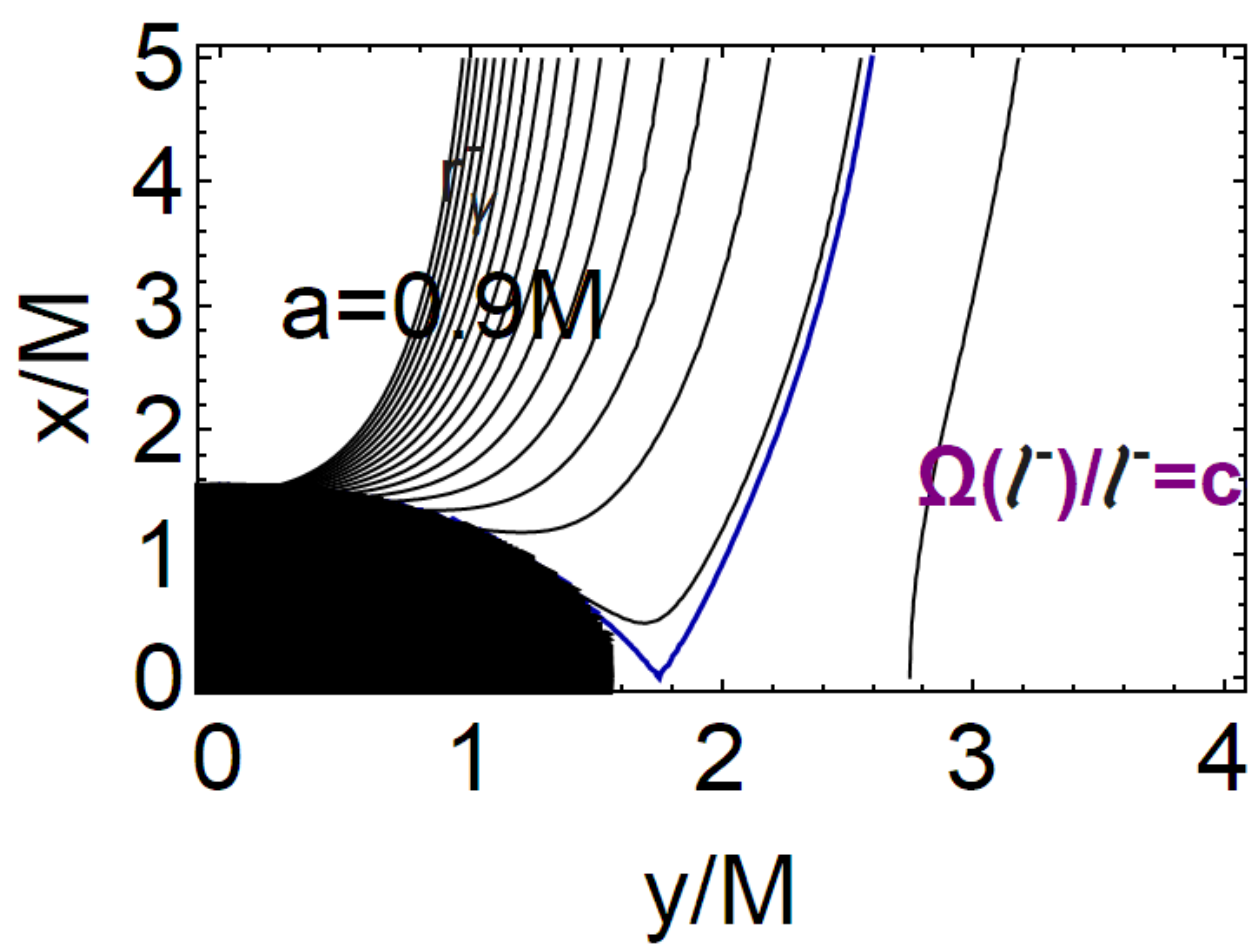}\\
      \includegraphics[width=5.6cm]{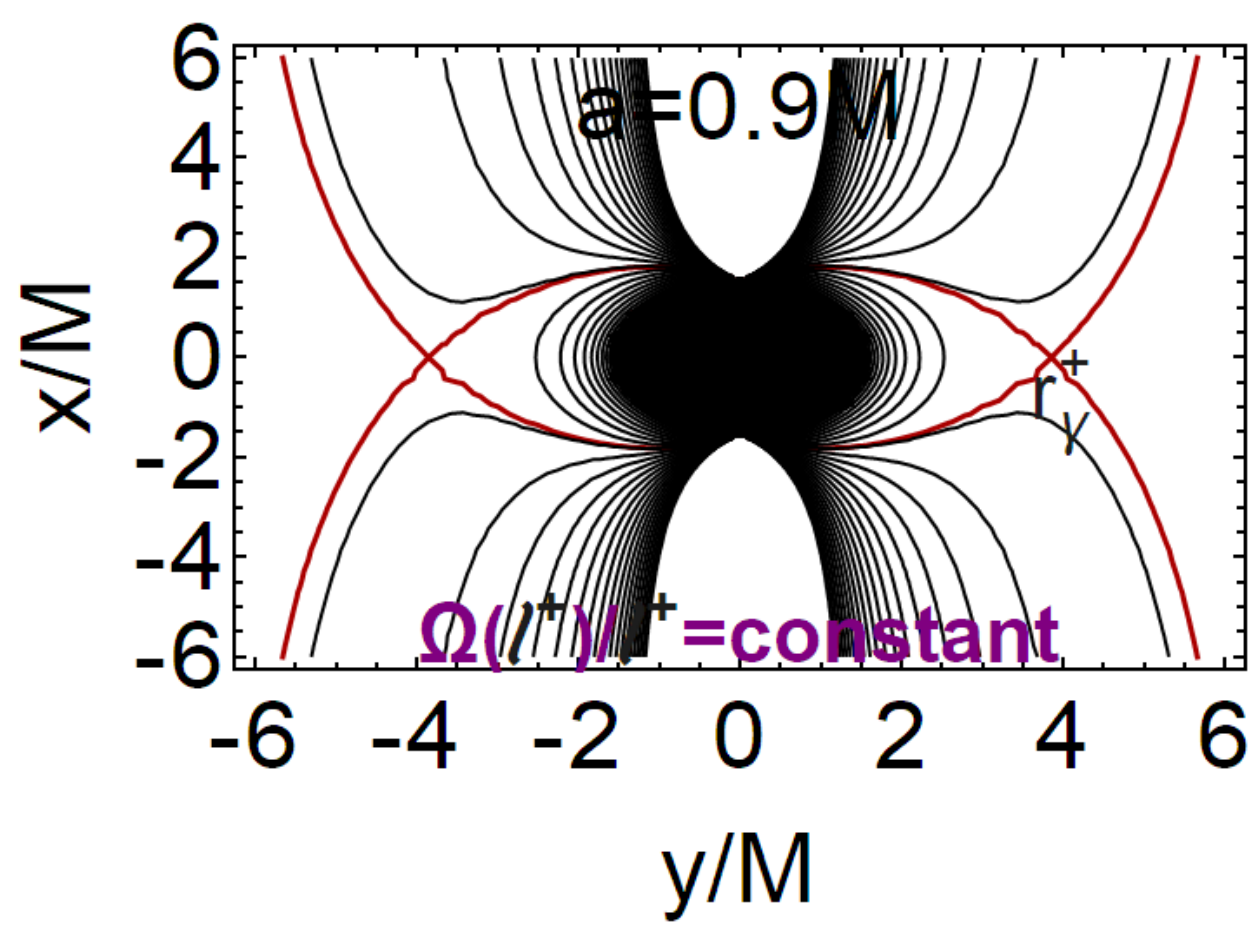}
      \includegraphics[width=5.6cm]{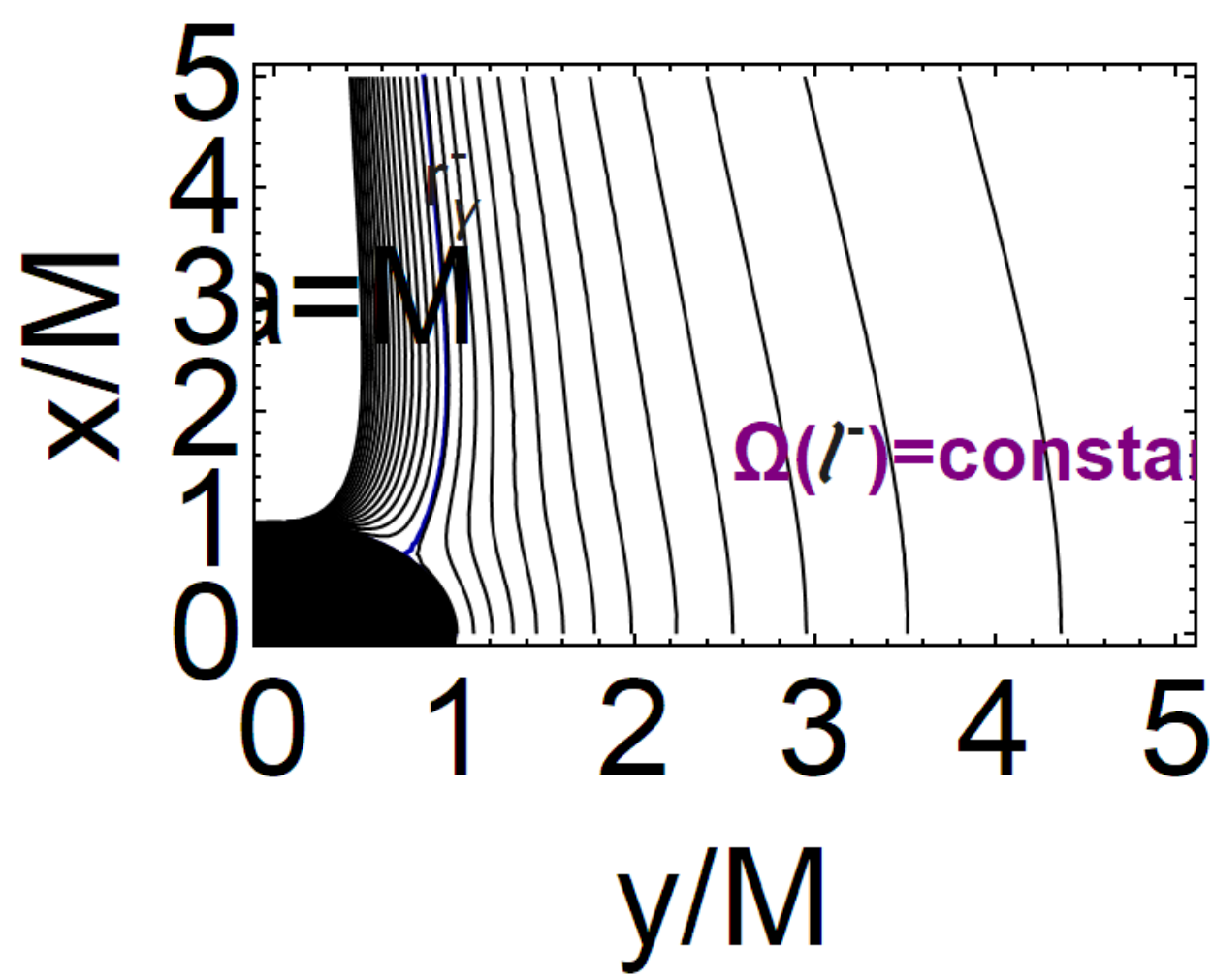}
      \includegraphics[width=5.6cm]{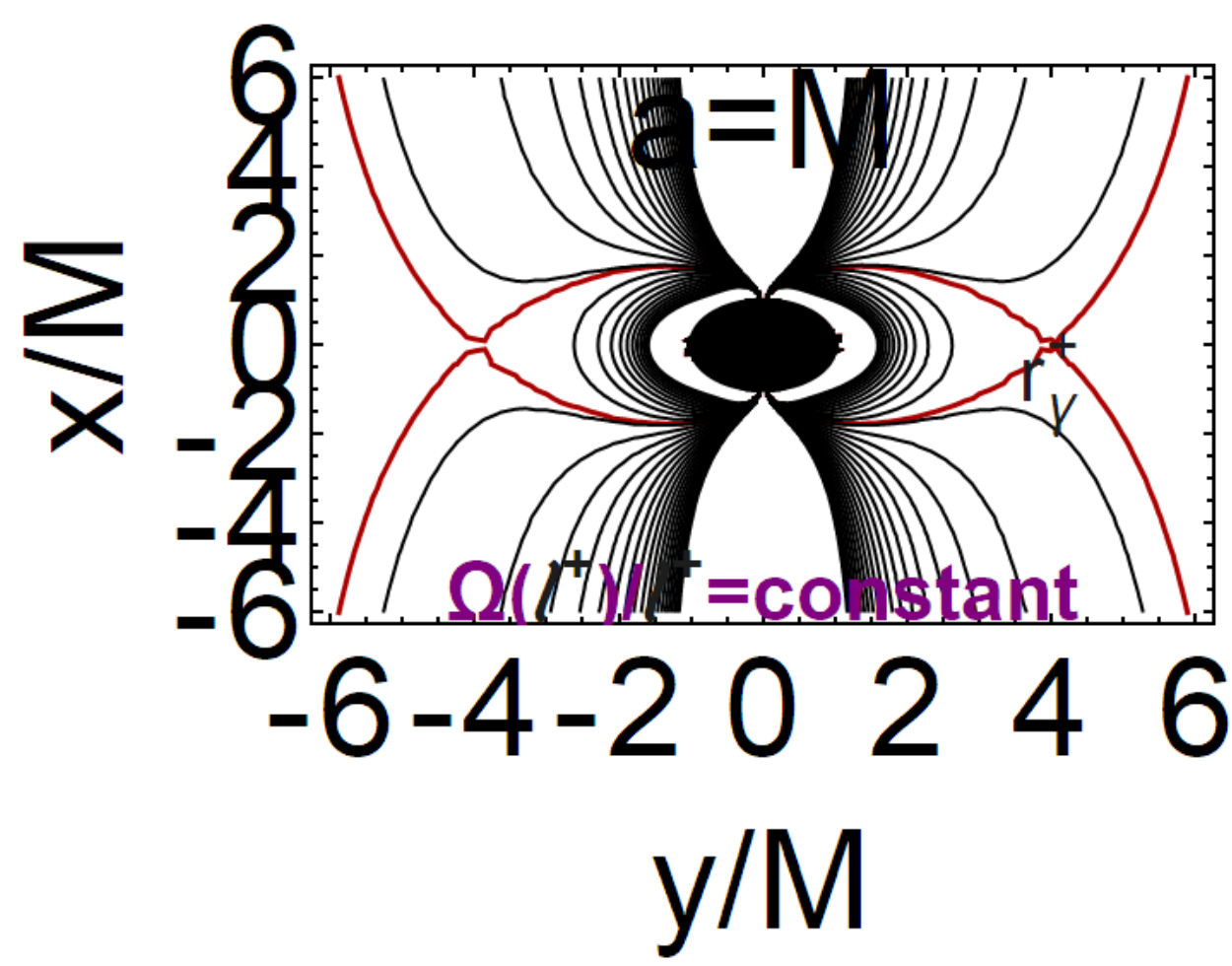}
  \caption{Von Zeipel surfaces for the static Schwarzschild spacetime $a=0$, for \textbf{BH} spin $a=0.9M$, for the extreme Kerr spacetime. Black region is the central \textbf{BH}. $r_{\gamma}^{\pm}$ is a photon orbit for corotating $(-)$ and counter-rotating orbit $(+)$. $\ell=$constant is the fluid specific angular momentum, $\Omega(\ell_{\pm})$ is the relativistic frequencies in Eqs\il(\ref{Eq:flo-adding}).}\label{Fig:giamm0}
\end{figure}
From  Fig.\il(\ref{Fig:giamm0}) it can be deduced  that $\ell_{\gamma}^{\pm}=1/(\Omega_{\pm}(r_{\gamma}^{\pm}))=
1/\omega_{\pm}=a_0$ respectively, where $a_0$ is the bundle origin, $\omega_{\pm}$ is the bundle characteristic frequency, and orbital frequency of the photon on $r_{\gamma}^{\pm}$ for the spacetime $a\neq a_0$  (on the equatorial plane).
\section{Adapted solution parameterization}\label{Sec:adaptedd}
In  \cite{proto-jet} the following variables were introduced:
\bea\label{Eq:anas}
&&\label{Eq:simm-MalD-BH}
\Delta_{\pm}\equiv\ell\pm a,\quad\mbox{where}\quad \Delta_- \Delta_+ >0,
\eea
where
\bea
&&
\ell=\frac{\Delta_-+\Delta_+}{2},\quad a=\frac{(\Delta_+-\Delta_-)}{2},
\quad
\mathcal{A}_{\pm}\equiv \frac{\Delta_{\pm}\mp \Delta_{\mp}}{2}: \quad
\mathcal{A}_+= a>0, \quad \mathcal{A}_-=\ell.
\eea

Then we can solve the Boyer problem  to find out the  tori in terms of the
$\Delta_{\pm}$ and not on the rotational law.  The critical points for the pressures where solutions of the Boyer problem exist are given by :
\bea
&&
\Delta_-= \frac{a (r-1) r^2-\sqrt{r^3 \Delta ^2}}{\Delta +X}, \quad{\mbox{and}}
\quad\Delta_+= -\frac{-\{2 a^3-a r [8+(r-7) r]\}+\sqrt{r^3 \Delta ^2}}{\Delta +X},
\\
&&
\Delta_-= \frac{a (r-1) r^2+\sqrt{r^3 \Delta ^2}}{\Delta +X} \quad{\mbox{and}}
\quad \Delta_+= \frac{2 a^3-a r [8+ r (r-7)]+\sqrt{r^3 \Delta ^2}}{\Delta +X},
\eea
where $X\equiv - r (r-2) (r-1)$ and $\Delta_+-\Delta_-=2a$.

We consider here a different problem,  re-phrasing the analysis above re-considering the  solutions for the Euler equation for the tori, by using new variables. We  focus on the potential  on the equatorial plane:
\bea&&\label{Eq:polic-s}
V_{eff}\equiv\sqrt{\frac{r \left[\left(\frac{\Delta_{-}+\Delta_{+}}{2}\right)^2+(r-2) r\right]}{2 \Delta_{-}^2+r^3+\Delta_{-} \Delta_{+}  r}},\quad \mbox{where}
\quad \ell=\frac{\Delta_{+}-\Delta_{-}}{2},\quad a=\frac{\Delta_{-}+\Delta_{+}}{2},\quad\{\Delta_{\mp}= a\mp\ell\}
\eea
%
We consider now the tori constrained   by the condition  $\partial_{\pm} V_{eff}\equiv\partial_{\Delta_\pm}V_{eff}=0$  where there is:
\bea
&&
\partial_-V=0;\quad \mbox{for}:
\quad (1)\Delta_{-}= \frac{\Delta_{+} r}{r-4}\quad or \quad(2)\Delta_{-}= -\Delta_{+}-\frac{2 (r-4) r}{\Delta_{+}};
 \eea
or  equivalently solving for $\Delta_+$ we find
 \bea
 (1)\Delta_+= \frac{\Delta_- (r-4)}{r},\quad\Delta_+=\Delta_{(s)}^{\pm}\equiv \frac{1}{2} \left[-\Delta_-\pm\sqrt{\Delta_-^2-8 (r-4) r}\right],
 \eea
 where in terms of specific angular momentum  there is
 \bea
 &&
  \ell_{(1)}= -\frac{2 a}{r-2},\quad \ell_{(2)}= \frac{4 r-a^2-r^2}{a},
  \eea
  there is $(1)=(2)$ for $r=r_+$, and $\ell= \sqrt{({a^4+4 \sqrt{1-a^2} a^2+8 \sqrt{1-a^2}+8})/{a^2}}-a$.

 On the other hand,  there is
\bea
&& \partial_+V=0,\quad \mbox{for}
\quad
\Delta_{+}= \frac{\Delta_{-} (r-4)}{r},\quad \Delta_+= -\Delta_--\frac{2 r^2}{\Delta_-}.
\eea
The last case is
$\Delta_{-}=\Delta_{\mathbf{d}}^{\pm}\equiv\left[-\Delta_{+}\pm\sqrt{\Delta_{+}^2-8 (r-4) r}\right]/2,
$;
in terms of specific angular momenta we find respectively
\bea
(1)\quad\ell= -\frac{2 a}{r-2},\quad (3)\quad \ell=\frac{a^2+r^2}{a}.
\eea
Note solution  (1) solved  as well as the derivative for $(a-\ell)$;  there is (1)=(3) for $r=r_+$.
On the other hand, the problem  $\ell^{\pm}=\ell$ is solved  for both corotating and counter-rotating fluids giving
the
solutions
\bea&&
\Delta_+=\Delta_-+\frac{2 \left(\Delta_-^3-\Delta_- r^2\right)}{r^3-\Delta_-^2}-2 \sqrt{\frac{r \left(r^3+\Delta_-^2 (r-2)\right)^2}{\left(r^3-\Delta_-^2\right)^2}},
\\
&&\Delta_+= \Delta_-+\frac{2 \left(\Delta_-^3-\Delta_- r^2\right)}{r^3-\Delta_-^2}+2 \sqrt{\frac{r \left(r^3+\Delta_-^2 (r-2)\right)^2}{\left(r^3-\Delta_-^2\right)^2}}.
\eea


\end{document}